\documentclass{msuphddissertationx}
\usepackage{amsmath,amssymb,amsthm,paralist}
\usepackage{slashed}
\usepackage{graphicx}
\usepackage{color}
\usepackage{lettrine}
\usepackage{url}
\author{Arsham Farzinnia} 
\title{Above and Beyond the Standard Model:
\protect \\
On Phenomenology of Lee--Wick Theory and Massive Vector Color--Octet} 
\field{Physics} 

\DeclareMathSizes{12}{12}{10}{8}

\begin{document}

\newcommand{\vect}[1]{\overrightarrow{#1}}
\newcommand{\smbox}[1]{\mbox{\scriptsize #1}}
\newcommand{\tanbox}[1]{\mbox{\tiny #1}}
\newcommand{\vev}[1]{\langle #1 \rangle}
\newcommand{\Tr}[1]{\mbox{Tr}\left[#1\right]}
\newcommand{\cosb}{c_{\beta}}
\newcommand{\sinb}{s_{\beta}}
\newcommand{\tanb}{t_{\beta}}
\newcommand{\picwidth}{3.4in}
\newcommand{\brac}[1]{\left( #1 \right)}

\maketitlepage 
\begin{abstract}

\begin{quote}
``\textit{A thesis has to be presentable... but don't attach too much importance to it. If you do succeed in the sciences, you will do later on better things, and then it will be of little moment. If you don't succeed in the sciences, it doesn't matter at all!}"
\begin{flushright}
|Paul Ehrenfest (1880 -- 1933)\\
\end{flushright}
\end{quote}

\vspace{\baselineskip}

The present Thesis is dedicated to a formal and phenomenological investigation of extensions to two separate sectors of the Standard Model of particle physics~(SM): the electroweak sector and the strong sector. The Thesis is divided into two main parts: Part~I focuses on the \textit{Lee-Wick Standard Model}~(LW~SM), which, by providing a solution to the Hierarchy problem, forms a natural extension of the electroweak sector, while Part~II studies the \textit{coloron theory}, arising from extending the strong sector gauge group.

Providing a general introduction about the current state of the SM and the associated challenges in Chapter~1, we proceed in Chapter~2 to analyze the tension between naturalness and isospin violation in the LW~SM. Chapter~3 discusses the global symmetries and the renormalizability of LW~scalar QED. A first complete calculation of QCD corrections to the production of a massive color-octet vector boson (colorons) is reported in Chapter~4. Finally, we conclude the Thesis in Chapter~5 by summarizing the discussed results and presenting an outlook for future research in the surveyed areas.

\end{abstract}


\begin{copyrt}
\end{copyrt}

\begin{dedication} \begin{center}
\vspace*{\fill}
\begin{center}
\noindent\includegraphics[width=\textwidth]{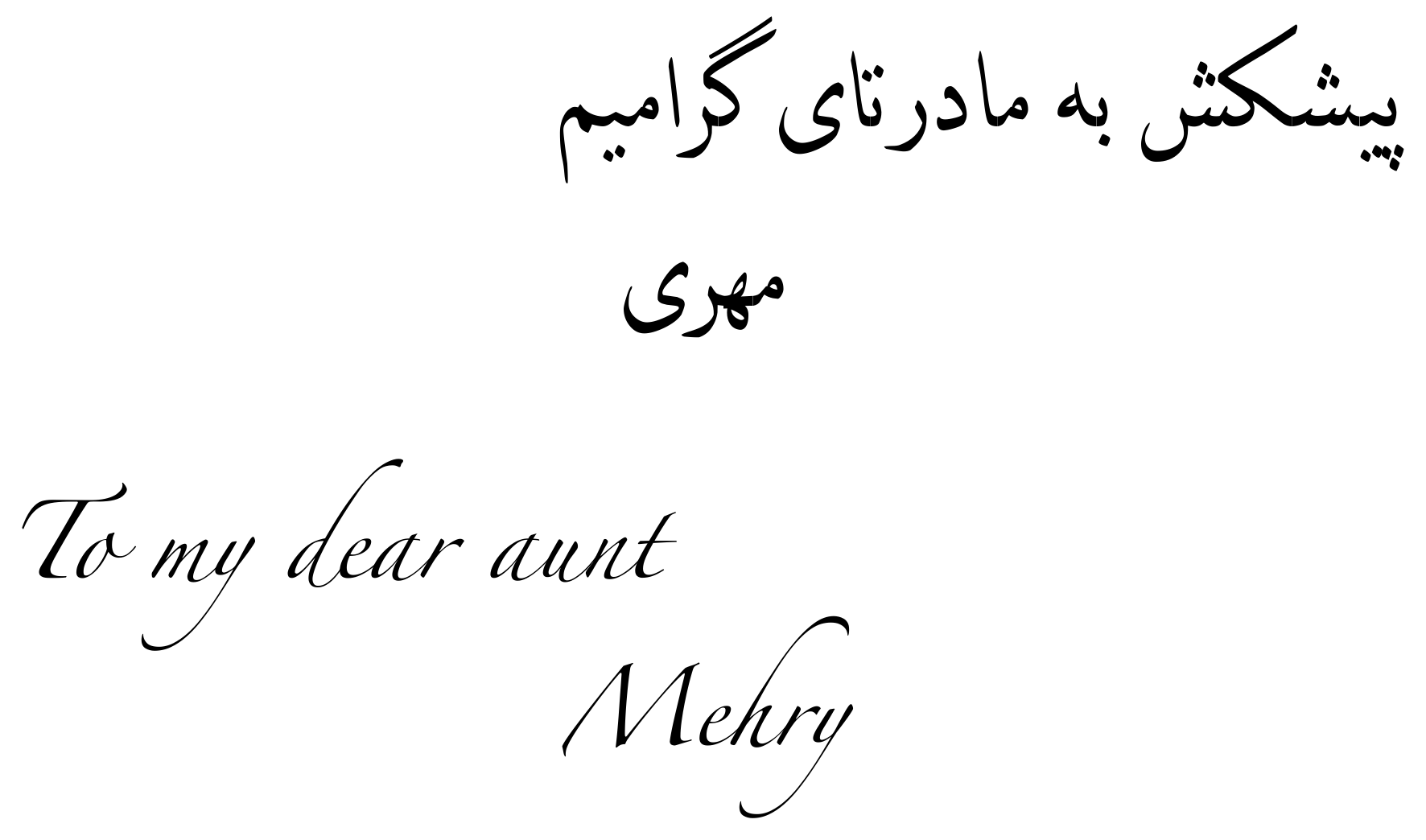}
\end{center}
\vfill
\end{center}\end{dedication}


\begin{acknowledgment}

My personal interest in physics and astronomy developed a well-defined shape first when I started reading the fascinating collection \textit{Library of the Universe} by Isaac Asimov, as a thirteen years old boy in my country of birth, Iran. I realized then where my true passion resided and what I wanted to pursue in the coming years academically. Seventeen years later and after traveling through several continents, I am finally on the verge of obtaining a \textit{Doctor of Philosophy} degree from Michigan State University in the United States of America.

On that account, several sincere gratitudes are in order. I would like to thank my dear aunt, \textit{Mehry}, for her unconditional love and encouragement throughout the difficult times in my personal and professional life. Without her continuous support, I might not have been where I am today. I would like to thank \textit{Dr.~Wayne~W.~Repko} for welcoming me into the captivating world of physics research, introducing me to a first flavor of what lay ahead. I owe my professional enthusiasm to his generous supervision. Finally I would like to thank my advisors \textit{Dr.~R.~Sekhar~Chivukula} and \textit{Dr.~Elizabeth~H.~Simmons} for their patience with me and their amazing professional guidance. I am truly honored to be their pupil.

Thinking about ``Mother Nature'' and trying to unlock her secrets is an incredibly exciting experience, and I am humbled by having the opportunity to contribute, no matter how small, to the world of human knowledge.

\end{acknowledgment}

\begin{preface}

\begin{quote}
``\textit{The world needs to wake up from its long nightmare of religious belief; and anything that we scientists can do to weaken the hold of religion should be done, and may in fact be our greatest contribution to civilization!}"
\begin{flushright}
|Steven Weinberg (1933 -- )\\
\end{flushright}
\end{quote}

\vspace{\baselineskip}

\textit{The scientific method} has been proven to be the only reliable method for unraveling the mysteries of the natural world. Since the dawn of \textit{Homo sapiens'} domination of the earth, curiosity has been the main thrust behind the slow but steady progress of this ``Wise Man". The primitive man, himself a product of a long evolutionary process on a planet which from the cosmological perspective is nothing but a speck of dust in the vast universe, looked with awe at the heavens and their apparent order. Being inescapably confronted with the ``big questions" of the origin and the nature of life and the universe, this creature's big brain saw initially no way out but to speculate about the possible answers, desperately trying to find temporary relief from curiosity. Indeed, curiosity was a curse from which there appeared to be no escape.

In the bumpy course of human history, since reaching full behavioral modernity around 50,000 years ago, various cultures and societies have tried to dogmatize their speculative (and often superstitious) resolutions in an effort to effectively terminate this curiosity ``disease". To some it seemed appropriate to go so far as attempting to prosecute and eradicate free-thought and intellectualism altogether. Despite ignorant opposition, a thirst for knowledge has always compelled many individuals not to succumb to superficial speculations, but rather to try to inquire about the natural world through objective observations and evidence-based empirical studies. Curiosity has played a central role in the development of this scientific method of studying natural phenomena.

Scientists across different cultures, throughout centuries, have accumulated the acquired knowledge of those before them and expanded upon it, ``standing on the shoulders of giants" as Isaac Newton put it. The scientific method stands in sharp contrast with the dogmatic speculations preceding it, and as Steven Weinberg phrased, it has no prophet or authority but it does have many \textit{heroes}. In our modern perspective, the evidence-based quest for truth about nature, inspired by the age-old curiosity and facilitated through falsifiable hypotheses and reproducible experiments, has culminated in what is called the Natural Sciences, containing physics as a branch.

The magnificent success of science in unlocking the secrets of the nature has had a tremendous impact on the development of the human intellect and its maturity. It has enabled us, to a certain extent, to leave behind childlike wishful-thinking and elect the rational deduction of facts as the correct approach towards understanding our own nature and our relationship with the world we live in. Objective observation, lying at the heart of science and the scientific method, provides us with the tools needed to tackle the global problems facing us as a species. It implies the necessity of employing rationalism as an alternative to tradition and the superstitious belief systems currently plaguing a substantial percentage of the human population across the planet, including in the developed world.

My hope is that one day humanity as a whole will achieve a level of maturity to break the millennia-old shackles of ignorance and embrace its own full intellectual potential which, if applied correctly, can construct a far better world than the one we are currently living in!

\newpage
\vspace*{\fill}
\begin{center}
\noindent\includegraphics[width=\textwidth]{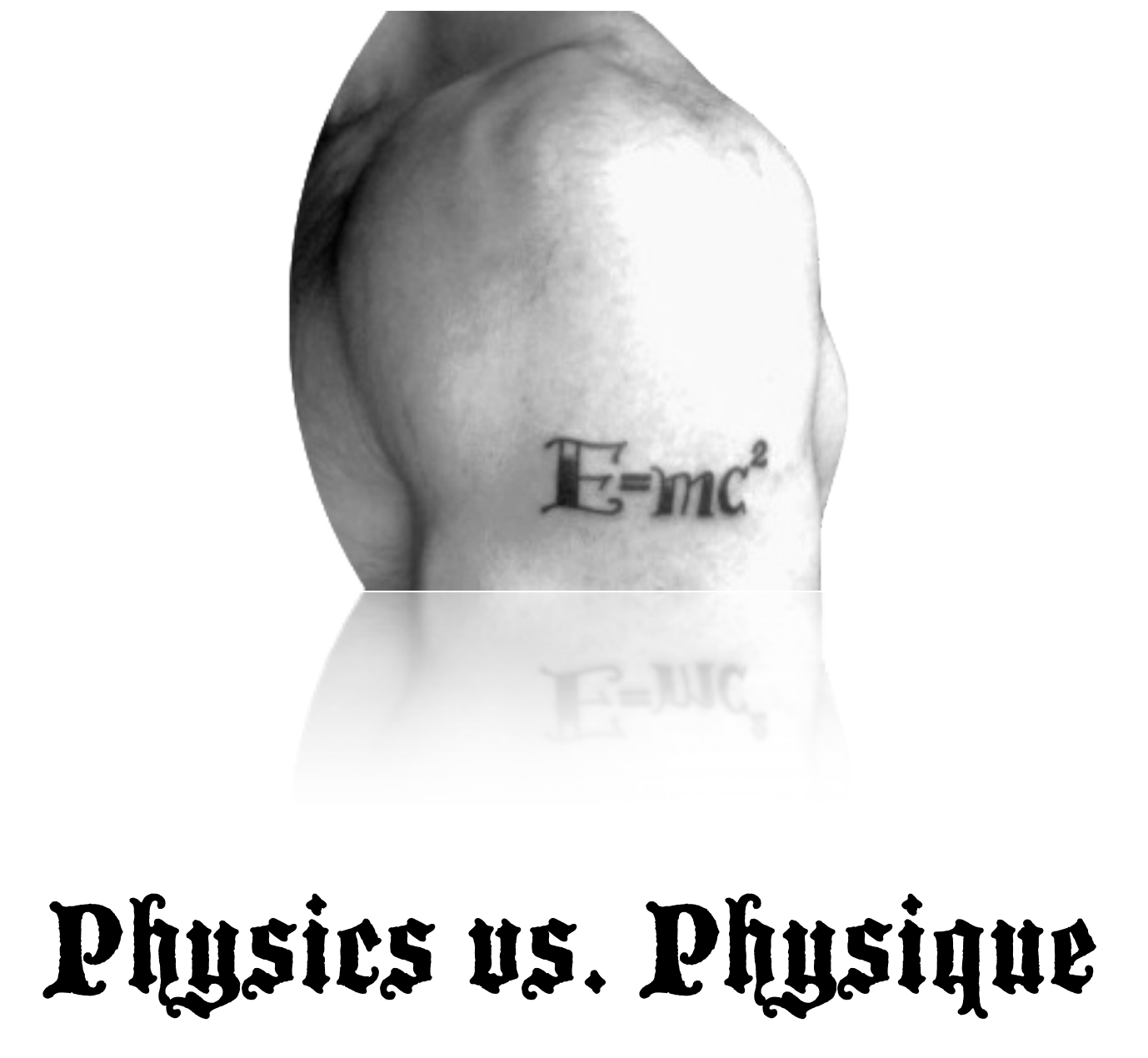}
\end{center}
\vfill

\end{preface}

\TOC
\LOT
\LOF


\newpage
\pagenumbering{arabic}
\begin{doublespace}

\begin{center}
\noindent\includegraphics[width=\textwidth]{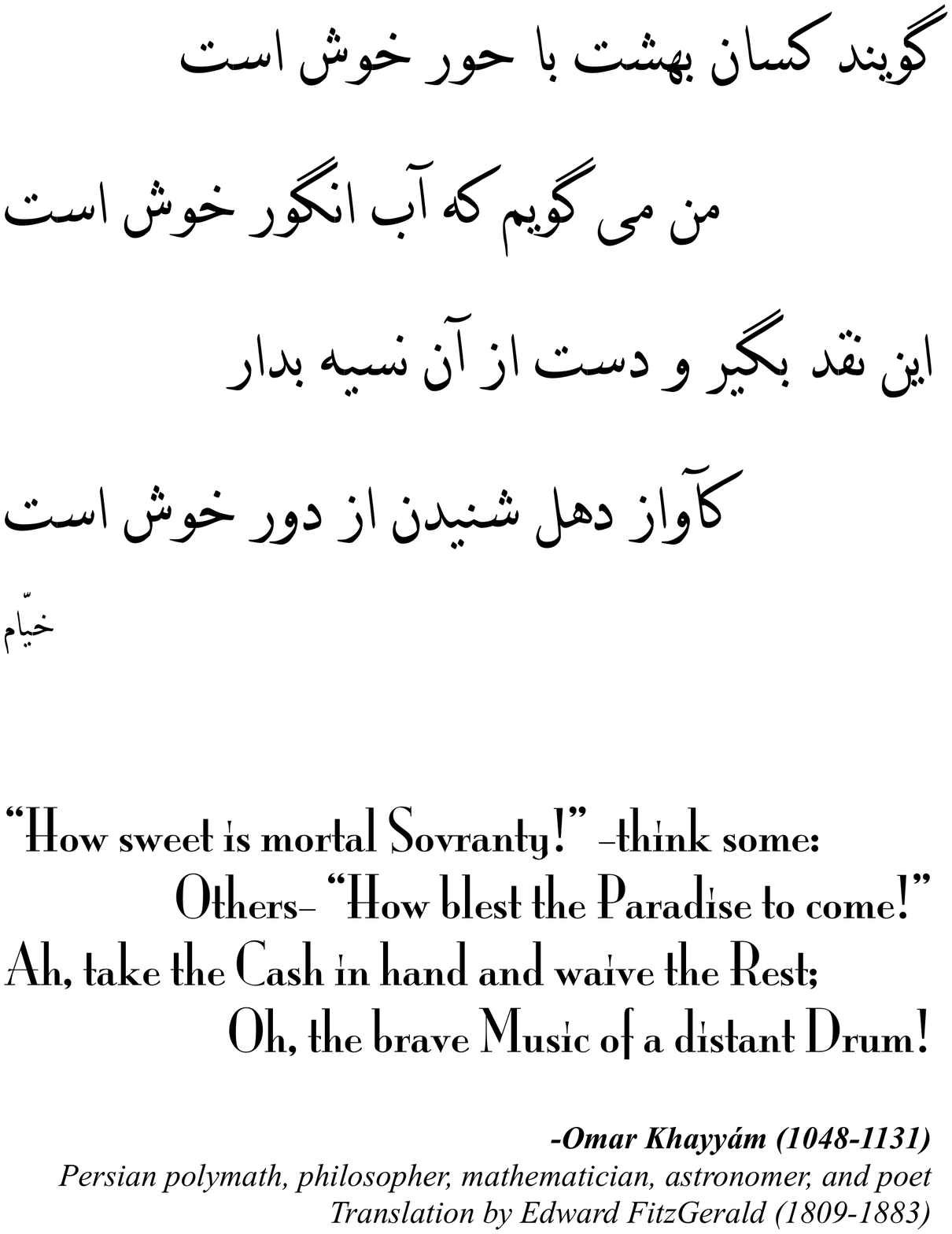}
\end{center}

\chapter{INTRODUCTION}\label{intro}

\begin{quote}
``\textit{The only thing that interferes with my learning is my education!}"
\begin{flushright}
|Albert Einstein (1879 -- 1955)\\
\end{flushright}
\end{quote}

\vspace{\baselineskip}

\lettrine[lines=1]{M}{odern particle physics} revolves around the idea of all matter and energy being composed of elementary undividable discrete constituents. In the modern physics nomenclature, this is referred to as ``quantization". This concept dates back to at least the 6th century BC, proposed first by ancient Greek philosophers such as Leucippus, Democritus, and Epicurus, ancient Indian philosophers such as Kanada, Dign\=aga, and Dharmakirti, and later studied by medieval Persian scientists Alhazen and Avicenna, among others. Democritus was the first to have coined the term \textit{\'atomos}, meaning ``indivisible", to describe these elementary constituents. The modern concept of atoms was introduced in the 19th century, through the work of John Dalton in chemistry, who thought of atoms as the fundamental particles of nature, and hence, adopted Democritus' terminology. By the early 20th century, however, it became evident through the experiments of Rutherford and others that atoms were not indivisible, but instead were composed of even smaller components themselves.\footnote{There are many books dedicated to reviewing the history of modern physics, with an emphasis on particle physics and gravity. The enthusiastic reader is encouraged to consult e.g. \cite{weinbergbook}\nocite{reality, veltmanboek}-\cite{heelal}, among many other excellent reviews.}

On the theoretical side, the special theory of relativity was discovered by Albert Einstein in 1905. Furthermore, in his account for the photoelectric effect, Einstein proposed a radical concept regarding the nature of light, describing it in terms of \textit{discrete} packets of energy (called \textit{photons}). These prominent theoretical breakthroughs led to a profound change in scientific understanding of the physical world in early 20th century. The notion of quantization of light, in particular, paved the way for the subsequent discovery and development of quantum mechanics through the work of Niels Bohr, Wolfgang Pauli, Werner Heisenberg, and Erwin Schr\"odinger, to name a few. Experimental discovery and identification of many subatomic particles and their properties, such as the electron, proton, and neutron, together with their quantum mechanical description, gave rise to ever more detailed models of atoms in the first half of the 20th century.

The high energy scattering experiments throughout the 1950s and 1960s uncovered the spectrum of a variety of new particles, initially all thought to be ``fundamental" exhibiting no further substructure, referred to as the ``particle zoo". Meanwhile, theoretical progress concerning the incorporation of the special theory of relativity into quantum mechanics, initiated by Paul Dirac in 1920s, gave birth to the development of a more comprehensive quantum theory by the mid 20th century, called the quantum field theory. Quantum field theory consistently describes all matter as being composed of point-like fundamental particles, and the interactions among them as being mediated by various quantized pockets of energy, themselves also identified as fundamental particles. In light of this theory, the particle zoo could be explained in terms of different bound-states of a limited number of fundamental particles and their quantized interactions; this dramatically simplified the high energy description of matter and interaction forces.\footnote{The gravitational force has not yet been described consistently by a quantized field theory.} 

Accordingly, many of the initially considered ``fundamental" particles in the zoo turned out to be described in terms of even more fundamental constituents, and hence, the notion of being fundamental was now reserved to apply only to a limited number of particles from which all others were composed. Initially, the existence of many of these truly fundamental particles were merely postulated in the quantum field theoretical description of the particle zoo; however, the experiments in the final decades of the 20th century unambiguously verified the existence of all but one of these postulated particles, confirming their lack of substructure up to the accessible energy scales. 

\begin{figure}[!t]\begin{center}
\includegraphics[width=\textwidth]{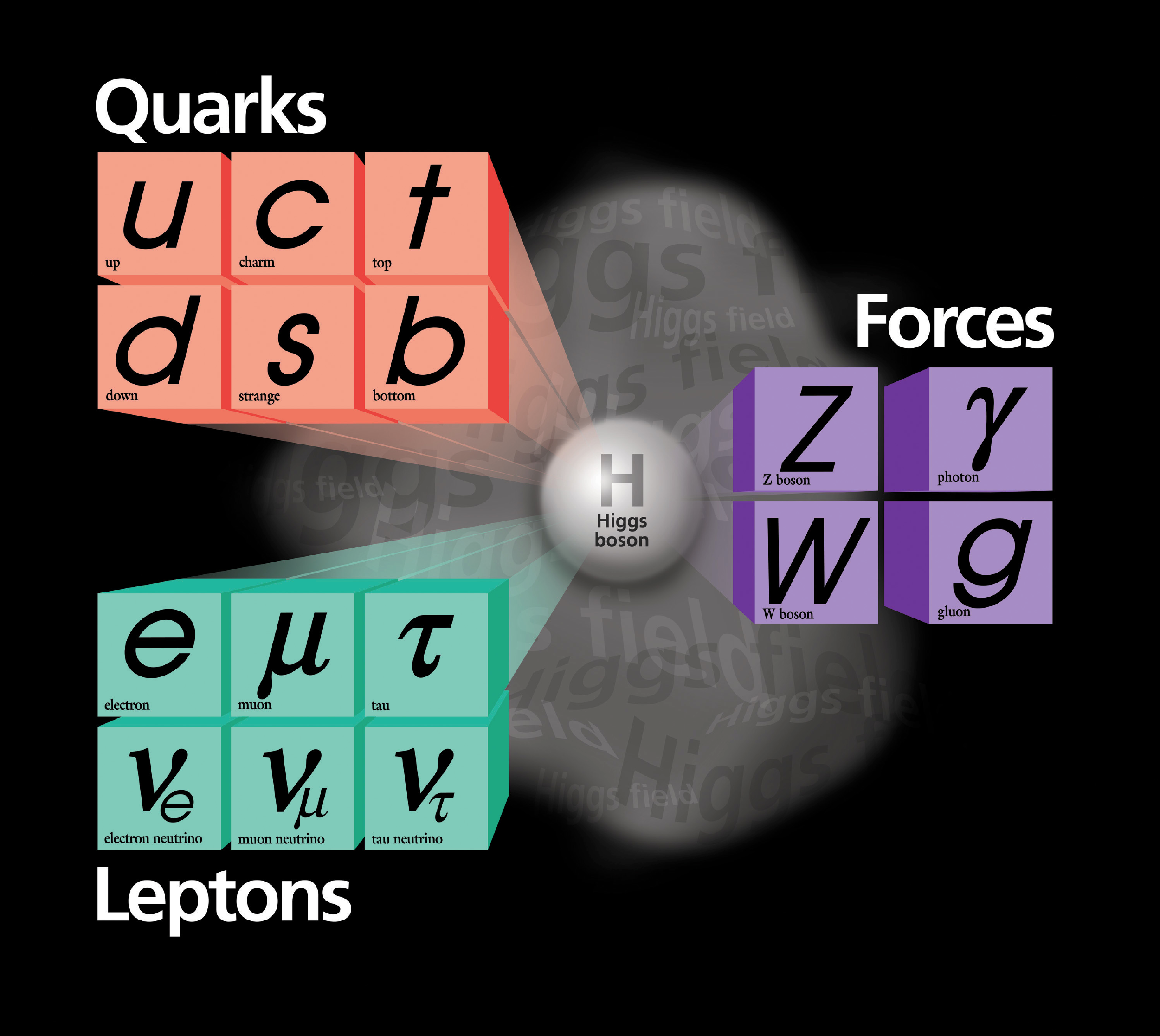}
\caption{An illustration of the Standard Model matter and force particle content and the Higgs boson. Quarks and leptons come in six different flavors, and are each organized in three generations. The existence of the Standard Model Higgs boson remains elusive as of yet. \textit{For interpretation of the references to color in this and all other figures, the reader is referred to the electronic version of this dissertation.}}\label{SM}
\end{center} (Figure courtesy of Fermilab Visual Media Services. The small text inside each box denotes the particle's name; it is not intended to be necessarily readable and is for visual reference only. \url{http://www-visualmedia.fnal.gov/VMS_Site/gallery/stillphotos/2005/0400/05-0440-01D.hr.jpg}) \end{figure}

According to the experiments, the currently accepted picture of the elementary constituents of matter, exhibiting no further substructure, contains the following: \textit{quarks} existing in six different varieties commonly referred to as ``flavors" (up, down, charm, strange, top, and bottom), and six \textit{leptons} consisting of electron, muon, and tau, each accompanied by a neutrino of the corresponding flavor\footnote{In light of the discovery of neutrino oscillations, the actual three neutrino particles have tiny masses, and are linear combinations of the flavor neutrinos described above.} (Fig.~\ref{SM}). Quarks and leptons are organized in three ``generations" or ``families", and are all spin--$\displaystyle{\frac{1}{2}}$ particles, which makes them \textit{fermions} (named after Enrico Fermi who made, among other things, important contributions towards understanding the behavior of these particles). A direct prediction of quantum field theory is that all of these particles have a partner with the same mass but opposite electric charge, forming the fundamental constituents of \textit{antimatter}.

Quarks tend to combine in special manners to form bound-states, called \textit{hadrons}. Hadron is a collective name for \textit{baryons} (composed of three quarks) and \textit{mesons} (composed of a quark and an antiquark). The proton (two up quarks and one down quark) and neutron (one up quark and two down quarks) are, therefore, baryons in this classification. Various numbers of protons and neutrons bind together to form the nuclei of atoms of different elements in nature, accompanied by an appropriate number of electrons in orbits around the nucleus to make the atom electrically neutral. Hence, all known forms of matter, from simple elements in the periodic table to complex molecules and structures made of them, can be traced back to the bound-states of a handful of elementary particles.

Next, let us take a look at the quantized interactions in nature among the matter particles. There are four known forces in nature: \textit{gravity} (attractive between masses), \textit{electromagnetism} (attractive between opposite charges and repulsive between like charges), the \textit{strong nuclear force} (binding quarks inside the proton and neutron, and binding protons and neutrons inside the nucleus), and the \textit{weak nuclear force} (responsible for certain radioactive decays). In the quantized description provided by quantum field theory, electromagnetism is mediated by a massless, electrically neutral particle, called the \textit{photon}, interacting only with electrically charged particles. The strong interaction has eight massless, electrically neutral mediators, called \textit{gluons}, and is felt only by particles carrying the \textit{color} quantum number, i.e. quarks and gluons. Color can be perceived as the analogue ``charge" appropriate for the strong interaction; hence leptons, being colorless, are not affected by the strong force. The weak interaction is mediated by three massive mediators: $W^+$, $W^-$, and $Z$ \textit{vector bosons}, where superscripts represent the appropriate electric charges. The weak force is felt by all the \textit{left-handed}\footnote{In the relativistic limit, a particle is called \textit{right-handed} if its direction of spin is parallel to the direction of motion, and \textit{left-handed} if the spin direction is anti-parallel to its motion.} quarks and leptons. All of these fundamental force-carriers are spin-1 particles, which makes them \textit{bosons} (named after Indian mathematician and physicist Satyendra Nath Bose, for his work related to Bose-Einstein statistics and the theory of the Bose-Einstein condensate).

A consistent quantum field theoretical description of gravity currently does not exist. A proposed quantized mediator of gravity takes the form of a massless, electrically neutral, spin-2 particle, called the \textit{graviton}, with a tensor nature (as opposed to the spin-1 vector mediators of the other three forces). This is motivated by the macroscopic properties of gravitational fields as described by the general theory of relativity. However, since the rest masses of the elementary particles are extremely small, and the gravitational interaction is much weaker than the other forces, gravity can be safely neglected in the relevant energy scales currently under study in particle physics.

The quantum field theoretical description of the electromagnetic, strong, and weak forces, and their interactions with quarks and leptons is collectively called the \textit{Standard Model} of particle physics~(SM). It represents the currently accepted theory explaining the particle zoo and other observed high energy phenomena with high precision. A final elementary particle postulated in the SM, but yet remaining undiscovered, is a massive, electrically neutral, spin-0 particle, called the \textit{Higgs boson} (named after Peter Higgs, one of the contributors to the description of the so-called ``Higgs mechanism" of spontaneous symmetry breaking). This scalar particle has been postulated in order to explain the origin of the masses of the quarks and leptons. Incorporation of the Higgs scalar into the SM leads, however, to the so-called \textit{Hierarchy Problem}. Solving the hierarchy problem is currently one of the main topics of research in particle physics. A potential solution to the Hierarchy problem will be discussed at length in Part~I of this Thesis.

At this point, let us turn to a more comprehensive description of the SM. In the 1950s and 1960s, it was realized by Chen Ning Yang, Robert Mills, Martinus Veltman, and many others that quantized ``gauge theories" provided a promising candidate for the quantum description of the forces encountered in nature. A familiar classical example of a gauge parameter would be the location where the gravitational potential field equals zero in the Newtonian gravity. This represents a \textit{global} free parameter in the theory, since its value is a constant,  independent of space-time. Depending on the problem at hand, the gauge parameter may be freely chosen without affecting the outcome of the predictions for physical observables, such as a particle's kinetic energy in this potential field. The same observation is true in quantum field theoretical description of interactions, although in a slightly more complicated context. A proper isolation and treatment of the gauge parameter turns out to be of extreme importance and leads to consistent theories which may describe a particular interaction with properties obtained from experiment.

Let us examine more carefully how gauge theories emerge in context of quantum field theory. Imagine a particular quantum field theory possessing a specific global \textit{continuous} mathematical symmetry; i.e. the theory remains invariant under a continuous symmetry transforming operation. The branch of mathematics dealing with symmetries is called group theory. As mentioned before, ``global" means that the mathematical parameter in group theory describing the symmetry is \textit{not} a function of space-time; in other words, this symmetry parameter is a constant across space-time of the theory. In that sense, the global continuous symmetry of our quantum field theory is described by the particular mathematical group associated with the symmetry under consideration, leaving the theory invariant. Given this, if the ground state (vacuum) of the system does not exhibit the same global continuous symmetry, the symmetry is said to be ``spontaneously broken" in this quantum field theory. In other words, non-invariance of the ground state serves as the condition for a spontaneous symmetry breaking (SSB), even though the Hamiltonian or Lagrangian of theory is still fully invariant under the symmetry transformation. Upon a spontaneous breaking of a global continuous symmetry, according to Goldstone's theorem \cite{springerlink:10.1007/BF02812722} (named after Jeffrey Goldstone who formulated and proved the theorem), a massless scalar particle, called the \textit{Nambu-Goldstone boson}, will be generated. This is a peculiarity of quantum field theories with spontaneously broken global continuous symmetries.

The situation turns out to be more interesting if our continuous symmetry is ``local"; i.e. the symmetry parameter \textit{is} a function of space-time of the theory. Such a quantum field theory is then called a \textit{gauge theory}. The reason for the terminology is that in order for the theory to be invariant under the local continuous symmetry transformation, a new massless vector degree of freedom, containing a free gauge parameter, must be introduced. Mathematically, this vector particle acts as if to ``connect" different space-time points that have different symmetry parameters.\footnote{Geometrically, the situation is analogous to ``parallel transporting" a vector on a curved surface, which leads necessarily to the introduction of the ``Christoffel symbols".} Hence, a quantum field theory possessing one or more local continuous symmetries necessarily contains massless vector bosons (one for each symmetry generator), which makes it quite tempting as a suitable candidate for describing quantized interactions in nature, mediated by massless gauge bosons, such as electromagnetism and the strong interaction.

What would happen if the local continuous symmetry of this gauge theory were spontaneously broken? According to Goldstone's theorem, again a massless Nambu-Goldstone boson is released. It turns out, however, that this Nambu-Goldstone boson combines with (or as is commonly referred to, is ``eaten" by) the vector boson of the gauge theory, making the latter massive. The previously massless gauge boson with only two (transverse) polarizations has now become massive and, consequently, has acquired a third (longitudinal) polarization component, which is exactly provided by the ``eaten" Nambu-Goldstone boson. Hence, spontaneously broken gauge theories contain \textit{massive} gauge bosons (one for each broken symmetry generator), making them again tempting as suitable candidates for describing quantized interactions in nature, mediated by massive gauge bosons, such as the weak interaction. Let us discuss how these ideas are adopted in practice.

The quantum field theoretical description of electromagnetism, involving photons, is called \textit{Quantum Electrodynamics} (QED), developed among others by Julian Schwinger \cite{Schwinger:1948fk, Schwinger:1949uq}, Sin-Itiro Tomonaga \cite{Tomonaga:1948fk}, Freeman Dyson \cite{Dyson:1949kx}, and Richard Feynman \cite{Feynman:1948vn}. Feynman called it ``the jewel of physics" for its extremely accurate predications of various physical observables, such as the anomalous magnetic moment of the electron and the so-called Lamb shift of the energy levels of hydrogen. It represents the first successful application of quantum field theory to the physics of elementary particles, describing the quantum nature of the interaction between light and electrically charged matter. QED is a gauge theory with a local continuous $U(1)$ symmetry, representing an unobservable (local) phase. The $U(1)$ symmetry has only one generator and is unbroken; therefore, QED contains one massless, electrically neutral gauge boson, identified with the photon. Since photons themselves do not carry electric charge, they do not interact with one another. The \textit{Abelian}\footnote{In group theory, a group is called \textit{Abelian} if its generators commute with one another; otherwise, it is referred to as \textit{non-Abelian}. $U(1)$ is a trivial example of an Abelian group, since it has only one generator.} nature of the $U(1)$ group mirrors directly this physical property of the theory. 

\begin{figure}[!t]\begin{center}
\includegraphics[width=\textwidth]{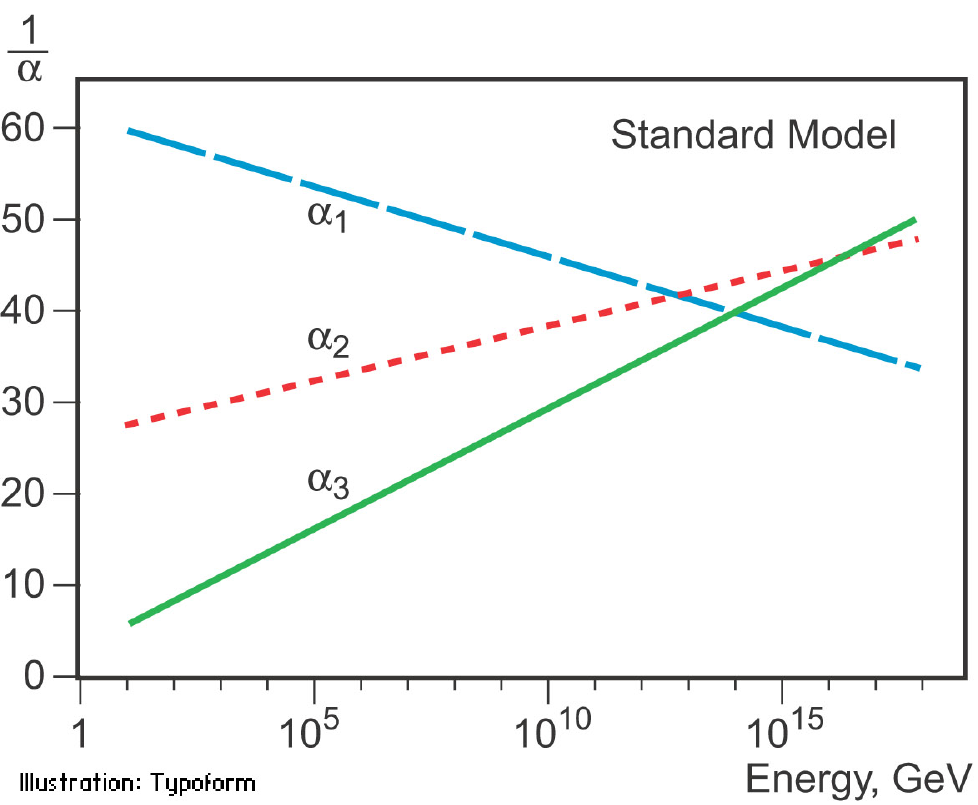}
\caption{Relative strengths of the Standard Model interactions as a function of energy. The electromagnetic coupling constant ($\alpha_1$) increases with energy scale, while the weak force ($\alpha_2$) and the strong force ($\alpha_3$) coupling constants decrease, reflecting the asymptoticly free nature of the latter interactions.}\label{runnings}
\end{center} (Figure courtesy of \copyright The Royal Swedish Academy of Sciences. ``The 2004 Nobel Prize in Physics - Popular Information". Nobelprize.org. 6 Apr 2012. \url{http://www.nobelprize.org/nobel_prizes/physics/laureates/2004/popular.html}) \end{figure}

The QED vacuum is a dynamical ``polarizable medium"; i.e. it contains virtual electron-positron pairs which are constantly created and annihilated for a short period of time, according to the Heisenberg uncertainty principle. In the presence of a ``bare" charge (which is formally infinite), these virtual pairs become polarized and induce a net screening of the bare charge at low energies, corresponding to a finite observed charge at long distance scales.\footnote{See the renormalization discussion later on in this chapter.} At high energies, shorter distance scales from the bare charge are probed, which are less screened by the dynamical vacuum. This results in a larger observed charge and a stronger perceived electromagnetic force. As a consequence, in QED the strength of the electromagnetic interaction \textit{increases} (logarithmically) with energy (see $\alpha_1$ in Fig.~\ref{runnings}); in other words, the electromagnetic force becomes stronger at shorter distance scales.

Strong interaction field theory is based on an $SU(3)$ unbroken local symmetry, and is called \textit{Quantum Chromodynamics} (QCD). It was developed by Chen Ning Yang and Robert Mills \cite{Yang:1954ys}, Murray Gell-Mann \cite{Gell-Mann:1962zr}, and many others. The $SU(3)$ group consists of the set of \textit{Special Unitary} $3 \times 3$ matrices, and is a \textit{non-Abelian} theory. It involves the color quantum number (analogous to the charge in QED), which is only present in quarks and gluons. The symmetry group is, therefore, formally designated as $SU(3)_C$  (subscript $C$ standing for color). The QCD's $SU(3)_C$ symmetry group contains eight generators, corresponding to eight massless gluons as force-carriers. In contrast with the photon, which itself does not carry electric charge, gluons do carry color quantum numbers and interact among themselves. This fact is reflected in the non-Abelian nature of the theory. Moreover, contrary to the Abelian QED, non-Abelian theories have as a peculiarity that the strength of the force \textit{decreases} logarithmically with increasing energy or equivalently at shorter distance scales (see $\alpha_3$ in Fig.~\ref{runnings}); in other words, quarks and gluons are practically free (i.e. non-interacting) inside hadrons at close distances. This phenomenon is referred to as \textit{Asymptotic Freedom} \cite{Politzer:1973ly, Gross:1973ve}, and makes a perturbative treatment of strong interaction possible at high energies. Although not yet formally proven, it is generally assumed that QCD predicts the appearance of only \textit{colorless} bound-states of quarks in nature; i.e. all hadrons are colorless, in agreement with observations. This is called \textit{color confinement}.

As mentioned previously, the asymptoticly free nature lying at the heart of the strong interaction makes it possible to invoke perturbation theory at high energies, since the strength of the interaction decreases with an increasing energy. To be precise, in a perturbative treatment the theory is expanded in terms of an asymptotic series as a function of its (small) coupling constant ($\alpha_3$ for the strong force in Fig.~\ref{runnings}). The \textit{leading order} term (LO) in the perturbative expansion corresponds to the classical level interaction, and is called the ``tree-level" interaction. The subsequent terms in the expansion form the so-called ``quantum corrections" to the tree-level interaction with an increasing level of complexity. Each of the higher order quantum corrections is, however, proportional to a higher power of the perturbative coupling constant, and is, therefore, increasingly suppressed. A perturbative analysis in quantum field theory is, hence, only meaningful so long as the coupling constant of the theory under consideration remains small, increasingly suppressing the higher powers of quantum corrections. Throughout this Thesis, we will extensively make use of perturbation theory, confining the analyses to the energy regions where such a treatment remains valid.

A field theoretical description of the weak force with three massive gauge bosons, two of which carry opposite electric charges, is quite subtle and hints at a gauge theory with spontaneously broken local continuous symmetries (to explain the masses), and involving some degree of mixture with the electromagnetic $U(1)$ symmetry (to explain the electric charges). Sheldon Glashow \cite{Glashow:1961qr}, Steven Weinberg \cite{Salam:1964hl}, and Abdus Salam \cite{Weinberg:1967qf} were independently able to show that a gauge theory based on an $SU(2)_L \times U(1)_Y$ symmetry, spontaneously broken to $U(1)_\text{EM}$, produces three massive gauge bosons (corresponding to the three broken generators) with positive, negative, and neutral electric charges, along with another gauge boson that is massless and electrically neutral (corresponding to the unbroken generator). As previously noted, the weak interactions affect only left-handed fermions, or equivalently, right-handed antifermions. The subscript $L$, thus, refers to the left-handed nature of the weak force. The subscript $Y$ represents ``hyper-charge", a $U(1)$ quantum number for the particles in the unbroken phase. Upon the spontaneous symmetry breaking, the surviving $U(1)$ symmetry is identified with the ordinary QED (hence, the subscript EM); the massless gauge boson associated with it is identified with the photon, while the three massive gauge bosons are the $W^+$, $W^-$, and $Z$ vector bosons, the force-carriers of the weak interaction. In that sense, $SU(2)_L \times U(1)_Y$ symmetry group provides a unified gauge description of weak and electromagnetic forces, and is, therefore, called \textit{Electroweak theory}. It should be emphasized that electroweak theory represents only a unified \textit{gauge description}, rather than a true physical unification of weak and electromagnetic interactions, since the weak and electromagnetic coupling strengths remain distinct (Fig.~\ref{runnings}).

\begin{figure}[!t]\begin{center}
\includegraphics[width=0.95\textwidth]{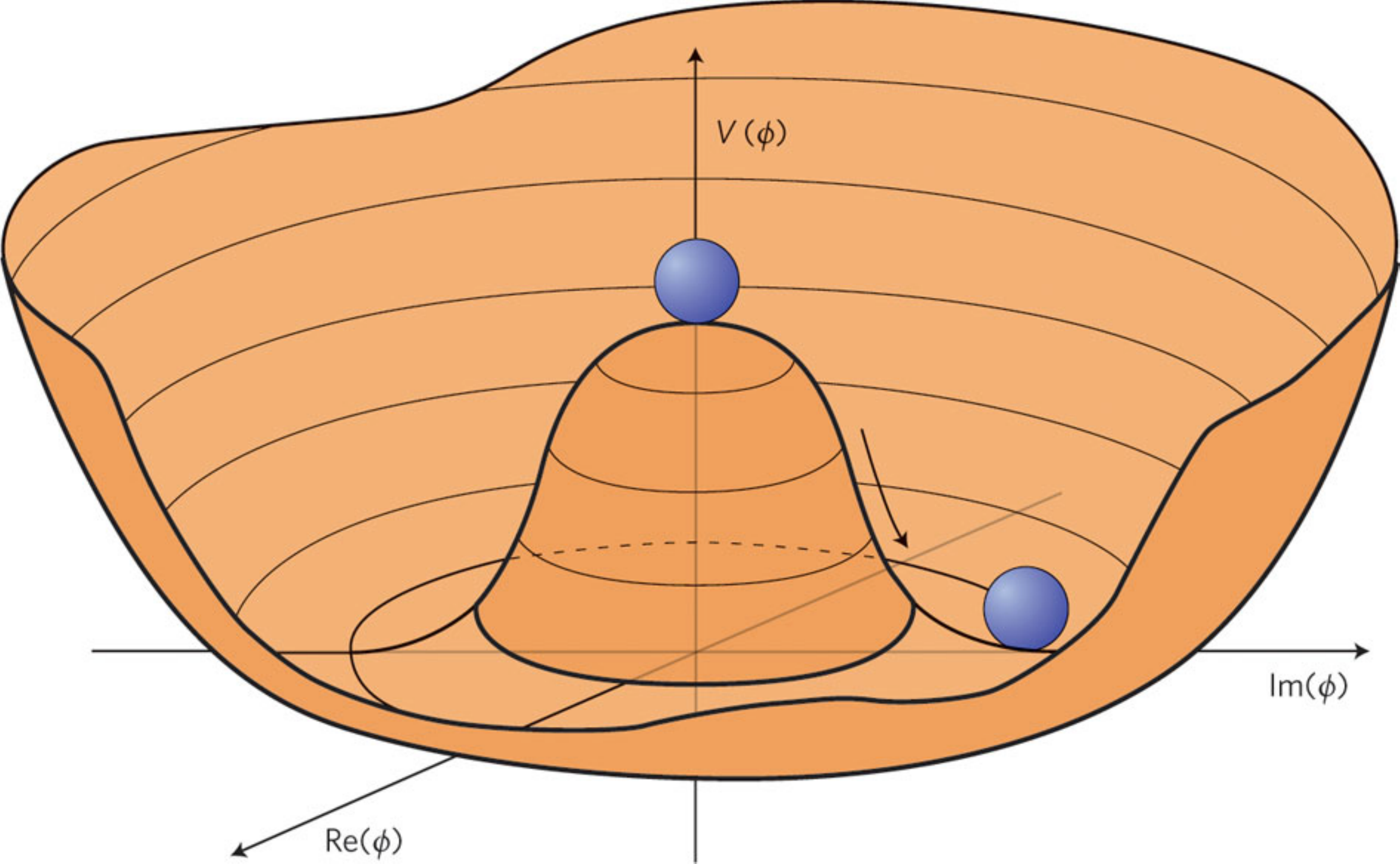}
\caption{Standard Model Higgs potential illustrated. Upon spontaneous symmetry breaking, the extremum at zero vacuum expectation value (VEV) becomes unstable and the system is forced to acquire a stable non-zero VEV at a randomly chosen point around the bottom of the hat. The degree of freedom along the radial direction (along the arrow) corresponds to the massive Higgs boson, while the degree of freedom perpendicular to it (on the equipotential circle at the bottom of the hat) is the massless Nambu-Goldstone boson.}\label{Higgs}
\end{center} (Figure courtesy of Nature Publishing Group. ``Eyes on a prize particle", Luis \'Alvarez-Gaum\'e \& John Ellis, Nature Physics 7, 2-3 (2011) $|$ doi:10.1038/nphys1874. \url{http://www.nature.com/nphys/journal/v7/n1/fig_tab/nphys1874_F1.html}) \end{figure}

At this stage, let us define more precisely the Higgs mechanism \cite{Anderson:1958rw}\nocite{Anderson:1963xe, Englert:1964qq, Guralnik:1964la, Higgs:1964cs}-\cite{Higgs:1966gd} responsible for electroweak symmetry breaking. In electroweak theory, the SSB is achieved by introducing a potential energy function for a complex scalar $SU(2)$ doublet (denoted as $\phi$ in Fig.~\ref{Higgs}), effectively adding four extra degrees of freedom to the theory. As depicted in Fig.~\ref{Higgs}, the potential develops a Mexican hat form, with the stable minima consisting of points along a circle at the bottom of the hat, representing an infinite number of degenerate vacua with \textit{non-zero} values (note that the vacuum with the \textit{zero} expectation value is unstable). One may choose any point on this circle\footnote{The different points correspond to different choices of gauge and are, therefore, physically equivalent.} as the true vacuum of the theory, thereby, allowing the development of a particular non-zero vacuum expectation value (VEV). This choice leads to an explicit spontaneous breaking of electroweak symmetry as described above. Fluctuations around this non-zero VEV redeploys three massless scalar degrees of freedom as the Nambu-Goldstone bosons (one for each broken symmetry generator), as well as one massive scalar state, which is called the \textit{Higgs boson}. In Fig.~\ref{Higgs}, the Nambu-Goldstone bosons correspond to fluctuations along the equipotential circle at the bottom, which is reflected in their massless nature. The Higgs boson, on the other hand, is generated by fluctuations in the radial direction where the potential changes, inducing a mass term for this degree of freedom.

In electroweak theory, additionally, the coupling of the scalar doublet to fermions is responsible for the fermion masses, with their masses proportional to the strength of the associated couplings. This type of coupling is called the Yukawa coupling (named after Hideki Yukawa who was the first to introduce a coupling between scalars and fermions in context of a different theory). The massive weak gauge bosons acquire their masses by eating the Nambu-Goldstone bosons, as discussed earlier. Therefore, in electroweak theory, gauge bosons and fermions are massless in the unbroken electroweak symmetry phase, and become massive in the broken phase. Even though the Higgs mechanism provides, among other things, an economical solution to mass generation for both fermions and weak gauge bosons, the Higgs boson itself remains elusive so far, and forms the last missing piece of the SM.\footnote{As of early 2012, the searches at the Large Hardon Collider (LHC) may indicate the existence of a SM Higgs boson at around $2\sigma$ statistical significance. This implies that the SM Higgs boson, if existed, is most likely to have a mass constrained to the range 116-131~GeV by the ATLAS experiment \cite{Aad:2012jt}\nocite{Aad:2012sf}-\cite{Aad:2012vl}, and 115-127~GeV by CMS \cite{Chatrchyan:2012zp}\nocite{Chatrchyan:2012ud, Chatrchyan:2012lr}-\cite{Chatrchyan:2012ao}. These findings are, however, statistically not strong enough to claim a discovery (the $5\sigma$ threshold) and await the accumulation of more data.}

\begin{figure}[!t]\begin{center}
\includegraphics[width=.89\textwidth]{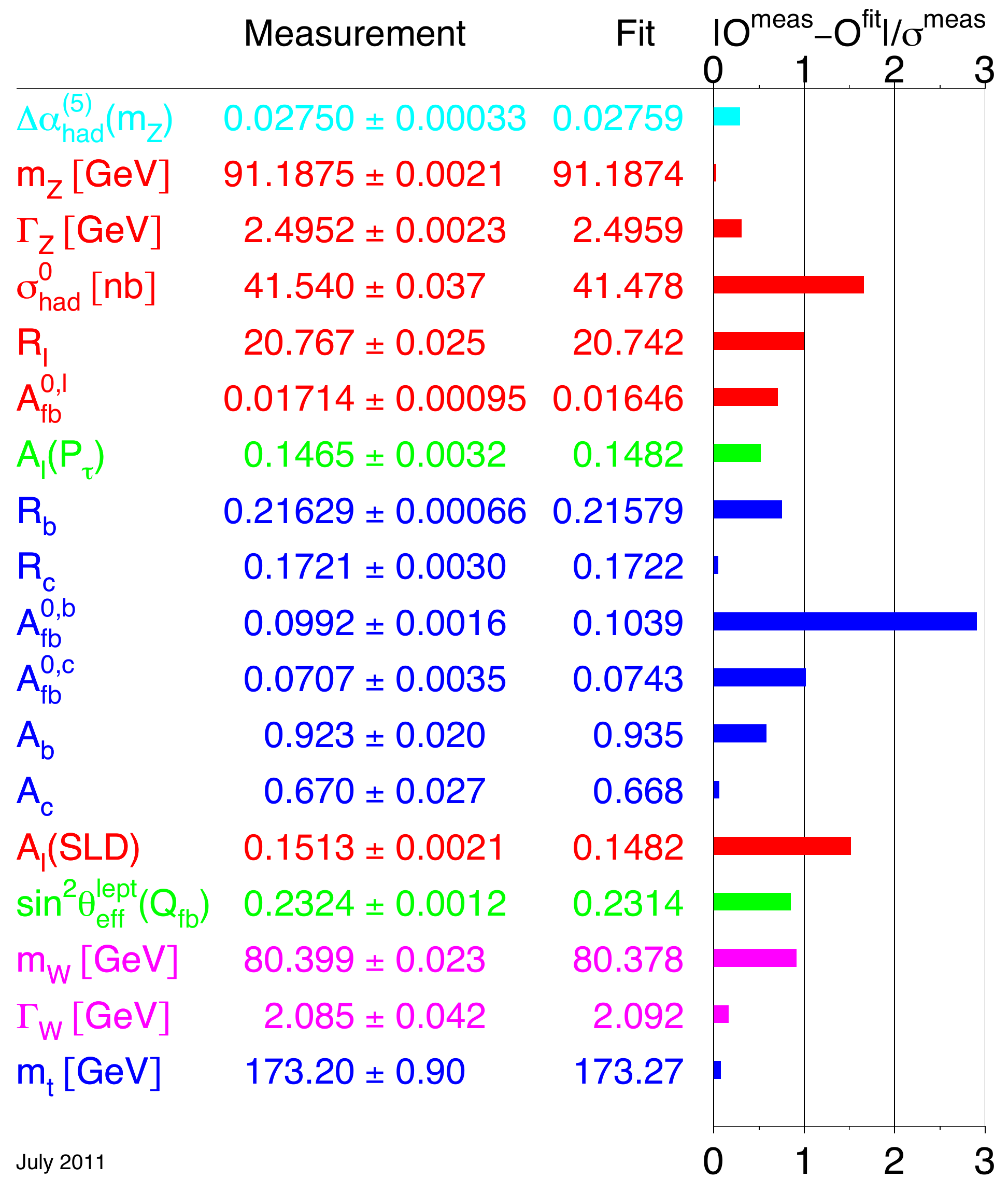}
\caption{Experimentally measured values of various SM parameters. Columns on right illustrate deviations from the SM theoretical prediction, ranging anywhere between 0 to 3 standard deviations.}\label{SMpar}
\end{center} (Figure courtesy of LEP/TEV Electroweak Working Group (EW WG). ``Preliminary constraints on the Standard Model" $|$ preprint: arXiv:1012.2367 [hep-ex],
updated for 2012 winter conferences. \url{http://lepewwg.web.cern.ch/LEPEWWG/plots/summer2011/s11_show_pull_18.pdf}) \end{figure}

To summarize, the SM has an $SU(3)_C \times SU(2)_L \times U(1)_Y$ gauge structure with a Higgs mechanism for the SSB, and encompasses all known subatomic particles and their interactions. Furthermore, using the experimentally determined values of a number of parameters as input, the SM can, given its renormalizable nature, make specific predictions regarding the outcome of many other experiments with, in principle, an arbitrary degree of accuracy. In order to test the SM predictions in the electroweak sector, a number of high precision experiments, the so-called \textit{electroweak precision tests}, have been conducted. Data collected through these experiments can be parametrized in various manners,\footnote{See Chapter~2 for an elaborate discussion on these electroweak parametrizations.} in order to make comparison with the theoretical predictions, and place tight constraints on the numerical values of many of the SM input parameters.\footnote{As we shall see, experimental deviations from the SM predicted value for these parameters can be attributed to new physics, allowing one to place (lower) bounds on the value of various beyond the Standard Model (BSM) variables.} Fig.~\ref{SMpar} summarizes the experimentally measured values of a number of parameters predicted in the SM, and compares them with the theoretical prediction, with the highest deviation lying within a $3\sigma$ bound. As one may appreciate, the SM is impressively successful in predicting the values of many observables, and passes the most stringent experimental tests.

As impressive as the SM might sound, there are, nevertheless, a number of theoretical and observational indications suggesting that it cannot represent a complete and final account of the quantum world, as the ultimate theory of nature. These indications demand explicit extensions and improvement of the SM, requiring inclusion of some new physics. An illustrative list of SM shortcomings contains the following observations:
\begin{itemize}
\item The SM does not contain a quantum field theoretical description of gravity;
\item As indicated in Fig.~\ref{runnings}, the strength of the electromagnetic, strong, and weak interactions do not meet at one single energy. If, based on theoretical considerations, one is to take the idea of unifying the interactions seriously, the SM does not unify the coupling constants of the three forces, neither does it provide a unified description of all three as different manifestations of one fundamental interaction;
\item There are over twenty free parameters in the SM, the origin of which remains unexplained within the framework of the theory, and their numerical values need to be considered as input to the SM;
\item As presented in Fig.~\ref{SMpar}, the SM predicted outcome for a number of experimental parameters deviates up to $3\sigma$ from the experimentally determined value, entertaining the possibility of some new physics substantially influencing those experiments;
\item \textit{Dark matter}, its existence inferred from various sources of astronomical and cosmological observations, is not explained within the context of the SM. In other words, the particle content of the SM (Fig.~\ref{SM}) cannot account for the observed dark matter in the universe;
\item The SM prediction for the value of \textit{dark energy}, presumably responsible for the observed accelerating expansion of the universe, is far too large by many orders of magnitude and is physically excluded;
\item The SM cannot account for the observed dominance of matter over antimatter in our universe;
\item The SM provides no justification for the observed homogeneity and isotropy of the universe at large distance scales;
\item Quantum corrections to the mass of the predicted Higgs boson are quadratically divergent (see below), making its mass highly sensitive to the ultra high energy behavior of the theory, around the Planck scale where gravity presumably becomes important. These corrections, therefore, tend to generate a large mass for the Higgs boson. In contrast, the actual mass of the Higgs boson must naturally lie near or below the electroweak symmetry breaking scale, many orders of magnitude smaller than the Planck scale. This requires a severe fine-tuning of the Higgs mass (and other SM parameters related to it), in order to remove the enormous contribution of the quadratically divergent quantum corrections, and produce a natural weak scale ``light" Higgs. This spectacular cancellation is known as the \textit{Hierarchy problem}. Another way to rephrase the Hierarchy problem is to contemplate why the weak energy scale is so much smaller than the Planck scale, or equivalently, why is gravity so much weaker than the other forces.
\end{itemize}
Let us elaborate on this last bullet note. Within the context of quantum field theory, quantum corrections often introduce divergent contributions to observables. This is a consequence of extrapolation of the quantum field theory to very high energies, where the current description presumably breaks down, and a new unknown and more fundamental theory takes over. Nevertheless, in order to obtain physically meaningful quantities with our current understanding, these unphysical infinities must be properly parametrized and isolated (the so-called \textit{regularization} procedure), and disposed of, by being reabsorbed into the definitions of physical quantities of the theory (the so-called \textit{renormalization} procedure). In other words, the unrenormalized theory contains \textit{bare} parameters (such as bare mass and charge) which are formally infinite due to the quantum corrections.

Upon renormalization, the properly regularized divergences of the theory are absorbed into the definitions of these bare parameters, rendering finite results for the physical observables (such as renormalized mass and charge). As noted before, in case of the Higgs mass the contributions from the quantum corrections are quadratically divergent. Reabsorbing these gigantic contributions into the definition of the bare Higgs mass through the renormalization procedure requires a cancellation to a remarkable degree of accuracy, in order to leave a finite, and in comparison tiny, renormalized weak scale mass for the Higgs. In other words, the renormalized Higgs mass is severely fine-tuned, leading to the mentioned Hierarchy problem.

Many theoretical extensions of the SM have been proposed in order to tackle and resolve one or more of the aforementioned SM issues. The main effort in particle physics community has been concentrated towards solving the Hierarchy problem, through somehow eliminating the quadratic divergences in the Higgs mass by construction. This is usually achieved by imposing some new symmetries, which result in a new set of (heavy) particles, inducing a cancellation among the quantum contributions originating from the ordinary particles and the new set of heavy particles. The first half of this Thesis is, accordingly, dedicated to elaborate on by far the simplest, although as we will see rather peculiar, solution of the Hierarchy problem; namely, the \textit{Lee-Wick Standard Model}~(LW~SM), as discussed below.

There are various methods of regularization in use in particle physics, each possessing certain advantages and shortcomings, depending on specific circumstances. A method suggested by Wolfgang Pauli and Felix Villars, published in 1949 \cite{Pauli:1949ix}, involves introducing a set of auxiliary ``fictitious" particles (see below) into the theory, the quantum contribution of which is subtracted from that of ordinary particles. The mass parameter of these auxiliary fields acts as the regulator, parameterizing the divergences of the original theory. Taking the infinite limit of this mass parameter removes the fictitious particles from the theory, and the original (divergent) theory is recovered.

In the original formulation of the Pauli-Villars (P-V) regularization method, the auxiliary particles were strictly introduced as a mathematical tool into the theory, in order to regularize the infinities associated with the quantum corrections of ordinary particles. In 1969 Tsung-Dao Lee and Gian-Carlo Wick explored the possibility of these auxiliary fields' being actual degrees of freedom, and the theoretical implications of this consideration \cite{Lee:1969fe}. They were motivated by the following observation: by construction, the divergences of the theory are cancelled among the quantum contributions from ordinary particles and their auxiliary ``partners". This feature made the resulting theory, from a formal perspective, highly attractive. At that time, their considerations led to a finite theory of QED \cite{Lee:1970hb}.\footnote{Modern finite theory of QED is based upon the renormalization formalism, in which the infinities are reabsorbed in the definitions of the bare parameters, rather than being cancelled by introduction of heavy partners.}

P-V auxiliary fields inherently contain an overall negative sign as part of their description, originally inserted to cancel divergences of ordinary fields. Therefore, promoting these auxiliary fields from fictitious to real particles turns them, by construction, into so-called ``ghosts". Ghost fields, carrying this extra negative sign, contribute a \textit{negative probability} to the processes in particle physics, and hence, are considered unphysical. The P-V ghosts as real degrees of freedom, therefore, violate \textit{unitarity}, indicating that the sum of the probabilities of all processes in the theory would not equal one, and lead to an unphysical theory. Coping with this problem, Lee and Wick were able to show that the issue could be averted by relaxing \textit{causality} at very high energies, having as a consequence that these auxiliary partners decay \textit{before} they are produced.\footnote{For an elaborate discussion on peculiarities related to the Lee-Wick theory, see e.g. \cite[p.~282]{colemanLW}.} As a result, the Lee-Wick~(LW)~theory inherently violates causality at small distance scales corresponding to very high energies, beyond the reach of current collider probes.\footnote{Odd as the theory might sound, one should bear in mind that causality in physics is an \textit{axiom}, and ultimately it is up to experiment to determine the energy bounds within which this axiom is valid.}

Lee and Wick also showed that, from a theoretical point-of-view, the addition of this ``auxiliary sector" to the original Lagrangian would be equivalent to invoking a ``higher-derivative term". In the latter approach, the kinetic term in the Lagrangian, ordinarily prescribed by a quadratic derivative term, is viewed as the first entity in an infinite expansion, and is extended to contain the next entity in the expansion series,\footnote{From a formal perspective, the inclusion of more LW~partners per SM field is equivalent to adding more higher-derivative expansion terms.} which will be of a quartic form.\footnote{Odd terms in the expansion are excluded by Lorentz invariance.} Using the equations of motion, one can prove the equivalence of the ``auxiliary-field formulation" and ``higher-derivative term formulation" of the theory. The Lagrangian with a higher-derivative term naturally produces a propagator with higher powers of momentum in the denominator for the ordinary fields, which, in calculations of quantum corrections, leads to a softening or complete removal of any divergences. This is in accordance with the auxiliary-field formulation, in which the auxiliary fields soften or cancel the infinities induced by the ordinary fields.

In 2008 Benjamin Grinstein and Mark Wise examined the possibility of applying LW~theory to the SM \cite{Grinstein:2008uk}, since the resulting theory would be naturally free of any quadratic divergences; hence, solving the Hierarchy problem in an economical way. As mentioned above, the higher-derivative terms added to various sectors of the SM produce the higher momentum propagators, softening or removing all SM divergences. Equivalently, in the auxiliary-field formulation, all bosons of the SM are accompanied by a massive ``Lee-Wick partner", while all left-handed and right-handed fermions have separately two corresponding massive left-handed and right-handed LW~partners, accomplishing the same result as the higher-derivative formulation of theory.

In Part~I of this Thesis, we take a deeper look into the LW~SM, which, in the context of resolving the Hierarchy problem, forms a natural extension to the SM electroweak sector. As alluded to before, the electroweak precision data may be parametrized in particular manners, in order to make comparison with theory possible. One particular way of parametrizing these data involves determining the quantum loop correction contributions to the self-energy of massive gauge bosons (also called \textit{vacuum polarization amplitude} (VPA)); i.e. $W^{\pm}$ and $Z$. The electroweak precision data place tight bounds on the value of the parametrizations associated with these contributions. At low energies, the postulated existence of the heavy LW~partners introduces new (previously non-existing) corrections to these parameters; for example, due to the LW~partners manifesting themselves as ``virtual particles" running in quantum correction loops. As the contributions of LW~particles are related to their masses, one can utilize the tight constraints on the VPA electroweak parametrizations, deduced from the electroweak precision tests, to place lower bounds on the LW~masses. This analysis \cite{Chivukula:2010if} will be the subject of Chapter~2.

In 1971 Gerard 't Hooft proved that the SM is a renormalizable theory \cite{Hooft:1971eq}. This implies that all divergences of the SM to all orders in perturbation theory are absorbed by renormalizing a limited number of the SM parameters, obtaining finite, physically meaningful quantities for the observables. Generally, the renormalizability of a quantum field theory may be qualitatively examined using a technique called \textit{power counting}. In this technique, one determines the powers of momentum in quantum correction loops, which in turn may be used as an indication of the number of divergent quantum correction amplitudes (to all orders) of the quantum field theory. If this number is finite, the theory under consideration is renormalizable and all of its divergences may be absorbed in a redefinition of a finite number of its bare parameters. If, however, by power counting the number of divergent amplitudes is infinite, the theory is non-renormalizable.

The addition of higher-derivative terms to the SM Lagrangian, as prescribed by LW~theory, might naturally raise concerns regarding the renormalizability of the resulting LW~SM, since it potentially alters the powers of momentum in the quantum correction loops. Renormalizability has been previously explored in the higher-derivative formulation of the theory by means of power counting arguments. It is, however, not clear \textit{a priori} how renormalizability manifests itself in the auxiliary-field formulation. In Chapter~3, studying a LW~scalar QED theory as a toy model, we directly examine the symmetries and renormalizability of this class of Abelian LW~theories in the auxiliary-field formulation, identifying the relevant symmetries which lead to a renormalizable theory.\footnote{The analysis performed in the context of LW~scalar QED is expected to generalize to non-Abelian LW~theories, and in particular to the LW~SM, although a formal proof is yet to be provided.}

In Part~II of the Thesis, we turn our attention to investigating the collider phenomenology of an extension to the strong sector of the SM described by QCD. As explained previously, QCD is based upon an $SU(3)_C$ local continuous symmetry. Since the late 1980s, it has been of theoretical and phenomenological interest to extend the SM strong sector to an $SU(3)_{1C} \times SU(3)_{2C}$ gauge theory, spontaneously broken to QCD's $SU(3)_C$ \cite{Frampton:1987wo}\nocite{Bagger:1988le, Hill:1991tx, Hill:1994lk, Popovic:1998bq, Braam:2008bx, Chivukula:1996qp, Simmons:1997hh, Martynov:2009sy, Davoudiasl:2001qd}-\cite{Lillie:2007ca}. This class of extensions represents an integral feature of theories in which the electroweak symmetry breaking is induced by the so-called \textit{strong dynamics}, where a new type of strongly-coupled gauge interaction forms a ``composite Higgs" out of colored fermions.\footnote{The situation is analogous to formation of ``Cooper pairs" from electrons in a superconducting medium below the critical temperature.} In accordance with Goldstone's theorem, the spontaneous symmetry breaking releases eight Nambu-Goldstone bosons, which are subsequently eaten by eight out of the sixteen originally massless gauge bosons. Consequently, we obtain, in addition to eight massless colored gauge bosons identified with ordinary QCD gluons, eight massive colored vector bosons which we generically refer to as the \textit{colorons}.

From the phenomenological point-of-view, the main coloron production channel in a hadron collider is the quark-antiquark annihilation process. To date, all theoretical and phenomenological analyses of colorons have been performed only at tree-level in perturbation theory (LO). The \textit{next-to-leading order} corrections (NLO) consist of the first order quantum corrections in the perturbative expansion, and additionally, the emissions of real soft and collinear particles which are undetected due to the limited resolution of the detectors. Chapter~4 is devoted to the first complete and comprehensive study of coloron production at NLO, taking into account the full corrections arising from the real emission of gluons and light quarks, in addition to first order quantum corrections \cite{Chivukula:2012fk}. The NLO study dramatically improves upon the previous LO calculations, and makes it possible to predict new coloron kinematic variables, which emerge only after a complete NLO analysis is performed.

Finally, we conclude the Thesis in Chapter~5, by summarizing the main contributions examined in the pervious chapters, and providing an outlook for future research in these areas.

\newpage
\vspace*{\fill}
\begin{center}
\Huge \textbf{PART I}
\end{center}
\vfill

\chapter{CUSTODIAL ISOSPIN VIOLATION IN THE LEE-WICK STANDARD MODEL\protect \footnote{This chapter is based on the paper first published in \cite{Chivukula:2010if}.}}\label{custodial}

\begin{quote}
``\textit{If the facts don't fit the theory, change the facts!}"
\begin{flushright}
|Albert Einstein (1879 -- 1955)\\
\end{flushright}
\end{quote}

\section{Introduction}

\lettrine[lines=1]{T}{he Lee-Wick Standard Model}\footnote{A Lee-Wick extension of the Higgs sector had been previously proposed in \cite{Jansen:1993uq}-\nocite{Jansen:1993kx}\cite{Jansen:1993vn}.} (LW~SM) \cite{Grinstein:2008uk} forms a natural and simple extension of the ordinary Standard Model, which solves the Hierarchy problem in an economical way. As discussed in Chapter~1, the Standard Model~(SM) suffers from the quadratic divergences brought forth by quantum corrections to the mass of the SM Higgs boson. Absorbing these contributions by means of renormalizing the bare Higgs mass requires a severe fine-tuning, and introduces the Hierarchy problem. The LW~SM, as an extension to the SM, introduces new fermions, among other particles, with exotic properties (see below) | the Lee-Wick~(LW)~fermions. Within the context of the LW~SM, the Hierarchy problem is remedied by an induced cancellation among the ordinary fermion quantum contributions and those generated by their LW~counterparts. The bare Higgs mass will then contain no quadratic divergences and can be renormalized in the usual way as explained in Chapter~1, remaining insensitive to the high energy behavior of the theory.

In the SM, the electroweak sector at one-loop may be fully parametrized using the following five observables: the mass of the $Z$ vector boson $m_Z$, the so-called \textit{Fermi constant} $G_F$ as a measure of the symmetry breaking vacuum expectation value (VEV), the so-called \textit{Weinberg angle} $\theta_W$ indicating the degree of mixture between the electromagnetic and weak forces, the mass of the Higgs boson $m_h$, and the mass of the top quark $m_t$. The values of these five \textit{electroweak observables} are extremely well-measured.

In order to extract the values of the electroweak observables from the electroweak precision data, convenient parametrizations have been introduced in terms of one-loop fermionic corrections to the electroweak gauge bosons' vacuum polarization amplitudes (VPA) (see e.g. Fig.~\ref{fig:loops}). In four-fermion scattering processes, \textit{oblique corrections} are defined as quantum corrections to the gauge bosons' VPA which do not depend on the identities of the initial and final state fermions. If such a flavor dependence does exist, the corrections are referred to as \textit{non-oblique corrections}.

In general, the oblique corrections are parametrized using the Peskin-Takeuchi $S$ and $T$ (or $\Delta \rho$) parameters \cite{Peskin:1992kl}\nocite{Altarelli:1991tg}-\cite{Altarelli:1992hc}. $S$ measures, for example, the size of the electroweak symmetry breaking sector, while $T$ tracks isospin violation (defined below).\footnote{Within the context of the standard electroweak theory, the $\rho$ parameter is defined as the zeroth-order ratio: $\rho = \displaystyle{\frac{m_{W}^{2}}{m_{Z}^{2} \, \cos^{2}\theta_{W}}}=1$. Small experimentally measured deviations from this value is attributed to the higher-order quantum corrections, and can place tight bounds on the new physics contributions.}  If both non-oblique and oblique contributions are present, one must employ, instead, the Barbieri {\em et al.} \cite{Barbieri:2004cr, Chivukula:2004nx} post-LEP electroweak parameters $\hat{S}$, $\hat{T}$, $W$, and $Y$.\footnote{For the precise definition of these parameters, see Sec.~\ref{treeBarbieri}.} The well-determined experimental values of precision electroweak observables can subsequently be used to place tight constraints on any contribution from new non-SM physics to the gauge bosons' VPA.

Since none of the LW~degrees of freedom have been observed in experiments, they must, if they exist, be heavy with masses beyond the reach of previous collider searches.\footnote{As we will show in this chapter, a lower bound for the LW~particle masses will be of the order of a few TeV, in which case, they might lie within the reach of the LHC.} As mentioned before, addition of these new heavy exotic particles to the spectrum in virtual form risks running into conflict with precise experimental data on the electroweak observables: virtual LW~fermions, running in quantum loops, contribute to the gauge bosons' VPA; thereby, modifying the values of the aforementioned (non-)oblique corrections. In conjunction with the electroweak experimental data, this information can be used to deduce lower bounds for the LW~particle masses. In what follows, we will discuss how the exotic properties of the LW~fermions solve the Hierarchy problem, and show, in detail, how corrections to the post-LEP electroweak parameters can be utilized to constrain the LW~masses.

The LW~SM features higher-derivative kinetic terms for each SM field.\footnote{See Sec.~\ref{sec:model} for the exact formal definition of the LW~SM.} As a consequence, the field propagators fall off to zero with momentum more rapidly than the ordinary SM propagators, and the infinities associated with ultraviolet quantum fluctuations either become less severe or are removed from the theory. In a scalar field theory all amplitudes turn out to be finite by power counting. In a gauge theory the higher-derivative kinetic terms generate new momentum-dependent interactions that prevent the theory from being finite; however, a simple power counting argument shows that all possible divergences are logarithmic. Thus, the LW~SM offers a potential solution to the hierarchy problem. This was the main motivation for studying the model \cite{Grinstein:2008uk} and analyzing its phenomenological implications \cite{Rizzo:2007ys}-\nocite{Rizzo:2008zr}\nocite{Alvarez:2009ly}\cite{Krauss:2008ve}.

If a higher-derivative kinetic term is added to the Lagrangian, the propagator of a LW~SM field displays two poles,\footnote{Note that in quantum field theory, the pole of a propagator (i.e. the zero-value of its denominator) is identified with the mass of the particle the propagator is representing.} the lighter one corresponding to a SM-like particle, and the heavier one corresponding to a new degree of freedom, the LW~partner. An equivalent formulation consists of separating the poles in such a way that to each field there corresponds only one pole and one mass. The LW~poles are then characterized by a negative residue, and, thus, act as Pauli-Villar regulators. However, unlike mere regulators, the LW~fields nontrivially participate in gauge and Yukawa interactions.

In electroweak sector of the SM, the left-handed top and bottom quarks, $t_L$ and $b_L$, (and their replicas in the other two quark generations, as well as the corresponding leptons) form a doublet under the $SU(2)_L \times U(1)_Y$ group, while their right-handed partners, $t_R$ and $b_R$, are singlets. The doublet structure is referred to as \textit{isospin}, in analogy with the ordinary spin doublet for the spin--$\displaystyle{\frac{1}{2}}$ fermions. In contrast with the ordinary spin doublet, however, members of these isospin doublets are different flavors and have different masses, leading to a violation of the isospin symmetry. The mass difference among members of the isospin doublet is most prominent in this third quark generation (top-bottom), compared to the other two generations, with the top quark being about two orders of magnitude heavier than the bottom quark. Hence, the isospin symmetry in the SM is broken, with the largest violation in the top-bottom doublet.

Likewise, in the LW~SM the largest one-loop contribution to the Higgs mass comes from an isospin violating sector of the theory: the top Yukawa coupling. There are two heavy partners of the top quark in the LW~SM, one associated with the left-handed top-bottom doublet, with mass $M_q$, and the other with the right-handed top, with mass $M_t$. The contributions to the Higgs mass involving a single LW~top are opposite in sign to those from a single SM top, so they cancel the quadratic divergence associated with the Higgs mass renormalization, $\delta m_h^2$. The net contribution is still logarithmically divergent, and for degenerate LW~top quarks, $M_q = M_t$, is of the form
\begin{equation}
\delta m_h^2=\frac{3\lambda_t^2}{8\pi^2}M_q^2 \log\frac{\Lambda^2}{M_q^2} \ ,
\end{equation} 
where $\Lambda$ is the cutoff, i.e. the highest energy scale marking the limit of the validity of the theory. In the limit $M_q \rightarrow \infty$ the ordinary quadratic divergence reappears, with $M_q$ acting as a cutoff. Therefore, as already pointed out in Ref.~\cite{Alvarez:2008fk}, in order to avoid fine-tuning the value of $M_q$ cannot be too large.

Because the dominant correction to the Higgs mass is associated with an isospin violating sector of the theory, it is important to check whether the LW~tops cause a large contribution to the electroweak observables, which are usually protected by the so-called \textit{custodial symmetry}: $\Delta \rho$, and, for theories with heavy replicas of the top quark, the $Zb_L\bar{b}_L$ coupling \cite{Agashe:2006bh, Chivukula:2009dq}. Large contributions to these quantities would lead to a stringent lower bound on $M_q$, which would result in large corrections to $m_h^2$ and thus the necessity of fine-tuning the scalar sector of the theory.\footnote{The electroweak gauge structure $SU(2)_L\times U(1)_Y$ may also be realized from a more general approximate global symmetry structure $SU(2)_L \times SU(2)_R$, in which only $SU(2)_L$, and $U(1)_R$ subgroup of $SU(2)_R$ are gauged (i.e. are local symmetries) and are identified with the electroweak theory. The latter $SU(2)_R$ approximate global symmetry is called the custodial symmetry, and protects the electroweak observables, such as the $W$-$Z$ mass ratio and the isospin, from large quantum corrections.}

In this chapter\footnote{Throughout this chapter, the timeline for the depicted Feynman diagrams is from left to right.} we analyze the potential conflict between naturalness and isospin violation, by computing the contribution of the top quark sector to the $\rho$ parameter and to the $Zb_L\bar{b}_L$ coupling. Furthermore, we compute the Barbieri {\em et al.} \cite{Barbieri:2004cr, Chivukula:2004nx} post-LEP electroweak parameters ($\hat{S}$, $\hat{T}$, $W$, and $Y$) to check for additional constraints. In terms of the post-LEP parameters, we find a simple picture for the constraints on the LW~SM. The dominant contributions to $\hat{T}$ come from the fermion sector at one-loop, and limits on this parameter provide the strongest constraints on the top quark sector.\footnote{The dominant contributions to $\hat{S}$, likewise, come from the fermion sector at one-loop, but they are too small to provide strong constraints on the top quark sector.} In contrast, the dominant contributions to $Y$ and $W$ arise from the gauge sector at tree-level, and limits on these parameters, therefore, provide the strongest constraints on the gauge sector. These results imply that the bounds on the LW~fermions, coming almost entirely from $\hat{T}$, are essentially independent of the LW~gauge masses.

Our results differ from those in Refs.~\cite{Alvarez:2008fk, Carone:2008oq} because their one-loop analysis of the effects of LW~top quarks on electroweak observables rests on the incorrect assumption that the corrections are purely oblique \cite{Peskin:1992kl}\nocite{Altarelli:1991tg}-\cite{Altarelli:1992hc}. As discussed in Ref.~\cite{Underwood:2009ij} important non-oblique corrections arise at tree-level in the LW~SM, in the form of non-zero values for $W$ and $Y$. Therefore, one must use the Barbieri {\em et al.} parameters to compare the LW~SM with experiment.

In Sec.~\ref{sec:model} we review the structure of the LW~SM \cite{Grinstein:2008uk} and establish notation. In Sec.~\ref{sec:effective} we present an effective field theory analysis of the LW~corrections to $\Delta \rho$ and to the $Zb_L\bar{b}_L$ coupling. In Sec.~\ref{sec:parameters} we present our analysis of the post-LEP electroweak parameters and the resulting constraints on the LW~SM, while the constraints from the $Zb_L\bar{b}_L$ coupling are analyzed in Sec.~\ref{sec:Zbb}. The leading logarithmic contributions to the electroweak observables in the full theory and the effective theory have to match; thus the results of Sec.~\ref{sec:effective} provide an important check for those of Sec.~\ref{sec:parameters} and \ref{sec:Zbb}.

Global symmetries and renormalizability of LW~theories will be discussed in detail in Chapter~3, while questions concerning unitarity~\cite{Cutkosky:1969bs}, causality~\cite{Grinstein:2009fv}, and Lorentz invariance in LW~theories, although potentially important, will not be considered in this analysis. A complete analysis of the one-loop renormalization of the LW~SM can be found in~\cite{Grinstein:2008dz}.

\section{The Lee-Wick Standard Model}\label{sec:model}

It is straightforward to write a higher-derivative extension of the SM electroweak Lagrangian
\cite{Grinstein:2008uk}. Adopting a non-canonical normalization for the gauge fields, the gauge Lagrangian reads
\begin{equation}\label{eq:gaugelagrg} \begin{split}
{\cal L}_\text{gauge}^\text{hd} =\ & -\frac{1}{4g_1^2} \, \hat{B}_{\mu\nu}\hat{B}^{\mu\nu}
-\frac{1}{2 g_2^2} \, \text{Tr}\left[\hat{W}_{\mu\nu}\hat{W}^{\mu\nu}\right] \\
&+ \frac{1}{2 g_1^2 M_1^2} \, \partial^\mu \hat{B}_{\mu\nu} \partial_\lambda \hat{B}^{\lambda\nu}
+\frac{1}{g_2^2 M_2^2} \, \text{Tr}\left[\hat{D}^\mu \hat{W}_{\mu\nu} \hat{D}_\lambda \hat{W}^{\lambda\nu}\right] \ ,
\end{split} \end{equation}
where
\begin{equation*}
\hat{B}_{\mu\nu} \equiv \partial_{\mu} \hat{B}_{\nu} - \partial_{\nu} \hat{B}_{\mu} \ , \quad
\hat{W}_{\mu\nu} \equiv \left( \partial_{\mu} \hat{W}_{\nu}^{a} - \partial_{\nu} \hat{W}_{\mu}^{a} + f^{abc} \, \hat{W}_{\mu}^{b} \hat{W}_{\nu}^{c} \right) \tau^{a} \ ,
\end{equation*}
with $\tau^{a} \equiv \displaystyle{\frac{\sigma^{a}}{2}}$, and $\sigma^{a}$ the $SU(2)$ generators (Pauli matrices). The ``hat" notation indicates that the field's propagator contains not only the ordinary SM poles but also the LW~poles. For example, in the limit of unbroken electroweak phase the $\hat{B}_\mu$ propagator has a massless pole, corresponding to the ordinary $B_\mu$ gauge field, and a mass-$M_1$ pole, corresponding to its LW~counterpart. Notice also that additional dimension-six operators could, in principle, be added to this Lagrangian. However, these would lead to scattering amplitudes for the longitudinally polarized gauge bosons growing like $E^2$, where $E$ is the center-of-mass energy, and thus to a rather early violation of unitarity \cite{Grinstein:2008fu}. We, therefore, do not include them in this analysis. Notice also that we only include one higher-derivative term per SM field, which introduces a single corresponding LW~pole. This is certainly fine for our purposes, since in this analysis we focus on the low momentum regime, where additional higher-derivative terms are negligible. However, at large momenta additional poles in the propagator can have important implications \cite{Carone:2009kl, Carone:2009qa}.
 
The higher-derivative extension of the Higgs sector is 
\begin{equation}
{\cal L}_\text{Higgs}^\text{hd} = |\hat{D}_\mu \hat{\phi}|^2
-\lambda\left(\hat{\phi}^\dagger \hat{\phi} -\frac{v^2}{2}\right)^2 
-\frac{1}{M_h^2} |\hat{D}^2 \hat{\phi}|^2 \ ,
\label{eq:higgslagrg}
\end{equation}
where as usual the Higgs doublet may be written in component form as
\begin{equation}
\hat{\phi} = \frac{1}{\sqrt{2}}
\begin{pmatrix} i\sqrt{2}\hat{\phi}^+ \\ v+\hat{h}-i\hat{\phi}^0 \end{pmatrix} \ .
\end{equation}
Here and in Eq.~(\ref{eq:gaugelagrg}) the covariant derivative written with a hat is built with the hatted gauge fields. We will find it convenient to have a compact way of denoting $i \sigma^2 \hat{\phi}^\ast$ as we build operators that couple the Higgs to the right-handed top quark. Hence, we make the definition
\begin{equation}
\hat{\varphi} \equiv (i \sigma^2 \hat{\phi}^\ast) =  
\frac{1}{\sqrt{2}}
\begin{pmatrix} v+\hat{h}+i\hat{\phi}^0\\
i\sqrt{2}\hat{\phi}^- \end{pmatrix} \ .
\end{equation}
The field $\hat{\phi}$ contains both the ordinary Higgs doublet and a massive doublet with mass~$\sim~M_h$.\footnote{If $M_h$ is smaller than all other LW~mass parameters, in a certain energy regime the model behaves like a two-Higgs doublet model, although one doublet is of LW~type. This scenario was analyzed in \cite{Carone:2009mi}.}

In the fermion sector we focus only on the third quark generation, since this is the dominant source of isospin violation and gives the largest correction to the Higgs mass.\footnote{Inclusion of the remaining flavors would introduce new mixing matrices, and, without the assumption of minimal flavor violation, potential sources of flavor changing neutral currents (FCNC). However, in Ref. \cite{Dulaney:2008pi} it was shown that for LW~fermion masses in the TeV range the FCNC are sufficiently small to satisfy experimental constraints.} The higher-derivative extension of the fermion Lagrangian is
\begin{equation}\label{eq:qhigher} \begin{split}
{\cal L}_\text{quark}^\text{hd} = \ &\bar{\hat{q}}_L i \hat{\slashed{D}} \hat{q}_L
+ \bar{\hat{t}}^\prime_R  i \hat{\slashed{D}} \hat{t}^\prime_R
+ \bar{\hat{b}}^\prime_R  i \hat{\slashed{D}} \hat{b}^\prime_R \\
&+ \frac{1}{M_q^2} \bar{\hat{q}}_L  i \hat{\slashed{D}}^3 \hat{q}_L
+\frac{1}{M_t^2} \bar{\hat{t}}^\prime_R  i \hat{\slashed{D}}^3 \hat{t}^\prime_R
+\frac{1}{M_b^2} \bar{\hat{b}}^\prime_R  i \hat{\slashed{D}}^3 \hat{b}^\prime_R \ ,
\end{split} \end{equation}
where $\hat{q}_L=(\hat{t}_L,\hat{b}_L)$. Notice that the right handed fields have been primed because, for example, $\hat{t}_L$ and $\hat{t}^\prime_R$ are not left and right component of the same Dirac spinor. In the unbroken electroweak phase $\hat{t}_L$ ($\hat{t}^\prime_R$) contains the ordinary massless SM left-handed (right-handed) top as well as a massive Dirac fermion of mass $M_q$ ($M_t$). 

Finally we consider the Yukawa Lagrangian, which in the SM has no derivatives. Therefore, we write
\begin{equation}
{\cal L}_\text{Yukawa} = -y_t\,\bar{\hat{q}}_L\, \hat{\varphi}\,\hat{t}^\prime_R \,+\,\text{h.c.} \ ,
\label{eq:yukhigher}
\end{equation}
where the bottom Yukawa coupling has been ignored, since $y_b\ll y_t$.

As explained in the introduction, this ``higher-derivative" formulation of the theory, in which both the ordinary pole and the LW~pole are contained in the same field, is equivalent to an ``ordinary formulation" in which, as follows: (i) the two poles belong to two different fields, and (ii) the kinetic and mass terms for the LW~fields have the ÔÔwrongÕÕ sign. This alternative formulation is especially useful for calculating loop diagrams. In this chapter we will compute loop diagrams with the top and bottom quarks in the loop. Thus we will find it helpful to replace the higher-derivative fermion and Yukawa Lagrangians with the ordinary formulation Lagrangians
\begin{equation}\label{eq:lower} \begin{split}
{\cal L}_\text{quark} =\ & \bar{q}_L i \hat{\slashed{D}} q_L
+ \bar{t}^\prime_R i \hat{\slashed{D}} t^\prime_R
+ \bar{b}^\prime_R i \hat{\slashed{D}} b^\prime_R \\
&- \bar{\tilde{q}}\left(i\hat{\slashed{D}}-M_q\right)\tilde{q}
-\bar{\tilde{t}}^\prime\left(i\hat{\slashed{D}}-M_t\right)\tilde{t}^\prime
-\bar{\tilde{b}}^\prime\left(i\hat{\slashed{D}}-M_b\right)\tilde{b}^\prime \ ,
\end{split} \end{equation}
and
\begin{equation}
{\cal L}_\text{Yukawa} = -y_t\,\left(\bar{q}_L-\bar{\tilde{q}}_L\right)\,\hat{\varphi}\ 
\left(t^\prime_R-\tilde{t}^\prime_R\right) \,+ \,\text{h.c.} \ ,
\label{eq:yuklower}
\end{equation}
where
\begin{equation}
\hat{q}_L \equiv q_L - \tilde{q}_L\ , \quad\quad \hat{t}^\prime_R\equiv t^\prime_R-\tilde{t}^\prime_R\ , \quad\quad 
\hat{b}^\prime_R\equiv b^\prime_R-\tilde{b}^\prime_R \ ,
\end{equation}
and where the fields with (without) a tilde are LW~(SM) fields. The equivalence between the higher-derivative formulation, Eqs.~\eqref{eq:qhigher} and \eqref{eq:yukhigher}, and the ordinary formulation, Eqs.~\eqref{eq:lower} and \eqref{eq:yuklower}, can be easily proved; see, for example, \cite{Grinstein:2008uk}. Notice that the wrong sign in front of the kinetic and mass terms makes the LW~(tilde) fields act like Pauli-Villars regulators, with the difference that they also participate nontrivially in gauge and Yukawa interactions.

\section{Effective Field Theory for $\Delta \rho$ and $Zb\bar{b}$}\label{sec:effective}

The appearance of the LW~fields in the Yukawa interactions, Eq.~(\ref{eq:yuklower}), suggest the presence of non-standard sources of custodial isospin violation at energies below the LW~scale. Dimension-six custodial violating operators can potentially arise from tree-level exchanges, and from loop diagrams with one or more LW~fermions in the loop. The leading contribution to these operators, in inverse powers of the LW~fermion masses, can be found by integrating out the LW~fermions at tree-level and computing loops in the resulting effective field theory. For LW~fermion masses much larger than both the Higgs vacuum expectation value (VEV) and the external momenta, the effective Lagrangian can be computed in powers of $\hat{\varphi}/M_{q,t}$ and $\hat{\slashed{D}}/M_{q,t}$. Including the leading non-standard corrections, this leads to
\begin{equation}\label{eq:efflagr0} \begin{split}
{\cal L}_\text{eff} =\ & \bar{q}_L i \hat{\slashed{D}} q_L
+ \bar{t}_R i \hat{\slashed{D}} t_R
+ \bar{b}_R i \hat{\slashed{D}} b_R
- y_t\left( \bar{q}_L \hat{\varphi}\,t_R + \bar{t}_R \hat{\varphi}^\dagger q_L \right) \\
&- \frac{y_t^2}{M_t^2}\,\bar{q}_L i \hat{\varphi}\hat{\slashed{D}}\left(\hat{\varphi}^\dagger q_L \right)
- \frac{y_t^2}{M_q^2}\,\bar{t}_R  i \hat{\varphi}^\dagger \hat{\slashed{D}}\left(\hat{\varphi} \, t_R \right) \ .
\end{split} \end{equation}
Notice that the primes have been removed from the right-handed fermion fields, because now left-handed and right-handed components are Dirac partners. Notice also that this Lagrangian assumes $M_q$ and $M_t$ to be of the same order, with no hierarchy between them. The leading logarithmic correction to observables will, therefore, be proportional to $\log M_q^2/v^2\sim \log M_t^2/v^2$. In what follows, we compute these leading-log corrections by constructing the operators which arise in
the effective theory appropriate for energy scales below $M_t\simeq M_q$, in which the LW~partners have
been ``integrated out" but the top quark remains in the spectrum.

After electroweak symmetry breaking, the higher-derivative operators lead to a renormalization of the fermion kinetic terms. An alternative approach consists of redefining $q_L$ and $t_R$ to make their kinetic terms canonically normalized in both the broken and the unbroken electroweak phase. This is achieved by the replacements
\begin{equation}
q_L  \to \left[1+\frac{y_t^2}{2 M_t^2} \hat{\varphi}\hat{\varphi}^\dagger 
+ {\cal O}(1/M_t^3)\right] q_L \ , \quad t_R  \to \left[1+\frac{y_t^2}{2 M_q^2} \hat{\varphi}^\dagger\hat{\varphi} 
+ {\cal O}(1/M_q^3)\right] t_R \ ,
\end{equation}
which leads to a new Lagrangian, equivalent to ${\cal L}_\text{eff}$
\begin{align}
{\cal L}^\prime_\text{eff} =\ & \bar{q}_L i \hat{\slashed{D}} q_L
+ \bar{t}_R i \hat{\slashed{D}} t_R
+ \bar{b}_R i \hat{\slashed{D}} b_R 
- y_t\,\bar{q}_L \hat{\varphi}\left[1+\frac{y_t^2}{2}\left(\frac{1}{M_q^2}+\frac{1}{M_t^2}\right)
\hat{\varphi}^\dagger \hat{\varphi}\right]t_R + \text{h.c.} \notag \\
&+ \frac{y_t^2}{2M_t^2}\,\bar{q}_L i \left[(\hat{D}_\mu\hat{\varphi})\hat{\varphi}^\dagger
-\hat{\varphi}(\hat{D}_\mu\hat{\varphi})^\dagger\right] \gamma^\mu q_L \label{eq:efflagr} \\
&+ \frac{y_t^2}{2M_q^2}\,\bar{t}_R  \gamma^\mu t_R\,i \left[(\hat{D}_\mu\hat{\varphi})^\dagger \hat{\varphi}
-\hat{\varphi}^\dagger(\hat{D}_\mu\hat{\varphi})\right] \notag \ .
\end{align}
As expected, there are custodial symmetry violating dimension-six operators. However, at tree-level there is no non-standard contribution to $\Delta\rho$ or the $Z b_L \bar{b}_L$ coupling.

\begin{figure}\begin{center}
\includegraphics[width=\textwidth]{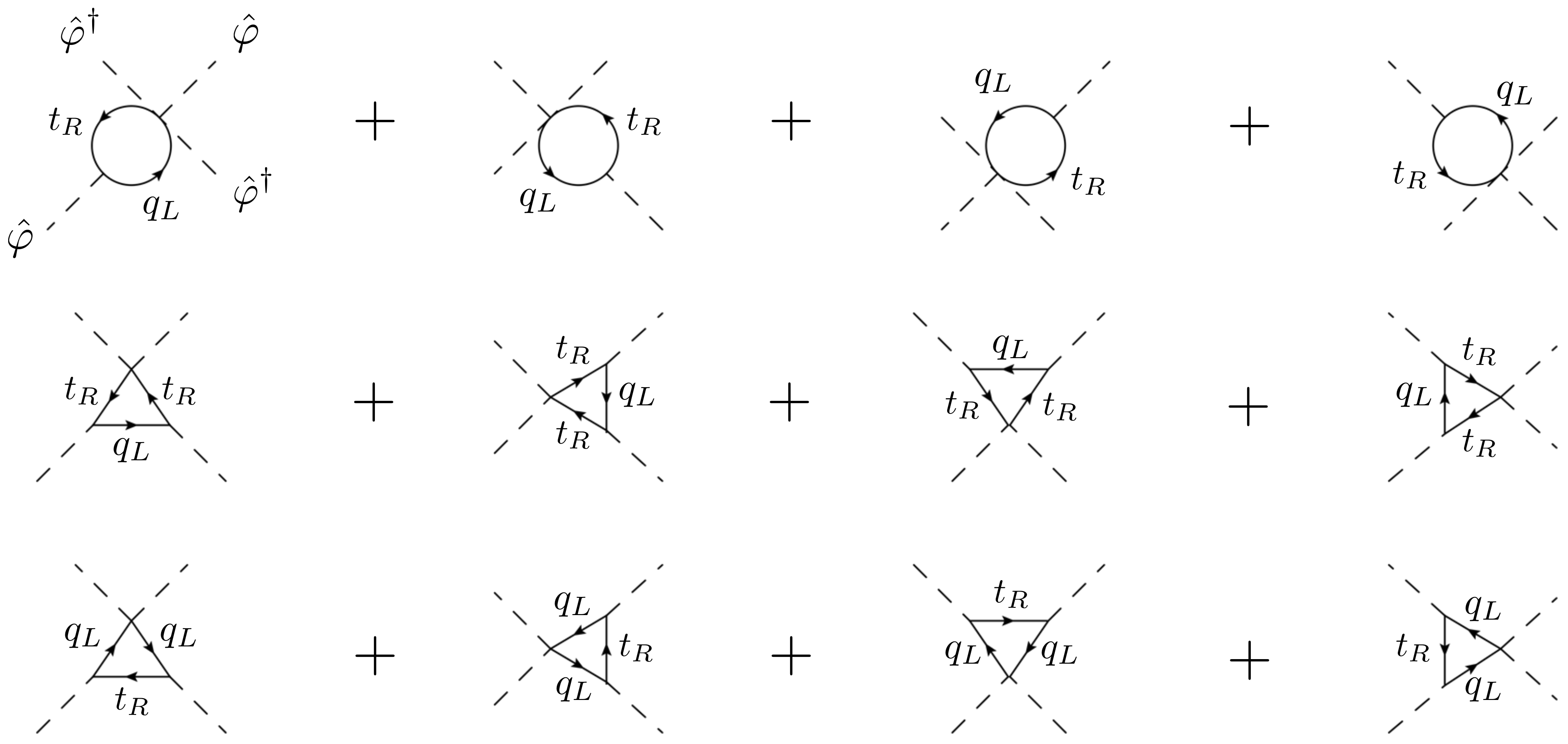}
\caption{Diagrams contributing to the dimension-six four-$\hat{\phi}$ operators in the effective theory, with the LW~fermions integrated out at tree-level.}
\label{fig:effectiverho}
\end{center}\end{figure}

${\cal L}^\prime_\text{eff}$ features terms coupling one, two, or three $\hat{\varphi}$ fields to a pair of fermions. Therefore, dimension-six four-$\hat{\varphi}$ operators arise both from vacuum polarization amplitudes and triangle diagrams, as shown in Fig.~\ref{fig:effectiverho}. The log-divergent parts of these diagrams (which yield the $\log(M^2_{t,q}/m^2_t)$ contributions) can be computed in the unbroken electroweak phase, with fermions in the loop. The logarithmically divergent part of the amplitude is reproduced by the operators\footnote{There are also quadratic divergences which are completely absorbed by a counterterm of the form $|\hat{\varphi}|^4$, with no derivatives.}
\begin{equation}
\frac{3y_t^4}{16\pi^2}\left[\frac{2}{M_t^2}+\frac{1}{M_q^2}\right]|\hat{D}\hat{\phi}|^2 |\hat{\phi}|^2 \cdot \frac{1}{\epsilon}
+\frac{3y_t^4}{16\pi^2}\left[\frac{1}{M_t^2}+\frac{2}{M_q^2}\right]|\hat{\phi}^\dagger\hat{D}\hat{\phi}|^2 \cdot \frac{1}{\epsilon} \ ,
\end{equation}
where as usual $\epsilon=2-d/2$ in dimensional regularization. The first operator respects custodial symmetry, but the second operator does not, since it contributes only to the $Z$~boson mass. The second operator gives the dominant contribution to $\Delta\rho$, which is, therefore, of the order
\begin{equation}
(\Delta\rho)_\text{LW} \sim -\frac{3}{16\pi^2}\frac{2m_t^4}{v^2}\left[\frac{1}{M_t^2}+\frac{2}{M_q^2}\right]\log{\frac{M_q^2}{m_t^2}} \ ,
\label{eq:rholeading}
\end{equation}
where the $1/\epsilon$ is replaced by the large $\log$ which arises in the effective theory scaling from the scale $M_q\sim M_t$ to the weak scale $m_t\sim v$. For LW~fermions lighter than 1 TeV this is a large \textit{negative} isospin violating effect. For example, taking $M_q=M_t=500$ GeV gives $\Delta\rho\sim -1.4\%$. Furthermore, since $\Delta\rho$ is always negative, a heavy Higgs is strongly disfavored in the LW~SM.

\begin{figure}\begin{center}
\includegraphics[width=\textwidth]{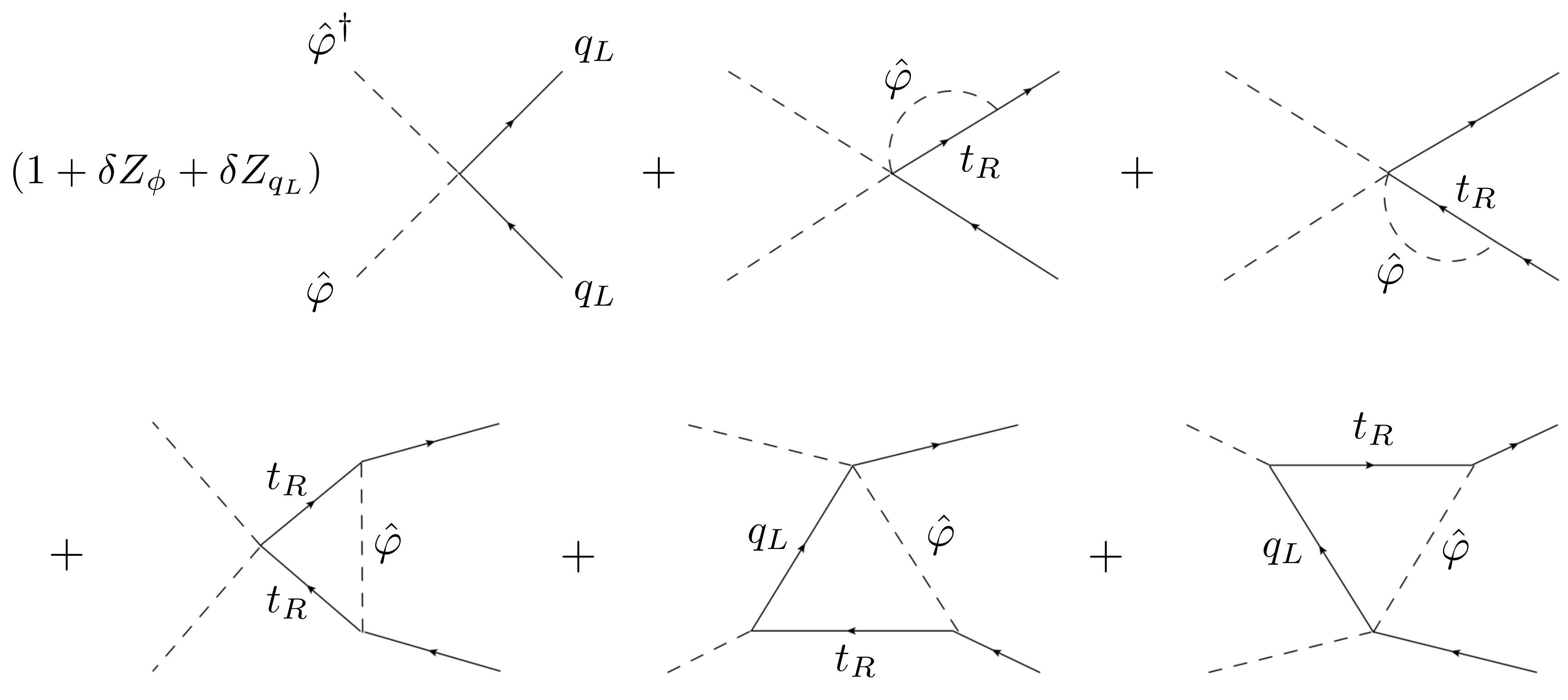}
\caption{Diagrams contributing to the dimension-six operators with two external $q_L$ and two $\hat{\varphi}$ fields. The triangle diagrams lead to the second operator of Eq.~(\ref{eq:Zbbeffective}), which contains non-universal corrections to the $Z b_L \bar{b}_L$ coupling.}
\label{fig:effectiveZbb}
\end{center}\end{figure}

The diagrams contributing to the left-handed fermionic gauge couplings up to one-loop order are shown in Fig.~\ref{fig:effectiveZbb}. The tree-level diagram (corrected by the field strength renormalizations) corresponds to the custodial violating operator proportional to $y_t^2/2M_t^2$, in Eq.~(\ref{eq:efflagr}). This operator contributes to the $Z t_L \bar{t}_L$ coupling, not to $Z b_L \bar{b}_L$.\footnote{Including the bottom Yukawa coupling would lead to a tree-level operator contributing to $Z b_L \bar{b}_L$. However, the top loop contribution is dominant, since $16\pi^2 y_b^2\sim 0.1$.} The remaining diagrams contain non-standard logarithmic divergences which are reproduced by the operators
\begin{equation}\label{eq:Zbbeffective} \begin{split}
&\frac{y_t^4}{16\pi^2}\,  \frac{1}{4M_q^2} \, i \left[\bar{q}_L \gamma^\mu \hat{D}_\mu q_L - \bar{q}_L \overleftarrow{\hat{D}^\dagger_\mu} \gamma^\mu q_L\right] \hat{\varphi}^\dagger \hat{\varphi} \cdot \frac{1}{\epsilon} \; + \\
&\frac{y_t^4}{16\pi^2} \left[\frac{1}{M_t^2}+\frac{1}{4 M_q^2}\right] \bar{q}_L \gamma^\mu q_L \, i \left[(\hat{D}_\mu\hat{\varphi})^\dagger \hat{\varphi}
-\hat{\varphi}^\dagger(\hat{D}_\mu\hat{\varphi})\right]  \cdot \frac{1}{\epsilon} \ .
\end{split} \end{equation}
In this expression the first (custodially symmetric) operator amounts to a renormalization of the standard gauge interactions, and does not contribute to non-standard fermionic gauge couplings. The second operator violates custodial symmetry, and is only due to the triangle diagrams in Fig.~\ref{fig:effectiveZbb}. This contributes both to the $Z t_L \bar{t}_L$ coupling and the $Z b_L \bar{b}_L$ coupling. Expressing the latter in the form
\begin{equation}\label{eq:totalZbb} \begin{split}
&\frac{e}{c_w s_w} \, g_L^{b\bar{b}}\, Z_\mu\,  \bar{b}_L\gamma^\mu b_L \\
&\equiv \frac{e}{c_w s_w} \left[-\frac{1}{2}  + \frac{1}{3}\sin ^2 \theta_W 
+ (\delta g_L^{b\bar{b}})_\text{SM} + (\delta g_L^{b\bar{b}})_\text{LW} \right] Z_\mu\,  \bar{b}_L\gamma^\mu b_L \ ,
\end{split}\end{equation}
where $(\delta g_L^{b\bar{b}})_\text{SM}$ includes all higher order SM corrections, and replacing the $1/\epsilon$ poles with the large log arising from scaling in the
theory, we find that the second 
operator of Eq.~(\ref{eq:Zbbeffective}) gives the dominant non-universal LW~contribution to $g_L^{b\bar{b}}$
\begin{equation}
(\delta g_L^{b\bar{b}})_\text{LW} \sim -\frac{m_t^4}{32\pi^2 v^2}\left[\frac{4}{M_t^2}+\frac{1}{M_q^2}\right]\log\frac{M_q^2}{m_t^2} \ .
\label{eq:Zbbleading}
\end{equation}
The SM prediction is already 1.96$\sigma$ below the observed central value. Hence, the additional 
 negative correction in the LW~theory goes in the direction opposite 
to what is favored by experiment.

In the next two sections we compute perturbatively (in $v^2/M_q^2$ and $v^2/M_t^2$) and numerically
the values of  $\Delta\rho$ and the $Z b_L \bar{b}_L$ coupling in the full LW~theory. Our effective field
theory results, Eq.~(\ref{eq:rholeading}) and Eq.~(\ref{eq:Zbbleading}), provide a check of these full 
calculations, since the leading logarithmic contributions have to match. More generally, below we compute the top sector one-loop contribution to all Barbieri {\em et al.} \cite{Barbieri:2004cr, Chivukula:2004nx} electroweak parameters, and provide lower bounds on $M_q$ and $M_t$ from comparison with experiment.

\section{Constraints from Post-LEP Parameters}\label{sec:parameters}

In the language of Barbieri {\em et al.} \cite{Barbieri:2004cr,Chivukula:2004nx}, the observables $\hat{S}$, $\hat{T}$, $Y$, and $W$ parametrize the flavor-universal deviations from the SM at low energies. We now analyze the tree-level and the fermionic one-loop contributions to these parameters and use them to obtain constraints on the masses of the LW~states.

\subsection{Tree-Level Contributions}\label{treeBarbieri}

\begin{figure}\begin{center}
\includegraphics[width=\textwidth]{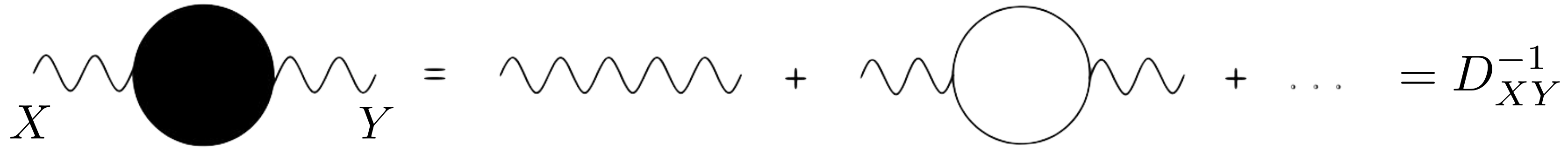}
\caption{The $XY$ VPA, defined as inverse of the full $XY$ propagator, $D_{XY}$ (see footnote~15 for details).}
\label{pLEP}
\end{center}\end{figure}

At tree-level, it is straightforward to read from Eq.~(\ref{eq:gaugelagrg}) the vacuum polarization amplitudes\footnote{The full Lorentz structure of a gauge boson's VPA contains a term proportional to $g^{\mu \nu}$ and another term proportional to the external momentum $q^{\mu}q^{\nu}$. For a massless gauge boson, these two coefficients are the same, and lead to a transverse VPA Lorentz structure. This is, however, generally not the case for a massive gauge boson, such as $W^{\pm}$ and $Z$, where the two coefficients are different. In \eqref{eq:twopoint}, $\Pi_{XY}(q^{2})$ represents the coefficient of $g^{\mu \nu}$ in the $XY$ VPA, since only this piece is relevant for our calculation. $B$ and $W^{3}$ are $U(1)_{Y}$ and the third component of $SU(2)_{L}$ group generators, respectively; within the context of the electroweak theory, they mix to produce massive $Z$ and massless photon.}
\begin{equation}\label{eq:twopoint} \begin{split}
&\Pi_{\hat{W}^+ \hat{W}^-}(q^2) = \frac{q^2}{g_2^2} -\frac{(q^2)^2}{g_2^2\,M_2^2}-\frac{v^2}{4} \ , \quad\Pi_{\hat{W}^3 \hat{W}^3}(q^2) = \frac{q^2}{g_2^2} -\frac{(q^2)^2}{g_2^2\,M_2^2}-\frac{v^2}{4} \ , \\
&\Pi_{\hat{W}^3 \hat{B}}(q^2) =  \frac{v^2}{4} \ , \qquad\qquad\qquad\qquad\quad\; \Pi_{\hat{B} \hat{B}}(q^2) = \frac{q^2}{g_1^2} -\frac{(q^2)^2}{g_1^2\,M_1^2}-\frac{v^2}{4} \ .
\end{split} \end{equation}
Following \cite{Barbieri:2004cr}, we see that there is no tree-level correction to the Fermi constant
\begin{equation}
\frac{1}{\sqrt{2}G_F} = - 4 \Pi_{\hat{W}^+ \hat{W}^-}(0) = v^2 \ .
\label{eq:GF}
\end{equation}
Barbieri {\em et al.} define the approximate electroweak gauge couplings
\begin{align}
\frac{1}{g^2} \equiv \ & \Pi^\prime_{\hat{W}^+ \hat{W}^-}(0) \ , \label{eq:g}\\
\frac{1}{g^{\prime 2}} \equiv \ & \Pi^\prime_{\hat{B} \hat{B}}(0) \ , \label{eq:ggprime} 
\end{align}
which in the LW~SM gives $g^{\prime}= g_1$ and $g = g_2$. We then compute the tree-level electroweak parameters~\cite{Underwood:2009ij},
\begin{align}
\hat{S} \equiv \ & g^2\,\Pi^\prime_{\hat{W}^3 \hat{B}}(0) = 0 \ ,  \label{eq:treelevelShat}\\
\hat{T} \equiv \ & g^2\left[\Pi_{\hat{W}^3 \hat{W}^3}(0)-\Pi_{\hat{W}^+ \hat{W}^-}(0)\right] = 0 \ , \\
Y \equiv \ & \frac{1}{2}g^{\prime 2}m_W^2\,\Pi^{\prime\prime}_{\hat{B} \hat{B}}(0)=-\frac{m_W^2}{M^{2}_1} \ , \label{eq:treelevelY} \\
W \equiv \ & \frac{1}{2}g^2m_W^2\,\Pi^{\prime\prime}_{\hat{W}^3 \hat{W}^3}(0)=-\frac{m_W^2}{M_2^2} \label{eq:treelevelW} \ ,
\end{align}
where in each equation the first equality is the definition of the corresponding post-LEP parameter \cite{Barbieri:2004cr}. The VPAs ($\Pi_{XY}$) are defined as inverse of the full $XY$ propagator (Fig.~\ref{pLEP}).

\subsection{Fermionic One-Loop Contributions}

\begin{figure}[!t]\begin{center}
\includegraphics[width=\textwidth]{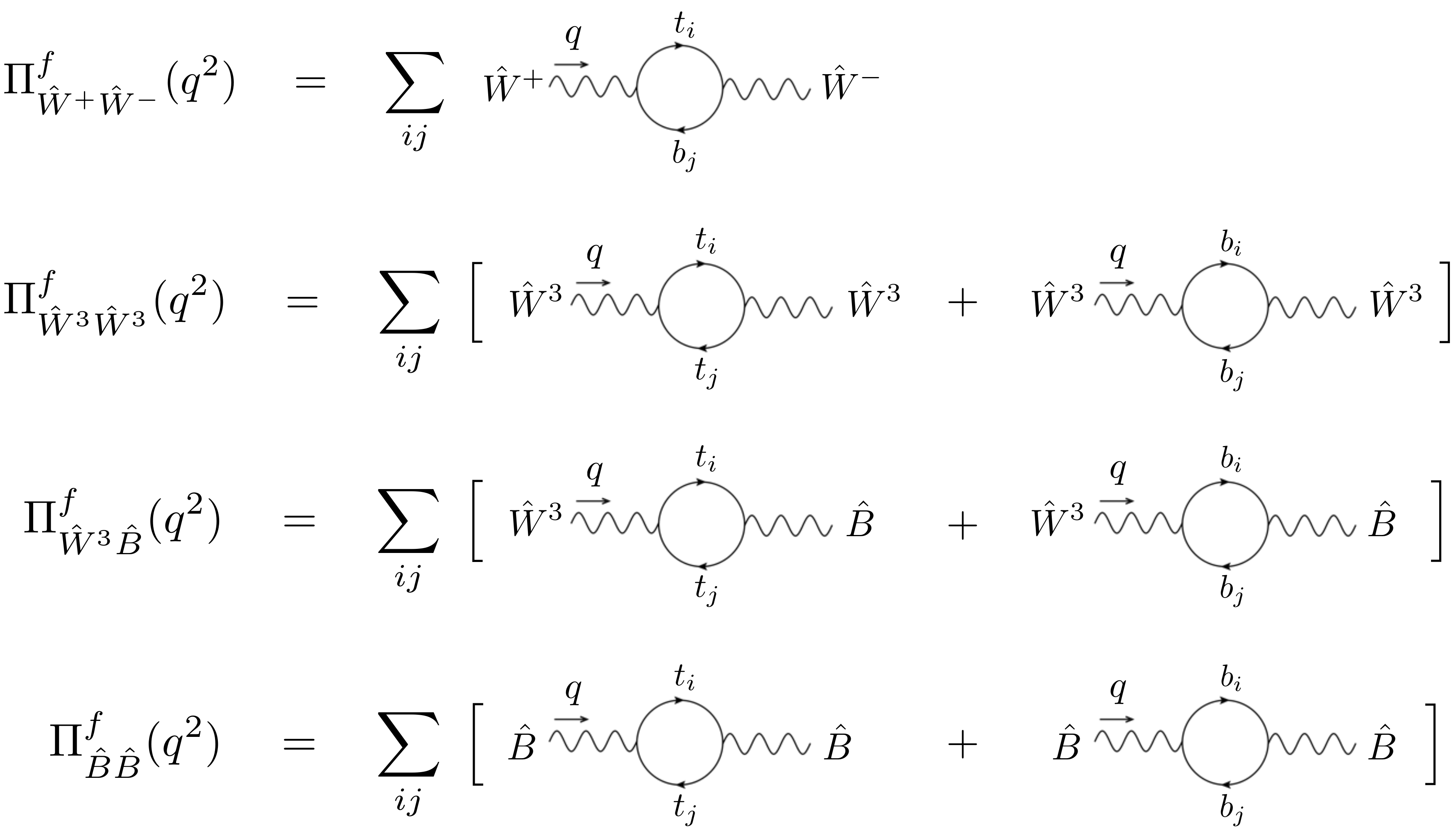}
\caption{Dominant vacuum polarization amplitudes for the LW~SM gauge fields. These include the ordinary ($t_0$ and $b_0$) and the LW~third generation quarks ($t_1$, $t_2$, $b_1$, and $b_2$) in the loop. These amplitudes contribute to the two-point functions of Eq.~(\ref{eq:twopoint}).}
\label{fig:loops}
\end{center}\end{figure}

The gauge current correlators receive important loop corrections from the top-bottom sector, through the diagrams shown in Fig.~{\ref{fig:loops}}. These vacuum polarization amplitudes contain two infinities, which are absorbed in the definitions of $g$ and $g^\prime$ given in Eqs.~(\ref{eq:g}) and (\ref{eq:ggprime}), respectively. 
As a consequence the non-canonical normalization adopted in Eq.~(\ref{eq:gaugelagrg}) forces us to define renormalized LW~gauge masses. A convenient scheme consists of defining $M$ and $M^\prime$ by
\begin{equation}
 -\,\frac{2}{g^2 M^2} \equiv \Pi^{\prime \prime}_{\hat{W}^+ \hat{W}^-}(0) \ , \qquad \quad -\,\frac{2}{g^2 M^{\prime 2}} \equiv \Pi^{\prime \prime}_{\hat{B}\hat{B}}(0) \ ,
\label{eq:newdef}
\end{equation}
which simplify the one-loop calculations below. At tree-level, from Eq.~(\ref{eq:twopoint}), we see that $M=M_2$ and $M^\prime=M_1$, and both are related to the masses of the LW~partners of the gauge bosons. Because of the power counting properties of LW~theories, after the usual\footnote{Notice that the vacuum polarization diagrams involving only one LW~fermion carry an overall negative sign. In fact this happens to make all zero-derivative functions, at $q^2=0$, finite. For this reason there is actually one less infinity compared to the ordinary SM, and the bare $v$ is finite \protect\cite{Grinstein:2008uk}.} coupling-constant and mass renormalizations, all physical quantities remain finite \cite{Grinstein:2008uk}. Hence, $M$ and $M^{\prime}$ remain finite at one-loop (and beyond). However, since they are defined by the zero-momentum properties of the gauge boson two-point functions, their values only approximately equal the masses of the LW~partners of the gauge bosons. This suffices for our purposes, since we are interested in low-energy observables; if we were studying quantities measured at higher energies, we would want to define $M$ and $M^{\prime}$ based on propagators renormalized at high $q^2$ instead.

The propagators in the loops of Fig.~\ref{fig:loops} correspond to mass eigenstates, where the masses are obtained by diagonalizing the mass matrices by means of symplectic rotations: in this way the LW~fields maintain the ``wrong-sign'' kinetic and mass terms. A perturbative diagonalization in $v^2/M_q^2$ and $v^2/M_t^2$~\cite{Alvarez:2008fk} requires considering two different scenarios: non-degenerate LW~masses, $|M_q^2-M_t^2|\sim M_q^2$, and (near) degenerate LW~masses, $|M_q^2-M_t^2|\ll M_q^2$. For non-degenerate LW~top quarks the contributions to the electroweak parameters are quite lengthy. To leading order we obtain
\begin{equation}\label{eq:nondeg} \begin{split}
\hat{S} = \ & -\frac{g^2 m_t^2}{48\pi^2 M_q^2}\Bigg[
\left(2+\frac{1}{r_t^2}\right)\log\frac{M_q^2}{m_t^2}
+\frac{1-3r_t^2+6r_t^4-r_t^6+3r_t^8}{r_t^2(1-r_t^2)^5}\log r_t^2 \\
&\quad - \frac{5-17r_t^2+4r_t^4+12r_t^6-23r_t^8+7r_t^{10}}{2r_t^2(1-r_t^2)^4}\Bigg] \ , \\
\hat{T} = \ & -\frac{3 g^2 m_t^4}{32\pi^2 m_W^2 M_q^2}\Bigg[
\left(2+\frac{1}{r_t^2}\right)\log\frac{M_q^2}{m_t^2}
+\frac{1-3r_t^2+6r_t^6}{r_t^2(1-r_t^2)^5}\log r_t^2 \\
&\quad - \frac{9-12r_t^2-21r_t^4+46r_t^6-68r_t^8+22r_t^{10}}{6r_t^2(1-r_t^2)^4}\Bigg] \ , \\
Y = \ & -\frac{m_W^2}{M^{\prime 2}} \ , \qquad \quad W =\ -\frac{m_W^2}{M^2}+\frac{g^2 m_W^2}{640\pi^2 M_q^2}\Bigg[-7+\frac{3}{r_b^2}-\frac{9}{r_t^2}\Bigg] \ ,
\end{split} \end{equation}
where
\begin{equation}
r_t\equiv M_t/M_q \ , \quad r_b \equiv M_b/M_q \ .
\end{equation}
The electroweak parameters in the (near) degenerate case cannot simply be obtained by taking the $r_t,r_b\to 1$ limit in Eqs.~(\ref{eq:nondeg}), since the corresponding expressions diverge. Instead, we must diagonalize the mass matrices perturbatively in $1/M_q^2$ (or $1/M_t^2$) and $|M_q^2-M_t^2|/M_q^2$, and then compute the electroweak parameters. For exact degeneracy, $M_q=M_t$, this gives
\begin{equation}\label{eq:deg} \begin{split}
\hat{S} = \ & -\frac{g^2 m_t^2}{16\pi^2 M_q^2}\Bigg[\log\frac{M_q^2}{m_t^2}-\frac{12}{5}\Bigg] \ , \\
\hat{T} = \ & -\frac{3 g^2 m_t^4}{32\pi^2 m_W^2 M_q^2}\Bigg[3\log\frac{M_q^2}{m_t^2}-\frac{141}{20}\Bigg] \ , \\
Y = \ & -\frac{m_W^2}{M^{\prime 2}} \ , \\
W = \ & -\frac{m_W^2}{M^2}-\frac{7 g^2 m_W^2}{640\pi^2 M_q^2} \ .
\end{split} \end{equation}
Note that the absence of fermionic one-loop corrections to the tree-level value of $Y$ is a direct consequence of the second definition in Eq.~(\ref{eq:newdef}): a different scheme choice would lead to an additional contribution. In the same way, changing the definition of $M$ would lead to a different fermionic one-loop expression for $W$; in any case, the second term\footnote{Note that the first definition in Eq.~(\protect\ref{eq:newdef}) pertains to
$\Pi^{\prime \prime}_{\hat{W}^+\hat{W}^-}$, whereas $W$ is defined in terms of $\Pi^{\prime \prime}_{\hat{W}^3 \hat{W}^3}$.}
 in $W$ is numerically very small and can be neglected. We, therefore, conclude that the leading contributions to $Y$ and $W$ are those arising from the LW~gauge-sector at tree-level, Eqs.~(\ref{eq:treelevelY}, \ref{eq:treelevelW}).
 
Since the tree-level values of $\hat{S}$ and $\hat{T}$ vanish, the leading LW~contributions to both $\hat{S}$ and $\hat{T}$ arise from the top quark sector at one-loop. In the case of $\hat{T}$ this is not surprising since the dominant locus of isospin violation in the model is the splitting between the top and bottom quark masses. Because $\hat{T}$ is the same as $\Delta\rho$~\cite{Chivukula:2004nx}, we may compare the leading logarithmic correction in Eq.~(\ref{eq:nondeg}) with the result obtained in the effective theory, Eq.~(\ref{eq:rholeading}); we see that they agree. In the case of $\hat{S}$, the situation is more subtle. The LW~gauge-eigenstate fermion partners, being massive, are not chiral and therefore, in the absence of electroweak contributions to the masses that mix them with the light chiral gauge-eigenstates, their contribution to $\hat{S}$ vanishes. Hence, the dominant LW~contributions to $\hat{S}$ also arise predominantly from the top sector of the theory.

Therefore, at tree-level plus one fermion loop we obtain a very simple conclusion: the fermion sector contributes to $\hat{S}$ and $\hat{T}$ only, while the gauge sector contributes to $Y$ and $W$ only.  It is true that when gauge loops are included, there will be additional contributions.  However, the gauge loop contributions are generally sub-dominant compared to the quantities we have already calculated. The only potential exception is $\hat{S}$, for which the fermionic one-loop contribution is small. However, in Ref.~\cite{Carone:2009kl} a numerical computation shows that the gauge loop contribution to $S$ is suppressed (see also \cite{Alvarez:2008fk}), since $| S_\text{tree} - S_\text{loop} | \lesssim 0.01$. Using the results of Ref.~\cite{Chivukula:2004nx}, this allows us to estimate the gauge loop contribution to $\hat{S}$ to be $10^3 \hat{S} \lesssim 0.1$, which is negligibly small. Thus, our existing results suffice for extracting constraints on the LW~fermions from the experimental data.

\subsection{Comparison with Data}

\begin{figure}[t!]\begin{center}
\includegraphics[width=.9\textwidth]{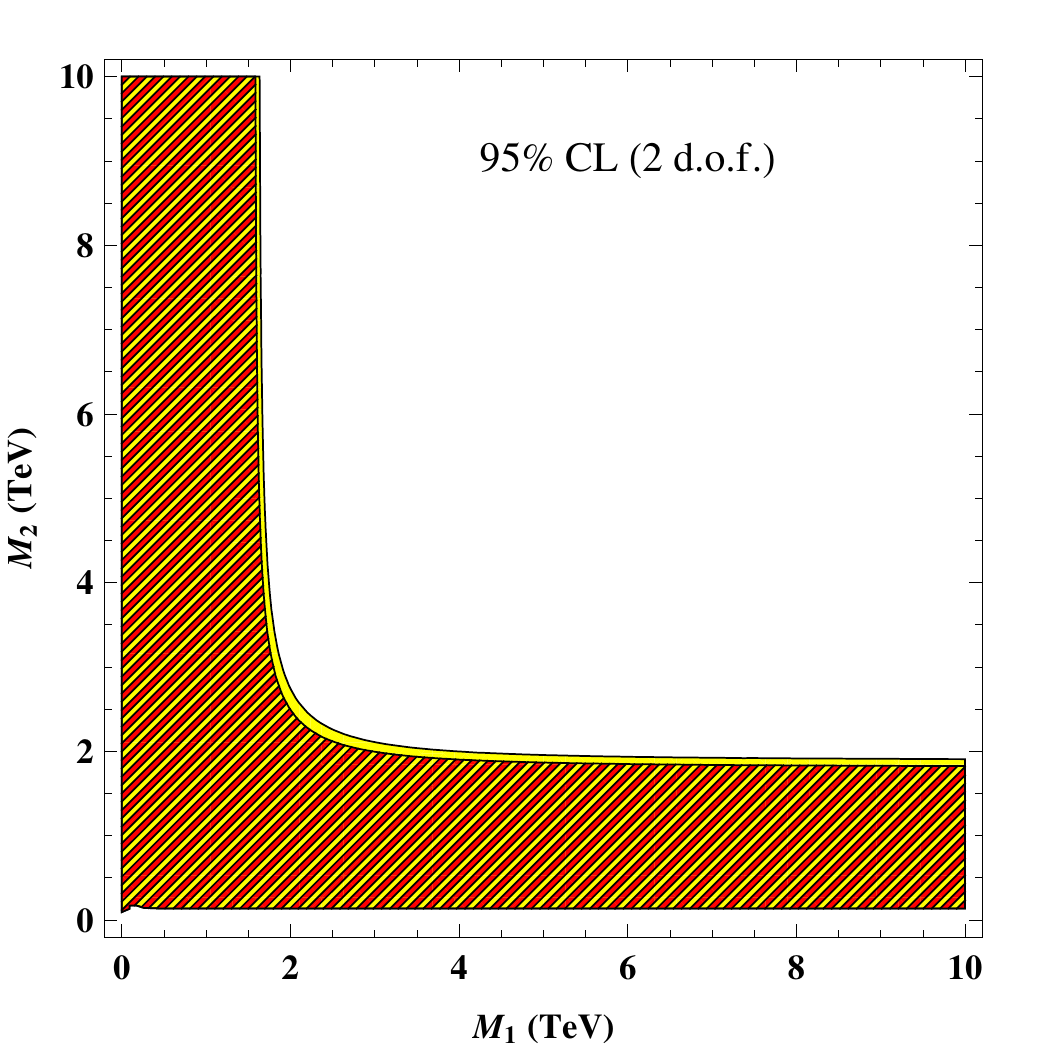}
\caption{Exclusion plot for the LW~gauge-field masses $M_2$ and $M_1$. These bounds are due to the constraints on $Y$ and $W$, as shown by Eq.~(\ref{eq:treelevelY}) and Eq.~(\ref{eq:treelevelW}). For a light Higgs ($m_h=115$ GeV) the striped region to the left of both curves is excluded. For a heavy Higgs ($m_h=800$ GeV) the additional yellow strip between the curves is excluded as well.}
\label{fig:gauge1}
\end{center}\end{figure}

\begin{figure}[t!]\begin{center}
\includegraphics[width=.9\textwidth]{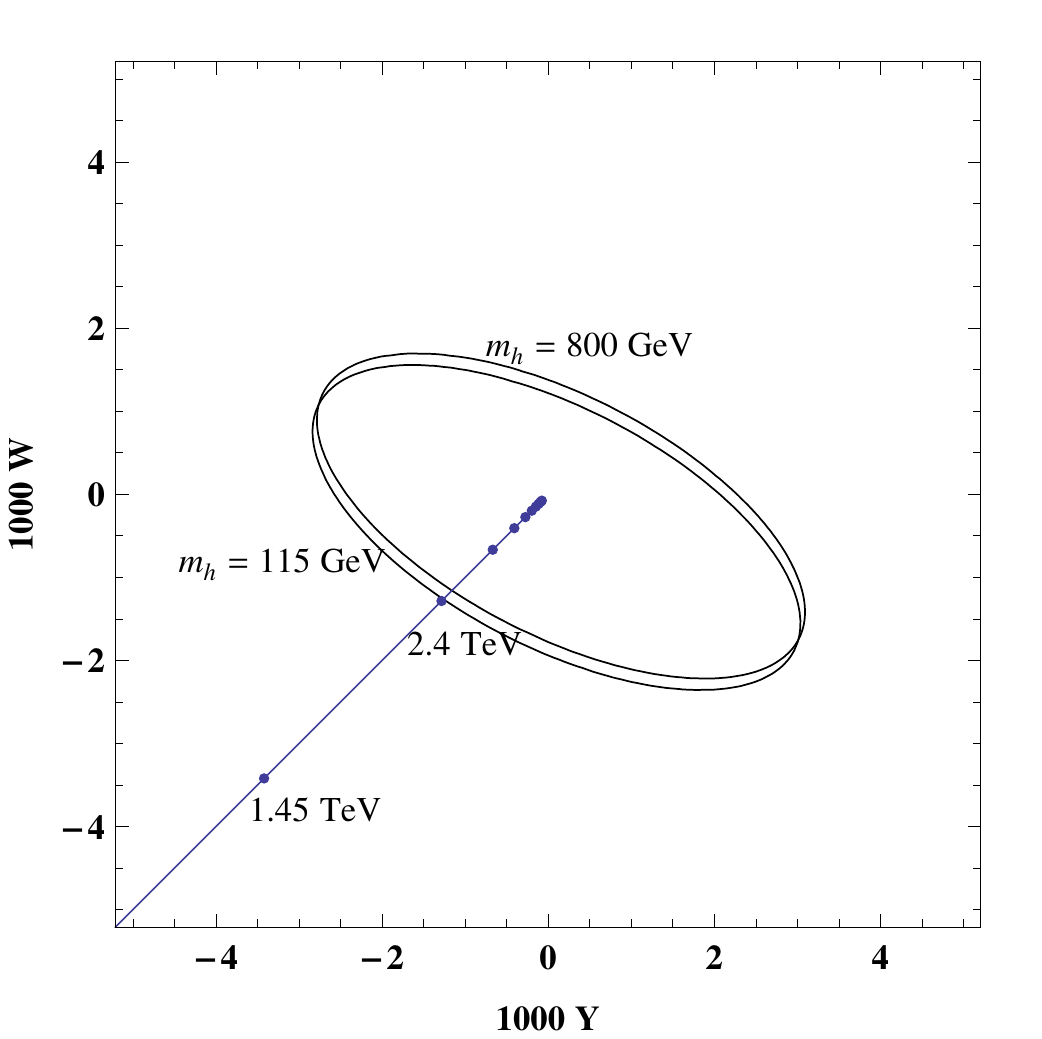}
\caption{The 95\% C.L. ellipses in the ($Y$,$W$) plane, and the LW~prediction for degenerate masses, $M_1=M_2$. The parametric plot is for  $0.5\,\text{TeV} < M_1 = M_2 < 10\,\text{TeV}$, and the dots are equally spaced in mass every 0.95TeV (only two dots labeled). The lower bound on $M_1 = M_2$ is approximately 2.4 TeV for a light Higgs.}
\label{fig:gauge2}
\end{center}\end{figure}

We begin with constraints on the masses of the LW~partners of the gauge bosons.  The previous subsection found that  the only post-LEP parameters affected by the LW~gauge boson masses are $W$ and $Y$, and also that the tree-level expressions for $W$ and $Y$, Eqs.~(\ref{eq:treelevelY}, \ref{eq:treelevelW}), suffice for comparison with data.   The experimental constraints on $Y$ and $W$ are rather tight and almost independent of the value of the Higgs mass~\cite{Barbieri:2004cr}. These translate into the 95\% C.L. lower bounds on $M_2$ and $M_1$ shown\footnote{These bounds are derived using the errors and correlation matrix given in Ref.~\cite{Barbieri:2004cr}.} in Figs.~\ref{fig:gauge1} and \ref{fig:gauge2}. Fig.~\ref{fig:gauge1} shows the bounds for arbitrary values of $M_1$ and $M_2$: for $m_h=115$ GeV the striped region is excluded, while for $m_h=800$ GeV the additional narrow yellow region is excluded as well. Fig.~\ref{fig:gauge2} shows the 95\% C.L. ellipses in the $(Y,W)$ plane from the global fit to data~\cite{Barbieri:2004cr}, for $m_h=115$ and $m_h=800$, as well as the LW~prediction for degenerate LW~masses, $M_1=M_2$. All this is in agreement with the results of Ref.~\cite{Underwood:2009ij} and gives the constraints $M_1$, $M_2\,\gtrsim 2.4$ TeV. 

Next, we seek constraints on the masses of the LW~partners of the top quark.  The previous subsection found that the post-LEP parameters sensitive to the LW~fermion masses are $\hat{S}$ and $\hat{T}$, which do not depend on the LW~gauge masses at the one-loop level. We should also note that, for a light Higgs, the LW~prediction of $\hat{S}$ is very close to its central value, $\hat{S}\simeq0$. Furthermore, from the global fit to the experimental data in Ref.~\cite{Barbieri:2004cr}, we conclude that $\hat{T}$ is only mildly correlated to $Y$ and $W$, the parameters that are most sensitive to the LW~gauge boson masses in the LW~SM. This confirms that the bounds on the LW~fermions should be essentially independent of the LW~gauge masses, and should come almost entirely from $\hat{T}$. 

\begin{figure}[!t]
\begin{center}
\includegraphics[width=\textwidth]{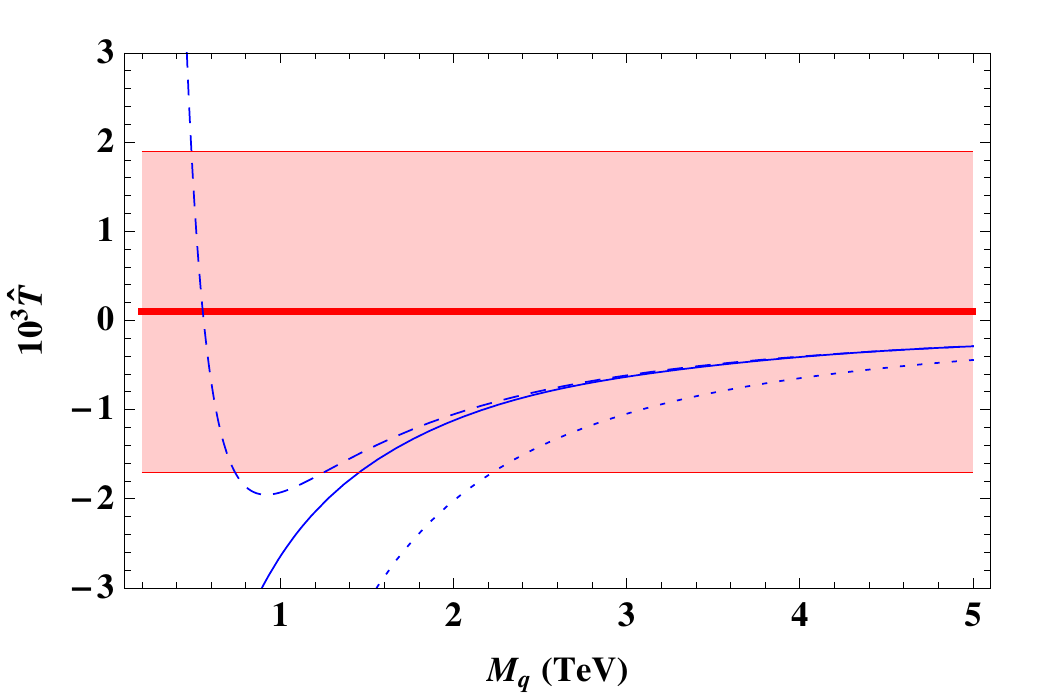}
\caption{$\hat{T}$ as a function of $M_q$ in the degenerate case, $M_q=M_t$. 
The experimental mean value for $\hat{T}$ is shown by the thick red line, along with
the $\pm 2 \sigma$ allowed region. Also shown are the all-order (in $v^2/M_q^2$) LW~prediction (solid blue curve), the leading order LW~prediction, Eq.~(\protect\ref{eq:deg}) (dashed blue curve), and the leading-log curve, Eq.~(\protect\ref{eq:rholeading})  (dotted blue curve), as functions of $M_q$, in the degenerate case.  Note that the leading order prediction is not valid below $M_q \sim 1$ TeV. (See text for details)}
\label{fig:That}
\end{center}
\end{figure}

In Fig.~\ref{fig:That} we show the experimental mean value for $\hat{T}$ (thick red line), the $\pm 2 \sigma$ allowed region, the all-order (in $v^2/M_q^2$) LW~prediction (solid blue curve), the leading order LW~prediction from Eq.~(\ref{eq:deg}) (dashed blue curve), and the leading-log approximation (dotted blue curve), as functions of $M_q$, in the degenerate case. This figure reveals the bound $M_q=M_t\gtrsim 1.6$ TeV on the LW~fermion masses in the degenerate case. Note that although Eq.~(\ref{eq:deg}) appears to predict a positive $\hat{T}$ for small $M_q$ (dashed blue curve), the complete numerical evaluation (solid blue curve) shows that $\hat{T}$ is always negative, as Fig.~\ref{fig:That} shows explicitly; below $M_q=1$ TeV the perturbative diagonalization of the mass matrix is no longer valid, rendering the leading order LW~prediction unreliable in that mass regime.

\begin{figure}[t!]\begin{center}
\includegraphics[width=.9\textwidth]{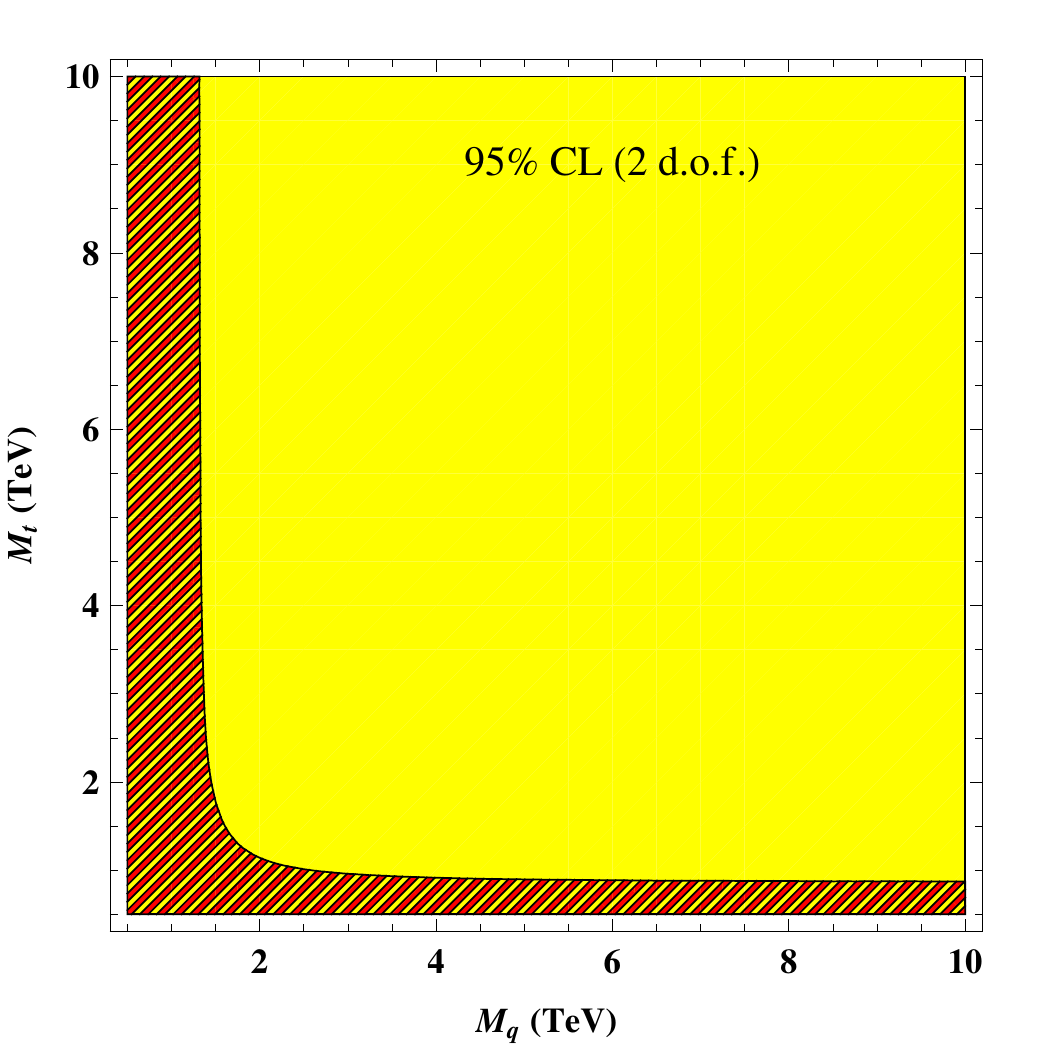}
\caption{The 95\% C.L. exclusion plots for the LW~fermion masses $M_q$ and $M_t$. These bounds come almost entirely from the experimental constraints on $\hat{T}$. For a light Higgs the striped region to the left of the curve is excluded, while a heavy Higgs is completely excluded.}
\label{fig:fermionbounds1}
\end{center}\end{figure}

If we relax the requirement of degenerate LW~fermion masses, we obtain the 95\% C.L. bounds on $M_q$ and $M_t$ shown in Fig.~\ref{fig:fermionbounds1}. For a light Higgs the striped region in Fig.~\ref{fig:fermionbounds1} is excluded, while for a heavy Higgs the whole (yellow) region is excluded.  Note from Figs.~\ref{fig:gauge1} and \ref{fig:fermionbounds1} that the mildest constraints on the LW~masses are obtained in the fully degenerate case, $M=M^\prime$ and $M_q=M_t$. 

\begin{figure}[t!]\begin{center}
\includegraphics[width=.9\textwidth]{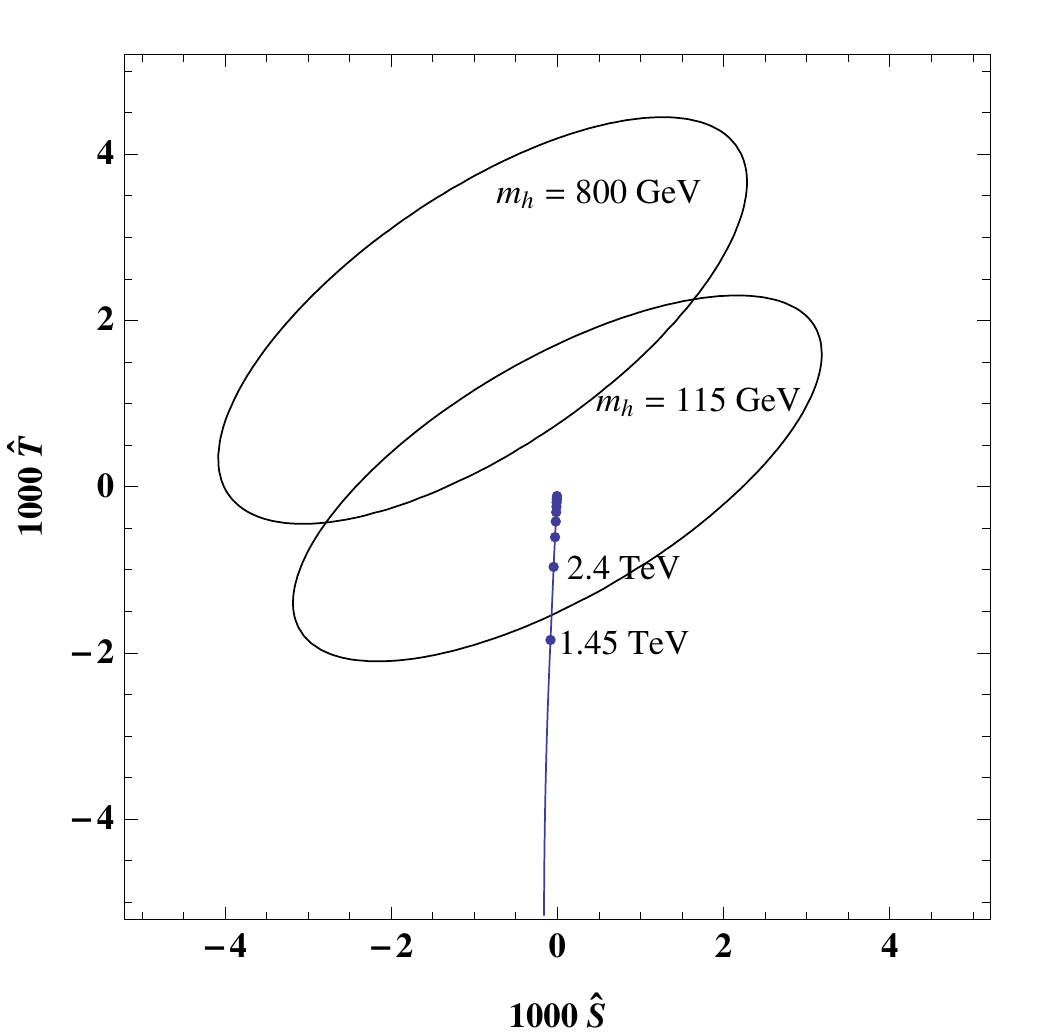}
\caption{The 95\% C.L. ellipses in the ($\hat{S}$,$\hat{T}$) plane, and the LW~prediction for degenerate masses, $M_q=M_t$. The parametric plot is for  $0.5\,\text{TeV} < M_q < 10\,\text{TeV}$ and the dots are equally spaced in mass every 0.95TeV (only two dots labeled). The lower bound on $M_q$ is approximately 1.6 TeV for a light Higgs.}
\label{fig:fermionbounds2}
\end{center}\end{figure}

Returning to the case of degenerate LW~fermion masses, we show in Fig.~\ref{fig:fermionbounds2} the values of $\hat{S}$ and $\hat{T}$ as a function of $M_q=M_t$  for $0.5\,\text{TeV} < M_q < 10\,\text{TeV}$; the dots representing different values of $M_q$ are placed at regular intervals. The 95\% C.L. ellipses from the global fit to the data~\cite{Barbieri:2004cr} confirm the constraint $M_q\gtrsim 1.6$ TeV for a light Higgs, while a heavy Higgs scenario is disfavored for any LW~fermion mass. In fact, for a heavy Higgs the $\hat{T}$ parameter is expected to be positive, while the LW~SM predicts a negative $\hat{T}$. This is a direct consequence of the negative sign in the LW~fermion propagators, which results in an overall negative sign from the (dominant) diagrams involving a single LW~fermion in the loop.   

Our results disagree with those of \cite{Alvarez:2008fk, Carone:2008oq} in two ways: their bounds on the LW~fermion masses appear more stringent for a light Higgs and their limits appear to depend on the masses of the LW~gauge boson partners.  The disagreement arises because their study of one-loop electroweak corrections in the LW~SM assumes the corrections  to be purely oblique and derives constraints by comparing the Peskin-Takeuchi $S$ and $T$~\cite{Peskin:1992kl} parameters to data. However, as clearly discussed in Ref.~\cite{Underwood:2009ij}, and confirmed above in Eqs.~(\ref{eq:treelevelY}) and (\ref{eq:treelevelW}), the LW~SM features large non-oblique corrections, in the form of  non-zero values for $Y$ and $W$ at tree-level.  Hence, one must use the Barbieri {\em et al.}  parameters to compare the LW~SM with experiment, as we have done.

\section{Constraints from the $Z b_L \bar{b}_L$ Coupling}\label{sec:Zbb}

\begin{figure}[!t]\begin{center}
\includegraphics[width=0.85\textwidth]{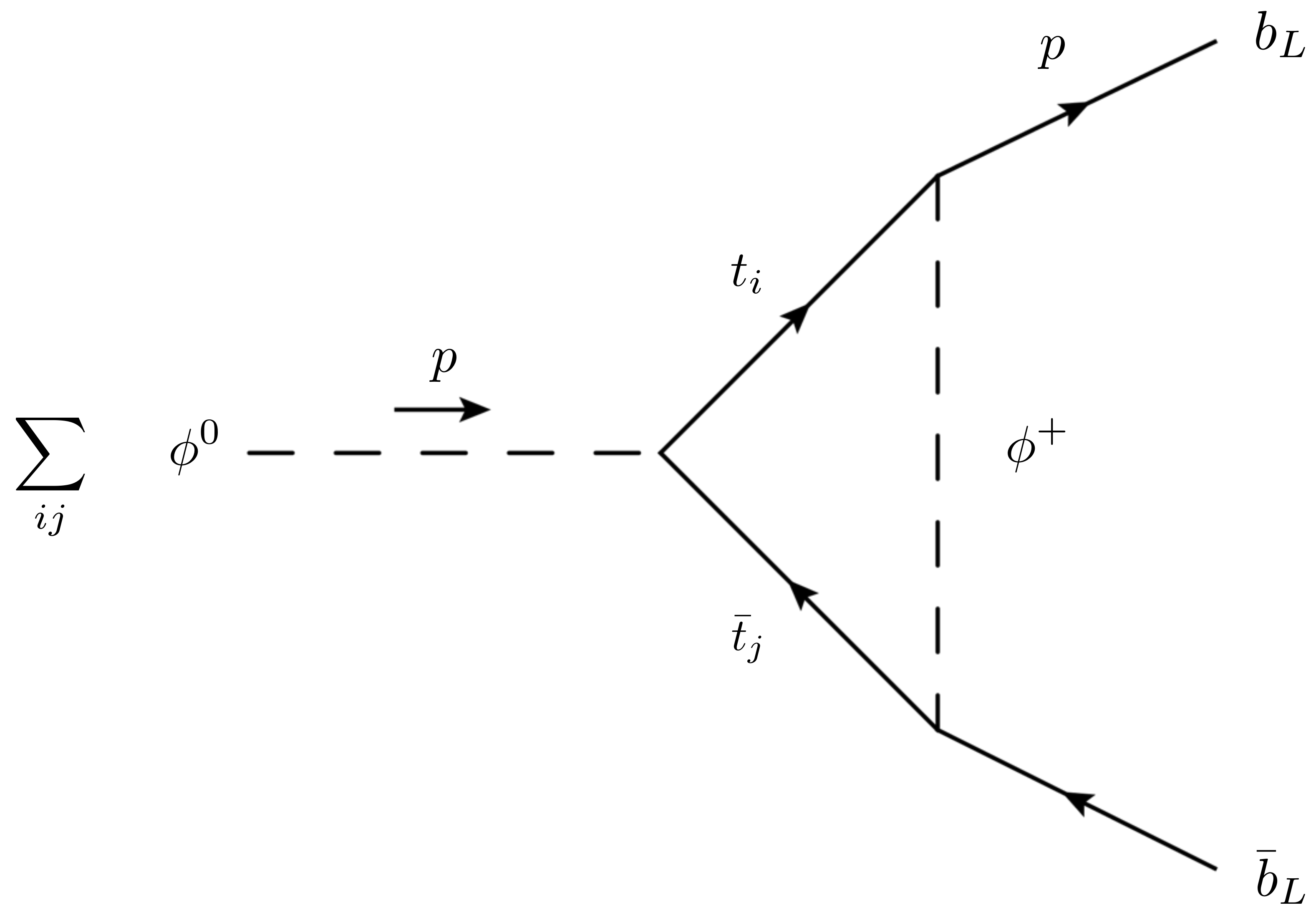}
\caption{Diagram giving the largest contribution to the $Z b_L \bar{b}_L$ coupling. The latter is related through the Ward identity to the $\phi^0 b_L \bar{b}_L$ coupling | where $\phi^0$ is the Nambu-Goldstone boson eaten by the $Z$ boson. The top quarks running in the loop are both ordinary and LW.}
\label{fig:Zbb}
\end{center}\end{figure}

The leading contribution to the $Z b_L \bar{b}_L$ coupling (in the gauge coupling expansion) can be obtained in the gaugeless limit from the  $\phi^0 b_L \bar{b}_L$ coupling~\cite{Barbieri:1992ff}\nocite{Barbieri:1993lh}\nocite{Oliver:2003fu}-\cite{Abe:2009ye}, where $\phi^0$ is the Nambu-Goldstone boson eaten by the $Z$. The loop diagram giving the largest correction involves the SM and LW~top quarks\footnote{In the gaugeless limit of the LW~SM, as in the SM itself, all external $b$-quark wavefunction renormalization corrections are proportional to $y_b^2$ and are, therefore, negligible. This should be contrasted with the situation in the three-site Higgsless model~\cite{Abe:2009ye}.} and is shown in Fig.~\ref{fig:Zbb}. A detailed computation of the loop integral, valid for arbitrary models with heavy replicas of the top quark, is given in Appendix~\ref{phibb}. At zero external momentum the amplitude corresponding to the diagram has the form
\begin{equation}
iM = - A\, \slashed{p} P_L \ ,
\label{eq:MZbb}
\end{equation}
where $P_L\equiv (1-\gamma^5)/2$ is the left-handed projector, $p$ is the incoming $\phi^0$ momentum, and the external fermion wavefunctions have been omitted. Then to leading order in $g$ the correction to the $Z b_L \bar{b}_L$ coupling is~\cite{Barbieri:1992ff}\nocite{Barbieri:1993lh}\nocite{Oliver:2003fu}-\cite{Abe:2009ye}
\begin{equation}
\delta g_L^{b\bar{b}} = \frac{v}{2} A \ . 
\label{eq:generalZBB}
\end{equation}

Expanding the amplitude in powers of $m_t^2/M_q^2$, we obtain
\begin{equation}\label{eq:Zbbnondeg} \begin{split}
(\delta g_L^{b\bar{b}})_\text{LW} = \ & -\frac{m_t^4}{32\pi^2 v^2 M_q^2}\Bigg[\left(\frac{4}{r_t^2}+1\right)\log\frac{M_q^2}{m_t^2} \\
&\quad + \frac{4-11 r_t^2+9 r_t^4}{r_t^2(1-r_t^2)^3}\log r_t^2 -\frac{6-10 r_t^2 + 2r_t^4}{r_t^2(1-r_t^2)^2}\Bigg] \ ,
\end{split} \end{equation}
for non-degenerate LW~fermion masses, and
\begin{equation}
(\delta g_L^{b\bar{b}})_\text{LW}=-\frac{m_t^4}{32\pi^2 v^2 M_q^2}\Bigg[5\log\frac{M_q^2}{m_t^2}-\frac{49}{6}\Bigg] \ ,
\label{eq:Zbbdeg}
\end{equation}
for degenerate LW~masses. Both of these expressions agree with the dominant contribution found in the effective theory, Eq.~(\ref{eq:Zbbleading}).  

\begin{figure}[!t]\begin{center}
\includegraphics[width=\textwidth]{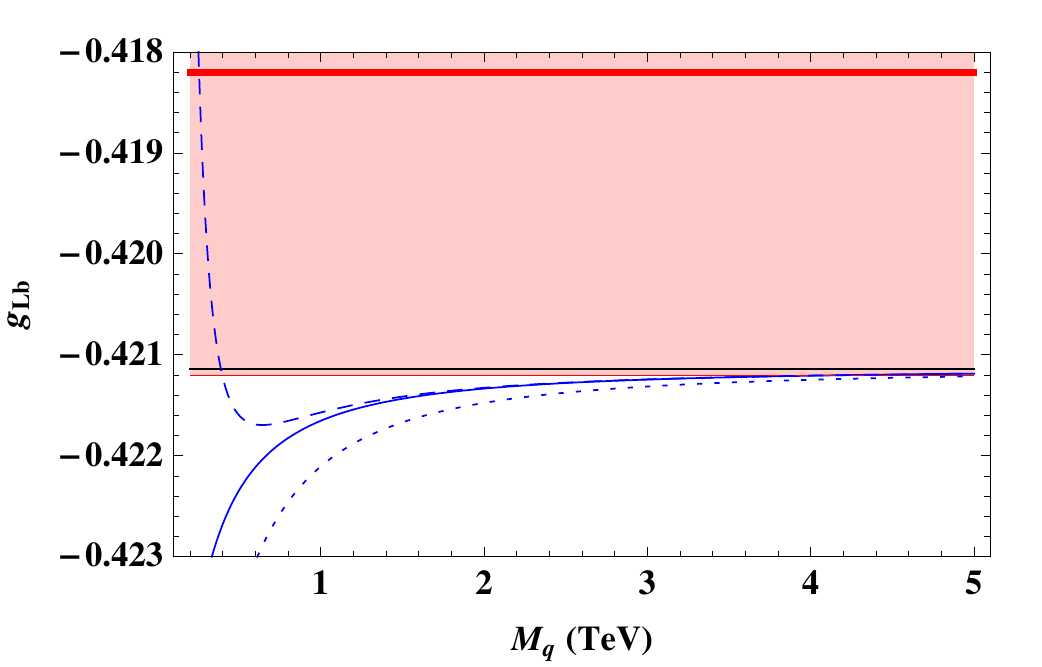}
\caption{Constraints from the $Z b_L \bar{b}_L$ coupling. This graph features the experimental mean value (thick horizontal red line), the 2$\sigma$ allowed region below the mean value, the SM prediction (solid horizontal black line), the all-order (in $v^2/M_q^2$) LW~prediction (solid blue curve), the leading order LW~prediction, Eq.~(\ref{eq:Zbbdeg}) (dashed blue curve), and the leading-log approximation (dotted blue curve), as functions of $M_q$, in the degenerate case.}
\label{fig:Zbbexclusion}
\end{center}\end{figure}

The experimental value of $g_L^{b\bar{b}}$ is derived from measurements of $R_b$, the ratio of the $Z\to b \bar{b}$ width to the width of the hadronic decays, and $A_b$, the forward-backward asymmetry for $Z$ decays into $b\bar{b}$~\cite{ALEPH:2006il}
\begin{equation}
(g_L^{b\bar{b}})_\text{exp}=-0.4182\pm 0.0015 \ .
\end{equation}
The SM value was computed using ZFITTER~\cite{Bardin:2001qo,Arbuzov:2006tw} in Ref.~\cite{Chivukula:2009dq}, leading to 
\begin{equation}
(g_L^{b\bar{b}})_\text{SM}=-\frac{1}{2}  + \frac{1}{3}\sin ^2 \theta_W 
+ (\delta g_L^{b\bar{b}})_\text{SM}=-0.42114 \ ,
\end{equation}
while the LW~prediction is given by Eq.~(\ref{eq:Zbbnondeg}) and Eq.~(\ref{eq:Zbbdeg}). In Fig.~\ref{fig:Zbbexclusion} we show the experimental mean value (thick horizontal red line), the 2$\sigma$ allowed region below the mean value, the SM prediction (solid horizontal black line), the all-order (in $v^2/M_q^2$) LW~prediction (solid blue curve), the leading order LW~prediction, Eq.~(\ref{eq:Zbbdeg}) (dashed blue curve), and the leading-log approximation (dotted blue curve), as functions of $M_q$, in the degenerate case. Note that the dashed curve and Eq.~(\ref{eq:Zbbdeg}) are not reliable for $M_q\lesssim 1$ TeV, because the perturbative diagonalization of the mass matrix is not valid in that mass regime.

As anticipated by the effective field theory calculation, the LW~correction is always negative: this is essentially due to the negative sign in front of the dominant non-standard triangle diagrams with one LW~top and one SM top. It is large (for small values of $M_q$) because of the explicit breaking of custodial isospin symmetry. Since the SM value is already $1.96 \sigma$ below the experimental mean value, this correction goes in the direction opposite to what is needed. Agreement at the 2$\sigma$ level requires $M_q\gtrsim 4$ TeV; at $2.5 \sigma$ this bound relaxes to $M_q\gtrsim 700$ GeV.

\section{Conclusion \& Discussion}

There is significant tension between naturalness and isospin violation in the LW~SM. While corrections to the Higgs mass are smallest when the LW~partners of the gauge bosons and fermions are light, isospin violation that must be present in the top sector to account for the large splitting between $m_t$ and $m_b$ tends to constrain the LW~partners to have masses over a TeV. We have performed an effective field theory analysis of the corrections to $\hat{T}$ and the $Zb_L\bar{b}_L$ coupling in the LW~SM, and used it to confirm our full calculation of the LW~effects on $\hat{S}$, $\hat{T}$, $W$, $Y$, and $g_L^{b\bar{b}}$, including tree-level and fermionic one-loop contributions. The post-LEP parameters yield a simple picture in the LW~SM: the gauge sector contributes to $Y$ and $W$ only, with leading contributions arising at tree-level, while the fermion sector contributes to $\hat{S}$ and $\hat{T}$ only, with leading corrections arising at one-loop.

In agreement with \cite{Underwood:2009ij}, we find that experimental limits on $W$ and $Y$ jointly constrain the masses of the LW~gauge bosons to satisfy $M_1, M_2 \gtrsim 2.4$ TeV at 95\% C.L., with relatively little sensitivity to the Higgs mass.

We also conclude that the experimental limits on $\hat{T}$ require the masses of the LW~fermions to satisfy $M_q, M_t \gtrsim 1.6$ TeV at 95\% C.L. if the Higgs mass is light and tend to exclude the LW~SM for any LW~fermion masses if the Higgs mass is heavy. This is because a model containing a heavy Higgs can be rendered consistent with the data only if some other sector of the model provides a large positive correction to $\hat{T}$. However, in the LW~SM, the fermionic loops that provide the dominant contribution to $\hat{T}$ always make $\hat{T}$ more negative, due to the negative sign in the LW~fermion propagators. The LW~fermions simply cannot compensate for the presence of a heavy Higgs. Our results differ from those in Refs.~\cite{Alvarez:2008fk, Carone:2008oq} because their analysis incorrectly assumes that the electroweak corrections due to LW~states are purely oblique. As explained in Ref.~\cite{Underwood:2009ij} the LW~states actually induce important non-oblique corrections, and one must, therefore, use the Barbieri {\em et al.} \cite{Barbieri:2004cr, Chivukula:2004nx} post-LEP parameters to compare the LW~SM with experiment, as we have done.

Weak isospin violation in the top quark sector also manifests itself through corrections to the $Zb_L\bar{b}_L$ coupling. The SM prediction for $g_L^{b\bar{b}}$ lies at the lower end of the range allowed by experiment at 95\% C.L., so that new physics making negative contributions to the value of $g_L^{b\bar{b}}$ would decrease the agreement with the data. As in the case of $\hat{T}$, however, we find that the negative sign in the LW~fermion propagators translates into a negative contribution to $g_L^{b\bar{b}}$; the lighter the LW~fermions, the greater the disagreement between prediction and data. We find that the $Zb_L\bar{b}_L$ coupling places a lower bound of 4 TeV on the LW~fermion masses at 95\% C.L.

On this note, we conclude our phenomenological analysis of the LW~SM, reflecting various constraints from the current experimental data. Having completed our phenomenological study of the LW~SM, we are going to focus, in the next chapter, on some of the theoretical aspects concerning the global symmetries and renormalizability of LW~theories, since they are in general poorly understood. Understanding renormalizability of this class of theories, in particular, is of great importance in order for them to be considered as reliable beyond the Standard Model (BSM) candidates with accurate all-order predictions for the experiments. We will examine these issues in both the higher-derivative and auxiliary-field formulations of the theory by means of a relatively simple Abelian example.

\chapter{GLOBAL SYMMETRIES AND RENORMALIZABILITY OF LEE-WICK THEORIES\protect \footnote{This chapter is based on the paper first published in \cite{Chivukula:2010oj}.}}\label{symmetry}

\begin{quote}
``\textit{Science is a way of trying not to fool yourself. The first principle is that you must not fool yourself | and you are the easiest person to fool!}"
\begin{flushright}
|Richard Feynman (1918 -- 1988)\\
\end{flushright}
\end{quote}

\section{Introduction}

\lettrine[lines=1]{R}{enormalizabilty of Lee-Wick~theories} (LW) is an important theoretical aspect which needs to be well-stablished, in order for them to be considered as reliable beyond the Standard Model (BSM) candidates. In Chapters~1 and 2, we explained how the LW~theory may be formulated in terms of two separate but equivalent formal descriptions. Depending on the problem at hand, one may choose to utilize either of the two formulations, in order to facilitate the calculations and make certain formal aspects of the theory more transparent. Renormalizability has been previously explored in the higher-derivative formulation of the theory (see below), but its precise theoretical description in the auxiliary-field formulation remained unknown. In this chapter, we address this nontrivial issue along with global symmetries possessed by LW~theories in the auxiliary-field formulation.

To summarize our discussion so far, we saw that in the higher-derivative formulation, the higher-derivative kinetic term gave rise to propagators that fall off with momentum more rapidly than the ordinary Standard Model~(SM) field propagators, thereby reducing the degree of divergence of loop diagrams. On the other hand, higher covariant derivatives also introduce new momentum-dependent interactions, which raise the degree of divergence of quantum fluctuations. Power counting arguments \cite{Grinstein:2008uk} show that these two competing effects conspire to make all loop diagrams at most logarithmically divergent. If the scale associated with the higher-derivative terms is of the order of the electroweak scale, then the latter becomes stable against radiative corrections: no quadratic divergences are present at any order in perturbation theory, and no unnatural fine-tuning of parameters is required. In addition, power counting arguments \cite{Grinstein:2008uk} show that this higher-derivative formulation of the theory is renormalizable.

The higher-derivative kinetic terms in the Lee-Wick Standard Model~(LW~SM) result in propagators with more than one pole. In minimal (so-called $N = 2$) LW~theories\footnote{We will focus on $N = 2$ theories throughout this Thesis, though our results can be potentially generalized to arbitrary-$N$ LW~theories, with $N-1$ higher-derivative kinetic terms associated with each field.} \cite{Carone:2009qa}, there is only one higher-derivative kinetic term for each field, corresponding to two-pole propagators. In the $N = 2$ LW~SM, the lighter pole can be identified as a SM-like state already seen in experiment, while the heavier pole corresponds to a new LW~ghost state with the same quantum numbers and negative norm. As explained in Chapter~1, this is dangerous, since it would lead to a violation of unitarity. In order to avoid this scenario, the LW~ghosts must appear only as virtual states \cite{Lee:1969fe}. Furthermore, the integration contour in momentum integrals involving ghost propagators must be modified so as to preserve unitarity \cite{Cutkosky:1969bs}. The price to be paid for these modifications is the presence of unobservable acausal effects in scattering processes \cite{Lee:1969fe, Grinstein:2009fv}.

Alternatively, in the auxiliary-field form of the theory, in addition to the SM fields there are ``LW~fields" with kinetic energy terms with the opposite sign from their SM counterparts. The opposite sign for the kinetic energy terms enforces the cancellations that soften the divergences in the theory. The main advantage of the auxiliary-field approach lies in the computation of loop diagrams, since aside from the overall sign the propagators are just ordinary propagators. There are, however, a number of open field-theoretic issues with this formulation, which require clarification.

In this chapter\footnote{Throughout this chapter, the timeline for the depicted Feynman diagrams is from left to right.} we clarify two issues in the auxiliary-field description of the theory in the context of a simple but nontrivial theory | LW~scalar quantum electrodynamics (QED).\footnote{Our analysis extends immediately to LW~QED with an arbitrary number of matter fields, either scalars or fermions.} First, the interaction terms involving the LW~fields have a very particular form, which is not the most general one allowed by gauge invariance. For example, the couplings of the LW~vector fields are identical to the gauge couplings of the corresponding SM gauge fields. This equality and others are essential if the cancellations softening or removing the infinities are to hold. On the one hand, it is not clear why, \textit{a priori}, this special form of the interactions should be preserved to all orders in perturbation theory. On the other hand, we know that it must be preserved since power counting shows that the equivalent higher-derivative theory is free of quadratic divergences.\footnote{In the auxiliary-field formulation power counting is more difficult because of the cancellations involved between different diagrams.} Here we identify approximate $SO(1,1)$ global symmetries of the auxiliary-field description of the theory that allow us to understand its structure.

Second, we clarify the renormalizability of LW~scalar QED in the presence of the massive ghost LW~vector field. Because of the $q^\mu q^\nu / M_A^2$ term in a heavy vector boson propagator (where $M_A$ is the heavy vector mass), power counting in the auxiliary-field formalism is difficult. We will identify two $SO(1,1)$ symmetric gauge-fixing conditions that simplify the auxiliary-field LW~analysis. In one case (ordinary), the gauge-fixing forces the $q^\mu q^\nu / M_A^2$ terms to appear with canceling signs in the gauge-LW~propagator matrix. In the other case (``no-mixing"), the gauge-fixing eliminates the $q^\mu q^\nu / M_A^2$ term in the vector field propagators. Working in the no-mixing gauge allows us to show that the number of superficially divergent amplitudes in an Abelian gauge theory is finite, and the theory is, therefore, renormalizable.

Finally, to illustrate the renormalizability of LW~scalar QED, we explicitly carry out the one-loop renormalization program and demonstrate how the counterterms required are constrained by the joint conditions of gauge and $SO(1,1)$ invariance. As a by-product of these discussions, we compute the one-loop beta functions in LW~scalar QED and contrast them with those of ordinary scalar QED.

In Sec.~\ref{sec:phi4} we introduce and illustrate the $SO(1,1)$ symmetries of a LW~theory in the context of $\phi^4$ theory. In Sec.~\ref{sec:lgrng} we consider LW~scalar QED and derive the equivalent auxiliary-field description. We then analyze the global symmetries of the theory and explain how these protect the form of the Lagrangian against radiative corrections. In Sec.~\ref{sec:fixing} we show how gauge-fixing can be implemented in an $SO(1,1)$ invariant fashion and derive the corresponding propagators. In Sec.~\ref{sec:renormalization} we show that the number of superficially divergent amplitudes is finite, and the theory is, therefore, renormalizable. Then we illustrate these results at one-loop by carrying out the renormalization program and computing the beta functions. Finally, in Sec.~\ref{sec:concl} we offer our conclusions and we sketch why a modified approach is needed for the case of non-Abelian gauge theories.

\section{Lee-Wick $\phi^4$ Theory}\label{sec:phi4}

We first consider LW~$\phi^4$ theory for a complex scalar field in order to introduce the auxiliary-field formalism and the $SO(1,1)$ global symmetry of the model, as well as to set our notational conventions.\footnote{In this chapter we follow closely the conventions of Ref. \protect\cite{Grinstein:2008uk}.} Lee-Wick $\phi^4$ theory is defined by the higher-derivative (hd) Lagrangian
\begin{equation}
{\cal L}_{\text{hd}} = |\partial_\mu \hat{\phi}|^2-\frac{1}{\hat{M}^2}|\partial^2 \hat{\phi}|^2
-\hat{m}^2|\hat{\phi}|^2 -\frac{\hat{\lambda}}{4}|\hat{\phi}|^4 \ ,
\label{eq:hdphi4}
\end{equation}
where $\hat{\phi}$ is a complex scalar field, and the Lee-Wick scale $\hat{M}$ parameterizes the energy at which the model deviates substantially from the standard $\phi^4$ model. As we will see, $\hat{M}$ also characterizes the mass scale of the LW~ghosts, so long as $\hat{m} \ll \hat{M}$.  This Lagrangian is equivalent to one in which we introduce an ``auxiliary" complex scalar field $\tilde{\phi}^\prime$ (the reason for the ``prime" will become clear in what follows)
\begin{equation}
{\cal L} = |\partial_\mu \hat{\phi}|^2 + \hat{M}^2 |\tilde{\phi}^\prime|^2 + 
\partial_\mu \hat{\phi} \, \partial^\mu \tilde{\phi}^{\prime \ast}
+\partial_\mu \hat{\phi}^{\ast} \partial^\mu \tilde{\phi}^\prime-\hat{m}^2|\hat{\phi}|^2 -\frac{\hat{\lambda}}{4}|\hat{\phi}|^4 \ .
\label{eq:auxphi4}
\end{equation}
Making the change of variable
\begin{equation}
\hat{\phi} = \phi^\prime - \tilde{\phi}^\prime \ ,
\end{equation}
we find
\begin{equation}
{\cal L} = 
|\partial_\mu \phi^\prime|^2-|\partial_\mu \tilde{\phi}^\prime|^2 +\hat{M}^2 |\tilde{\phi}^\prime|^2 
-\hat{m}^2|\phi^\prime-\tilde{\phi}^\prime|^2 - \frac{\hat{\lambda}}{4} |\phi^\prime-\tilde{\phi}^\prime|^4 \ .
\end{equation}

The symplectic rotation\footnote{Note the contrast with an orthogonal rotation, $\begin{pmatrix}\cos\theta&-\sin\theta\\\sin\theta&\cos\theta\end{pmatrix}$.}
\begin{equation}
\begin{pmatrix}\phi^\prime \\ \tilde{\phi}^\prime \end{pmatrix}
=\begin{pmatrix} \cosh\theta&\sinh\theta \\ \sinh\theta&\cosh\theta \end{pmatrix}
\begin{pmatrix}\phi \\ \tilde{\phi} \end{pmatrix} \ ,
\label{eq:symplectic}
\end{equation}
where
\begin{equation}
\tanh 2\theta = \frac{-2\hat{m}^2/\hat{M}^2}{1-2\hat{m}^2/\hat{M}^2} \ ,
\end{equation}
diagonalizes the scalar field mass terms while preserving the symplectic structure of the kinetic terms~\cite{Grinstein:2008uk}. Hence, we arrive at the auxiliary-field description of the LW~$\phi^4$ theory
\begin{equation}\label{eq:phi4} \begin{split}
{\cal L}_{\phi^4} = \ &  |\partial_\mu \phi|^2 - |\partial_\mu \tilde{\phi}|^2 +M^2 |\tilde{\phi}|^2 -m^2|\phi|^2 
- \frac{\lambda}{4} |\phi-\tilde{\phi}|^4 \\
= \ & |\partial_\mu \phi|^2 - |\partial_\mu \tilde{\phi}|^2 +M^2 |\tilde{\phi}|^2 -m^2|\phi|^2 - \frac{\lambda}{4} |\phi|^4 + \frac{\lambda}{2} |\phi|^2\left(\phi\tilde{\phi}^\ast+\phi^\ast\tilde{\phi}\right) \\
&- \lambda |\phi|^2 |\tilde{\phi}|^2 - \frac{\lambda}{4} \left(\phi^2\tilde{\phi}^{\ast 2}+\phi^{\ast 2}\tilde{\phi}^2\right)
+ \frac{\lambda}{2} |\tilde{\phi}|^2\left(\phi\tilde{\phi}^\ast+\phi^\ast\tilde{\phi}\right)
- \frac{\lambda}{4} |\tilde{\phi}|^4 \ ,
\end{split} \end{equation}   
where
\begin{equation}\label{eq:relabel} \begin{split}
M^2 = \ & \cosh^2\theta\,\hat{M}^2-e^{-2\theta}\,\hat{m}^2 \ , \\
m^2 = \ & e^{-2\theta}\,\hat{m}^2 - \sinh^2\theta\,\hat{M}^2 \ , \\
\lambda = \ & e^{-4\theta}\,\hat{\lambda} \ .
\end{split} \end{equation}
Note that the kinetic term of the $\tilde{\phi}$ field has the opposite sign to the usual one, and, hence, the corresponding particle has negative norm and is the LW~ghost field. Furthermore, the mass of the LW~ghost $M$ is, in the limit $\hat{m} \ll \hat{M}$, approximately the LW~scale $\hat{M}$ introduced in Eq.~(\ref{eq:hdphi4}).

This  theory has an exact global $U(1)$ symmetry, but is not the most general $U(1)$ symmetric renormalizable Lagrangian that can be built out of the ordinary field $\phi$ and the ghost field $\tilde{\phi}$ charged under the $U(1)$ symmetry. In particular, the six interaction terms in the second line can in principle have six independent couplings.  However, the dimension-four terms in Eq.~(\ref{eq:phi4}) do have an additional  $SO(1,1)$ symmetry, under which the fields transform as
\begin{equation}
\begin{pmatrix}\phi \\ \tilde{\phi} \end{pmatrix}
\to \begin{pmatrix} \cosh\beta&\sinh\beta \\ \sinh\beta&\cosh\beta \end{pmatrix}
\begin{pmatrix}\phi \\ \tilde{\phi} \end{pmatrix} \ ,
\end{equation}
so long as we also promote $\lambda$ to a spurion field\footnote{A spurion is a (fictitious) field that parametrizes the symmetry breaking. Its initial symmetry invariant transformation can be used to construct the invariant operators of the theory. Setting this field equal to its actual constant value will, subsequently, capture all of the symmetry-breaking operators.} that
transforms as
\begin{equation}
\lambda\to e^{4\beta}\,\lambda \ .
\end{equation}

Furthermore, the Lagrangian of Eq.~(\ref{eq:phi4}) is the most general renormalizable and $U(1)$-symmetric Lagrangian with $SO(1,1)$-symmetric dimension-four terms. The different mass terms for $\phi$ and $\tilde{\phi}$ break the $SO(1,1)$
symmetry, but do so only softly. They are also the only $U(1)$-preserving soft terms that break $SO(1,1)$. Thus in the  LW~$\phi^4$ theory, the global $SO(1,1)$ symmetry of the dimension-four terms implies that loop corrections can only modify the structure of the mass terms, introducing a mixing term between $\phi$ and $\tilde{\phi}$ with infinite coefficient. This can always be diagonalized via a symplectic rotation (of the form given in Eq.~(\ref{eq:symplectic})), which leaves the rest of the Lagrangian unchanged, except for a redefinition of the coupling. Hence, Lee-Wick $\phi^4$ theory is renormalizable by power-counting.

LW~$\phi^4$ theory  is rather simple, because aside from mass renormalization the theory is finite. The LW~scenario is, however, much less trivial in LW~gauge theories, because of the new momentum-dependent interactions in the higher-derivative formulation. In this case global symmetries  are important to understand the full structure of the theory. In the following we will show that Abelian $N=2$ LW~theories have a softly broken $SO(1,1)^{m+1}$ symmetry, where $m$ is the number of matter fields, and the remaining $SO(1,1)$ transformation acts on the vector fields. Since the $SO(1,1)^{m+1}$ breaking is soft, the special relation between the LW~couplings and the ordinary couplings is protected against radiative corrections.

\section{Global Symmetries of Lee-Wick Scalar QED}\label{sec:lgrng}

Let us now study an $N=2$ LW~theory of scalar electrodynamics.  In the higher-derivative formulation, the Lagrangian is\footnote{In non-Abelian theories there can be additional dimension-six higher-derivative operators, which lead to heavy vector scattering amplitudes growing like $E^2$, where $E$ is the center-of-mass energy~\cite{Grinstein:2008fu}. For $N\geq 2$ LW~theories see, for example, Ref.~\cite{Carone:2009kl}.}
\begin{equation}
{\cal L}_\text{hd} = -\frac{1}{4}\hat{F}_{\mu\nu}^2
+\frac{1}{2 M_A^2} \left(\partial^\mu \hat{F}_{\mu\nu}\right)^2
+|\hat{D}_\mu \hat{\phi}|^2 - \frac{1}{\hat{M}^2} |\hat{D}^2 \hat{\phi}|^2
-\hat{m}^2|\hat{\phi}|^2 - \frac{\hat{\lambda}}{4} |\hat{\phi}|^4 \ ,
\label{eq:higher}
\end{equation}
where
\begin{equation}
\hat{D}_\mu \equiv \partial_\mu  - ig \, \hat{A}_\mu  \ .
\label{eq:covariant}
\end{equation}
The scalar sector is simply that of $\phi^4$ theory as shown in Eq.~(\ref{eq:hdphi4}), and, hence, our analysis of this Lagrangian will parallel the discussion of Sec.~\ref{sec:phi4}. Introducing auxiliary fields, now for both the vector and the scalar and using the notation described above, we see that the Lagrangian of Eq.~(\ref{eq:higher}) is equivalent to
\begin{equation}\label{eq:lower1} \begin{split}
{\cal L} = \ & -\frac{1}{4}\hat{F}_{\mu\nu}^2 - \partial^\mu \tilde{A}^\nu\, \hat{F}_{\mu\nu} -\frac{M_A^2}{2} \tilde{A}_\mu^2
+|\hat{D}_\mu \hat{\phi}|^2 +\hat{M}^2 |\tilde{\phi}^\prime|^2 \\
&+ \hat{D}_\mu\hat{\phi} \hat{D}^\mu\tilde{\phi}^{\prime\ast}
+ \hat{D}_\mu\hat{\phi}^\ast \hat{D}^\mu\tilde{\phi}^\prime 
-\hat{m}^2|\hat{\phi}|^2 - \frac{\hat{\lambda}}{4} |\hat{\phi}|^4 \ ,
\end{split} \end{equation}
to all orders in perturbation theory. Changing variables from $\hat{A}_\mu$, $\tilde{A}_\mu$, $\hat{\phi}$, $\tilde{\phi}^\prime$ to $A_\mu$, $\tilde{A}_\mu$, $\phi^\prime$, $\tilde{\phi}^\prime$, where
\begin{align}
\hat{A}_\mu = \ & A_\mu - \tilde{A}_\mu \ , \label{eq:AAtilde} \\
\hat{\phi} = \ & \phi^\prime - \tilde{\phi}^\prime \ ,
\end{align}
and substituting in Eqs.~(\ref{eq:covariant}) and (\ref{eq:lower1}), gives
\begin{equation} \begin{split}
{\cal L} = \ & -\frac{1}{4}F_{\mu\nu}^2 + \frac{1}{4}\tilde{F}_{\mu\nu}^2 -\frac{M_A^2}{2} \tilde{A}_\mu^2
+|D_\mu \phi^\prime|^2-|D_\mu \tilde{\phi}^\prime|^2 +\hat{M}^2 |\tilde{\phi}^\prime|^2 
-\hat{m}^2|\phi^\prime-\tilde{\phi}^\prime|^2 \\
&- \frac{\hat{\lambda}}{4} |\phi^\prime-\tilde{\phi}^\prime|^4 - ig\tilde{A}_\mu \left(\phi^\prime D^\mu \phi^{\prime\ast} - \phi^{\prime\ast} D^\mu \phi^\prime \right)
+ ig\tilde{A}_\mu \left(\tilde{\phi}^\prime D^\mu \tilde{\phi}^{\prime\ast} - \tilde{\phi}^{\prime\ast} D^\mu \tilde{\phi}^\prime \right) \\
&+ g^2 \tilde{A}_\mu^2\left(|\phi^\prime|^2-|\tilde{\phi}^\prime|^2\right) \ ,
\end{split} \end{equation}
where now the covariant derivative is in terms of $A_\mu$,
\begin{equation}
D_\mu = \partial_\mu- ig\, A_\mu \ .
\end{equation}
The symplectic rotation
of Eq.~(\ref{eq:symplectic}) again diagonalizes the scalar field mass terms while preserving the symplectic structure of the kinetic terms~\cite{Grinstein:2008uk}. Since the gauge interactions stem from kinetic terms, and the $\phi^4$-interaction has a symplectic structure as well, it follows that Eq.~(\ref{eq:symplectic}) only diagonalizes the mass terms leaving the rest of the Lagrangian invariant in form. In terms of $\phi$ and $\tilde{\phi}$ the Lagrangian now reads
\begin{equation}\label{eq:Ltwofields} \begin{split}
{\cal L} = \ & -\frac{1}{4}F_{\mu\nu}^2 + \frac{1}{4}\tilde{F}_{\mu\nu}^2 -\frac{M_A^2}{2} \tilde{A}_\mu^2
+|D_\mu \phi|^2-|D_\mu \tilde{\phi}|^2 +M^2 |\tilde{\phi}|^2 
-m^2|\phi|^2 \\
&- \frac{\lambda}{4} |\phi-\tilde{\phi}|^4 + ig\tilde{A}_\mu \left(\phi\,D^\mu \phi^\ast - \phi^\ast D^\mu \phi \right)
- i g\tilde{A}_\mu \left(\tilde{\phi}\,D^\mu \tilde{\phi}^\ast - \tilde{\phi}^\ast D^\mu \tilde{\phi} \right) \\
&+ g^2 \tilde{A}_\mu^2\left(|\phi|^2-|\tilde{\phi}|^2\right) \ ,
\end{split} \end{equation}
where we redefine parameters as in Eq.~(\ref{eq:relabel}).

The Lagrangian of Eq.~(\ref{eq:Ltwofields}) has an exact $U(1)$ gauge symmetry. In the limit $\lambda\to 0$ the global symmetry is promoted to $U(1)\times U(1)$, because the $\phi$ and $\tilde{\phi}$ fields can now rotate independently, and only the diagonal $U(1)$ subgroup is gauged. Thus, we expect loop corrections to generate $U(1)$-symmetric terms | some with infinite coefficients |  that will be $U(1)\times U(1)$-symmetric in the $\lambda\to 0$ limit. Eq.~(\ref{eq:Ltwofields}) is not the most general renormalizable Lagrangian with this symmetry structure; for example, the coefficients of the interactions involving $\tilde{A}_\mu$ could be arbitrary. Notice, however, that this Lagrangian can be re-arranged in the form
\begin{align}
{\cal L} = \ & -\frac{1}{4}F_{\mu\nu}^2 + \frac{1}{4}\tilde{F}_{\mu\nu}^2 -\frac{M_A^2}{2} \tilde{A}_\mu^2
+|\partial_\mu \phi|^2-|\partial_\mu \tilde{\phi}|^2 +M^2 |\tilde{\phi}|^2
-m^2|\phi|^2 - \frac{\lambda}{4} |\phi-\tilde{\phi}|^4 \notag \\
& - ig(A_\mu-\tilde{A}_\mu) \left(\phi\,\partial^\mu \phi^\ast- \tilde{\phi}\,\partial^\mu \tilde{\phi}^\ast - \text{h.c.}\right) -g^2 (A_\mu-\tilde{A}_\mu)^2\left(|\phi|^2-|\tilde{\phi}|^2\right) \label{eq:Ltwofields2} \ .
\end{align}
In the limit $M_A\to 0$, and treating the gauge coupling as a spurion field, the Lagrangian respects a global $SO(1,1)$ symmetry under which
\begin{equation}
\begin{pmatrix} A_\mu \\ \tilde{A}_\mu \end{pmatrix}
\to \begin{pmatrix} \cosh\alpha&\sinh\alpha \\ \sinh\alpha&\cosh\alpha \end{pmatrix}
\begin{pmatrix} A_\mu \\ \tilde{A}_\mu \end{pmatrix}, \quad g \to e^{\alpha}\ g \ .
\label{eq:SOgauge}
\end{equation}
As mentioned in Sec.~\ref{sec:phi4}, an additional $SO(1,1)$ global symmetry for the scalar field arises in the limit $M\to m$, when $\lambda$ is treated as a spurion field
\begin{equation}
\begin{pmatrix}\phi \\ \tilde{\phi} \end{pmatrix}
\to \begin{pmatrix} \cosh\beta&\sinh\beta \\ \sinh\beta&\cosh\beta \end{pmatrix}
\begin{pmatrix}\phi \\ \tilde{\phi} \end{pmatrix}, \quad \lambda\to e^{4\beta}\,\lambda \ .
\label{eq:SOscalar}
\end{equation}
In Sec.~\ref{sec:renormalization} we will argue that this theory is renormalizable, because the number of superficially divergent amplitudes is finite, and no operators of dimension greater than four are present in the auxiliary-field formulation. We may wonder whether radiative corrections require introducing dimension-four $SO(1,1)\times SO(1,1)$-breaking counterterms. However, as in the LW~$\phi^4$ theory described above, the answer is no: since $SO(1,1)\times SO(1,1)$ is only softly broken by mass terms,\footnote{The renormalizability of massive Abelian gauge theory \protect\cite{Kroll:1967zt} | arising from the coupling of the Abelian gauge boson to a conserved current | insures that the gauge boson mass term is ``soft".} the $SO(1,1)\times SO(1,1)$-breaking corrections to the renormalizable terms are finite. Furthermore, Eq.~(\ref{eq:Ltwofields2}) is the most general $U(1)$ gauge Lagrangian with dimension-four $SO(1,1)\times SO(1,1)$-symmetric terms. Since renormalizability prevents higher dimensional operators from being generated, we conclude that the form of the Lagrangian is protected to all orders against radiative corrections, with the exception of the scalar field mass terms. However, as we have already seen, these can be diagonalized with a symplectic rotation, without affecting the rest of the Lagrangian. In the simple example we have shown there is only one matter field: for an arbitrary number $m$ of matter fields the global symmetry is promoted to $SO(1,1)\times SO(1,1)^m$, since each field is acted upon with a different $SO(1,1)$ symmetry transformation, and the conclusions about renormalizability persist, \textit{mutatis mutandis}.

\section{Gauge-Fixing}\label{sec:fixing}

In order to quantize the electromagnetic field, one must introduce a gauge-fixing term. To facilitate our subsequent analyses of divergences and renormalizability, we will find it most convenient to employ gauge-fixing functions that respect the $SO(1,1)$ symmetry; otherwise it can be unnecessarily difficult to recognize when significant cancellations occur. For example, Ref.~\cite{Grinstein:2008uk} employs an $SO(1,1)$ violating gauge-fixing term $-\displaystyle{\frac{1}{2\xi}}\left(\partial^\mu A_\mu\right)^2$
which leads to diagonal propagators of the form
\begin{equation*}
P_{AA}^{\mu\nu} = \frac{-i}{q^2}\left[g^{\mu\nu}-(1-\xi)\frac{q^\mu q^\nu}{q^2}\right] \ , \quad
P_{\tilde{A}\tilde{A}}^{\mu\nu} = \frac{i}{q^2-M_A^2}\left[g^{\mu\nu}-\frac{q^\mu q^\nu}{M_A^2}\right] \ .
\end{equation*}
Because the $q^\mu q^\nu/M_A^2$ term is only present in the LW~photon propagator, there is no simple cancellation of the badly-behaved terms and the theory appears to suffer from quadratic divergences and non-renormalizability at one-loop. The reason that Ref.~\cite{Grinstein:2008uk} found no quadratic divergences when computing the self-energy amplitudes for a massless scalar field at zero momentum is that the quadratic divergence vanishes in the limit $m\to 0$ and $q\to 0$, since it is necessarily of the form $\Lambda^2 m^2/M_A^2$ or $\Lambda^2 q^2/M_A^2$. These quadratic divergences are ``gauge artifacts" \cite{Grinstein:2008dz} in the sense that they contribute to both scalar wavefunction and mass renormalization in such a way that the pole mass of the scalar is \textit{not} quadratically sensitive to the cutoff.

To ensure that the symmetry will be preserved, it is sufficient to write the gauge-fixing term in terms of $\hat{A}_\mu/\sqrt{\xi}$, where $\xi$ is the gauge-fixing parameter. If one treats $\xi$ as a spurion field, the $SO(1,1)$ transformation
\begin{equation}
\hat{A}_\mu \to e^{-\alpha}\,\hat{A}_\mu, \quad \xi \to e^{-2\alpha}\,\xi \ ,
\end{equation}
clearly leaves $\hat{A}_\mu/\sqrt{\xi}$ invariant. We will consider two different $SO(1,1)$ symmetric gauge-fixing scenarios that are each convenient in different circumstances, and will denote them as ordinary and ``no-mixing" gauge-fixing.

\subsection{Ordinary Gauge-Fixing}

First, let us consider a gauge-fixing function of the typical form $G(\hat{A})=\partial^\mu \hat{A}_\mu$. In $R_\xi$ gauge, this amounts to adding to the Lagrangian the $SO(1,1)$ symmetric gauge-fixing term
\begin{equation}
{\cal L}^\text{ordinary}_\text{fixing} = -\frac{1}{2\xi}\left(\partial^\mu \hat{A}_\mu\right)^2 \ .
\end{equation}
Using Eq.~(\ref{eq:AAtilde}) to rewrite this in terms of $\tilde{A}^\mu$ and $A^\mu$, one obtains 
\begin{equation}
{\cal L}^\text{ordinary}_\text{fixing} = -\frac{1}{2\xi}\left(\partial^\mu A_\mu\right)^2 - \frac{1}{2\xi}\left(\partial^\mu \tilde{A}_\mu\right)^2
+\frac{1}{\xi}\partial^\mu A_\mu\, \partial^\nu \tilde{A}_\nu \ .
\end{equation}
With the gauge-fixing included, and after integrating by parts, the gauge field Lagrangian reads
\begin{equation} \begin{split}
{\cal L}^\text{ordinary}_\text{gauge} = \ & \frac{1}{2}A_\mu \left[g^{\mu\nu}\partial^2-(1-1/\xi)\partial^\mu\partial^\nu\right]A_\nu \\
&-\frac{1}{2}\tilde{A}_\mu \left[g^{\mu\nu}(\partial^2-M_A^2)-(1+1/\xi)\partial^\mu\partial^\nu\right]\tilde{A}_\nu
- A_\mu \frac{1}{\xi}\, \partial^\mu \partial^\nu \tilde{A}_\nu \ .
\end{split} \end{equation}

We can invert the diagonal terms, in momentum space, to find the partial propagators
\begin{equation}
D_{AA}^{\mu\nu} = \frac{-i}{q^2}\left[g^{\mu\nu}-(1-\xi)\frac{q^\mu q^\nu}{q^2}\right], \ 
D_{\tilde{A}\tilde{A}}^{\mu\nu} = \frac{i}{q^2-M_A^2}\left[g^{\mu\nu}-(1+\xi)\frac{q^\mu q^\nu}{q^2+\xi M_A^2}\right] \ .
\end{equation}
Then the full tree-level photon and LW~photon propagators, as well as the mixed-propagators, can be computed by resumming the Dyson series to obtain
\begin{equation}\label{eq:proplower} \begin{split}
P_{AA}^{\mu\nu} = \ & \frac{-i}{q^2}\left[g^{\mu\nu}-(1-\xi)\frac{q^\mu q^\nu}{q^2} + \frac{q^\mu q^\nu}{M_A^2}\right], \quad 
P_{\tilde{A}\tilde{A}}^{\mu\nu} = \frac{i}{q^2-M_A^2}\left[g^{\mu\nu}-\frac{q^\mu q^\nu}{M_A^2}\right], \\ 
P_{\tilde{A}A}^{\mu\nu} = \ & P_{A\tilde{A}}^{\mu\nu}  = -\frac{i\,q^\mu q^\nu}{q^2 M_A^2} \ .
\end{split} \end{equation}
Notice that only the photon propagator depends on the gauge-fixing parameter $\xi$, since the photon is the only true gauge field in this theory. 

Up to an overall sign, the $\tilde{A}_\mu$ propagator is identical to the unitary-gauge propagator of a massive gauge boson, in a spontaneously broken gauge theory. In particular, it contains the $q^\mu q^\nu/M_A^2$ term which would apparently render  the theory non-renormalizable and reintroduce quadratic divergences. However, the photon propagator and the mixed-propagators contain the same term; when the $P_{AA}$, $P_{\tilde{A}\tilde{A}}$, and $P_{A\tilde{A}}$ propagators are all included in loop integrals, the badly behaved terms cancel, and quadratic divergences are avoided. This can be seen even more clearly from the form of the $P^{\mu\nu}_{\hat{A}\hat{A}}$ propagator.  Recalling that all gauge interactions depend on $A_\mu-\tilde{A}_\mu=\hat{A}_\mu $ and working in Feynman gauge  ($\xi = 1$), we obtain
\begin{equation}
P^{\mu\nu}_{\hat{A}\hat{A}} = P^{\mu\nu}_{AA}+P^{\mu\nu}_{\tilde{A}\tilde{A}} - 2 P^{\mu\nu}_{A\tilde{A}}
=\frac{-i}{q^2-q^4/M_A^2}\left[g^{\mu\nu}-\frac{q^\mu q^\nu}{M_A^2}\right] \ ,
\label{eq:hatprop}
\end{equation}
which decays like $1/q^2$ for large values of the momentum.

\subsection{No-Mixing Gauge-Fixing}

Next, we consider the alternative\footnote{In non-Abelian theories this introduces a $q^2/M_A^2$ expansion in the gauge-ghost interaction, which renders it less interesting. In Abelian gauge theories, however, the ghosts are decoupled. } gauge-fixing function $G(\hat{A})=\left(1+\partial^2/M_A^2\right)^{1/2}\partial^\mu \hat{A}_\mu$. The resulting $SO(1,1)$-symmetric gauge-fixing Lagrangian is
\begin{equation}
{\cal L}^\text{no-mixing}_\text{fixing} = -\frac{1}{2\xi}\left(\partial^\mu \hat{A}_\mu\right)^2
+\frac{1}{2 \, \xi M_A^2} \left(\partial^\mu\partial^\nu \hat{A}_\nu \right)^2 \ .
\end{equation}
Adding this to the original higher-derivative gauge Lagrangian gives, after integration by parts,
\begin{equation}
{\cal L}^\text{no-mixing}_\text{gauge} = \frac{1}{2}\hat{A}_\mu \left[g^{\mu\nu}\partial^2-(1-1/\xi)\partial^\mu\partial^\nu\right]
\left(1+\partial^2/M_A^2\right)\hat{A}_\nu \ .
\label{eq:no-mixing-hd-gauge}
\end{equation}
This is equivalent to
\begin{equation}\label{eq:no-mixing-gauge} \begin{split}
{\cal L}^\text{no-mixing}_\text{gauge} = \ & \frac{1}{2}\hat{A}_\mu \left[g^{\mu\nu}\partial^2-(1-1/\xi)\partial^\mu\partial^\nu\right]\hat{A}_\nu \\
&+ \hat{A}_\mu \left[g^{\mu\nu}\partial^2-(1-1/\sqrt{\xi})\partial^\mu\partial^\nu\right]\tilde{A}_\nu -\frac{M_A^2}{2}\tilde{A}_\mu^2 \ .
\end{split} \end{equation}
in the sense that solving the equations of motion for $\tilde{A}_\mu$ and inserting the solution in (\ref{eq:no-mixing-gauge}) recovers the form of (\ref{eq:no-mixing-hd-gauge}).  

At this point we can eliminate the $\hat{A}^\mu$ field from Eq.~(\ref{eq:no-mixing-gauge}) via  Eq.~(\ref{eq:AAtilde}); the Lagrangian will include both diagonal and mixing terms in $A^\mu$ and $\tilde{A}^\mu$, and the full tree-level propagators can again be computed by summing the Dyson series. However, for $\xi=1$ the mixing term vanishes,\footnote{As an alternative to the $\xi=1$ gauge, one could replace Eq.~(\ref{eq:AAtilde}) with
\begin{equation*}
\hat{A}_\mu = A_\mu - \left[\delta_\mu^\nu - (1-\sqrt{\xi})\frac{q_\mu q^\nu}{q^2}\right]\tilde{A}_\nu \ ,
\end{equation*}
in momentum space. This cancels the off-diagonal terms for any value of $\xi$, at the price of introducing non-local interactions in coordinate space for $\xi\neq1$. } and we obtain the simpler diagonal Lagrangian 
\begin{equation}
{\cal L}^\text{no-mixing}_{\rm gauge, \xi=1} = \frac{1}{2}A_\mu \partial^2 A^\mu
-\frac{1}{2}\tilde{A}_\mu (\partial^2-M_A^2) \tilde{A}^\mu \ .
\end{equation}
The corresponding propagators
\begin{equation}
P_{AA}^{\mu\nu} = \frac{-i\,g^{\mu\nu}}{q^2}\ , \quad 
P_{\tilde{A}\tilde{A}}^{\mu\nu} = \frac{i\,g^{\mu\nu}}{q^2-M_A^2} \ ,
\label{eq:prophigher}
\end{equation}
have no $q^\mu q^\nu$ terms so they are well-behaved at high energies.  This is equally clear if we construct the $\hat{A}$ propagator,
\begin{equation}
P^{\mu\nu}_{\hat{A}\hat{A}} = P^{\mu\nu}_{AA}+P^{\mu\nu}_{\tilde{A}\tilde{A}} - 2 P^{\mu\nu}_{A\tilde{A}}
=\frac{-i g^{\mu\nu}}{q^2-q^4/M_A^2} \ ,
\label{eq:hatprophigher}
\end{equation}
which falls off like $q^{-4}$ in the ultraviolet.

We will now use the two convenient gauges introduced in this section to explore LW~scalar QED at one-loop.

\section{One-Loop Renormalization}\label{sec:renormalization}

 We will start by establishing an upper bound on the superficial degree of divergence of Feynman diagrams in $N=2$ LW~scalar QED. For specificity, we work in the auxiliary-field formulation and employ the no-mixing $\xi=1$ gauge. Recalling that each loop integral introduces four powers of momentum in the numerator, each trilinear gauge-scalar-scalar vertex introduces one power of momentum in the numerator, and the propagator has two powers of momentum in the denominator, we arrive at
\begin{equation}
D \leq 4L - 2P_A - 2P_{\tilde{A}} - 2P_\phi - 2P_{\tilde{\phi}} + V_{gss} \ ,
\label{eq:divergence}
\end{equation}
where $L$ is the number of loops, $P_f$ is the number of propagators of the $f$ field, and $V_{gss}$ is the number of trilinear gauge-scalar-scalar vertices. The number of loop integrals is, in turn, given by the total number of propagators (each carrying its own momentum-space integral) minus the total number of vertices (each carrying a momentum-space delta function) plus one, since an overall delta function ensures momentum conservation for the external fields. Therefore, denoting the number of quartic gauge-gauge-scalar-scalar vertices by $V_{ggss}$ and the number of four-point scalar vertices by $V_{ssss}$, we have
\begin{equation}
L = P_A + P_{\tilde{A}} + P_\phi + P_{\tilde{\phi}} - V_{gss} - V_{ggss} - V_{ssss} + 1 \ .
\label{eq:elloop}
\end{equation}
Finally we can relate the number of lines attached to a vertex to the number of propagators (each connecting two vertices) and the number of external lines $N_f$
\begin{equation}\label{eq:verte} \begin{split}
V_{gss} + 2 V_{ggss} = \ &  2P_A + 2P_{\tilde{A}} + N_A + N_{\tilde{A}}\ , \\
2V_{gss} + 2V_{ggss} + 4 V_{ssss} = \ & 2P_\phi + 2P_{\tilde{\phi}} + N_\phi + N_{\tilde{\phi}} \ ,
\end{split} \end{equation} 
where the first relation deals with gauge lines and the second with scalar lines. Inserting Eqs.~(\ref{eq:elloop}) and (\ref{eq:verte}) in Eq.~(\ref{eq:divergence}) yields
\begin{equation}
D \leq 4 - N_A - N_{\tilde{A}} - N_{\phi} - N_{\tilde{\phi}} \ .
\label{eq:D}
\end{equation}
This equation tells us that the number of superficially divergent amplitudes is finite; since no operators of dimension greater than four are present in the auxiliary-field Lagrangian, we conclude that the theory is renormalizable.

In order to confirm renormalizability explicitly and to verify how the $SO(1,1)\times SO(1,1)$ structure of the theory is protected against radiative corrections, we will now compute the infinite\footnote{If we were to compute the finite part as well, we would need to employ the Cutkosky-Landshoff-Olive-Polkinghorne prescription in order to avoid unitarity violation~\cite{Cutkosky:1969bs}. However, the infinite part is not affected by this subtlety~\cite{Grinstein:2008dz}.} part of the divergent one-particle irreducible (1PI) diagrams at one-loop.   As a way of checking our results and exploring the detailed symmetry structure, we will compute the diagrams in both the ordinary and no-mixing gauges, with $\xi = 1$.  As we shall see, in the no-mixing  gauge only the vacuum polarization and self-energy amplitudes are infinite, while in the ordinary gauge infinities also arise in the vertex corrections. Therefore, the way the counterterms preserve the symplectic structure of the theory is different in the two gauges.

\subsection{Counterterms}

Radiative corrections renormalize the fields and mix the ordinary fields with the LW~partners. This not only preserves the $U(1)$ gauge symmetry, but also does not generate any hard breaking of the global $SO(1,1)\times SO(1,1)$ symmetry. Let us derive the most general relation between bare and renormalized fields satisfying these requirements. We will employ the standard QED nomenclature for the counterterms by using the subscript ``3'' for the photon wavefunction renormalization, ``2'' for the matter field wavefunction renormalization, and ``1'' for gauge vertex renormalization.

For the vector fields we have in general
\begin{equation*}
\begin{pmatrix} A^\mu \\ \tilde{A}^\mu \end{pmatrix}
= \begin{pmatrix} \sqrt{Z_3} & \sqrt{Z_3^\prime} \\ \sqrt{Z_3^{\prime\prime}} & \sqrt{\widetilde{Z}_3} \end{pmatrix}
\begin{pmatrix} A_r^\mu \\ \tilde{A}_r^\mu \end{pmatrix} \ .
\end{equation*}
Gauge invariance requires $Z_3^{\prime\prime} = 0$, lest a photon mass term be generated. Preserving the form of the symplectic combination $A_\mu-\tilde{A}_\mu$ demands
\begin{equation*}
\sqrt{Z_3^\prime}=\sqrt{\widetilde{Z}_3}-\sqrt{Z_3} \ .
\end{equation*}
Finally, substituting in the kinetic term Lagrangian and imposing the  $SO(1,1)$ symmetry on the counterterms gives
\begin{equation*}
\widetilde{Z}_3 = \frac{1}{Z_3} \ .
\end{equation*}
Therefore, the relation between bare and renormalized vector fields consistent with the symmetries of the theory is
\begin{equation}
\begin{pmatrix} A^\mu \\ \tilde{A}^\mu \end{pmatrix}
= \sqrt{Z_3} \begin{pmatrix} 1 & \ \ \ Z_3^{-1}-1 \\ 0 & Z_3^{-1} \end{pmatrix}
\begin{pmatrix} A_r^\mu \\ \tilde{A}_r^\mu \end{pmatrix}, \quad Z_3\equiv 1+\delta_3 \ .
\label{eq:renormA}
\end{equation}
Similarly, in order to preserve the $SO(1,1)$ symmetry on the scalar fields, the relation between bare and renormalized scalar fields must be a symplectic rotation times a wavefunction renormalization
\begin{equation}
\begin{pmatrix}\phi \\ \tilde{\phi} \end{pmatrix}
= \sqrt{Z_2} \begin{pmatrix} \cosh\eta&\sinh\eta \\ \sinh\eta&\cosh\eta \end{pmatrix}
\begin{pmatrix}\phi_{r} \\ \tilde{\phi}_{r} \end{pmatrix}, \quad Z_2\equiv 1+\delta_2 \ .
\label{eq:renormphi}
\end{equation}

Substituting Eqs.~(\ref{eq:renormA}) and (\ref{eq:renormphi}) in the Lagrangian, Eq.~(\ref{eq:Ltwofields2}), and denoting the gauge-scalar-scalar vertex and gauge-gauge-scalar-scalar vertex renormalizations, respectively, by $Z_1 \equiv 1 + \delta_1$ and $Z_1^\prime = 1 + \delta_1^\prime$ leads to
\begin{equation}\label{eq:Lrenorm} \begin{split}
{\cal L} = \ & -\frac{1}{4}F_{r\mu\nu}^2 + \frac{1}{4}\tilde{F}_{r\mu\nu}^2 -\frac{M_{Ar}^2}{2} \tilde{A}_{r\mu}^2
+|\partial_\mu \phi_r|^2-|\partial_\mu \tilde{\phi}_r|^2 +M_r^2 |\tilde{\phi}_r|^2
-m_r^2|\phi_r|^2 \\
&- \frac{\lambda_r}{4} |\phi_r-\tilde{\phi}_r|^4
- ig_r\, (A_{r\mu}-\tilde{A}_{r\mu}) \left(\phi_r\, \partial^\mu \phi_r^\ast- \tilde{\phi}_r\, \partial^\mu \tilde{\phi}_r^\ast - \text{h.c.}\right) \\
&- g_r^2 (A_{r\mu}-\tilde{A}_{r\mu})^2\left(|\phi_r|^2-|\tilde{\phi}_r|^2\right) + {\cal L}_\text{ct} \ ,
\end{split} \end{equation}
where the counterterm (ct) Lagrangian is
\begin{equation}\label{eq:Lct} \begin{split}
{\cal L}_\text{ct} = \ & -\frac{\delta_3}{4}F_{r\mu\nu}^2 + \frac{\delta_3}{2}F_{r\mu\nu} \tilde{F}_r^{\mu\nu}
-\frac{\delta_3}{4}\tilde{F}_{r\mu\nu}^2 -\frac{\delta_{M_A^2}}{2} \tilde{A}_{r\mu}^2 \\
&+\delta_2\, |\partial_\mu \phi_r|^2-\delta_2\, |\partial_\mu \tilde{\phi}_r|^2 +\delta_{M^2} |\tilde{\phi}_r|^2
-\delta_{m^2}|\phi_r|^2 -\delta_{mM}\left(\phi_r^\ast\tilde{\phi}_r+\phi_r\tilde{\phi}_r^\ast\right) \\
&- \frac{\delta_\lambda}{4} |\phi_r-\tilde{\phi}_r|^4 - i \delta_1g_r\, (A_{r\mu}-\tilde{A}_{r\mu}) 
\left(\phi_r\, \partial^\mu \phi_r^\ast- \tilde{\phi}_r\, \partial^\mu \tilde{\phi}_r^\ast - \text{h.c.}\right) \\
&- \delta_1^\prime\, g_r^2 (A_{r\mu}-\tilde{A}_{r\mu})^2\left(|\phi_r|^2-|\tilde{\phi}_r|^2\right) \ .
\end{split} \end{equation}

The renormalized trilinear and quartic gauge-scalar couplings are related to the bare couplings by
\begin{equation}
g\, \sqrt{Z_3}\, Z_2 = g_r\, Z_1 \ , \qquad\quad g^2\, Z_3\, Z_2 = g_r^2\, Z_1^\prime\ ,
\end{equation}
where gauge invariance guarantees
\begin{equation}
Z_1 = Z_1^\prime = Z_2 \ ,
\label{eq:ward}
\end{equation}
to all orders in perturbation theory. The renormalized mass parameters are related to the bare masses by
\begin{equation}\label{eq:ctrel} \begin{split}
Z_2 \left[(\cosh\eta)^2\,m^2 - (\sinh\eta)^2\,M^2\right] = \ &m_r^2 + \delta_{m^2} \ , \\
Z_2 \left[(\cosh\eta)^2\,M^2 - (\sinh\eta)^2\,m^2\right] = \ & M_r^2 + \delta_{M^2} \ , \\
Z_2 (M^2-m^2)\cosh\eta\, \sinh\eta =\ & - \delta_{mM} \ ,
\end{split} \end{equation}
whereas the renormalized and bare four-scalar couplings are related by
\begin{equation}
\lambda\, Z_2^2\, e^{-4\eta} = \lambda_r + \delta_\lambda \ .
\end{equation}
The vector field kinetic terms in the counterterm Lagrangian are now mixed. However, it can be easily shown that these are still invariant under an $SO(1,1)$ transformation, provided that $\delta_3$ is promoted to a spurion field.

We shall now prove that this set of counterterms is sufficient to absorb all infinities at one-loop. In the process, the $SO(1,1)\times SO(1,1)$ global symmetry leads to cancellation of the quadratic divergences in the scalar field self-energy amplitudes. In order to simplify our notation we will drop the subscript $r$ everywhere, but it should be kept in mind that all fields and parameters involved in the calculations below are the renormalized ones.
\\
\\

\subsection{Vacuum Polarization Amplitudes}

\begin{figure}
\begin{center}
\includegraphics[width=0.85\textwidth]{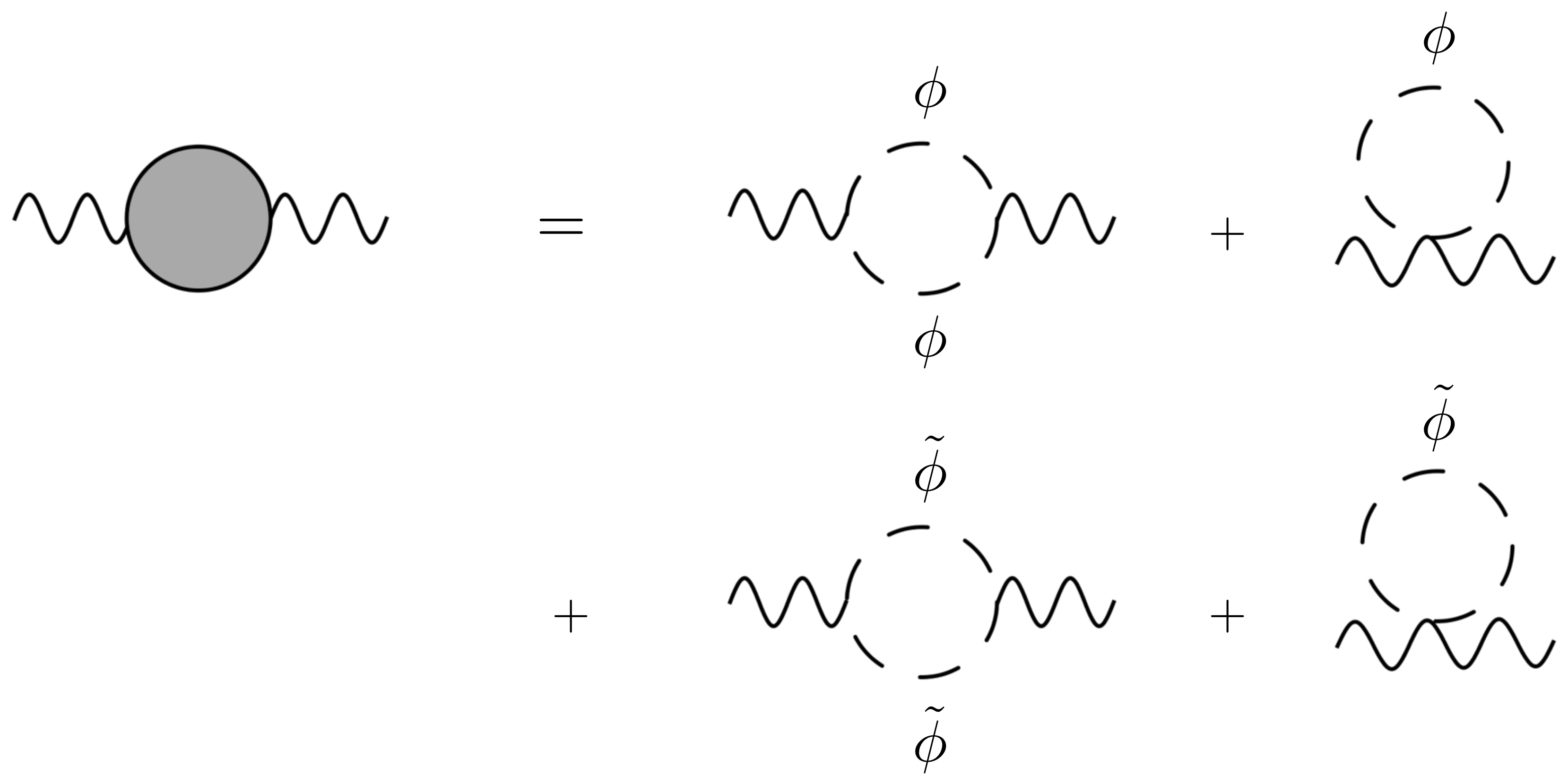}
\end{center}
\caption{One-loop contribution to the vacuum polarization amplitudes. Each external vector field is either a photon or a LW~photon.}
\label{fig:VPA}
\end{figure}

We begin our examination of the infinite part of the divergent 1PI diagrams of LW~scalar QED by computing the one-loop contributions to the vacuum polarization amplitudes for the vector fields.  The relevant diagrams are illustrated in Fig.~\ref{fig:VPA}, where each external field is either a photon or a LW~photon. Since no gauge field propagators are involved in the one-loop diagrams, the results are manifestly gauge independent. We find
\begin{equation}
i\Pi_{AA}^{\mu\nu}=i \Pi_{\tilde{A}\tilde{A}}^{\mu\nu}=-i\Pi_{A\tilde{A}}^{\mu\nu} =i\Pi(q^2)\,(q^2 g^{\mu\nu}-q^\mu q^\nu) \ ,
\label{eq:VPA}
\end{equation}
where, in dimensional regularization,
\begin{equation}
\Pi(q^2)= - 2\times \frac{e^2}{48\pi^2}\, \frac{1}{\epsilon}\, +\, \text{finite\ terms} \ ,
\label{eq:Pi}
\end{equation}
with $\epsilon\equiv 2-d/2$ as usual.  The explicit factor of two arises from the presence of the LW~scalar loops, and the remaining factor is the ordinary scalar QED contribution. Since $\Pi_{\tilde{A}\tilde{A}}^{\mu\nu}$ contains no mass term, we have
\begin{equation}
\delta_{M_A^2} = 0 \ .
\end{equation}
The relevant counterterm contributions from the field-strength terms in Eq.~(\ref{eq:Lct}) are
\begin{equation}
i\delta\Pi_{AA}^{\mu\nu}=i \delta\Pi_{\tilde{A}\tilde{A}}^{\mu\nu}
=-i\delta\Pi_{A\tilde{A}}^{\mu\nu} =-i \delta_3\, (q^2 g^{\mu\nu}-q^\mu q^\nu) \ ,
\end{equation}
which are precisely of the form required to cancel the infinities in Eq.~(\ref{eq:VPA}). In the minimal subtraction scheme we obtain
\begin{equation}
\delta_3 = - \frac{e^2}{24\pi^2}\, \frac{1}{\epsilon}\ .
\label{eq:delta3}
\end{equation}

\subsection{Self-Energy Amplitudes}

\begin{figure}
\begin{center}
\includegraphics[width=0.9\textwidth]{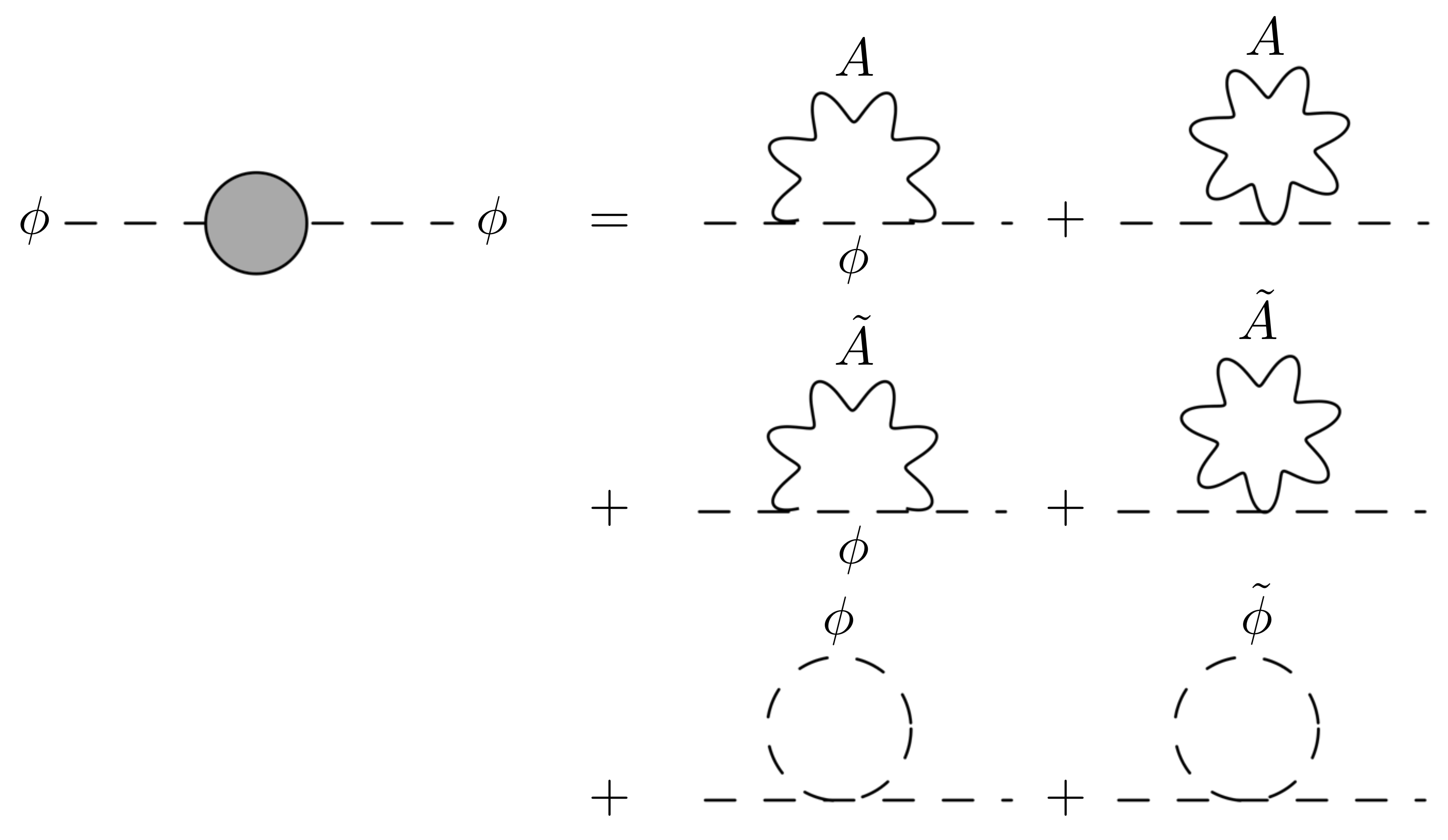}
\end{center}
\caption{One-loop contribution to the 1PI self-energy amplitude $\Sigma_{\phi\phi}$ in the no-mixing $\xi=1$ gauge. In the ordinary gauge there are also diagrams involving internal mixed gauge propagators, $P_{A\tilde{A}}$.}
\label{fig:LWself}
\end{figure}

We will calculate the one-loop contribution to the self-energy amplitude $\Sigma$ for a scalar field in the no-mixing $\xi=1$ gauge and then will repeat the calculation in the ordinary $\xi=1$ gauge as a check. The relevant diagrams for the $\phi$ field, in the no-mixing gauge, are shown in Fig.~\ref{fig:LWself}. Those for the $\tilde{\phi}$ field are obtained by replacing $\phi$ with $\tilde{\phi}$; given the form of the Lagrangian (\ref{eq:Ltwofields2}), we expect that the contributions of the diagrams involving internal gauge bosons will change sign. The mixed self-energy amplitude $\Sigma_{\phi\tilde{\phi}}$ explicitly breaks the $U(1)\times U(1)$ symmetry to diagonal $U(1)$, and must vanish in the limit $\lambda\to 0$; therefore, only the diagrams with scalar loops will contribute to $\Sigma_{\phi\tilde{\phi}}$.

We begin our calculation of $\Sigma_{\phi\phi}$ in the no-mixing $\xi=1$ gauge by considering potential quadratic divergences.  First, we examine the gauge-scalar diagrams on the top and middle lines of Fig.~\ref{fig:LWself}. The first two diagrams correspond to the gauge-sector contribution in ordinary scalar QED, which is quadratically divergent. That quadratic divergence is canceled by the $\tilde{A}$ diagrams, as we now demonstrate. The $SO(1,1)$ symmetry acting on the vector fields guarantees that: (i) the gauge-gauge-scalar and gauge-gauge-scalar-scalar couplings involving the photon and LW~photon are identical (up to an unphysical minus sign in the gauge-gauge-scalar coupling\footnote{This minus sign comes from the $A_\mu-\tilde{A}_\mu$ dependence. However, we can always redefine $\tilde{A}_\mu$ to $-\tilde{A}_\mu$, which turns $A_\mu-\tilde{A}_\mu$ into $A_\mu+\tilde{A}_\mu$. This, for example, is the convention adopted in Ref.~\cite{Grinstein:2008uk}.}), and (ii) the LW~photon propagator has a minus sign, relative to the photon propagator. As a result, each diagram with an internal LW~photon is opposite in sign to its counterpart with an ordinary photon, and in the UV (where the LW~photon mass becomes irrelevant), there is an exact cancellation of the quadratic divergences.  Likewise, moving to the diagrams in the bottom row of Fig.~\ref{fig:LWself}, we recognize that the first diagram is familiar from the ordinary $\phi^4$ theory, and is of course quadratically divergent. The second diagram exactly cancels the quadratic divergence, as the $SO(1,1)$ symmetry acting on the scalar fields guarantees the equality of the $|\phi|^4$ and $|\phi|^2|\tilde{\phi}|^2$ couplings, as well as the negative sign in the $\tilde{\phi}$ propagator.

Having established that $\Sigma$ is free from quadratic divergences, we may proceed to complete the one-loop calculation in no-mixing $\xi=1$ gauge. In dimensional regularization, near $d=4$, the result is
\begin{equation}\label{eq:self1} \begin{split}
-i\Sigma_{\phi\phi} = \ & -i \, \frac{\lambda(M^2-m^2)+3 g^2 M_A^2}{16\pi^2}\, \frac{1}{\epsilon}\, +\, \text{finite\ terms}\ , \\
-i\Sigma_{\tilde{\phi}\tilde{\phi}} = \ &  
-i\, \frac{\lambda(M^2-m^2)-3 g^2 M_A^2}{16\pi^2}\, \frac{1}{\epsilon} \, +\, \text{finite\ terms}\ , \\
-i\Sigma_{\phi\tilde{\phi}} = \ & i\, \frac{\lambda(M^2-m^2)}{16\pi^2}\, \frac{1}{\epsilon} \, +\, \text{finite\ terms} \ .
\end{split} \end{equation}
Notice that there is mass renormalization but not wavefunction renormalization, as the $1/\epsilon$ coefficients are $q^2$ independent in this gauge.  Moreover, in the limit of exact $SO(1,1)\times SO(1,1)$ symmetry (where $M\to m$ and $M_A\to 0$) the self-energy amplitudes vanish exactly in the no-mixing $\xi = 1$ gauge, because the LW~and ordinary propagators are then of equal magnitude and opposite sign.  The theory is still not finite, because, for example, the vacuum polarization amplitudes do not vanish.

From  Eq.~(\ref{eq:Lct}) we find that the relevant counterterm contributions are of the form
\begin{equation}\label{eq:self1ct} \begin{split}
-i \delta\Sigma_{\phi\phi} = \ & i \delta_2\, q^2 - i \delta_{m^2}\ , \\
-i \delta\Sigma_{\tilde{\phi}\tilde{\phi}} = \ & - i \delta_2\, q^2 + i \delta_{M^2}\ , \\
-i \delta\Sigma_{\phi\tilde{\phi}} = \ & - i \delta_{mM}\ ,
\end{split} \end{equation}
and in the minimal subtraction scheme, we conclude
\begin{equation}
\delta_2 = 0 \ ,
\label{eq:delta2higher}
\end{equation}
and
\begin{equation}\label{eq:deltam} \begin{split}
\delta_{m^2} = \ & -\frac{\lambda(M^2-m^2)+3 g^2 M_A^2}{16\pi^2}\, \frac{1}{\epsilon} \ , \\
\delta_{M^2} = \ & \frac{\lambda(M^2-m^2)-3 g^2 M_A^2}{16\pi^2}\, \frac{1}{\epsilon} \ , \\
\delta_{mM} = \ & \frac{\lambda(M^2-m^2)}{16\pi^2}\, \frac{1}{\epsilon} \ .
\end{split} \end{equation}
Inserting these results into Eq.~(\ref{eq:ctrel}) yields the following expression for the mixing angle
\begin{equation}
\eta=-\frac{\lambda}{16\pi^2}\, \frac{1}{\epsilon}  \ .
\label{eq:varepsilon}
\end{equation}
Note that this vanishes in the $\lambda\to 0$ limit, as expected:  for $\lambda\to 0$ the theory acquires a global $U(1)\times U(1)$ symmetry (with the diagonal $U(1)$ gauged) under which $\phi$ and $\tilde{\phi}$ rotate independently. This prevents $\phi$-$\tilde{\phi}$ mixing terms from being radiatively generated.

The calculation in the ordinary $\xi=1$ gauge proceeds somewhat differently because mixed gauge propagators ($P_{A\tilde{A}}$) are present.  Once their effects are included, the quadratic divergences still cancel among the diagrams involving internal gauge propagators.   A direct computation of the self-energy functions then yields
\begin{equation}\label{eq:self2} \begin{split}
-i\Sigma_{\phi\phi} = \ & 
-i\, \frac{\lambda(M^2-m^2)+g^2(3 M_A^2 + m^2 - q^2)}{16\pi^2}\,\frac{1}{\epsilon} \, +\, \text{finite\ terms}\ , \\
-i \Sigma_{\tilde{\phi}\tilde{\phi}} = \ & 
-i\, \frac{\lambda(M^2-m^2)-g^2(3 M_A^2 + M^2 - q^2)}{16\pi^2}\, \frac{1}{\epsilon} \, +\, \text{finite\ terms}\ , \\
-i \Sigma_{\phi\tilde{\phi}} = \ & i\, \frac{\lambda(M^2-m^2)}{16\pi^2}\, \frac{1}{\epsilon} \, +\, \text{finite\ terms} \ .
\end{split} \end{equation}
In this gauge, both mass renormalization and wavefunction renormalization are present. The counterterm contributions are, of course, still given by Eq.~(\ref{eq:self1ct}), which have the right form to cancel both the $q^2$-dependent and $q^2$-independent infinities in Eq.~(\ref{eq:self2}). In minimal subtraction scheme, one obtains the relationships
\begin{equation}\begin{split}
\delta_{m^2} = \ & -\frac{\lambda(M^2-m^2)+g^2(3 M_A^2+m^2)}{16\pi^2}\, \frac{1}{\epsilon} \ , \\
\delta_{M^2} = \ & \frac{\lambda(M^2-m^2)-g^2(3M_A^2+M^2)}{16\pi^2}\, \frac{1}{\epsilon} \ , \\
\delta_{mM} = \ & \frac{\lambda(M^2-m^2)}{16\pi^2}\, \frac{1}{\epsilon} \ ,
\end{split} \end{equation}
which lead to the same expression for the mixing angle as in the no-mixing $\xi=1$ gauge, Eq.~(\ref{eq:varepsilon}). However, this time the wavefunction renormalization counterterm is
\begin{equation}
\delta_2 = -\frac{g^2}{16\pi^2}\, \frac{1}{\epsilon} \ ,
\label{eq:delta2lower}
\end{equation}
which differs from the result in the other gauge.

Since universality of the $U(1)$ gauge coupling insures that the scalar field wavefunction renormalizations are always exactly compensated by the vertex corrections, as Eq.~(\ref{eq:ward}) shows explicitly, all that remains to show at one-loop, is the cancellation of all $SO(1,1)$ breaking amplitudes in the $\phi^4$ sector.

\subsection{$\phi^4$ Vertex}

There are many diagrams contributing to the 1PI amplitudes with four external scalar fields. As with the self-energy amplitudes, there are significant $SO(1,1)\times SO(1,1)$ cancellations involved. However, now that these are well understood, we can reduce the number of diagrams by using the hat-field propagators on all internal lines:  the $\hat{A}$ propagator of Eq.~(\ref{eq:hatprop}) when working in the ordinary $\xi=1$ gauge, the $\hat{A}$ propagator of Eq.~(\ref{eq:hatprophigher}) when employing the no-mixing $\xi=1$ gauge, and the gauge-independent $\hat{\phi}$ propagator  which is constructed by summing the simple propagators of the $\phi$ and $\tilde{\phi}$ fields
\begin{equation}
P_{\hat{\phi}\hat{\phi}} = P_{\phi\phi} + P_{\tilde{\phi}\tilde{\phi}} = \frac{i\,(M^2-m^2)}{(q^2-m^2)(M^2-q^2)} \ .
\end{equation}
When we use the hat-field propagators, the expected cancellations occur within single diagrams, and are due to denominators with higher powers of loop momenta. 

\begin{figure}[!t]
\begin{center}
\includegraphics[width=\textwidth]{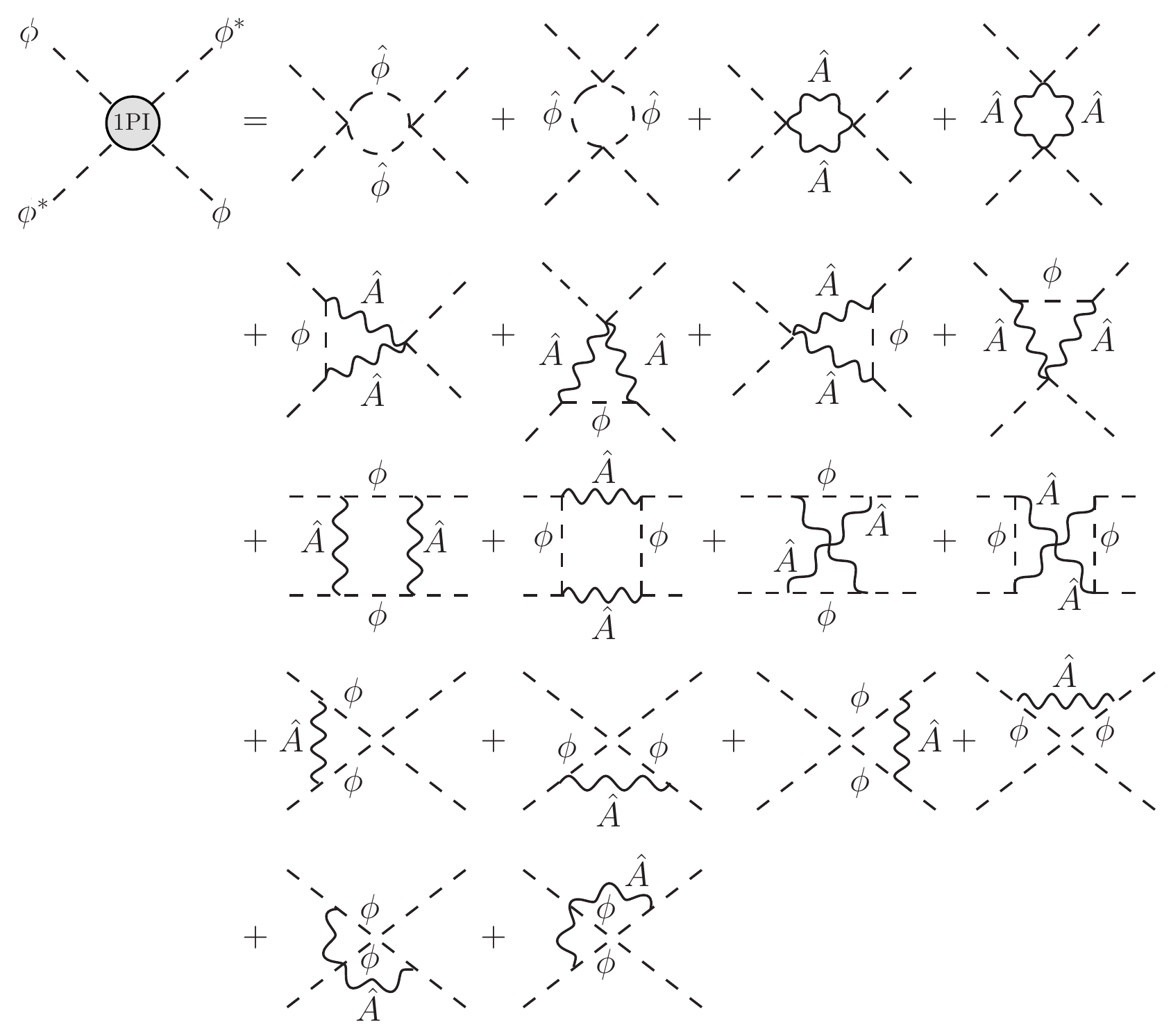}
\end{center}
\caption{One-loop contribution to the 1PI amplitude with four external $\phi$ fields. The number of diagrams is reduced by employing the hat-field propagators. (See text for details)}
\label{fig:phi4}
\end{figure}

Let us denote by $\Gamma_{f_1 f_2 f_3 f_4}$ the one-loop contribution to the 1PI amplitude with external scalar fields $f_1$, $f_2$, $f_3$, and $f_4$ and start by working within the no-mixing $\xi = 1$ gauge.  As a concrete example, we will consider $\Gamma_{\phi\phi\phi^\ast\phi^\ast}$, for which the relevant diagrams are those shown in Fig.~\ref{fig:phi4}. The first two diagrams are entirely due to the $\phi^4$ interaction, and involve only internal $\hat{\phi}$ fields. These diagrams are, thus, finite by power counting, since the $\hat{\phi}$ propagator decays like $q^{-4}$, at large momenta, and the vertices are momentum independent. The remaining two diagrams of the first row, as well as the diagrams of the second and third row, are entirely due to the gauge-scalar interactions, which depend on $A_\mu-\tilde{A}_\mu\equiv \hat{A}_\mu$. Although the trilinear gauge-scalar-scalar vertices are momentum-dependent, in the no-mixing gauge all these diagrams are finite by power counting. Finally, there are the diagrams in the last two rows, which involve both $\phi^4$ and gauge vertices. Once again these are finite by power counting in the no-mixing gauge. The same reasoning holds for all of the $U(1) \times U(1)$ symmetric amplitudes; for the  $U(1)\times U(1)$ violating amplitudes $\Gamma_{\phi\phi\phi^\ast\tilde{\phi}^\ast}$, $\Gamma_{\phi\tilde{\phi}\tilde{\phi}^\ast\tilde{\phi}^\ast}$, and $\Gamma_{\phi\phi\tilde{\phi}^\ast\tilde{\phi}^\ast}$, the only difference is that the diagrams involving only gauge-scalar vertices do not contribute.  We conclude that the amplitudes are purely finite in the no-mixing gauge and 
\begin{equation}
\delta_\lambda = 0 \ .
\label{eq:deltalambdahigher}
\end{equation}

The situation is different in the ordinary $\xi=1$ gauge.  If we, again, start by considering $\Gamma_{\phi\phi\phi^\ast\phi^\ast}$, we still conclude that the first two diagrams, which involve only  the $\phi^4$ interaction, and involve only internal $\hat{\phi}$ fields are finite.  The ten diagrams involving only gauge-scalar interactions are another story.  Power counting now  predicts a logarithmic divergence for each  of these diagrams, because of the $q^\mu q^\nu/M_A^2$ term in the $\hat{A}_\mu$ propagator.  However, we expect that these divergences must cancel against one another for symmetry reasons.  Recall that  the $U(1)\times U(1)$ violating amplitudes $\Gamma_{\phi\phi\phi^\ast\tilde{\phi}^\ast}$, $\Gamma_{\phi\tilde{\phi}\tilde{\phi}^\ast\tilde{\phi}^\ast}$, and $\Gamma_{\phi\phi\tilde{\phi}^\ast\tilde{\phi}^\ast}$ receive no contribution at all from the diagrams with only gauge vertices. Then those diagrams cannot make an infinite contribution to the $U(1)\times U(1)$-symmetric amplitudes like $\Gamma_{\phi\phi\phi^\ast\phi^\ast}$ either, since this would correspond to a hard breaking of the $SO(1,1)$ symmetry acting on the scalar fields. Explicit calculation confirms that the infinities arising from the diagrams with an even number of gauge vertices (last two diagrams of the first row, and diagrams of the third row) are precisely cancelled by the infinities from the diagrams with an odd number of gauge vertices (diagrams of the second row).  Therefore, the only possible infinite contribution to $\Gamma_{\phi\phi\phi^\ast\phi^\ast}$ and the other 1PI amplitudes in this gauge must arise from the diagrams in the last two rows, which involve both $\phi^4$ and gauge-scalar vertices.  Note that this implies that in the $\lambda\to 0$ limit all amplitudes with four external scalar fields are finite.
 
Computing the $\phi^4$ vertex correction diagrams from the last two rows of  Fig.~\ref{fig:phi4} in ordinary $\xi = 1$ gauge yields the following infinite contributions
\begin{equation}\label{eq:1PI} \begin{split}
&i\Gamma_{\phi\phi\phi^\ast\phi^\ast} = i X + \text{finite}, 
\ i \Gamma_{\phi\tilde{\phi}\phi^\ast\tilde{\phi}^\ast} =  i\,4 X + \text{finite} , \ 
i \Gamma_{\tilde{\phi}\tilde{\phi}\tilde{\phi}^\ast\tilde{\phi}^\ast} = i X + \text{finite} , \\
&i \Gamma_{\phi\phi\phi^\ast\tilde{\phi}^\ast} = - i\, 2 X + \text{finite} , \ 
i\Gamma_{\phi\tilde{\phi}\tilde{\phi}^\ast\tilde{\phi}^\ast} = - i\, 2 X + \text{finite} , \ 
i \Gamma_{\phi\phi\tilde{\phi}^\ast\tilde{\phi}^\ast} = iX + \text{finite},
\end{split} \end{equation}
where
\begin{equation}
X \equiv  -\frac{\lambda\,e^2}{8\pi^2}\, \frac{1}{\epsilon} \ .
\end{equation}
\\%
At the same time, Eq.~(\ref{eq:Lct}) provides the counterterm contributions
\begin{equation}\label{eq:1PIct} \begin{split}
&i \delta\Gamma_{\phi\phi\phi^\ast\phi^\ast} = -i\delta_\lambda \ , 
\quad i \delta\Gamma_{\phi\tilde{\phi}\phi^\ast\tilde{\phi}^\ast} = - i\, 4 \delta_\lambda \ , \quad
i\delta\Gamma_{\tilde{\phi}\tilde{\phi}\tilde{\phi}^\ast\tilde{\phi}^\ast} = - i\delta_\lambda  \ , \\
&i \delta\Gamma_{\phi\phi\phi^\ast\tilde{\phi}^\ast} =  i\, 2 \delta_\lambda  \ , \quad 
i\delta\Gamma_{\phi\tilde{\phi}\tilde{\phi}^\ast\tilde{\phi}^\ast} =  i\, 2  \delta_\lambda  \ , \quad
i \delta\Gamma_{\phi\phi\tilde{\phi}^\ast\tilde{\phi}^\ast} = - i \delta_\lambda  \ ,
\end{split} \end{equation}
which are precisely of the form required to cancel the infinities in Eq.~(\ref{eq:1PI}).  In the minimal subtraction scheme we obtain
\begin{equation}
\delta_\lambda = - \frac{\lambda\, e^2}{8\pi^2}\, \frac{1}{\epsilon} \ .
\label{eq:deltalambdalower}
\end{equation}

\subsection{Running of $g$ and $\lambda$}

We will now determine the $\beta$ functions of the LW~theory and compare them with the results for ordinary scalar QED. To lowest order, the $\beta$ functions for the electromagnetic and $\phi^4$ couplings in the LW~theory are given by
\begin{equation}
\beta_g = \frac{g^2}{2}\,\mu\frac{\partial}{\partial\mu}\delta_3 \ , \qquad\quad 
\beta_\lambda = \mu\frac{\partial}{\partial\mu}\left(-\delta_\lambda+2 \lambda\, \delta_2\right) \ ,
\end{equation}
where $\mu$ is the scale we must introduce in dimensional regularization to make the log arguments dimensionless, and $1/\epsilon$ in the counterterms is interpreted as
\begin{equation}
\frac{1}{\epsilon} \rightarrow \log\frac{\Lambda^2}{\mu^2} \ .
\end{equation}

Since we are employing a mass-independent renormalization scheme, below the LW~mass scale ($M_\text{LW}$) we must impose the decoupling theorem and integrate out the LW~fields.  Since what remains is identical to ordinary scalar QED, the counterterms are
\begin{equation}
\delta_3 = - \frac{g^2}{48\pi^2}\, \frac{1}{\epsilon}\ , \quad\quad \delta_2 = \frac{g^2}{8\pi^2}\,\frac{1}{\epsilon} \ , \quad\quad
\delta_\lambda  = \left[\frac{\lambda^2}{8\pi^2} - \frac{\lambda\, g^2}{8\pi^2}\right] \frac{1}{\epsilon} \, .
\end{equation}
These, in turn, yield the low-energy leading order $\beta$ functions
\begin{equation}
\beta_g (\mu < M_\text{LW}) = \frac{g^3}{48\pi^2} \ , \quad\quad  \beta_\lambda (\mu < M_\text{LW}) = \frac{\lambda^2}{4\pi^2} - \frac{3\lambda\, g^2}{4\pi^2}\ ,
\end{equation}
that are characteristic of ordinary scalar QED.

Above the LW~mass scale, the appropriate counterterms are those we have derived for the full LW~theory.   For the vector coupling, the counterterm value in Eq.~(\ref{eq:delta3}) leads to
\begin{equation}
\beta_g (\mu > M_\text{LW}) = \frac{g^3}{24\pi^2} \ ,
\end{equation}
which is twice the ordinary scalar QED $\beta_g$ function. In other words, the contribution from loops of the LW~scalar is identical to that from loops of the ordinary scalar; since there are no internal gauge fields, the calculation is manifestly gauge invariant.  

For the $\beta_\lambda$ function above the LW~scale we consider the no-mixing and ordinary gauges separately. If we employ the no-mixing $\xi=1$ gauge, then  Eqs.~(\ref{eq:delta2higher}) and (\ref{eq:deltalambdahigher}) tell us that $\delta_2$ and $\delta_\lambda$ are each separately zero. In this gauge the LW~scalar and vector fields make contributions to the counterterms that are equal and opposite to those of the ordinary scalar and vector fields.  As a result, we find
\begin{equation}
\beta_\lambda (\mu > M_\text{LW}) = 0 \ .
\end{equation}
In the ordinary $\xi=1$ gauge, the values of $\delta_2$ are $\delta_\lambda$ are non-zero, as shown, respectively, by Eqs.~(\ref{eq:delta2lower}) and (\ref{eq:deltalambdalower}); however, the final result for $\beta_\lambda$ is the same, which provides a useful check of our calculations.

\section{Conclusion \& Discussion}\label{sec:concl}

In this chapter we have discussed the global symmetries and the renormalizability of Lee-Wick scalar QED. The combination of $SO(1,1)$ global symmetry, $U(1)$ gauge invariance, and an $SO(1,1)$ invariant gauge-fixing condition allows us to show directly in the auxiliary-field formalism that the number of superficially divergent amplitudes in a LW~Abelian gauge theory is finite. To illustrate the renormalizability of the theory, we have explicitly carried out the one-loop renormalization program in LW~scalar QED and demonstrated how the counterterms are constrained by the joint conditions of gauge and $SO(1,1)$ invariance. We have also computed the one-loop beta functions in LW~scalar QED.

It would be interesting to generalize the discussion presented here to the case of non-Abelian gauge theories. However, this is not immediately possible. Notice that the $SO(1,1)$ transformation of Eq.~\eqref{eq:SOgauge} mixes a gauge field, $A_\mu$, with a non-gauge vector field, $\tilde{A}_\mu$. In an Abelian theory, we have the freedom to promote $A^\prime_\mu \equiv \cosh \alpha A_\mu + \sinh \alpha \tilde{A}_\mu$ to a gauge field for two reasons. First, the requirement that $A^\prime_\mu - \tilde{A}^\prime_\mu$ transform like a gauge field gives us the freedom to choose which field should bear the transformation. Second, all gauge interactions depend solely on $e(A_\mu - \tilde{A}_\mu)$, which is an $SO(1,1)$ invariant. That these conditions are satisfied in Abelian gauge theory is perhaps not surprising, given that a massive Abelian gauge theory is renormalizable \cite{Kroll:1967zt}.

In a non-Abelian gauge theory, however, interactions do not depend solely on $g(A^a_\mu - \tilde{A}^a_\mu)$, and the $SO(1,1)$ symmetry is violated by the gauge interactions themselves. To see this consider the generalization of the gauge kinetic energy terms of Eq.~\eqref{eq:lower1} to non-Abelian interactions,
\begin{equation}
{\cal L}_\text{gauge} = -\frac{1}{2}\, \text{Tr}\, \hat{F}_{\mu\nu}^2 -2\, \text{Tr}\,  D^\mu\tilde{A}^\nu \hat{F}_{\mu\nu} \ ,
\end{equation}
where
\begin{equation}
D^\mu\tilde{A}^\nu = \partial^\mu \tilde{A}^\nu - i g [\hat{A}^\mu,\tilde{A}^\nu] \ .
\end{equation}
An $SO(1,1)$ transformation on the hat and tilde fields, and the gauge coupling, reads
\begin{equation}
\hat{A}_\mu \to e^{-\alpha}\, \hat{A}_\mu \ , \quad \tilde{A}_\mu \to \sinh\alpha \, \hat{A}_\mu + e^\alpha\, \tilde{A}_\mu
\ ,\quad g \to e^\alpha\, g \ .
\end{equation} 
Applying this to ${\cal L}_\text{gauge}$ gives
\begin{equation}
{\cal L}_\text{gauge} \to {\cal L}_\text{gauge} + ig\, \sinh\alpha \, e^{-\alpha}\, \text{Tr}\, \hat{F}^{\mu\nu}[\hat{A}_\mu,\hat{A}_\nu] \ .
\label{eq:breaking}
\end{equation}
Thus the $SO(1,1)$ symmetry associated to the vector fields is explicitly broken by dimension-four gauge boson self-interactions.

In principle we would, therefore, expect the $SO(1,1)$ breaking to propagate to other sectors of the theory, and spoil the special relations between couplings that guarantee the cancellation of quadratic divergences. However, both power counting in the higher-derivative formulation \cite{Grinstein:2008uk} and the high energy behavior of massive vector meson scattering in Lee-Wick gauge theory \cite{Grinstein:2008fu} suggest that the number of superficially divergent diagrams remains finite and that non-Abelian LW~gauge theory may be renormalizable. A more thorough understanding of non-Abelian LW~gauge theories is, therefore, necessary in order to extend the results demonstrated here for Abelian theories.

As we have seen, Part~I of this Thesis has focused on an extension of the electroweak sector of the SM in which new (heavy) fermions, among other degrees of freedom, are introduced in order to tackle and resolve the Hierarchy Problem, canceling the quadratic divergences associated with the SM Higgs boson's mass. We investigated in detail some of the interesting and important theoretical aspects of this BSM theory, and analyzed how current experimental data constrains the new heavy degrees of freedom. In Part~II of this Thesis, we proceed to discuss a different type of BSM theory and some of its thought-provoking field theoretical issues. This type of BSM theory would form an extension to the strong sector of the SM, one in which new (heavy) colored gauge bosons (rather than new fermions) appear. Once more, we will carefully explore both the formal and phenomenological aspects of this BSM theory in detail.

\newpage
\vspace*{\fill}
\begin{center}
\Huge \textbf{PART II}
\end{center}
\vfill

\chapter{PRODUCTION OF MASSIVE COLOR-OCTET VECTOR BOSONS AT NEXT-TO-LEADING ORDER\protect \footnote{This chapter is based on the paper first published in \cite{Chivukula:2012fk}.}}\label{coloron}

\begin{quote}
``\textit{Everything should be made as simple as possible, but no simpler!}"
\begin{flushright}
|Albert Einstein (1879 -- 1955)\\
\end{flushright}
\end{quote}

\section{Introduction}

\lettrine[lines=1]{M}{assive color-octet vector bosons} are present in theories which constitute an extension of the QCD sector of the Standard Model~(SM). As explained in Chapter~1, since late 1980s it has been of theoretical and phenomenological interest to extend the SM strong sector in order to accommodate, among others, theories in which the electroweak symmetry breaking is induced by strong dynamics, where a new type of strongly-coupled gauge interaction forms composite Higgs out of colored fermions. Just as in the case of ordinary QCD, one may apply perturbation theory to study this class of theories at high energies. This may be again achieved through an asymptotic expansion of the theory in terms of the strong coupling, $\alpha_{s}$,\footnote{In previous chapters, we defined $\alpha_{3}$ to indicate the strong coupling. In this chapter, however, we employ the more conventional term $\alpha_{s}$. Both conventions are interchangeably used in the literature.} since the coupling decreases in strength as a function of increasing energy, reflecting the asymptotically free nature of the theory.

The analyses of the production of the massive colored gauge bosons to date, however, capture only the leading order (LO) in perturbation theory. In this chapter, we extend the production analysis to the next-to-leading order (NLO) in perturbation theory, thereby, improving dramatically upon the previous LO studies, in addition to predicting new kinematic variables important for comparison with experiments.

Massive color-octet vector bosons are predicted in a variety of models, including axigluon models \cite{Frampton:1987wo,Bagger:1988le}, topcolor models \cite{Hill:1991tx}\nocite{Hill:1994lk,Popovic:1998bq}-\cite{Braam:2008bx},
technicolor models with colored technifermions \cite[p.~352-382]{Chivukulabook}, flavor-universal \cite{Chivukula:1996qp,Simmons:1997hh} and
chiral \cite{Martynov:2009sy} coloron models, and extra-dimensional models with KK gluons \cite{Lillie:2007ca,Davoudiasl:2001qd}. These states have also recently been considered
as a potential source  \cite{Ferrario:2009jl,Frampton:2010gb} of the top quark forward-backward asymmetry observed by the CDF collaboration \cite{Aaltonen:2008mb,Aaltonen:2011cq}.\footnote{Note, however, that the observation of a top quark forward-backward asymmetry is not confirmed by results of the D0 collaboration \protect\cite{Abazov:2008kh,Abazov:2011rq}. Furthermore, if the observed top quark forward-backward asymmetry is confirmed, explaining this using color-octet vector bosons is problematic given the tight constraints on flavor-changing neutral-currents \protect\cite{Chivukula:2010fc}.}
Recent searches for resonances in the dijet mass spectrum at the LHC imply that the lower bound on such a boson is now 2-3 TeV \cite{Chatrchyan:2011bd},\cite{Han:2010qc}\nocite{Haisch:2011mw}-\cite{Aad:2012ud}.\footnote{At least for the fermion charge assignments considered, and in the case where the resonance is narrow compared to the djiet mass resolution of the detector.} If there are color-octet vector bosons associated with the electroweak symmetry breaking sector, as suggested by several of  the models discussed above, their presence should be uncovered by the LHC in the future.

In this chapter\footnote{Throughout this chapter, the timeline for the depicted Feynman diagrams is from bottom to top, with exception of the fermion self-energy (Fig.~\ref{fig:self}) and the gauge boson VPA (Figs.~\ref{fig:CC}-\ref{fig:CCGCfermions}) diagrams where it is from left to right.} we report the first complete calculation\footnote{As this work was being completed, a computation of
the NLO virtual corrections of top quark pair production via a heavy color-octet vector boson has been reported in \protect\cite{Zhu:2011if}. That work is complementary to ours in that it does not employ the narrow width approximation for the color-octet boson, but neither does it include real gluon or quark emission. After this work was submitted for publication,
real emission has also been considered by those authors \cite{Zhu:2012kc}.} of QCD corrections to the production of a massive color-octet vector boson. We will refer to these massive color-octet vector states generically as ``colorons". We treat the coloron as an asymptotic state in our calculations, employing the narrow width approximation. Our next-to-leading order (NLO) calculation includes both virtual corrections, as well as corrections arising from the emission of gluons and light quarks, and we demonstrate the reduction in factorization-scale dependence relative to the leading order (LO)
approximation used in previous hadron collider studies.

The QCD NLO calculation of coloron production reported here differs substantially from the classic computation of the QCD NLO
corrections to Drell-Yan production \cite{Altarelli:1979hs}, because the final state is colored. In particular, Drell-Yan
production involves the coupling of the light quarks to a conserved (or, in the case of $W$- or $Z$-mediated processes,
conserved up to quark masses) current. Hence, in computing the NLO corrections to Drell-Yan processes, the current conservation Ward identity insures a cancellation between the UV divergences arising from virtual
quark wavefunction and vertex corrections. These cancellations do not occur in the calculation of the NLO corrections to coloron production, because of vertex corrections involving the three-point non-Abelian colored-boson vertices. As we describe in Section \ref{sec:virual}, we use the ``pinch technique" \cite{Binosi:2009lo} to divide the problematic non-Abelian vertex corrections into two pieces | a ``pinched" piece whose UV divergence contributes to the renormalization of the coloron wavefunction (and, ultimately, a renormalization of the coloron coupling) and an ``unpinched" part whose UV divergence (when combined with an Abelian vertex correction)
cancels against the UV divergences in quark wavefunction renormalization. As we show, once the UV divergences are properly accounted for, the IR divergences cancel in the usual way: the IR divergences arising from real quark or gluon emission cancel against the IR divergences in the virtual corrections, and the IR divergences arising from collinear quarks or gluons in the initial state are absorbed in the properly defined parton distribution functions (PDFs).

We compute the gauge-, quark-, and self-couplings of the coloron from a theory with
an extended $SU(3)_{1C} \times SU(3)_{2C} \to SU(3)_C$ gauge structure, where $SU(3)_C$ is identified with QCD. The calculation yields
the minimal coupling of gluons to colorons, and
allows for the most general couplings of quarks to colorons.  The cancellation of UV divergences described above, however,
occurs only when the three-coloron coupling has the strength that arises from the dimension-four gauge-kinetic energy terms of
the extended $SU(3)_{1C} \times SU(3)_{2C}$ gauge structure. Our computation applies directly
to any theory with this structure, i.e. to massive color-octet vector bosons in axigluon, topcolor, and coloron models.
In general, the triple coupling of KK gluons in extra-dimensional models, or of colored technivector mesons in
technicolor models, will not follow this pattern. However, our results apply approximately to these cases
as well, to the extent that the $SU(3)_{1C} \times SU(3)_{2C}$
model is a good low-energy effective theory for the extra-dimensional model (a ``two-site" approximation in
the language of deconstruction \cite{Arkani-Hamed:2001sw,Hill:2001fh}) or for the technicolor theory (a hidden local symmetry approximation for
the effective technivector meson sector \cite{Bando:1985pr,Bando:1985sh}).\footnote{Arbitary three- and four-point coloron self-couplings
can be incorporated in the $SU(3)_{1C} \times SU(3)_{2C}$ by adding ${\cal O}(p^4)$ terms in the effective chiral Lagrangian of Eq.~(\protect\ref{eq:L}), and deviations in these couplings are, therefore, of ${\cal O}(M^2_C/\Lambda^2)$, where $\Lambda$ is the cutoff of the effective coloron theory. The three- and four-point self-couplings, however, are neither relevant to the leading order $q\bar{q}$, nor to
the IR divergent NLO coloron production contributions, and, therefore, are numerically insignificant.}

This chapter is structured as follows: in Sec.~\ref{formalsec} we introduce the formalism of a minimal vector coloron theory, deriving all the Feynman rules, and setting the stage for the subsequent calculations. In Sec.~\ref{LOsec} we review the leading order computations of the amplitude and cross section for coloron production due to $q\bar{q}$ pair annihilation. Sec.~\ref{sec:virual} describes in detail the one-loop virtual corrections to the $q\bar{q}$ pair annihilation process, elaborating on the contributions from the quark self-energy, coloron-coloron and gluon-coloron mixed vacuum polarization amplitudes, and the vertex corrections.
We employ the pinch technique \cite{Binosi:2009lo}, described above, in order to consistently treat the UV divergences, and obtain a gauge-invariant, mutually independent set of counterterms. The one-loop cross section is constructed, and the IR singularities of the virtual correction properly extracted. In Sec.~\ref{sec:real} we consider the real emission processes, consisting of real (soft and collinear) gluon and (collinear) quark emission. In Sec.~\ref{sec:NLOcs} we put all the pieces together, exhibiting the explicit cancellation of the IR divergences among the real and virtual corrections, and demonstrate the renormalization of the quark and gluon PDFs. We give a finite expression for the NLO-corrected production cross section.
Finally, in Sec.~\ref{sec:conclusion} we plot the cross section, demonstrate that the QCD NLO corrections are
as large as 30\%, and show that the residual factorization-scale dependence is at the 2\% level.
We also calculate the $K$-factor and the $p_T$ spectrum for coloron production, since these are
valuable for comparison with experiment. Appendix~\ref{FR} contains all the Feynman rules of the coloron theory.

\section{A Minimal Theory for Spin-One Colorons} \label{formalsec}

In this section, we introduce colorons\footnote{Colorons can, in principle, be introduced as matter fields in the adjoint of $SU(3)_C$. This approach, however, would lead to an early violation of tree-level unitarity, as the scattering amplitude of longitudinally polarized massive spin-one bosons can grow, by power counting, like $E^4$, where $E$ is the center-of-mass (CM) energy. The only way to avoid this is to ``promote'' the coloron to the status of gauge field of a spontaneously broken gauge theory: then the special relation between trilinear and quartic gauge couplings will lead to an exact cancellation of the terms growing like $E^4$, as happens in the standard electroweak theory.} as the massive color-octet bosons arising when an extended  $SU(3)_{1C}\times SU(3)_{2C}$ gauge symmetry is spontaneously broken by a non-linear sigma model field to its diagonal subgroup, $SU(3)_C$, which we identify with QCD. The symmetry breaking results in a low-energy spectrum that includes both a massless spin-one color-octet of gauge bosons, the gluons, and a massive spin-one color-octet of gauge bosons, the colorons.

In detail, we replace the QCD Lagrangian with
\begin{equation}\label{eq:L} \begin{split}
{\cal L}_\text{color} = \ & - \frac{1}{4}G_{1\mu\nu}G_1^{\mu\nu} - \frac{1}{4}G_{2\mu\nu}G_2^{\mu\nu}
+\frac{f^2}{4}\ \text{Tr} \, D_\mu\Sigma \,D^\mu\Sigma^\dagger \\
&+ {\cal L}_\text{gauge-fixing} + {\cal L}_\text{ghost} + {\cal L}_\text{quark} \ .
\end{split} \end{equation}
Here, $\Sigma$ is the non-linear sigma field breaking $SU(3)_{1C}\times SU(3)_{2C}$ to $SU(3)_C$,
\begin{equation}
\Sigma = \exp\left(\frac{2 i \pi^a t^a}{f}\right) \ , \quad a=1,\dots, 8 \ ,
\end{equation}
where $\pi^a$ are the Nambu-Goldstone bosons ``eaten'' by the coloron, $f$ is corresponding ``decay-constant", and $t^a$ are the Gell-Mann matrices, normalized as $\text{Tr}\,t^a t^b=\delta^{ab}/2$. The $\Sigma$ field transforms as the bi-fundamental of $SU(3)_{1C}\times SU(3)_{2C}$,
\begin{equation}
\Sigma\to u_1 \Sigma \, u_2^\dagger \ , \quad u_i=\exp\left(i \alpha_i^a t^a\right) \ ,
\end{equation}
where the $\alpha_i^a$ are the parameters of the $SU(3)_{iC}$ transformations. This leads to the covariant derivative
\begin{equation}
D_\mu \Sigma = \partial_\mu \Sigma - i g_{s_1} G^a_{1\mu} t^a \Sigma + i g_{s_2} \Sigma\,  G^a_{2\mu} t^a \ ,
\end{equation}
where $g_{s_i}$ is the gauge coupling of the $SU(3)_{iC}$ gauge group. Up to a total divergence, the quadratic terms in the Lagrangian are
\begin{equation} \begin{split}
{\cal L}^{(2)}_\text{color} = \ & \frac{1}{2} G^a_{i\mu}\left(g^{\mu\nu}\partial^2-\partial^\mu\partial^\nu\right)G^a_{i\nu}
+\frac{f^2}{8}\left(g_{s_1} G^a_{1\mu}-g_{s_2} G^a_{2\mu}\right)^2+\frac{1}{2}\left(\partial_\mu\pi^a\right)^2 \\
&- \frac{f}{2}\left(g_{s_1} G^a_{1\mu}-g_{s_2} G^a_{2\mu}\right)\partial^\mu\pi^a + {\cal L}^{(2)}_\text{gauge-fixing} + {\cal L}^{(2)}_\text{ghost} + {\cal L}^{(2)}_\text{quark} \ ,
\end{split} \end{equation}
where a sum over $i=1,2$ in the gauge kinetic terms is implied.

The gauge-Goldstone mixing term can be removed, up to a total divergence, by choosing the gauge-fixing Lagrangian to be
\begin{equation}
{\cal L}_\text{gauge-fixing} = -\frac{1}{2} \left({\cal F}_i^a\right)^2 \ ,
\end{equation}
where the gauge-fixing functions are
\begin{equation}
{\cal F}_1^a\equiv \frac{1}{\sqrt{\xi}}\left(\partial^\mu G^a_{1\mu} + \xi \frac{g_{s_1} f}{2}\pi^a\right) \ , \quad
{\cal F}_2^a\equiv \frac{1}{\sqrt{\xi}}\left(\partial^\mu G^a_{2\mu} - \xi \frac{g_{s_2} f}{2}\pi^a\right) \ .
\end{equation}
The Faddeev-Popov ghost Lagrangian is obtained by taking the functional determinant of $\delta {\cal F}_i^a/\delta\alpha_j^b$. This leads to
\begin{equation} \begin{split}
{\cal L}_\text{ghost} = \ & \bar{c}_i^a \Big[-\partial^\mu\left(\delta_{ij}\delta^{ab}\partial_\mu
-g_{s_i} f^{abc} \delta_{ij} G_{i\mu}^c\right) \\
&\quad - \xi \frac{g_{s_i}^2 f^2}{4}\left(\delta_{i1}-\delta_{i2}\right)\left(\delta_{1j}-\delta_{2j}\right)
\delta^{ab} + {\cal O}(\pi)\Big] c_j^b \ ,
\end{split} \end{equation}
where $f^{abc}$ are the $SU(3)$ structure constants, and a sum over $i,j=1,2$ is implied. Notice that we have included only the inhomogeneous terms in the transformation of the eaten Goldstone boson, whence, the unspecified ${\cal O}(\pi)$ term in the ghost Lagrangian which are unnecessary for our computation. Up to a total divergence, the quadratic Lagrangian now reads
\begin{align}
{\cal L}^{(2)}_\text{color} = \ & \frac{1}{2} G^a_{i\mu}\Bigg[\delta_{ij} g^{\mu\nu} \partial^2
-\delta{ij}\left(1-\frac{1}{\xi}\right)\partial^\mu\partial^\nu + \frac{g_{s_i}^2 f^2}{4}
\left(\delta_{i1}-\delta_{i2}\right)\left(\delta_{1j}-\delta_{2j}\right)\Bigg] G^a_{j\nu} \notag \\
&- \frac{1}{2}\pi^a\Bigg[\partial^2+\frac{\xi}{4}\left(g_{s_1}^2+g_{s_2}^2\right)f^2\Bigg]\pi^a \label{eq:quadratic} \\
&- \bar{c}_i^a \Bigg[\delta_{ij}\partial^2 + \xi\frac{g_{s_i}^2 f^2}{4}
\left(\delta_{i1}-\delta_{i2}\right)\left(\delta_{1j}-\delta_{2j}\right)\Bigg] c_j^a + {\cal L}^{(2)}_\text{quark} \notag \ .
\end{align}
Aside from a factor of the gauge-fixing parameter $\xi$, the gauge and ghost fields share the same mass matrix, as expected. This is diagonalized by
\begin{equation}
\begin{pmatrix} G^a_{1\mu} \\ G^a_{2\mu} \end{pmatrix}
= R \begin{pmatrix} G^a_\mu \\ C^a_\mu \end{pmatrix} \ ,
\quad
\begin{pmatrix} c^a_1 \\ c^a_2 \end{pmatrix}
= R \begin{pmatrix} c_G^a \\ c_C^a \end{pmatrix} \ ,
\label{eq:rotations}
\end{equation}
where
\begin{equation}
R\equiv \begin{pmatrix} \cos\theta_c & -\sin\theta_c \\ \sin\theta_c & \cos\theta_c \end{pmatrix} \ , \quad
\sin\theta_c \equiv \frac{g_{s_1}}{\sqrt{g_{s_1}^2+g_{s_2}^2}} \ .
\end{equation}
In Eq.~(\ref{eq:rotations}) $G^a_\mu$ is the gluon field and $C^a_\mu$ is the coloron field, whereas $c_G^a$ and $c_C^a$ are the corresponding ghost fields. Inserting these expressions in Eq.~(\ref{eq:quadratic}) gives, for the coloron mass,
\begin{equation}
M_C = \frac{\sqrt{g_{s_1}^2+g_{s_2}^2}\, f}{2}\equiv \frac{g_s\, f}{\sin 2\theta_c} \ ,
\end{equation}
where $g_s$ is the $SU(3)_C$ coupling,
\begin{equation}
\frac{1}{g_s^2} = \frac{1}{g_{s_1}^2}+\frac{1}{g_{s_2}^2}\ .
\end{equation}
The gluon ghost is massless, whereas both the coloron ghost and the eaten Goldstone boson have mass $\sqrt{\xi}M_C$. The interaction vertices and the corresponding Feynman rules can be found in Appendix~\ref{FR}.

We will leave the quark charge assignments under $SU(3)_{1C}\times SU(3)_{2C}$ arbitrary, for greater generality. In the mass eigenstate basis we write
\begin{equation} \label{Lferm}
\mathcal{L}_\text{quark} = \bar{q}^i i\left[\slashed{\partial}-i g_s \slashed{G}^a t^a -i \slashed{C}^a t^a \left(g_L P_L+g_R P_R\right)\right]q_i \ ,
\end{equation}
where $P_L$ and $P_R$ are the helicity projection operators,
\begin{equation}
P_L\equiv \frac{1-\gamma_5}{2}\ , \quad  P_R\equiv \frac{1+\gamma_5}{2} \ ,
\end{equation}
and $i$ is a flavor index.\footnote{Here we work in the broken electroweak phase, and only employ fermion mass eigenstates.} The coupling to the gluon is dictated by charge universality, whereas the $g_L$ and $g_R$ couplings to the coloron depend on the original charge assignments of the quarks. For example, if both left-handed and right-handed quarks are only charged under $SU(3)_{1C}$, then $g_L=g_R=-g_s\tan\theta_c$, while the axigluon \cite{Frampton:1987wo,Bagger:1988le} corresponds to
$g_L=-g_R=g_s$ (i.e. $\theta_c=\pi/4$). In general, $g_L$ and $g_R$ can each take on the values $-g_s\tan\theta_c$ or  $g_s\cot\theta_c$ in any specific model,\footnote{It is possible to generalize this setup to non-universal charge assignments: in this case flavor-diagonal chiral couplings to the coloron would depend on a generation index. Flavor-changing couplings are strongly constrained \protect\cite{Chivukula:2010fc}.}
\begin{equation} \label{gLgR}
g_L, g_R \in \left \{ -g_s \tan \theta_c, g_s \cot \theta_c \right \} \ .
\end{equation}

\section{LO Coloron Production} \label{LOsec}

\begin{figure}
\begin{center}
\includegraphics[width=.4\textwidth]{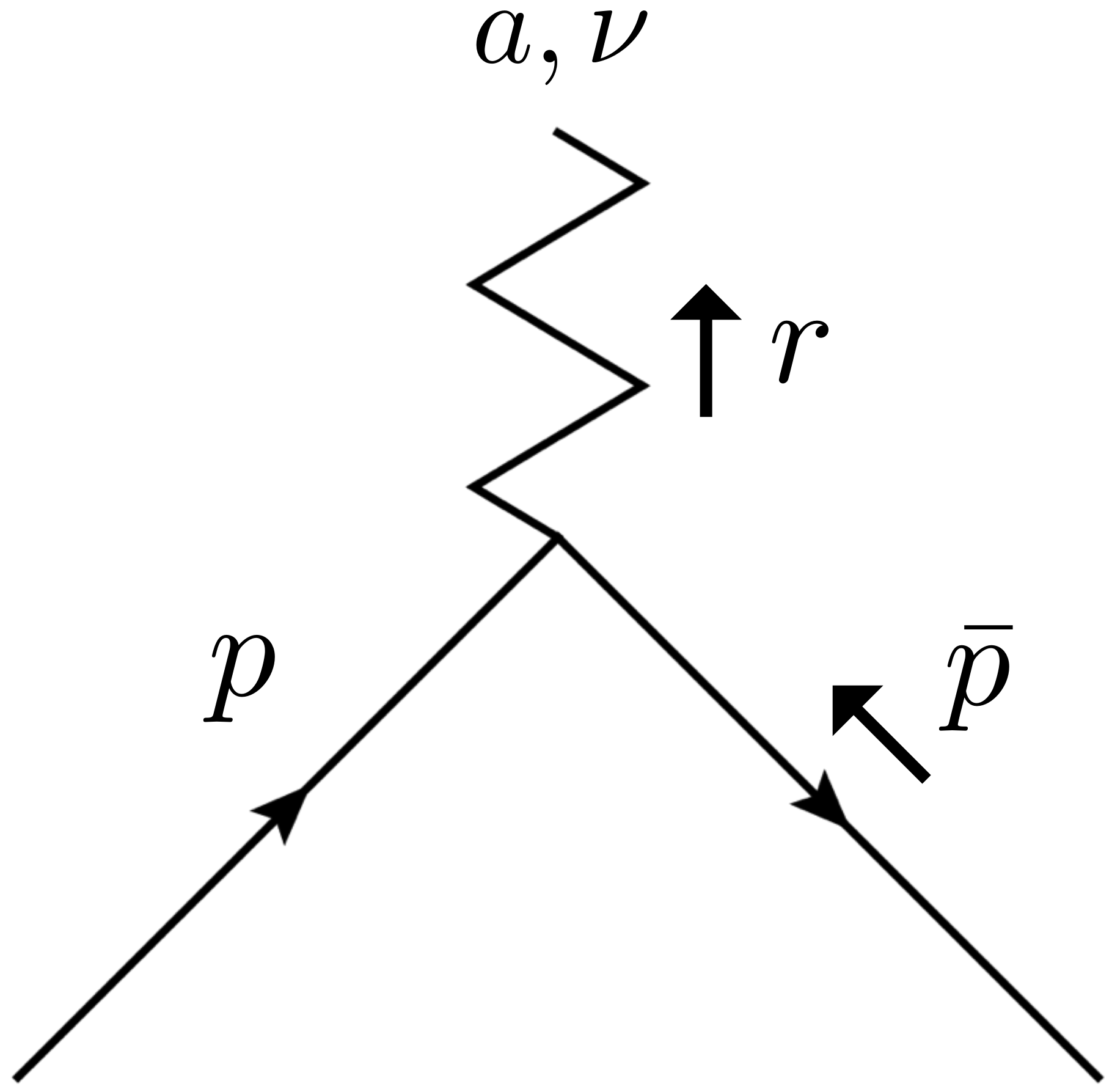}
\caption{Tree-level contribution to coloron production.  The coloron field, $C^a_\nu$, is represented by the zigzag line.}
\label{LOfig}
\end{center}
\end{figure}

The dominant channel for coloron production at a hadron collider is given by the tree-level diagram of Fig.~\ref{LOfig}, in which a $q\bar{q}$ pair annihilates into a coloron. The tree-level diagram with gluon-gluon fusion into a coloron does not exist in the Lagrangian of Eq.~(\ref{eq:L}): in general there are no dimension-four terms with two gauge bosons of an unbroken symmetry and a spin-one field charged under the same symmetry. We use the narrow width approximation for the coloron, take the quarks to be on-shell, and set their masses to zero:\footnote{Note that the Yukawa couplings of quarks to the eaten Goldstone bosons are proportional to the quark masses, and hence vanish in the zero-mass limit.} this is certainly a good approximation, as current experimental bounds \cite{Chatrchyan:2011bd},\cite{Han:2010qc}\nocite{Haisch:2011mw}-\cite{Aad:2012ud} constrain the coloron mass to be in the TeV range.

The leading order (LO) amplitude corresponding to the diagram of Fig.~\ref{LOfig} is
\begin{equation}
i \mathcal{M}_{q\bar{q}\to C}^{(0)}
= g_s\,\bar{v}^r(\bar{p}) \, i \gamma^{\mu} \left(r_L P_L+r_R P_R \right) t^a \, u^s (p) \, \varepsilon_\mu^{a\lambda\ast}(r) \ ,
\label{eq:tree}
\end{equation}
where the superscripts $r$ and $s$ denote quark spin projections, $\lambda$ is the coloron polarization, and
\begin{equation}
r_L\equiv \frac{g_L}{g_s}\ , \quad r_R\equiv \frac{g_R}{g_s} \ , \quad r_L, r_R \in \left \{-\tan \theta_c, \cot \theta_c \right \} \ .
\end{equation}
In $d=2(2-\epsilon)$ dimensions, the squared amplitude averaged over initial spins and colors, and summed over final polarization states, is
\begin{align}
\overline{|\mathcal{M}_{q\bar{q}\to C}^{(0)}|^2} \equiv \ & \left(\frac{1}{\text{dim}(r)}\right)^{2} \left(\frac{1}{2}\right)^2
\sum_{\text{spin \& color}} |\mathcal{M}_{q\bar{q}\to C}^{(0)}|^2 \notag \\
= \ & \frac{C_2(r)(1-\epsilon)}{2\,\text{dim}(r)}\,g_s^2 \left(r_L^2+r_R^2 \right) \hat{s} \label{m0sqav} \ ,
\end{align}
where $\text{dim}(r)=3$ and $C_2(r)=4/3$ are respectively the dimension and Casimir of the fundamental representation of $SU(3)$, and $\hat{s}\equiv (p+\bar{p})^2=2 \, p \cdot \bar{p}$ is the partonic center-of-mass (CM) energy. This gives the LO cross section \cite{Bagger:1988le} for $q\bar{q}\to C$,  
\begin{equation}
\hat{\sigma}_{q\bar{q}\to C}^{(0)} = \frac{\pi}{\hat{s}^2} \, \overline{|\mathcal{M}_{q\bar{q}\to C}^{(0)}|^2} \, \delta(1-\chi)
=\frac{\alpha_s \, A (r_L^2+r_R^2)}{\hat{s}}\, \delta(1-\chi)\ ,
\label{cstree}
\end{equation}
where $\alpha_s\equiv g_s^2/4\pi$,
\begin{equation}
A \equiv \frac{2\pi^2 C_2(r)(1-\epsilon)}{\text{dim}(r)} \ ,
\label{eq:defA}
\end{equation}
and
\begin{equation}
\chi \equiv \frac{M_C^2}{\hat{s}} \ .
\label{eq:chi}
\end{equation}

The full LO cross section for $pp\to C$ is given by the convolution of the LO partonic cross section $\hat{\sigma}_{q\bar{q}\to C}^{(0)}$ with the parton distribution functions (PDFs) for the quarks within the protons, and a sum over all quark flavors,
\begin{equation}
\sigma^\text{LO} = \int dx_1\int dx_2 \sum_q \Big[ f_q(x_1) f_{\bar{q}}(x_2) + f_{\bar{q}}(x_1) f_q(x_2) \Big]
\hat{\sigma}_{q\bar{q}\to C}^{(0)} \ ,
\label{CSfulldetect}
\end{equation}
where $f_q(x)$ is the PDF of parton $q$, and $x$ the momentum fraction of the corresponding parton. Taking the collision axis to be the 3-axis, the four-momenta
of the partons are
\begin{equation}
p=\frac{\sqrt{s}}{2}\left(x_1,0,0,x_1\right)\ ,\quad
\bar{p}=\frac{\sqrt{s}}{2}\left(x_2,0,0,-x_2\right) \ ,
\end{equation}
where $s$ is the CM energy of the colliding hadrons. This gives
\begin{equation}
\hat{s}=x_1 x_2\, s \ , \quad
\chi = \frac{M_C^2}{x_1 x_2 \, s} \ .
\label{eq:x1x2}
\end{equation}

\section{NLO Coloron Production: Virtual Corrections} \label{sec:virual}

In this section we compute the next-to-leading order (NLO) virtual QCD corrections to the $q\bar{q}\to C$ amplitude. These include one-loop wavefunction and vertex corrections, which we choose to compute in 't Hooft-Feynman gauge, $\xi=1$. The non-Abelian vertex corrections are computed by employing the \textit{pinch technique}: this allows us to obtain QED-like Ward identities, and absorb all UV infinities in the renormalization of the gauge field propagators. After inclusion of the counterterms, the virtual corrections are UV-finite, yet IR infinite. In Sec.~\ref{sec:NLOcs} we show that the IR divergences cancel once the real corrections, corresponding to the emission of soft  and collinear gluons and quarks, are included in the calculation of the inclusive production cross section. Our loop integrals are computed in dimensional regularization, with $d=2(2-\epsilon)$ dimensions. We first regulate the IR divergences by giving the gluon a small mass ($m_g \to 0^+$): in this way all infinities are in the UV, and regularization requires $\epsilon>0$. After all of the UV infinities are removed, by cancellation and inclusion of the counterterms, we let the gluon mass approach zero. This will make the virtual corrections IR divergent, with the infinities being regulated by taking $\epsilon<0$.

Since the quark couplings to the coloron are chiral, in general, we need a prescription for treating $\gamma_5$ in $d\neq 4$. Here we take $\gamma_5$ to always anticommute with $\gamma^\mu$. Choosing an alternative prescription, such as 't Hooft-Veltman in which $\gamma_5$ anticommutes with $\gamma^\mu$ for $\mu=0,1,2,3$ and commutes for other values of $\mu$, would lead to a cross section for $q\bar{q}\to C$ which differs from ours by only a finite renormalization of the coupling(s).

\begin{figure}[!t]
\begin{center}
\includegraphics[width=\textwidth]{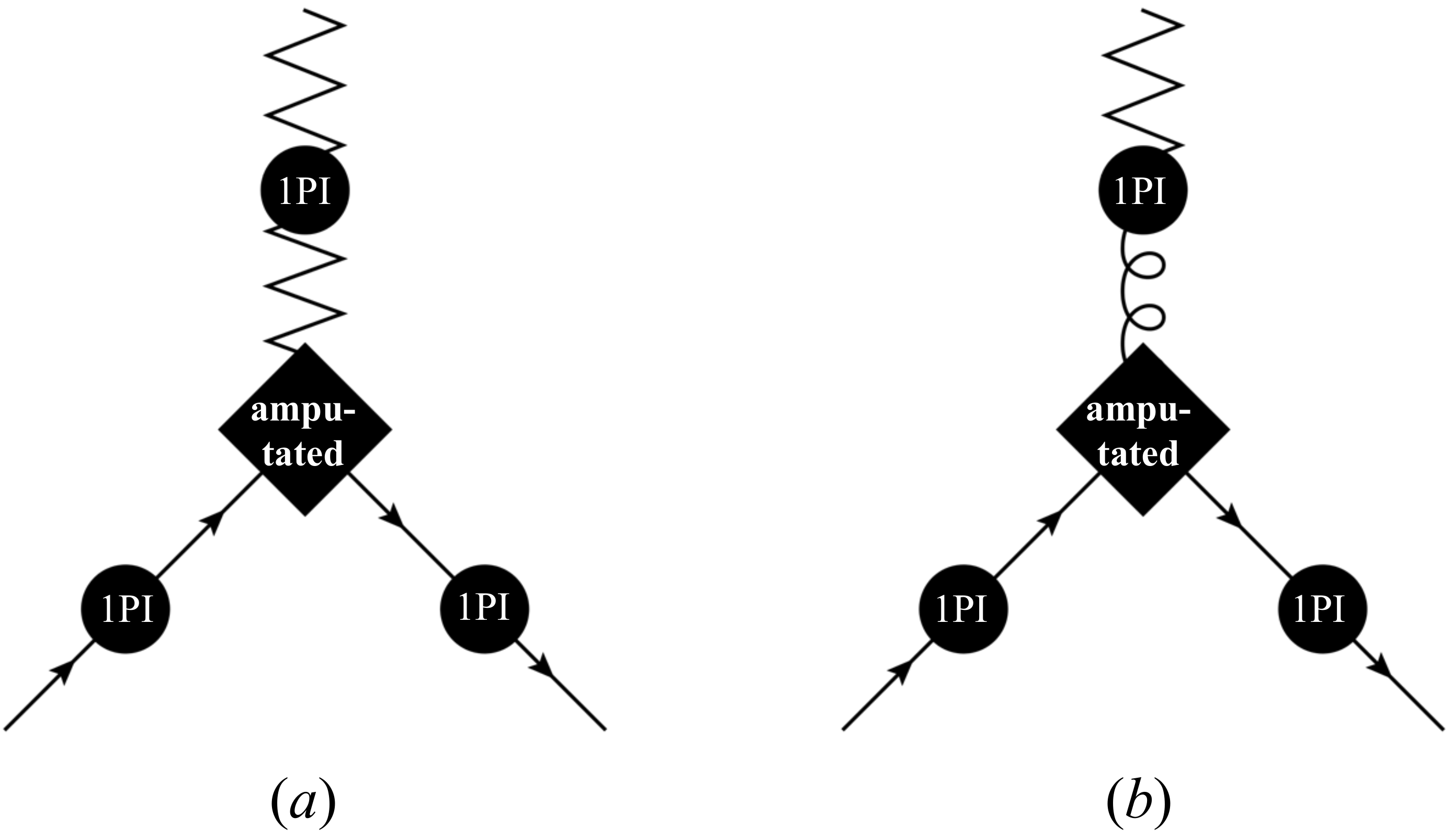}
\caption{Structure of $q\bar{q} \to C$ amplitude, to all orders in perturbation theory. Direct coloron production is illustrated on the left, while production via mixing
with the gluon is shown on the right.  The gluon field is, as usual, represented by the coiling line; the coloron field is represented by the zigzag line.}
\label{fig:lsz}
\end{center}
\end{figure}

The general structure of the $q\bar{q}\to C$ amplitude, illustrated in Fig.~\ref{fig:lsz}, is
\begin{equation}\label{eq:M}
i\mathcal{M}_{q\bar{q}\to C} = g_s\, \bar{v}^r(\bar{p})\, i\left[Z_C^{1/2}\Gamma_{qqC}^{a\mu}+\Gamma^{a\mu}_{qqG} \frac{\Pi_{GC}(\hat{s})}{\hat{s}} \right]
Z_q u^s (p) \, \varepsilon_\mu^{a\lambda\ast}(r) \ ,
\end{equation}
where $\Gamma^{a\mu}_{qqC}$ ($\Gamma^{a\mu}_{qqG}$) is the one-particle irreducible (1PI) quark-quark-coloron (quark-quark-gluon) vertex and $\Pi_{GC}$ is the coefficient of $g^{\mu\nu}$ in the gluon-coloron vacuum polarization mixing amplitude (VPA).  The factors $Z_q$ and $Z_C$ are, respectively, the residues of the full quark and coloron propagators at the mass pole; they are obtained from the quark self-energy amplitude, $\Sigma(\slashed{p})$, and the coefficient of $g^{\mu\nu}$  in the coloron-coloron VPA, $\Pi_{CC}(q^2)$, as follows
\begin{equation}
Z_q=\frac{1}{1-\Sigma^\prime(0)} \ , \quad
Z_C=\frac{1}{1-\Pi^\prime_{CC}(M^2_{C \,\text{phys}})} \ ,
\label{eq:Z}
\end{equation}
where the prime denotes a derivative with respect to the argument, and $M_{C \,\text{phys}}$ is the coloron's physical mass. To lowest order, $Z_q=1$, $Z_C=1$, $\Pi_{GC}=0$, and $i\Gamma^{a\mu}_{qqC}=\gamma^\mu \left(r_L P_L+r_R P_R \right) t^a$; inserting these expressions in Eq.~(\ref{eq:M}) recovers the tree-level amplitude of Eq.~(\ref{eq:tree}).
\\
\\

\subsection{Quark Self-Energy} \label{sec:self}

\begin{figure}[!t]
\begin{center}
\includegraphics[width=\textwidth]{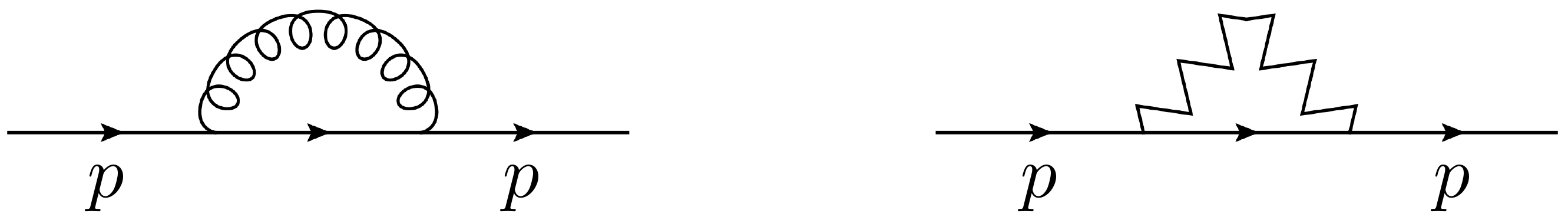}
\caption{Quark self-energy diagrams at one-loop.  Particle notation as defined in Fig.~\protect\ref{fig:lsz}.}
\label{fig:self}
\end{center}
\end{figure}

The NLO quark self-energy correction to the $q\bar{q}\to C$ amplitude is found, from Eqs.~(\ref{eq:M}) and (\ref{eq:Z}), to be
\begin{equation}
i Q = \bar{v}^r(\bar{p}) \, i \gamma^{\nu} \left(g_L P_L+g_R P_R \right) t^a\, \delta Z_q \, u^s (p) \, \varepsilon_\nu^{a\lambda\ast}(r) \ ,
\label{eq:Q1}
\end{equation}
where
\begin{equation}
\delta Z_q = \Sigma^\prime(0) \ .
\label{eq:Zq}
\end{equation}
At one-loop, the $\Sigma(\slashed{p})$ amplitude is given by the diagrams of Fig.~\ref{fig:self}. These lead to the expression
\begin{equation}\label{eq:Sigma} \begin{split}
\Sigma(\slashed{p}) = \ & -\slashed{p}\, \frac{g_s^2 \, C_2(r) \, 2(1-\epsilon)\Gamma(\epsilon)}{16\pi^2} \\
&\times \int_0^1 dx\,(1-x)
\left[\left(\frac{4\pi\mu^2}{\Delta_{Gq}}\right)^\epsilon  +
\left(\frac{4\pi\mu^2}{\Delta_{Cq}}\right)^\epsilon (r_L^2 P_L+r_R^2 P_R)\right] \ ,
\end{split} \end{equation}
where $\Gamma(\epsilon)$ is the Euler Gamma-function evaluated at infinitesimal $\epsilon$, and
\begin{equation}
\Delta_{Gq}=(1-x)m_g^2-x(1-x)p^2-i\eta \ , \quad
\Delta_{Cq}=(1-x)M_C^2-x(1-x)p^2-i\eta \ .
\end{equation}

The parameter $\mu$ is the mass scale introduced by the loop integral in $d$ dimensions, and $\eta$ is the positive infinitesimal parameter giving the appropriate prescription for computing the integral in momentum space. As previously anticipated, we have introduced a small gluon mass, $m_g$, in order to regulate the IR divergences and isolate the UV infinities: with $m_g\neq 0$, $\Sigma(\slashed{p})$ and $\Sigma^\prime(\slashed{p})$ contain only UV divergences. Inserting Eq.~(\ref{eq:Sigma}) in Eq.~(\ref{eq:Zq}) gives
\begin{equation}\label{eq:Qp} \begin{split}
\delta Z_q = \ & - \frac{g_s^2 \, C_2(r) \, 2(1-\epsilon)\Gamma(\epsilon)}{16\pi^2} \int_0^1 dx\,(1-x) \\
&\times
\left[\left(\frac{4\pi\mu^2}{(1-x)m_g^2-i\eta}\right)^\epsilon  +
\left(\frac{4\pi\mu^2}{(1-x)M_C^2-i\eta}\right)^\epsilon (r_L^2 P_L+r_R^2 P_R)\right] \ .
\end{split} \end{equation}
The amplitude of Eq.~(\ref{eq:Q1}) becomes
\begin{equation}\label{eq:Q} \begin{split}
i Q = \ & - \frac{\alpha_s}{4\pi}\,2 C_2(r)(1-\epsilon)\Gamma(\epsilon) \int_0^1 dx \int_0^{1-x} dy \\
&\ \Bigg[\left(\frac{4\pi\mu^2}{(1-x)m_g^2-i\eta}\right)^\epsilon i \mathcal{M}_{q\bar{q}\to C}^{(0)} +
\left(\frac{4\pi\mu^2}{(1-x)M_C^2-i\eta}\right)^\epsilon i \mathcal{M}_{q\bar{q}\to C}^{\prime(0)}\Bigg] \ ,
\end{split} \end{equation}
where $\mathcal{M}_{q\bar{q}\to C}^{(0)}$ is given by Eq.~(\ref{eq:tree}), and
\begin{equation}
i \mathcal{M}_{q\bar{q}\to C}^{\prime(0)} = g_s\, \bar{v}^r(\bar{p}) \, i \gamma^{\nu} \left(r_L^3 P_L+r_R^3 P_R \right) t^a \, u^s (p) \
\varepsilon_\nu^{a\lambda\ast}(r)  \ .
\label{eq:selfloop}
\end{equation}
For later convenience we have traded the $1-x$ factor, in Eq.~(\ref{eq:Qp}), for an integral over $dy$: this will allow us to directly add the self-energy correction to the vertex correction and explicitly show the cancellation of the UV divergences.

\subsection{Abelian Vertex Corrections} \label{sec:Abelian}

\begin{figure}[!t]
\begin{center}
\includegraphics[width=.8\textwidth]{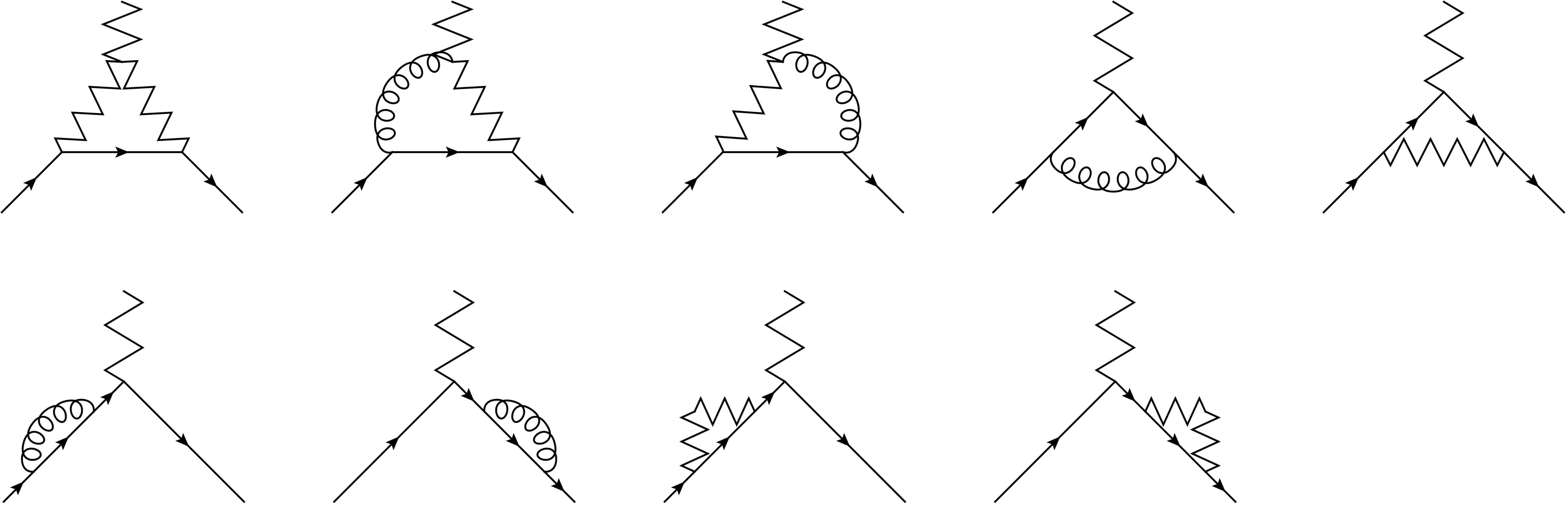}
\caption{One-loop Abelian vertex correction to the $q\bar{q}\to C$ amplitude. Particle notation is as defined in Fig.~\protect\ref{fig:lsz}.}
\label{fig:AV}
\end{center}
\end{figure}

The one-loop Abelian vertex correction to the $q\bar{q}\to C$ amplitude is given by the diagrams of Fig.~\ref{fig:AV}. These lead to the amplitude
\begin{align}
i V_\text{Abelian} = \ & \frac{\alpha_s}{4\pi} \left[2C_2(r)-C_2(G)\right] \Gamma(1+\epsilon) \int_0^1 dx \int_0^{1-x} dy \notag \\
& \Bigg\{\Bigg[
\frac{(1-\epsilon)^2}{\epsilon}-\left(x y \epsilon-(1-x)(1-y)\right)\frac{\hat{s}}{\Delta_{Gqq}}\Bigg]
\left(\frac{4\pi\mu^2}{\Delta_{Gqq}}\right)^\epsilon i\mathcal{M}_{q\bar{q}\to C}^{(0)} \label{eq:VA} \\
&\ + \Bigg[
\frac{(1-\epsilon)^2}{\epsilon}-\left(x y \epsilon-(1-x)(1-y)\right)\frac{\hat{s}}{\Delta_{Cqq}}\Bigg]
\left(\frac{4\pi\mu^2}{\Delta_{Cqq}}\right)^\epsilon i\mathcal{M}_{q\bar{q}\to C}^{\prime(0)} \Bigg\} \notag \ ,
\end{align}
where $C_2(G)=3$ is the Casimir of the adjoint representation, and
\begin{equation} \begin{split}
\Delta_{Gqq} = \ & (1-x-y)m_g^2 - x y \hat{s}- i \eta \ , \\
\Delta_{Cqq} = \ & (1-x-y)M_C^2 - x y \hat{s}- i \eta \ .
\end{split} \end{equation}
Once again, we have included a small gluon mass $m_g$ in order to regulate the IR divergences.
\\
\\

\subsection{Non-Abelian Vertex Corrections {\em a la} Pinch-Technique: Unpinched Diagrams} \label{sec:nonpinched}

\begin{figure}[!t]
\begin{center}
\includegraphics[width=\textwidth]{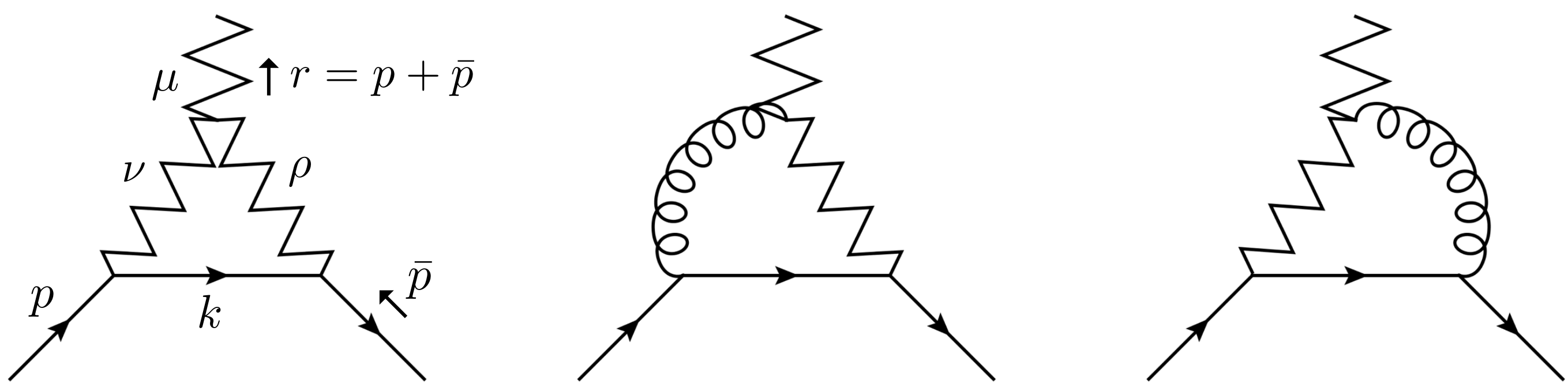}
\caption{One-loop non-Abelian vertex correction to the $q\bar{q}\to C$ amplitude.  Particle notation is as defined in Fig.~\protect\ref{fig:lsz}.  Each three-gauge boson vertex is a full non-Abelian vertex $\Gamma^{\mu\nu\rho}$ in Eq.~(\ref{eq:refa}).}
\label{fig:NAV}
\end{center}
\end{figure}

The non-Abelian vertex corrections are given by the diagrams of Fig.~\ref{fig:NAV}. When added to the overall Abelian vertex correction, Eq.~(\ref{eq:VA}), these give the one-loop total vertex correction to $q\bar{q}\to C$. Unlike in QED, the UV divergences in the vertex correction do not cancel the UV divergences arising from the self-energy amplitudes. The reason for this is that the QED Ward identity $\partial^\mu j_\mu=0$ is now replaced by its non-Abelian counterpart $D^\mu j^a_\mu=0$, which does not imply the equality of vertex and quark-wavefunction renormalization constants. It is possible, though, to recover QED-like Ward identities for the currents $j^a_\mu$ by employing the pinch technique. This consists of breaking up the gauge boson internal momenta of a Feynman diagram into ``pinching'' and ``non-pinching'' pieces. The pinching momenta are those which cancel some internal propagators, leading to a simpler diagram with the external-momentum structure of a propagator. The non-pinching momenta will instead give overall amplitudes satisfying QED-like Ward identities. A formal proof of these statements, for an arbitrary non-Abelian gauge theory, can be found in the review of Ref.~\cite{Binosi:2009lo} (and references therein).

\begin{figure}[!t]
\begin{center}
\includegraphics[width=\textwidth]{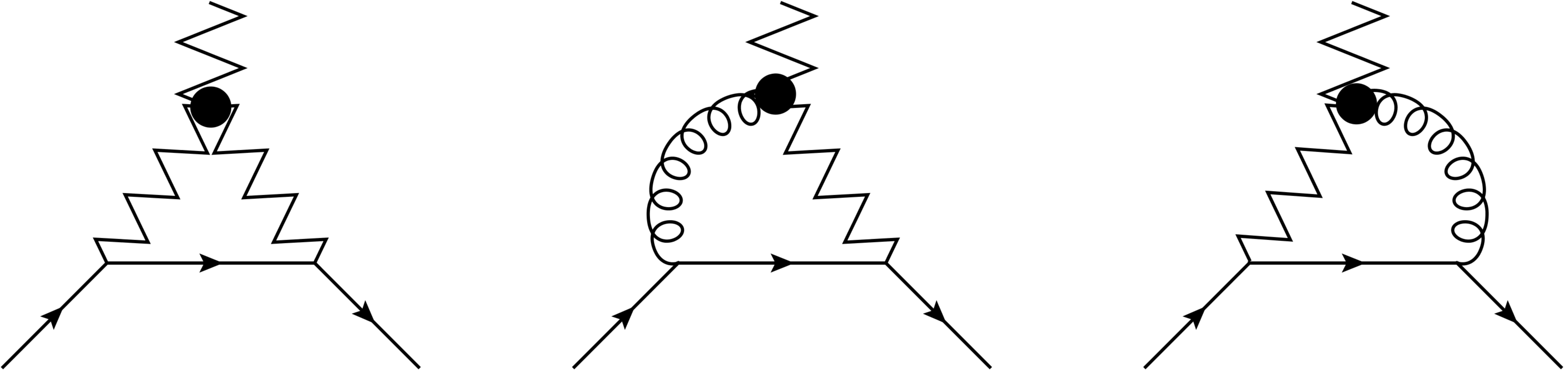}
\caption{Non-Abelian unpinched vertex-correction diagrams for the $q\bar{q}\to C$ amplitude at one-loop.  Particle notation is as defined in Fig.~\protect\ref{fig:lsz}. The black disk indicates that each three-point gauge boson vertex in these diagrams has been replaced by the non-pinched portion $\Gamma_F^{\mu\nu\rho}$, as described in Eqs.~(\protect\ref{eq:breakup}) and (\protect\ref{eq:RR}). }
\label{fig:NAVF}
\end{center}
\end{figure}

In our vertex computation the pinch technique works as follows. The non-Abelian vertex structure in each of the diagrams in Fig.~\ref{fig:NAV} is
\begin{equation}
\Gamma^{\mu\nu\rho}(k,p,\bar{p})=g^{\mu\nu}(-2p-\bar{p}+k)^\rho + g^{\nu\rho}(p-\bar{p}-2k)^\mu
+g^{\rho\mu}(k+p+2\bar{p})^\nu \ .
\label{eq:refa}
\end{equation}
We can break this into two parts,
\begin{equation}
\Gamma^{\mu\nu\rho}(k,p,\bar{p})=\Gamma_F^{\mu\nu\rho}(k,p,\bar{p}) + \Gamma_P^{\mu\nu\rho}(k,p,\bar{p}) \ ,
\label{eq:breakup}
\end{equation}
where
\begin{align}
\Gamma_F^{\mu\nu\rho}(k,p,\bar{p}) = \ &-2 g^{\mu\nu}(p+\bar{p})^\rho + 2 g^{\rho\mu}(p+\bar{p})^\nu
+ g^{\nu\rho}(p-\bar{p}-2k)^\mu \ , \label{eq:RR}\\
\Gamma_P^{\mu\nu\rho}(k,p,\bar{p}) = \ & g^{\mu\nu}(\bar{p}+k)^\rho + g^{\rho\mu}(k-p)^\nu \ .
\label{eq:breakup2}
\end{align}
Unlike $\Gamma^{\mu\nu\rho}(k,p,\bar{p})$, the $\Gamma_F^{\mu\nu\rho}(k,p,\bar{p})$ vertex satisfies a QED-like Ward identity for the $g C\to C$ and $C C\to C$ amplitudes,
\begin{equation}
(p+\bar{p})_\mu \Gamma_F^{\mu\nu\rho}(k,p,\bar{p}) = g^{\nu\rho} \Big[(p-k)^2-(\bar{p}+k)^2\Big] \ .
\label{eq:qedWI}
\end{equation}
As shown below, when $\Gamma_F^{\mu\nu\rho}(k,p,\bar{p})$  is used to compute the integral in momentum space (instead of  $\Gamma^{\mu\nu\rho}(k,p,\bar{p})$), its UV divergences, added to the UV divergences of the Abelian vertex corrections, exactly cancel the UV divergences of the quark self-energy amplitudes. As mentioned above, this occurs because a QED-like Ward identity for $q\bar{q}\to C$ holds, as one can prove by using the QED-like Ward identity for the $g C\to C$ and $C C\to C$ amplitudes given in Eq.~(\ref{eq:qedWI}). The three diagrams which correspond to using $\Gamma_F^{\mu\nu\rho}(k,p,\bar{p})$ instead of $\Gamma^{\mu\nu\rho}(k,p,\bar{p})$ are symbolically denoted with a black disk over the non-Abelian vertex, and are shown in Fig.~\ref{fig:NAVF}. These lead to the following contribution to the $q\bar{q}\to C$ amplitude
\begin{align}
& i V_\text{non-Abelian}= \frac{\alpha_s}{4\pi} \, C_2(G)\Gamma(1+\epsilon) \int_0^1 dx \int_0^{1-x} dy \notag \\
&\ \Bigg\{ \Bigg[\left(\frac{1-\epsilon}{\epsilon}-(x+y)\frac{\hat{s}}{\Delta_{GCq}}\right)\left(\frac{4\pi\mu^2}{\Delta_{GCq}}\right)^\epsilon +\left(\frac{1-\epsilon}{\epsilon}- (x+y)\frac{\hat{s}}{\Delta_{CGq}}\right)\left(\frac{4\pi\mu^2}{\Delta_{CGq}}\right)^\epsilon \notag \\
& \qquad - \left(\frac{1-\epsilon}{\epsilon}-(x+y)\frac{\hat{s}}{\Delta_{CCq}}\right)\left(\frac{4\pi\mu^2}{\Delta_{CCq}}\right)^\epsilon
\Bigg] i \mathcal{M}_{q\bar{q}\to C}^{(0)} \label{eq:NAV} \\
& \quad + \left(\frac{1-\epsilon}{\epsilon}-(x+y)\frac{\hat{s}}{\Delta_{CCq}}\right)\left(\frac{4\pi\mu^2}{\Delta_{CCq}}\right)^\epsilon
i \mathcal{M}_{q\bar{q}\to C}^{\prime(0)} \Bigg\} \notag \ ,
\end{align}
where
\begin{equation} \begin{split}
\Delta_{GCq} = \ & x m_g^2 + y M_C^2 - x y \hat{s} - i\eta \ , \\
\Delta_{CGq} = \ & x M_C^2 + y m_g^2 - x y \hat{s} - i\eta \ , \\
\Delta_{CCq} = \ & (x+y)M_C^2 - x y \hat{s} - i\eta \ .
\end{split} \end{equation}
In order to obtain Eq.~(\ref{eq:NAV}) we have used the equations of motion for the external spinors, together with the relations
\begin{equation}
2 \cot (2\theta_c)\, r_L = -1 + r_L^2 \ , \quad
2 \cot (2\theta_c)\, r_R = -1 + r_R^2 \ ,
\label{eq:couplingrel}
\end{equation}
which are true for any charge assignment of the quarks. As anticipated, $i Q+i V_\text{Abelian}+ i V_\text{non-Abelian}$ is free of UV divergences, as manifestly shown by adding together Eqs.~(\ref{eq:Q}), (\ref{eq:VA}) and (\ref{eq:NAV}). This part of the amplitude is, however, IR divergent in the limit of zero gluon mass. Setting $m_g= 0$ and $\epsilon<0$ gives
\begin{equation} \begin{split}
& i Q+i V_\text{Abelian}+ i V_\text{non-Abelian} \\
&= \frac{\alpha_s}{4\pi}\left[ C_2(r)\left(-\frac{2}{\epsilon^2}
-\frac{3+2i}{\epsilon}\right)
+ C_2(G) \frac{i\pi}{\epsilon} \right] i \mathcal{M}_{q\bar{q}\to C}^{(0)} \,+\, \text{finite} \ .
\end{split} \end{equation}

\begin{figure}
\begin{center}
\includegraphics[width=\textwidth]{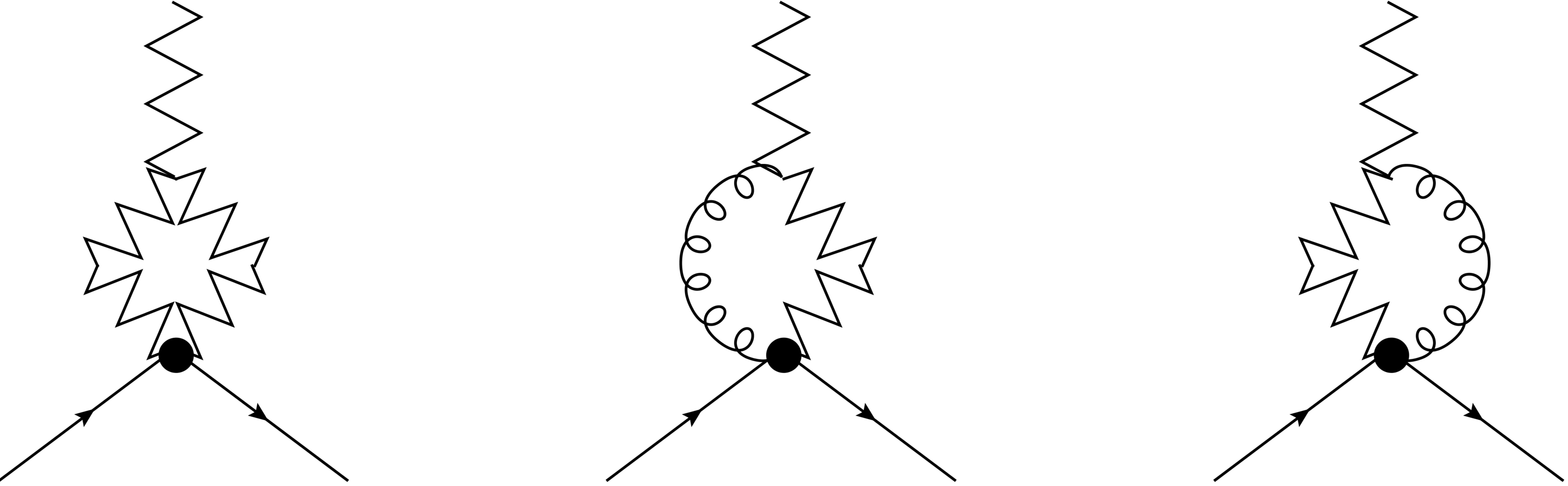}
\caption{Non-Abelian pinched vertex-correction diagrams for the $q\bar{q}\to C$ amplitude at one-loop. Particle notation is as defined in Fig.~\protect\ref{fig:lsz}.}
\label{fig:NAVP}
\end{center}
\end{figure}

Of course we still need to include the contribution from $\Gamma_P^{\mu\nu\rho}(k,p,\bar{p})$ (of Eq.~(\ref{eq:breakup2})) in the full non-Abelian vertex correction. This contains the pinching momenta: the action of $p$ and $\bar{p}$ on the external spinors gives zero, and the remaining piece cancels the internal fermion propagator in the diagram. Thus, the internal fermion line in each diagram is pinched away, leaving an effective diagram with a four-point coupling between fermions and gauge bosons as shown in Fig.~\ref{fig:NAVP}. The UV divergences of the pinched diagrams have the same group- and momentum-structure as those of the VPAs, and can be absorbed in the counterterms for the gauge field propagators. In order to see this clearly, we will now consider the form of the ``true'' propagator corrections to the $q\bar{q}\to C$ amplitude in the following subsection.

\subsection{Form of the Vacuum Polarization Amplitudes} \label{sec:VPA}

The NLO corrections to the $q\bar{q}\to C$ amplitude due to the VPAs are found, from Eqs.~(\ref{eq:M}) and (\ref{eq:Z}), to have the form
\begin{equation}
iP= i\mathcal{M}_{q\bar{q}\to C}^{(0)} \, \frac{\delta Z_C}{2} + i\mathcal{M}_{q\bar{q}\to C}^{\prime\prime(0)} \,
\frac{\Pi_{GC}(\hat{s})}{\hat{s}} \ ,
\label{eq:Pform}
\end{equation}
where
\begin{equation}
\delta Z_C = \Pi^\prime_{CC}(M^2_C) \ ,
\label{eq:derivative}
\end{equation}
and
\begin{equation}
i\mathcal{M}_{q\bar{q}\to C}^{\prime\prime(0)} = g_s\, \bar{v}^r(\bar{p}) \, i \gamma^{\mu}  t^a \, u^s (p) \, \varepsilon_\mu^{a\lambda\ast}(r) \ .
\end{equation}
In order to obtain the second term of Eq.~(\ref{eq:Pform}), we have replaced $\Gamma^{a\mu}_{qqG}$ with its LO component $i\gamma^\mu t^a$. Notice also that at this order we can swap $M^2_{C \text{phys}}$ for $M_C^2$.

\begin{figure}\begin{center}
\includegraphics[width=\textwidth]{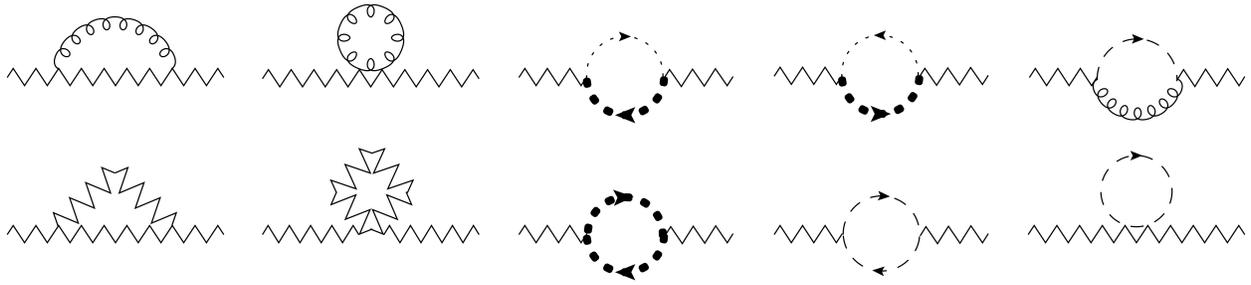}
\caption{Coloron-coloron vacuum polarization amplitude at one-loop. A gluon field is, as usual, represented by a coiling line; a coloron field is represented by a zigzag line. The coloron ghost is represented by a sequence of filled circles, and the eaten Goldstone bosons are represented by dashed lines.}
\label{fig:CC}
\end{center}\end{figure}
\begin{figure}\begin{center}
\includegraphics[width=\textwidth]{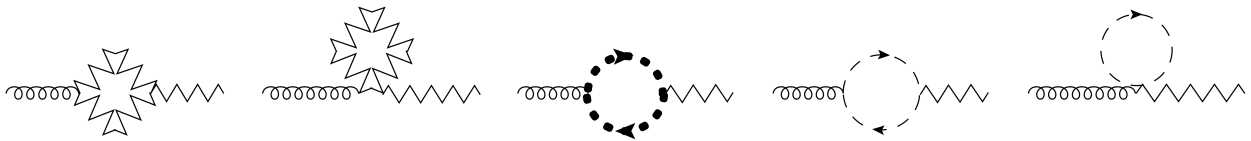}
\caption{Gluon-coloron mixing amplitude at one-loop. Particle notation is as defined in Fig.~\protect\ref{fig:CC}.}
\label{fig:GC}
\end{center}\end{figure}
\begin{figure}\begin{center}
\includegraphics[width=.7\textwidth]{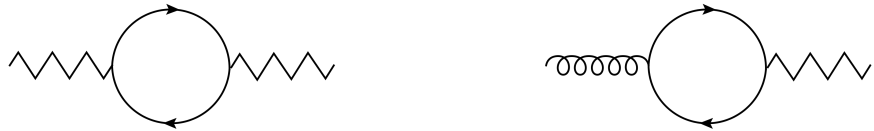}
\caption{Fermion contributions (solid lines) to coloron-coloron vacuum polarization amplitude and gluon-coloron mixing amplitude. Particle notation is as defined in Fig.~\protect\ref{fig:CC}.}
\label{fig:CCGCfermions}
\end{center}\end{figure}

At one-loop, $\Pi_{CC}(q^2)$ is given by the diagrams of Figs.~\ref{fig:CC} and \ref{fig:CCGCfermions}, in which the gluon ghost is represented by dotted lines, the coloron ghost by a sequence of filled circles, and the eaten Goldstone bosons are represented by dashed lines. There are poles at $d=2$ proportional to both $q^2$ and $M_C^2$. The latter correspond to quadratic divergences (renormalizing the coloron mass scale $f$), whereas the former can only be logarithmic by dimensional analysis (renormalizing the coloron field).\footnote{This situation parallels the renormalization of the electroweak chiral lagrangian \protect\cite{Appelquist:1980xd,Appelquist:1981le}.} The momentum-dependent part of the full coloron-coloron VPA is not transverse, as the coefficients of the $q^2$ and $q^\mu q^\nu$ terms are different. However, we have explicitly verified that the infinite part is transverse: this is necessary, because the corresponding Lagrangian counterterms {\em are} transverse. For small values of $\epsilon$ we obtain
\begin{align}
&\left(\frac{\alpha_s}{4\pi}\right)^{-1}\Pi_{CC}(q^2)g^{\mu\nu}+q^\mu q^\nu\text{--terms} \notag \\
&= C_2(G)\int_0^1 dx \Bigg\{
\Bigg[\left(\frac{\mu^2}{\Delta_{GC}}\right)^\epsilon 2\Big(1+4x(1-x)\Big)E+2(1-2x)^2\Bigg]\left(g^{\mu\nu}q^2-q^\mu q^\nu\right) \nonumber \\
&\quad +\Bigg[\left(\frac{\mu^2}{\Delta_{GC}}\right)^\epsilon \Big(1-x(4-3x)\Big) E-x(1-x)\Bigg] g^{\mu\nu}q^2 \notag \\
& \quad +\Bigg[\left(\frac{\mu^2}{\Delta_{GC}}\right)^\epsilon 2 x\, E +3-5x\Bigg]g^{\mu\nu} M_C^2 \Bigg\} \label{eq:PCC} \\
&\ +4\cot^2(2\theta_c) C_2(G) \int_0^1 dx \Bigg\{
\Bigg[\left(\frac{\mu^2}{\Delta_{CC}}\right)^\epsilon \Big(1+4x(1-x)\Big)E+(1-2x)^2\Bigg]\left(g^{\mu\nu}q^2-q^\mu q^\nu\right) \nonumber \\
&\quad +\Bigg[-\left(\frac{\mu^2}{\Delta_{CC}}\right)^\epsilon \frac{x(1-x)}{4} E-\frac{x(1-x)}{4}\Bigg] g^{\mu\nu} q^2
+\left(\frac{\mu^2}{\Delta_{CC}}\right)^\epsilon \frac{(1-2x)^2}{8}\, E\,q^\mu q^\nu \notag \\
&\quad +\Bigg[\left(\frac{\mu^2}{\Delta_{CC}}\right)^\epsilon \frac{5}{4}E+\frac{1}{4}\Bigg]g^{\mu\nu}M_C^2\Bigg\} \nonumber \\
&\ +(r_L^2+r_R^2)N_f \int_0^1 dx \left(\frac{\mu^2}{\Delta_{qq}}\right)^\epsilon \Big[-2x(1-x)\Big] E\left(g^{\mu\nu}q^2-q^\mu q^\nu\right) \notag \ ,
\end{align}
where our results depend only on the coefficient of $g^{\mu\nu}$,  the quantity $N_f$ is the number of quark flavors in the loop (see
Fig.~\ref{fig:CCGCfermions}),
\begin{equation}
E\equiv \frac{1}{\epsilon}-\gamma+\log 4\pi \ ,
\label{eq:E}
\end{equation}
and $\gamma$ is the Euler-Mascheroni constant. The $\Delta$ functions in Eq.~(\ref{eq:PCC}) are
\begin{equation} \begin{split}
\Delta_{GC} \equiv \ & x\, m_g^2+(1-x)M_C^2-x(1-x)q^2-i\eta \ , \\
\Delta_{CC} \equiv \ & M_C^2-x(1-x)q^2-i\eta \ , \\
\Delta_{qq} \equiv \ & -x(1-x)q^2 -i\eta \ .
\end{split} \end{equation}
Notice that the coloron-coloron VPA of Eq.~(\ref{eq:PCC}) is not IR divergent in the limit $m_g\to 0$, since there are no contributions with only massless (gluon) states. However, what enters in Eq.~(\ref{eq:Pform}) is the {\em derivative} of $\Pi_{CC}$ (see Eq.~(\ref{eq:derivative})), which {\em is} IR divergent in the limit $m_g\to 0$.

The momentum-dependent part of the gluon-coloron mixing amplitude (Figs.~\ref{fig:GC} and \ref{fig:CCGCfermions}) is found to be transverse, both in the infinite and the finite parts. For small values of $\epsilon$ we find
\begin{equation} \begin{split}
&\left(\frac{\alpha_s}{4\pi}\right)^{-1}\Pi_{GC}(q^2)g^{\mu\nu} \, +\, q^\mu q^\nu\text{--terms} = 2 \cot(2\theta_c) C_2(G) \int_0^1 dx \left(\frac{\mu^2}{\Delta_{CC}}\right)^\epsilon \\
& \quad \times \Bigg\{\Bigg[\Big(\frac{3}{4}+5x(1-x)\Big)E+(1-2x)^2\Bigg]\left(q^2 g^{\mu\nu}-q^\mu q^\nu\right) +E\, M_C^2  \Bigg\} \\
& \ +(r_L+r_R)N_f \int_0^1 dx \left(\frac{\mu^2}{\Delta_{qq}}\right)^\epsilon
\big[-2 x(1-x)\big] \left(q^2 g^{\mu\nu}-q^\mu q^\nu\right) \ .
\end{split} \end{equation}
There are no potential IR divergences hidden in $\Pi_{GC}$.

\subsection{Non-Abelian Vertex Corrections {\em a la} Pinch-Technique: Pinched Diagrams} \label{sec:pinched}

The pinched diagrams of Fig.~\ref{fig:NAVP} are obtained from the diagrams of Fig.~\ref{fig:NAV} by replacing the full non-Abelian vertex momentum structure $\Gamma^{\mu\nu\rho}(k,p,\bar{p})$  from Eq.~(\ref{eq:refa}), with $\Gamma_P^{\mu\nu\rho}(k,p,\bar{p})$ from Eq.~(\ref{eq:breakup2}). This  leads to the amplitude
\begin{equation}\label{eq:P} \begin{split}
i P_\text{pinched} = \ & \frac{\alpha_s}{4\pi} \, C_2(G) \int_0^1 dx \Bigg[
2\left(\frac{\mu^2}{\Delta_{GC}}\right)^\epsilon
+4\cot^2(2\theta_c) \left(\frac{\mu^2}{\Delta_{CC}}\right)^\epsilon
\Bigg] E\, \mathcal{M}_{q\bar{q}\to C}^{(0)} \\
&+ \frac{\alpha_s}{4\pi}\, 2\cot(2\theta_c) C_2(G)\int_0^1 dx \left(\frac{\mu^2}{\Delta_{CC}}\right)^\epsilon E\,
\mathcal{M}_{q\bar{q}\to C}^{\prime\prime(0)} \ ,
\end{split} \end{equation}
where we have used Eq.~(\ref{eq:couplingrel}) to rewrite the fermion couplings in terms of $\theta_c$. This contribution to the amplitude has the form of a VPA correction, like that in Eq.~(\ref{eq:Pform}). In fact, we can write
\begin{equation}
iP_\text{pinched}= i\mathcal{M}_{q\bar{q}\to C}^{(0)}  \frac{\widetilde{\Pi}^\prime_{CC}(M^2_C)}{2}
+ i\mathcal{M}_{q\bar{q}\to C}^{\prime\prime(0)} \frac{\widetilde{\Pi}_{GC}(\hat{s})}{\hat{s}} \ ,
\label{eq:Pformtilde}
\end{equation}
where
\begin{equation}\begin{split}
\left(\frac{\alpha_s}{4\pi}\right)^{-1}\widetilde{\Pi}_{CC}(q^2) = \ & C_2(G) \int_0^1 dx
\left(\frac{\mu^2}{\Delta_{GC}}\right)^\epsilon 4(q^2-M_C^2)E \\
&+4\cot^2(2\theta_c) C_2(G) \int_0^1 dx \left(\frac{\mu^2}{\Delta_{CC}}\right)^\epsilon 2(q^2-M_C^2)E \ ,
\end{split} \end{equation}
and
\begin{equation}
\left(\frac{\alpha_s}{4\pi}\right)^{-1}\widetilde{\Pi}_{GC}(q^2) = 2\cot(2\theta_c) C_2(G)\int_0^1 dx
\left(\frac{\mu^2}{\Delta_{CC}}\right)^\epsilon E\, q^2 \ .
\end{equation}

\subsection{Full Propagator Correction}

We have just seen that, due to the pinch technique, the coloron-coloron and gluon-coloron VPAs receive an additional contribution from the pinched non-Abelian vertex corrections. Combining the VPAs, the UV divergences can be removed by two wavefunction renormalization counterterms (which arise from renormalizing the gauge eigenstates $G_{1\mu}$ and $G_{2\mu}$) and one mass counterterm (which arises from renormalizing the vacuum expectation value~$f$), in the usual way. In the $\overline{\rm MS}$ scheme we obtain
\begin{align}
&\left(\frac{\alpha_s}{4\pi}\right)^{-1}\Big[\Pi_{CC}(q^2)+\widetilde{\Pi}_{CC}(q^2)\Big] \notag \\
&= C_2(G)\int_0^1 dx \Bigg\{
\Bigg[\left(\left(\frac{\mu^2}{\Delta_{GC}}\right)^\epsilon-1\right) 2\Big(3+4x(1-x)\Big)E+2(1-2x)^2\Bigg]q^2 \notag \\
&\quad +\Bigg[
\left(\frac{\mu^2}{\Delta_{GC}}\right)^\epsilon \Big(1-x(4-3x)\Big)E-x(1-x)\Bigg] q^2 \notag \\
&\quad +\Bigg[-\left(\left(\frac{\mu^2}{\Delta_{GC}}\right)^\epsilon-1\right)2(2-x)E +3-5x\Bigg] M_C^2 \Bigg\} \\
&\ +4\cot^2(2\theta_c) C_2(G) \int_0^1 dx \Bigg\{
\Bigg[\left(\left(\frac{\mu^2}{\Delta_{CC}}\right)^\epsilon-1\right) \Big(3+4x(1-x)\Big)E+(1-2x)^2\Bigg]q^2 \notag \\
&\quad -\Bigg[\left(\left(\frac{\mu^2}{\Delta_{CC}}\right)^\epsilon-1\right)E+1\Bigg]\frac{x(1-x)}{4}  q^2
+\Bigg[-\left(\left(\frac{\mu^2}{\Delta_{CC}}\right)^\epsilon-1\right) \frac{3}{4} \, E+\frac{1}{4}\Bigg]M_C^2\Bigg\} \notag \\
&\ +(r_L^2+r_R^2)N_f \int_0^1 dx \left(\left(\frac{\mu^2}{\Delta_{qq}}\right)^\epsilon-1\right) \big[-2x(1-x)\big]E\, q^2 \notag \ ,
\end{align}
and
\begin{equation} \begin{split}
& \left(\frac{\alpha_s}{4\pi}\right)^{-1}\Big[\Pi_{GC}(q^2) + \widetilde{\Pi}_{GC}(q^2)\Big] =2 \cot(2\theta_c) C_2(G) \int_0^1 dx \\
& \quad \Bigg\{
\left(\left(\frac{\mu^2}{\Delta_{CC}}\right)^\epsilon-1\right)
\Bigg[\Big(\frac{7}{4}+5x(1-x)\Big)q^2+M_C^2\Bigg]E +(1-2x)^2 q^2 \Bigg\} \\
&\ +(r_L+r_R)N_f \int_0^1 dx \
\left(\left(\frac{\mu^2}{\Delta_{qq}}\right)^\epsilon-1\right) \big[-2x(1-x)\big]E\, q^2 \ .
\end{split} \end{equation}

The overall UV-finite propagator correction to the $q\bar{q}\to C$ amplitude can be found by insering these expressions in
\begin{equation}\begin{split}
iP+iP_\text{pinched} =\ & i\mathcal{M}_{q\bar{q}\to C}^{(0)} \, \frac{\Pi^\prime_{CC}(M_C^2)+\widetilde{\Pi}^\prime_{CC}(M^2_C)}{2} \\
&+ i\mathcal{M}_{q\bar{q}\to C}^{\prime\prime(0)} \,\frac{\Pi_{GC}(\hat{s})+\widetilde{\Pi}_{GC}(\hat{s})}{\hat{s}} \ .
\end{split} \end{equation}
Letting $m_g\to 0$, we find that $P+P_\text{pinched}$ becomes IR divergent, with the divergence arising from $\Pi_{CC}^\prime$. Setting $m_g=0$ and $\epsilon<0$ gives
\begin{equation}
iP+iP_\text{pinched} = \frac{\alpha_s}{4\pi} C_2(G) \left(-\frac{1}{\epsilon}\right) i\mathcal{M}_{q\bar{q}\to C}^{(0)} \, +\, \text{finite} \ .
\end{equation}

\begin{figure}[t!]
\begin{center}
\includegraphics[width=\textwidth]{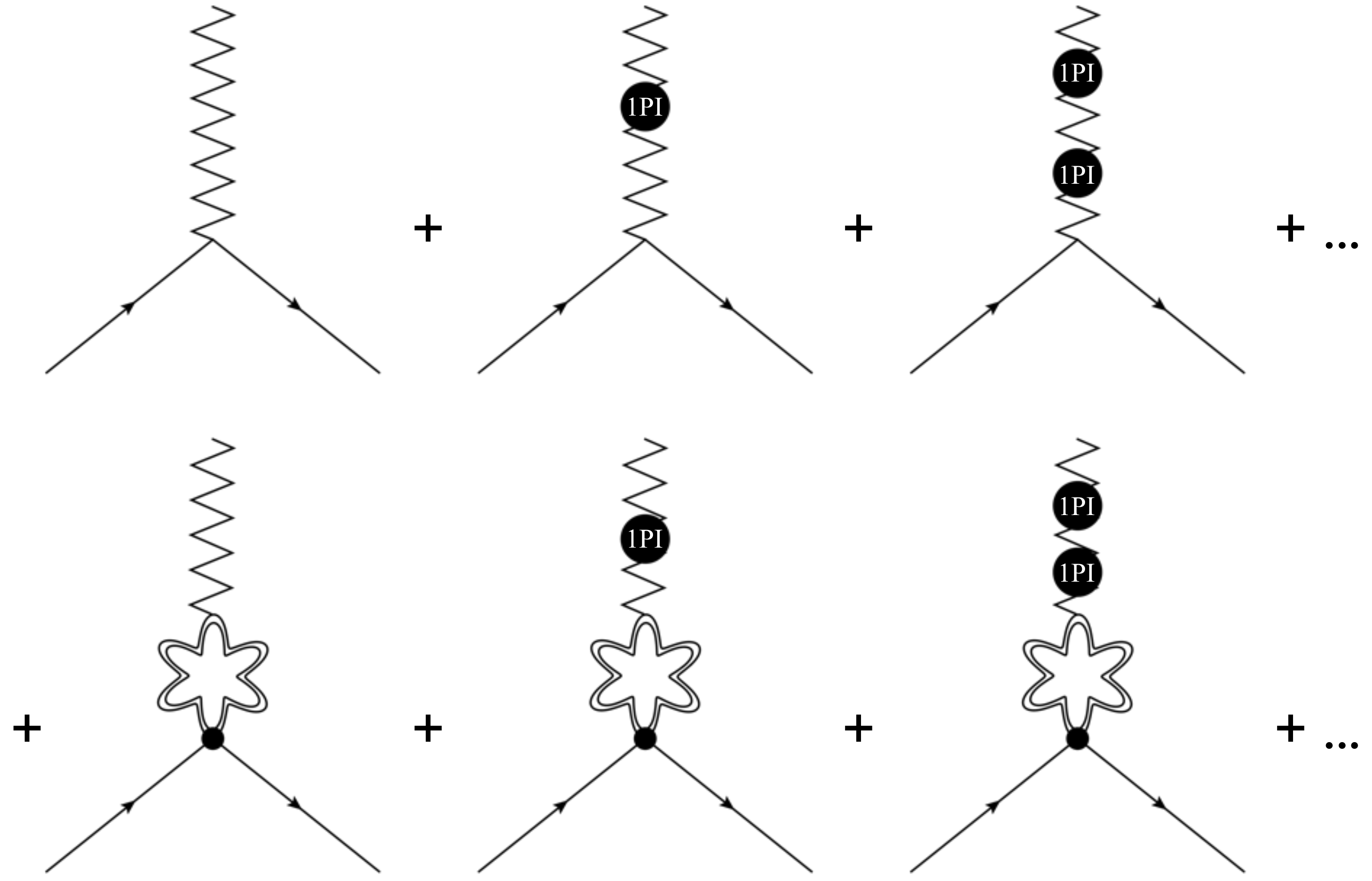}
\caption{The relevant contributions to the coloron Dyson series; as before, the zigzag lines represent colorons. The first row is the sum of the coloron VPA diagrams in the propagator, while the second row represents the sum of the VPA diagrams on top of the one-loop contribution from the pinched vertex correction (the double curly line illustrates generically all the allowed gauge bosons in the original non-Abelian vertices). The overall pinched amplitude factors out, and has no effect on the coloron pole mass.}
\label{fig:dyson}
\end{center}
\end{figure}

We have seen that the pinched diagrams contribute to the full propagators of the gluon-coloron system. This might seem in conflict with the expectation that the mass poles should be a property of freely propagating particles, and should not depend on any initial and/or final state. However, when we sum the Dyson series to obtain the full propagator, the pinched diagrams always appear as an overall prefactor, as pictorially shown in Fig.~\ref{fig:dyson}. This has an overall effect on the full propagators, which depend on the initial and final states, but has no effect on the propagator poles. Thus when we compute physical masses, we can do so by employing the true propagators in the computation, without the contribution from the pinched diagrams.

\subsection{Cross Section at One-Loop}

Adding up the tree-level contribution and the NLO contributions from $i Q+i V_\text{Abelian}+i V_\text{non-Abelian}$, and $iP+iP_\text{pinched}$, gives a $q\bar{q}\to C$ amplitude of the form
\begin{align}
i\mathcal{M}_{q\bar{q}\to C} = \ & i \mathcal{M}_{q\bar{q}\to C}^{(0)} + i Q+i V_\text{Abelian}
+i V_\text{non-Abelian} + iP+iP_\text{pinched} \notag \\
\equiv \ & i \mathcal{M}_{q\bar{q}\to C}^{(0)} + \frac{\alpha_s}{4\pi}\Big(
T\, i \mathcal{M}_{q\bar{q}\to C}^{(0)} + T^\prime\, i \mathcal{M}_{q\bar{q}\to C}^{\prime(0)}
+ T^{\prime\prime}\, i \mathcal{M}_{q\bar{q}\to C}^{\prime\prime(0)}
\Big) \ ,
\end{align}
where expressions for the real parts of $T$, $T^{\prime}$, and $T^{\prime \prime}$ are given below.
Averaging the squared amplitude over initial spins and colors, summing over final polarization states, and integrating over the phase space, gives the NLO result of the form
\begin{align}
\hat{\sigma}_\text{virt} \equiv \ & \hat{\sigma}_{q\bar{q}\to C}^{(0)}+\hat{\sigma}_{q\bar{q}\to C}^{(1)}  \label{eq:virtual} \\
= \ & \frac{\alpha_s \, A (r_L^2+r_R^2)}{\hat{s}} \, \delta(1-\chi) \Bigg[
1+\frac{\alpha_s}{2\pi}\Bigg(
\text{Re}\, T + \frac{r_L^4+r_R^4}{r_L^2+r_R^2}\, \text{Re}\, T^\prime + \frac{r_L+r_R}{r_L^2+r_R^2}\, \text{Re}\, T^{\prime\prime}\Bigg)
\Bigg] \notag \ .
\end{align}
At $\hat{s}=M_C^2$ it is possible to integrate over the Feynman parameter space in the expressions for $i Q+i V_\text{Abelian}+i V_\text{non-Abelian}$, and $iP+iP_\text{pinched}$. As we have seen, the UV infinities cancel in $i Q+i V_\text{Abelian}+i V_\text{non-Abelian}$ and are absorbed by propagator conterterms in $iP+iP_\text{pinched}$. Thus for $m_g\neq 0$ the overall amplitude is finite. Taking the $m_g\to 0$ limit leads to IR divergences in $\text{Re}\, T$, which are parametrized by taking $\epsilon<0$. For small and negative values of $\epsilon$ we obtain
\begin{align}
&\text{Re}\, T =  \left(\frac{4\pi\mu^2}{M_C^2}\right)^{\epsilon} \Gamma(1+\epsilon)
\Bigg[
-\frac{2}{\epsilon^2}-\frac{3}{\epsilon}-8+\frac{4\pi^2}{3}
\Bigg]C_2(r) \label{eq:T} \\
&\quad \; \; \, \quad \, +\Bigg[-E+\frac{61}{9}-\frac{5\pi}{2\sqrt{3}}-\frac{\pi^2}{3}-\frac{8}{3}\log\frac{M_C^2}{\mu^2}\Bigg]C_2(G)
\nonumber \\
&\quad \; \; \, \quad \, +\Bigg[
\frac{77}{48}-\frac{7\pi}{16\sqrt{3}}-\frac{29}{16}\log\frac{M_C^2}{\mu^2}\Bigg]4\cot^2(2\theta_c) C_2(G)
+ \Bigg[-\frac{1}{9}+\frac{1}{6}\log\frac{M_C^2}{\mu^2}\Bigg](r_L^2+r_R^2)N_f  \ , \nonumber \\
&\text{Re}\, T^\prime = \Bigg[
-\frac{11}{2}+\frac{2\pi^2}{3}
\Bigg]C_2(r)
+\Bigg[
1+\frac{5\pi}{2\sqrt{3}}-\frac{2\pi^2}{3}
\Bigg]C_2(G) \ ,  \notag \\
&\text{Re}\, T^{\prime\prime} = \Bigg[\frac{95}{9}-\frac{7\sqrt{3}\pi}{4}-\frac{43}{12}\log\frac{M_C^2}{\mu^2}\Bigg]2\cot (2\theta_c )C_2(G)
+\Bigg[-\frac{5}{9}+\frac{1}{3}\log\frac{M_C^2}{\mu^2}\Bigg](r_L+r_R)N_f \ . \notag
\end{align}
In the next section we will compute the corrections to the tree-level cross section due to the emission of soft and collinear gluons. We will show that the real emission cross section has IR divergences which exactly cancel the IR divergences contained in $\hat{\sigma}_\text{virt}$ (Eq.~(\ref{eq:virtual})), leading to a total cross section free of both UV and IR divergences.

\section{NLO Coloron Production: Real Corrections} \label{sec:real}

\begin{figure}[t]
\begin{center}
\includegraphics[width=\textwidth]{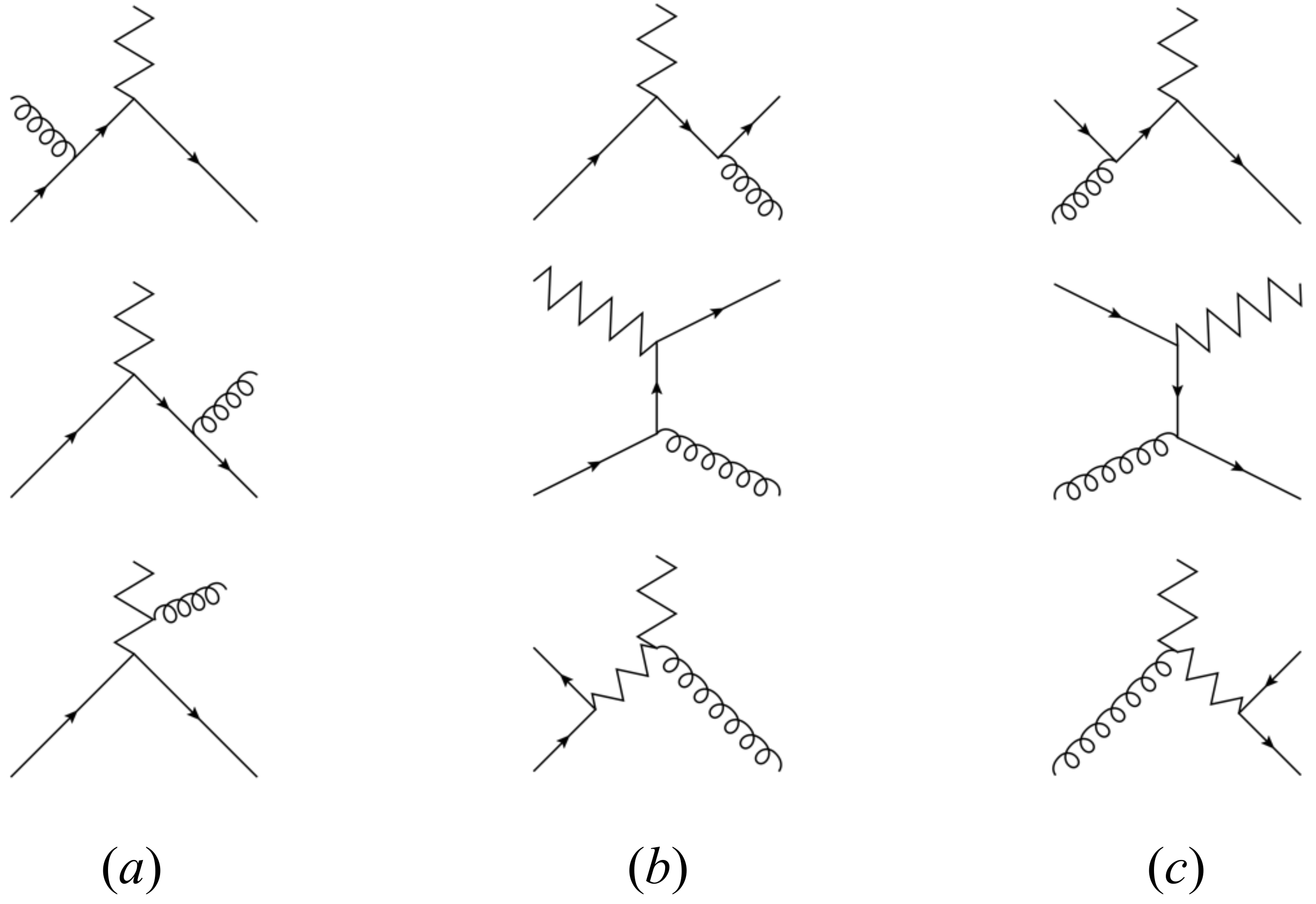}
\caption{Diagrams contributing to the real emission processes. A gluon field is, as usual, represented by a coiling line; a coloron field is represented by a zigzag line. $(a)$ Gluon emission. $(b)$ Quark emission. $(c)$ Antiquark emission.}
\label{fig:real}
\end{center}
\end{figure}

The real emission corrections, at NLO, are given by the diagrams of Fig~\ref{fig:real}. We first consider the diagrams with real emission of a gluon, shown in Fig~\ref{fig:real}(a). The squared amplitude, averaged over initial colors and spins, and summed over final colors and polarizations, is found to be, in $d=2(2-\epsilon)$ dimensions,

\begin{equation}\label{MRgsqavfin} \begin{split}
\overline{| \mathcal{M}_{q\bar{q} \rightarrow gC}^{(1)} |^{2}} = \ & \frac{C_{2}(r)\, g_{s}^{4} \, (r_{L}^{2}+r_{R}^{2}) }{\text{dim}(r)}\, \mu^{2\epsilon} \, (1-\epsilon) \\
&\times \left[ \frac{-1}{\omega(1-\omega)} \, C_{2}(r) + C_{2}(G) \right] \left[ \epsilon - \frac{1+\chi^{2}}{(1-\chi)^{2}} + 2\omega \, (1-\omega) \right]  \ ,
\end{split} \end{equation}
where
\begin{equation}
\omega\equiv \frac{1-\cos\theta}{2} \ .
\label{eq:omega}
\end{equation}
$\chi$ was defined in Eq.~(\ref{eq:chi}), and $\theta$ is the angle between the emitted gluon and the colliding quarks. The cross section for the real gluon emission is
\begin{equation} \label{csRg}
\hat{\sigma}_{q\bar{q}\to g C}^{(1)} = \frac{1}{2\hat{s}} \int d\Pi_{2} \, \overline{|\mathcal{M}_{q\bar{q}\to g C}^{(1)}|^{2}} \ ,
\end{equation}
where the integral is over the two-body Lorentz-invariant phase space in parton CM. In $d=2(2-\epsilon)$ dimensions,
\begin{equation} \label{LIPSd}
\int d\Pi_{2} = \frac{1}{8\pi} \, \frac{1-\chi}{\Gamma(1-\epsilon)} \left[ \frac{M_C^2(1-\chi)^{2}}{4\pi \chi} \right]^{-\epsilon} \int_{0}^{1}d \omega \left[ \omega \, (1-\omega) \right]^{-\epsilon} \ .
\end{equation}
This leads to the partonic cross section
\begin{equation} \begin{split}
\hat{\sigma}_{q\bar{q}\to g C}^{(1)} = \ & \frac{\alpha_s (r_L^2+r_R^2) A}{\hat{s}}\, \frac{\alpha_s}{2\pi}
\left(\frac{4\pi \mu^{2}}{M_C^2} \right)^\epsilon
\frac{\Gamma(1-\epsilon)}{\Gamma(1-2\epsilon)} \\
&\times \Bigg[-C_2(r) \, \frac{2}{\epsilon} \frac{\chi^\epsilon(1 + \chi^2)}{(1-\chi)^{1+2\epsilon}}
- C_2(G) \,  \frac{2}{3}\frac{\chi^\epsilon(1+\chi+\chi^2)}{(1-\chi)^{1+2\epsilon}} \Bigg] \ .
\label{eq:csreal}
\end{split} \end{equation}

Now $\chi$ is no longer constrained to be equal to one. Instead we must have $\chi\leq 1$, or else no on-shell coloron can be produced. The term proportional to $C_2(r)$ features a collinear singularity, parametrized by $\epsilon$, and a soft singularity, parametrized by $1-\chi$. The term proportional to $C_2(G)$ only features a soft singularity. The integral over $\chi$ in Eq.~(\ref{eq:csreal}) is finite for $\epsilon<0$, in spite of the singularity of the integrands.
For small and negative values of $\epsilon$ we can rewrite the $\chi$-dependence as follows
\begin{align}
& \frac{\chi^\epsilon(1 + \chi^2)}{(1-\chi)^{1+2\epsilon}} = -\frac{1}{\epsilon}\delta(1-\chi)
+\frac{1+\chi^2}{(1-\chi)_+}-\Bigg[2(1+\chi^2)\left(\frac{\log(1-\chi)}{1-\chi}\right)_{+}
-\frac{1+\chi^2}{1-\chi}\log\chi\Bigg] \epsilon \ , \nonumber \\
& \frac{\chi^\epsilon(1+\chi+\chi^2)}{(1-\chi)^{1+2\epsilon}}  = - \frac{3}{2\epsilon}\delta(1-\chi)
+\frac{1+\chi+\chi^2}{(1-\chi)_+} \ ,
\end{align}
where, as conventional, the ``+'' distributions are defined by

\begin{equation} \begin{split}
\int_0^1 d\chi \frac{f(\chi)}{(1-\chi)_+} \equiv \ & \int_0^1 d\chi \frac{f(\chi)-f(1)}{1-\chi} \ , \\
\int_0^1 d\chi\,f(\chi)\left(\frac{\log(1-\chi)}{1-\chi}\right)_{+} \equiv \ &
\int_0^1 d\chi \left[f(\chi)-f(1)\right]\frac{\log(1-\chi)}{1-\chi} \ .
\end{split} \end{equation}
The coefficients of the delta functions are found by integrating both sides of the equations. The partonic cross section becomes
\begin{equation}
\hat{\sigma}_{q\bar{q}\to g C}^{(1)} =
\frac{\alpha_s \, A (r_L^2+r_R^2)}{\hat{s}}\, \frac{\alpha_s}{2\pi}
\Bigg[\delta(1-\chi)\,R + R^\prime\Bigg] \ ,
\label{eq:csrealsplit}
\end{equation}
where, using Eq.~(\ref{eq:E}), and expanding for small values of $\epsilon$,
\begin{equation}\label{eq:RRp} \begin{split}
& R = \left(\frac{4\pi\mu^2}{M_C^2}\right)^\epsilon \frac{\Gamma(1-\epsilon)}{\Gamma(1-2\epsilon)}\Bigg\{
C_2(r)\Bigg[\frac{2}{\epsilon^2}+\frac{3}{\epsilon}\Bigg] + C_2(G)\frac{1}{\epsilon}\Bigg\}\ ,  \\
& R^\prime =-2\Bigg[E-\log\frac{M_C^2}{\mu^2}\Bigg] P_{q\to q}(\chi) \\
& \; \; \, \quad \ + C_2(r)\Bigg[4(1+\chi^2)\left(\frac{\log(1-\chi)}{1-\chi}\right)_+ -2\frac{1+\chi}{1-\chi}\log\chi\Bigg]
+C_2(G)\frac{2}{3}\frac{1+\chi+\chi^2}{(1-\chi)_+}  \ .
\end{split} \end{equation}

In the second equation $P_{q\to q}(\chi)$ is the Altarelli-Parisi splitting function for an on-shell quark to evolve into a virtual quark and a real gluon
\begin{equation}
P_{q\to q}(\chi) = C_2(r)\Bigg[\frac{1+\chi^2}{(1-\chi)_+}+\frac{3}{2}\delta(1-\chi)\Bigg] \ .
\end{equation}
Adding together $\hat{\sigma}_\text{virt}$, given by Eqs.~(\ref{eq:virtual}) and (\ref{eq:T}), and $\hat{\sigma}_{q\bar{q}\to gC}^{(1)}$, given by Eqs.~(\ref{eq:csrealsplit}) and \eqref{eq:RRp}, shows that the IR divergences proportional to $\delta(1-\chi)$ cancel. There is still a collinear singularity in $R^\prime$, proportional to the Altarelli-Parisi evolution $P_{q\to q}(\chi)$. This singularity arises from integrating over all collinear initial-state gluons.
As we will see in the next section, these collinear IR divergences will be absorbed through renormalization of the PDFs.

The real quark and antiquark emission diagrams are shown in Figs.~\ref{fig:real}(b) and \ref{fig:real}(c), respectively. The corresponding summed-averaged squared amplitudes in $d=2(2-\epsilon)$ are
\begin{equation}\label{MRfsqavfin} \begin{split}
\overline{| \mathcal{M}_{qg \rightarrow qC}^{(1)} |^{2}} = \ & \frac{C_{2}(r)\, g_{s}^{4} \, (r_{L}^{2}+r_{R}^{2}) }{\text{dim}(G)}\, \mu^{2\epsilon} \left[ C_{2}(r) + C_{2}(G) \frac{(1-\chi)(1-\omega)}{(1-(1-\chi)(1-\omega))^2} \right] \\
&\times \left[ 2\,(\epsilon+\chi) + \frac{1 - \epsilon -2\chi(1- \chi)}{(1 - \chi) \, (1-\omega)} +  (1-\epsilon)(1-\chi) \, (1-\omega) \right]  \ ,
\end{split} \end{equation}
and
\begin{equation}\label{MRfbsqavfin} \begin{split}
\overline{| \mathcal{M}_{\bar{q}g \rightarrow \bar{q}C}^{(1)} |^{2}} = \ & \frac{C_{2}(r)\, g_{s}^{4} \, (r_{L}^{2}+r_{R}^{2}) }{\text{dim}(G)}\, \mu^{2\epsilon} \left[ C_{2}(r) + C_{2}(G) \frac{(1-\chi)\, \omega}{(1-(1-\chi)\, \omega)^2} \right] \\
& \times \left[ 2\,(\epsilon+\chi) + \frac{1 - \epsilon -2\chi(1- \chi)}{(1 - \chi) \, \, \omega} +  (1-\epsilon)(1-\chi) \, \, \omega \right] \ ,
\end{split} \end{equation}
where $\text{dim}(G)\equiv 8$ is the dimension of the adjoint representation. Note that the amplitudes for quark
and antiquark emission are related by crossing, i.e. $\omega \leftrightarrow (1-\omega)$.
The integration over the two-body Lorentz-invariant phase space proceeds as in the gluon emission case, yielding
\begin{equation}
\hat{\sigma}_{q g\to q C}^{(1)} = \hat{\sigma}_{\bar{q} g\to \bar{q} C}^{(1)} = 
\frac{\alpha_s \, A (r_L^2+r_R^2)}{\hat{s}}\, \frac{\alpha_s}{2\pi}\, R^{\prime\prime} \ ,
\label{eq:refextra}
\end{equation}
where
\begin{align}
R^{\prime\prime} = \ & \frac{\text{dim}(r)}{\text{dim}(G)}
\Bigg\{C_2(r)\frac{3+2\chi-3\chi^2}{2}
+C_2(G)\Bigg[\frac{(1-\chi)(2+\chi+2\chi^2)}{\chi}+2(1+\chi)\log\chi\Bigg]\Bigg\} \nonumber \\
&-\Bigg[E-\log\frac{M_C^2}{\mu^2}-\log\frac{(1-\chi)^2}{\chi}+1\Bigg]P_{g\to q}(\chi) \ .
\label{eq:Rpp}
\end{align}
Here $P_{g\to q}(\chi)$ is the Altarelli-Parisi splitting function for an on-shell gluon to evolve to a virtual-real quark pair,
\begin{equation}
P_{g\to q}(\chi) =\frac{C_2(r)\cdot\text{dim}(r)}{\text{dim}(G)} \left[\chi^2+(1-\chi)^2\right] \ ,
\end{equation}
where $C_2(r)\cdot \text{dim}(r)/\text{dim}(G)= 1/2$. There is no soft singularity in $\hat{\sigma}_{q g\to q C}^{(1)}\equiv \hat{\sigma}_{\bar{q} g\to \bar{q} C}^{(1)}$, only a collinear singularity proportional to the Altarelli-Parisi evolution $P_{g\to q}(\chi)$. As noted above regarding $\hat{\sigma}_{q\bar{q}\to g C}^{(1)}$, this singularity will be canceled by renormalization of the PDFs when we compute the
total hadronic cross section.

\section{NLO Cross Section}\label{sec:NLOcs}

Our calculations in the previous sections have produced all of the relevant partonic cross sections at NLO and demonstrated them to be both UV and IR finite.\footnote{Note that the $g g\to C$ process vanishes at tree-level \cite{Chivukula:2001wb} and the one-loop contributions are small, less than of order 0.1\% of the $q\bar{q}$-initiated
leading order contribution \cite{Allanach:2010cj}; we, therefore, do not include this process in this work.}

The full NLO cross section for coloron production at the LHC is
\begin{align}
&\sigma^\text{NLO} = \int d x_1 \int d x_2 \notag \\
&\ \Bigg\{
\sum_q\Big[f^0_q(x_1) f^0_{\bar{q}}(x_2) + f^0_{\bar{q}}(x_1) f^0_q(x_2)\Big]\Big(\hat{\sigma}_{q\bar{q}\to C}^{(0)}
+\hat{\sigma}_{q\bar{q}\to C}^{(1)} + \hat{\sigma}_{q\bar{q}\to g C}^{(1)}\Big) \label{eq:barePDFcs} \\
&\quad + \sum_q\Big[f^0_q(x_1) f^0_g(x_2)
+ f^0_g(x_1) f^0_q(x_2)+ f^0_{\bar{q}}(x_1) f^0_g(x_2) + f^0_g(x_1) f^0_{\bar{q}}(x_2)\Big] \hat{\sigma}_{q g\to q C}^{(1)}
\Bigg\} \notag \ ,
\end{align}
where the partonic cross sections $\hat{\sigma}$ are given in Eqs.~(\ref{cstree}), (\ref{eq:virtual}), (\ref{eq:csrealsplit}), and (\ref{eq:refextra}), and
where the superscript ``0'' in the PDFs will be clear in a moment. We saw that all IR divergences contained in $\sigma$ cancel, except for a couple of collinear singularities proportional to Altarelli-Parisi evolutions. Such singularities arise because we integrated over all collinear quarks and gluons, even those which we should have included in the PDFs. Therefore, the corresponding IR singularities are absorbed by renormalizing the bare PDFs in Eq.~(\ref{eq:barePDFcs}). In the $\overline{\text{MS}}$ scheme,
\begin{equation}\label{eq:renormPDF} \begin{split}
f_i(x, \mu_F) =\ & f^0_i (x) \\
& - \frac{g_3^2}{8\pi^2} \brac{\frac{1}{\epsilon} - \gamma + \log (4\pi) - \log \frac{\mu_F^2}{\mu^2}} \int \frac{d\chi}{\chi} \sum_j \, f^0_j \brac{\frac{x}{\chi}} P_{j \rightarrow i} (\chi) \ ,
\end{split} \end{equation}
where $i,j=q,g$, and $\mu_F$ is the factorization scale. Exchanging the bare PDFs for the renormalized ones replaces $E$ with $\log\mu_F^2/\mu^2$ in Eqs.~(\ref{eq:RRp}) and (\ref{eq:Rpp}). The hadronic cross section becomes
\begin{align}
&\sigma^\text{NLO} = \frac{\alpha_s\, A\, H_1(\theta_c)}{s} \int \frac{d x_1}{x_1} \int \frac{d x_2}{x_2} \\ 
&\ \Bigg\{
\sum_q\Big[f_q(x_1,\mu_F) f_{\bar{q}}(x_2,\mu_F)+f_{\bar{q}}(x_1,\mu_F)f_q(x_2,\mu_F)\Big] \Big(\delta(1-\chi)+\frac{\alpha_s}{2\pi}\, {\cal F}^{qq}(\chi)\Big) \notag \\
&\quad + \sum_q\Big[f_q(x_1,\mu_F)  f_g(x_2, \mu_F) + f_g(x_1, \mu_F) f_q(x_2,\mu_F)
+( f_q \to f_{\bar{q}})
\Big] \, \frac{\alpha_s}{2\pi}\, {\cal F}^{qg}(\chi)\Bigg\}\notag \ , 
\end{align}
where the function $H_1(\theta_c)$ is defined below, in Eq.~(\ref{eq:H}), $A$ is defined in Eq.~(\ref{eq:defA}), and the partonic CM energy $\hat{s}$ has been traded for the hadronic one, as in Eq.~(\ref{eq:x1x2}). Notice that since the integrand is now finite, we can ignore the $1-\epsilon$ factor in $A$. The functions ${\cal F}^{qq}(\chi)$ and ${\cal F}^{qg}(\chi)$ are

\begin{equation}\label{eq:F} \begin{split}
{\cal F}^{qq}(\chi) = \ & 2\log\frac{M_C^2}{\mu_F^2} P_{q\to q}(\chi) +D_q(\chi) \ , \\
{\cal F}^{qg}(\chi) = \ & \log\frac{M_C^2}{\mu_F^2} P_{g\to q}(\chi) +D_g(\chi) \ ,
\end{split} \end{equation}
where
\begin{equation} \begin{split}
D_q(\chi) = \ & C_2(r)\Bigg[4(1+\chi^2)\left(\frac{\log(1-\chi)}{1-\chi}\right)_+ -2\,\frac{1+\chi}{1-\chi}\log\chi\Bigg] \\
&+ C_2(G)\, \frac{2}{3}\,\frac{1+\chi+\chi^2}{(1-\chi)_+}+\delta(1-\chi)\, Q  \ , \\
D_g(\chi) = \ & \frac{\text{dim}(r)}{\text{dim}(G)}\Bigg\{
C_2(r)\Bigg[\Big(\chi^2+(1-\chi)^2\Big)\Bigg(\log\frac{(1-\chi)^2}{\chi}-1\Bigg)+\frac{3}{2}+\chi-\frac{3}{2}\chi^2\Bigg] \\
&+C_2(G)\Bigg[\frac{(1-\chi)(2+\chi+2\chi^2)}{\chi} +2(1+\chi)\log\chi\Bigg]
\Bigg\} \ ,
\end{split} \end{equation}
and
\begin{align}
Q =\ & N_f\Bigg[\Bigg(-\frac{1}{9}+\frac{1}{6}\log\frac{M_C^2}{\mu^2}\Bigg)H_1(\theta_c)
+ \Bigg(-\frac{5}{9}+\frac{1}{3}\log\frac{M_C^2}{\mu^2}\Bigg)H_2(\theta_c)\Bigg] \notag \\
&+C_2(r)\Bigg[-8+\frac{2\pi^2}{3}+\Bigg(-\frac{11}{2}+\frac{2\pi^2}{3}\Bigg)H_3(\theta_c)\Bigg] \label{eq:QNLO} \\
&+C_2(G)\Bigg[\frac{61}{9}-\frac{5\pi}{2\sqrt{3}}-\frac{\pi^2}{3}-\frac{11}{3}\log\frac{M_C^2}{\mu^2}
+\Bigg(\frac{77}{12}-\frac{7\pi}{4\sqrt{3}}-\frac{29}{4}\log\frac{M_C^2}{\mu^2}\Bigg)\cot^2(2\theta_c) \notag \\
&\quad +\Bigg(1+\frac{5\pi}{2\sqrt{3}}-\frac{2\pi^2}{3}\Bigg)H_3(\theta_c) +\Bigg(\frac{190}{9}-\frac{7\sqrt{3}\pi}{2}-\frac{43}{6}\log\frac{M_C^2}{\mu^2}\Bigg)\cot(2\theta_c)H_4(\theta_c)
\Bigg] \notag \ .
\end{align}
The functions $H_i(\theta_c)$ are determined by the chiral couplings of the quarks to the colorons~(which depend on the charges of the quarks under the full $SU(3)_{1C} \times SU(3)_{2C}$ symmetry)
\begin{equation}\label{eq:H} \begin{split}
H_1(\theta_c) = \ &
\begin{cases}
2\tan^2\theta_c&r_L=r_R=-\tan\theta_c \\
\tan^2\theta_c + \cot^2\theta_c&r_L\neq r_R \\
2\cot^2\theta_c&r_L=r_R=\cot\theta_c
\end{cases} \ , \\
H_2(\theta_c) = \ &
\begin{cases}
2&r_L=r_R=-\tan\theta_c \\
\displaystyle{\frac{2(1+\cos(4\theta_c))}{3+\cos(4\theta_c))}}&r_L\neq r_R \\
2&r_L=r_R=\cot\theta_c
\end{cases} \ , \\
H_3(\theta_c) = \ &
\begin{cases}
\tan^2\theta_c&r_L=r_R=-\tan\theta_c \\
\displaystyle{\frac{\tan^4\theta_c+\cot^4\theta_c}{\tan^2\theta_c+\cot^2\theta_c}}&r_L\neq r_R \\
\cot^2\theta_c&r_L=r_R=\cot\theta_c
\end{cases} \ , \\
H_4(\theta_c) = \ &
\begin{cases}
-\cot\theta_c&r_L=r_R=-\tan\theta_c \\
\displaystyle{\frac{\sin(4\theta_c)}{3+\cos(4\theta_c)}}&r_L\neq r_R \\
\tan\theta_c&r_L=r_R=\cot\theta_c
\end{cases} \ .
\end{split} \end{equation}
At NLO the $\mu$ dependence is removed by trading the $\overline{\rm MS}$ couplings $g_{1s}$ and $g_{2s}$, or $g_s$ and $\theta_c$, for the corresponding running couplings. Since $\theta_c$ is a free parameter, we simply set $\mu\equiv M_C$, and express the cross section as a function of the $\overline{\rm MS}$ couplings. At the same time,
the NLO $\mu_F$ dependence weakens once the renormalized PDFs are employed, as $\sigma$ in Eq.~(\ref{eq:barePDFcs}) is independent of $\mu_F$
to this order in perturbation theory.
From these results we may also compute the transverse momentum distribution of the produced coloron, which  is given by
\begin{align}
\frac{d\sigma}{d p_T} = & \int d x_1 \int d x_2 \Bigg\{
\sum_q \Big[f_q(x_1,\mu_F) f_{\bar{q}}(x_2,\mu_F) + f_{\bar{q}}(x_1,\mu_F) f_q(x_2,\mu_F)\Big]
\, \frac{d\hat{\sigma}_{q\bar{q}\to g C}}{d p_T} \notag \\
&\ +\sum_q\Big[f_q(x_1,\mu_F) f_g(x_2,\mu_F)+ f_g(x_1,\mu_F) f_q(x_2,\mu_F) \label{eq:dsigmadpt} \\
&\qquad + f_{\bar{q}}(x_1,\mu_F) f_g(x_2,\mu_F) + f_g(x_1,\mu_F) f_{\bar{q}}(x_2,\mu_F)\Big] \,
\frac{d\hat{\sigma}_{q g\to q C}}{d p_T}
\Bigg\} \notag \ ,
\end{align}
where
\begin{align}
\frac{d\hat{\sigma}_{q\bar{q}\to g C}}{d p_T} =\ & \frac{1}{4\pi\hat{s}^2 (1-\chi^2)} \,
\frac{p_T}{\displaystyle{\sqrt{1-\frac{4p_T^2}{\hat{s}(1-\chi)^2}}}}
\times 2 \overline{|\mathcal{M}_{q\bar{q}\to g C}^{(1)}|^{2}} \ , \\
\frac{d\hat{\sigma}_{q g\to q C}}{d p_T} =\ & \frac{1}{4\pi\hat{s}^2 (1-\chi^2)} \,
\frac{p_T}{\displaystyle{\sqrt{1-\frac{4p_T^2}{\hat{s}(1-\chi)^2}}}} \notag \\
&\times \Bigg( \overline{|\mathcal{M}_{q g\to q C}^{(1)}|^{2}} + \overline{|\mathcal{M}_{q g\to q C}^{(1)}|^{2}}\restriction_{\omega\to 1-\omega} \Bigg) \ ,
\end{align}
and $\omega$ (Eq.~(\ref{eq:omega})) is given by
\begin{equation}
\omega=\frac{1-\displaystyle{\sqrt{1-\frac{4p_T^2}{\hat{s}(1-\chi)^2}}}}{2} \ .
\end{equation}
Note that this is the \textit{leading order} prediction for $d\sigma/dp_T$, and, therefore, this distribution is strongly
$\mu_F$-dependent.

\section{Conclusion \& Discussion}\label{sec:conclusion}

\begin{figure}[t]\begin{center}
\includegraphics[width=\textwidth]{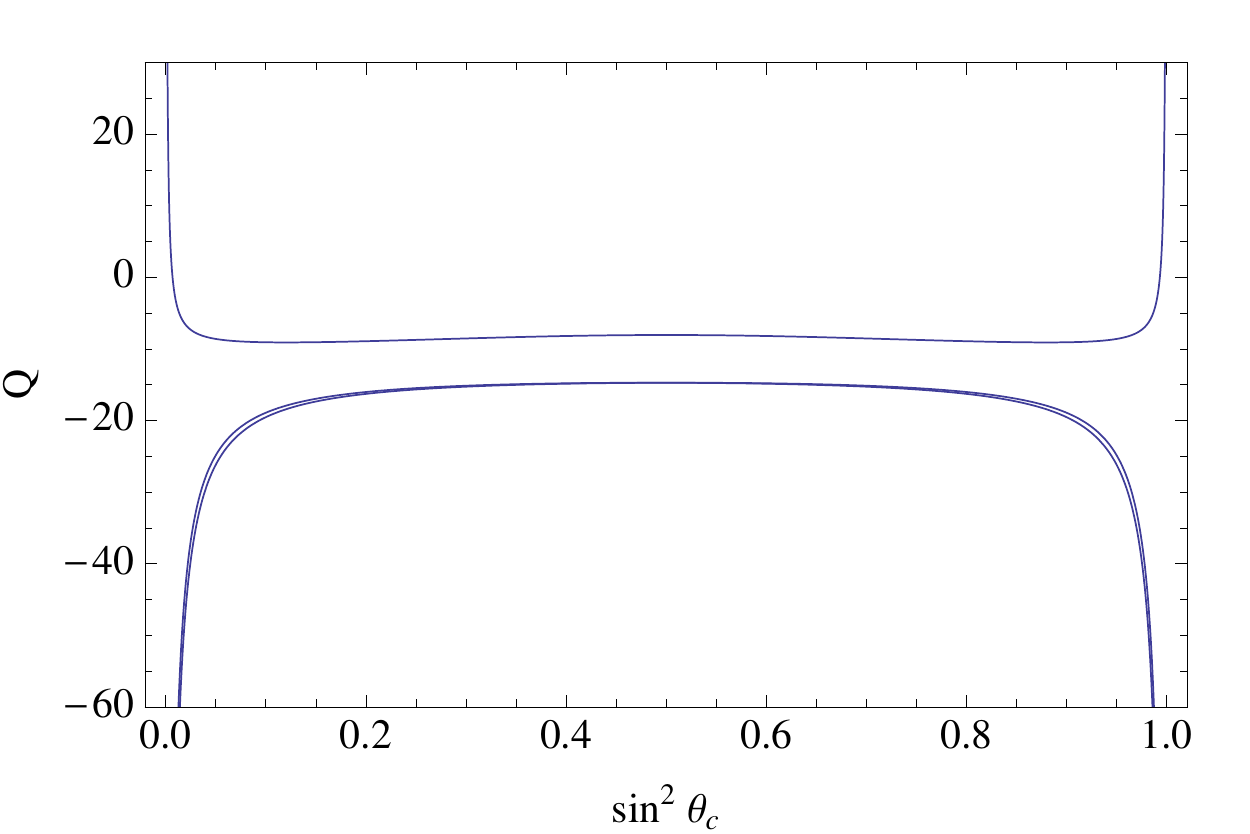}
\caption{Behavior of the $Q$ function defined in Eq.~(\ref{eq:QNLO}), for $\mu=M_C$: this gives the contribution from the virtual corrections to the NLO cross section for coloron production. The upper curve is for the $r_L\neq r_R$ scenario, whereas the almost identical lower curves are for $r_L=r_R=-\tan\theta_c$, and $r_L=r_R=\cot\theta_c$.
Note that $Q$, and, therefore, the NLO corrections, become very large when $\sin^2\theta_c$ is either too small or too large.}
\label{fig:Q}
\end{center}\end{figure}

\begin{figure}\begin{center}
\includegraphics[width=.8\textwidth,height=2.7in]{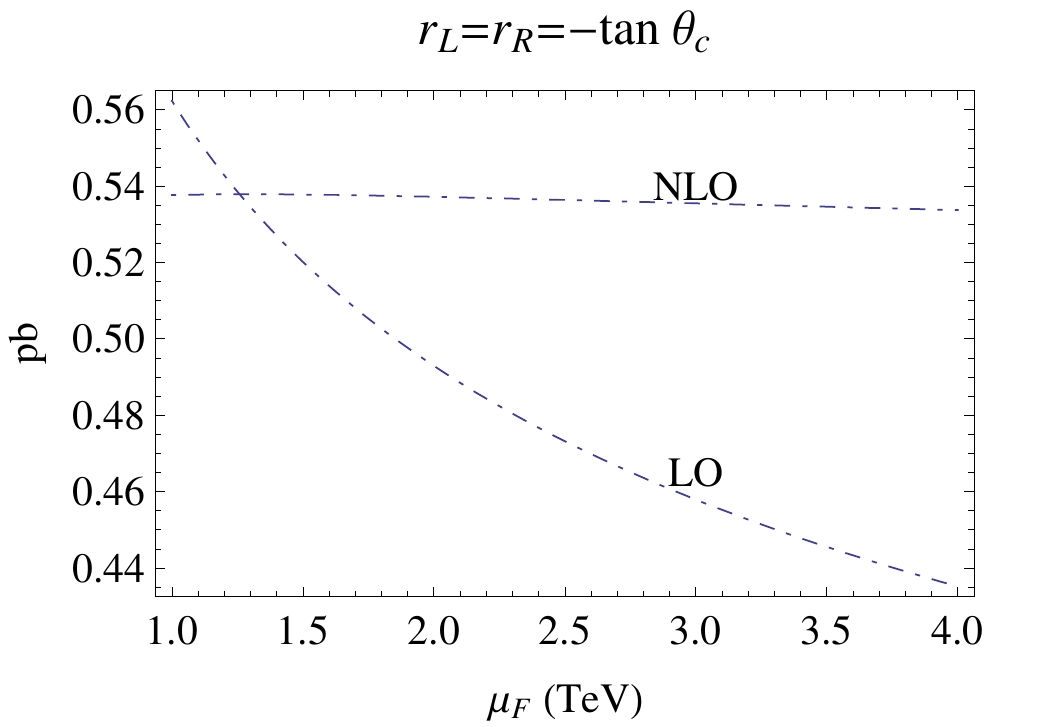}
\includegraphics[width=.8\textwidth,height=2.7in]{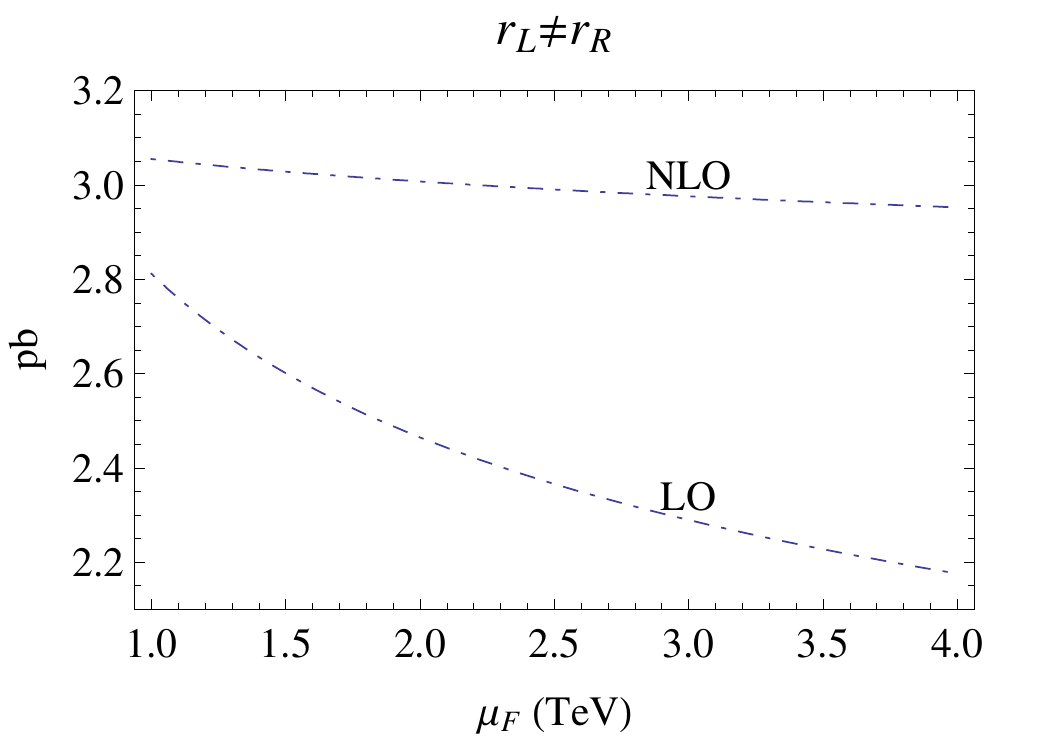}
\includegraphics[width=.8\textwidth,height=2.7in]{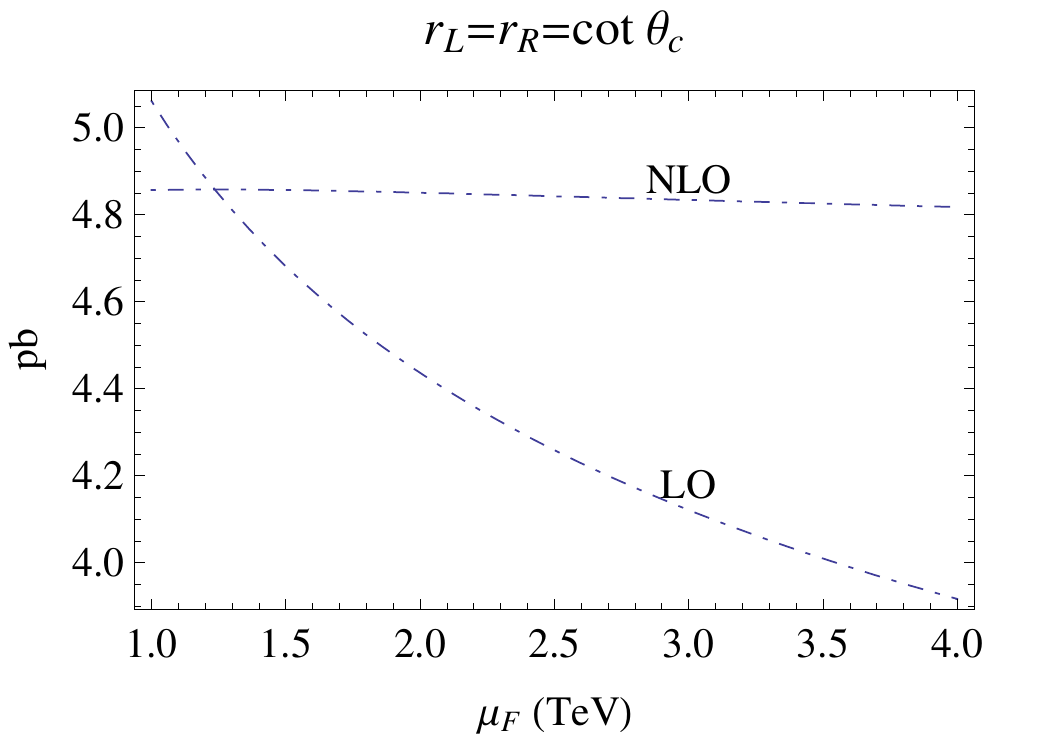}
\caption{Dependence of LO and NLO cross sections at the LHC ($\sqrt{s}=$ 7 TeV), as a function of factorization scale $\mu_F$
for $M_C$ = 2.0 TeV, $\sin^2\theta_c\vert_{\mu = 2.0\,\text{TeV}} = 0.25$, and the three possible flavor-universal scenarios for the quark charge assignments. As expected, the NLO cross section has a much weaker (formally, two-loop) residual scale-dependence.}
\label{fig:scale-dependence}
\end{center}\end{figure}

\begin{figure}\begin{center}
\includegraphics[width=.8\textwidth,height=2.35in]{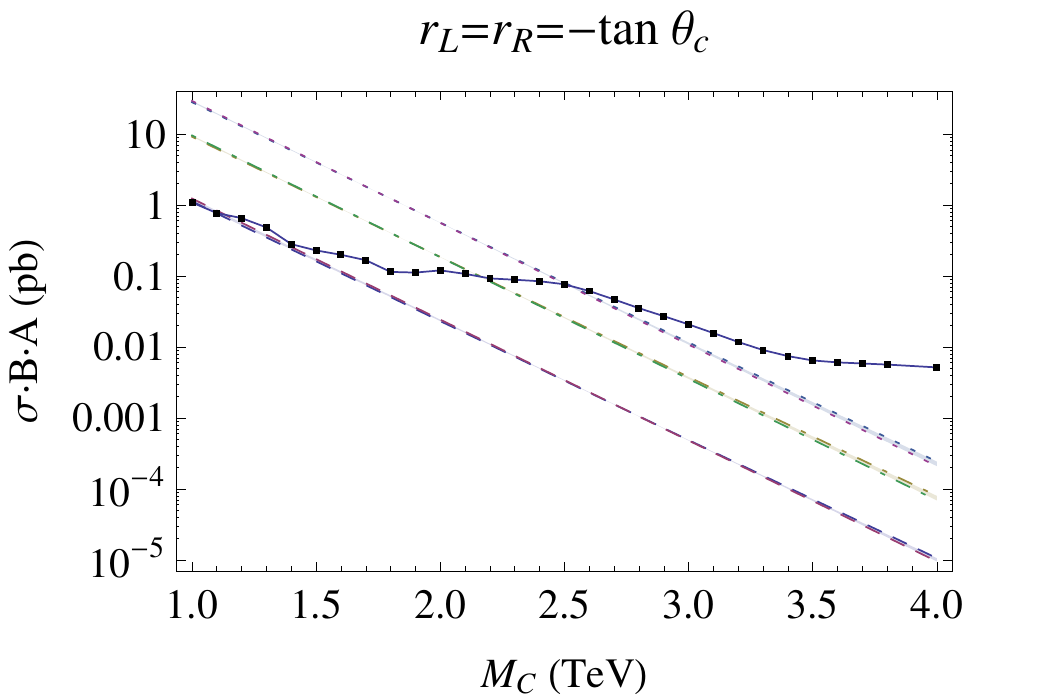}
\includegraphics[width=.8\textwidth,height=2.35in]{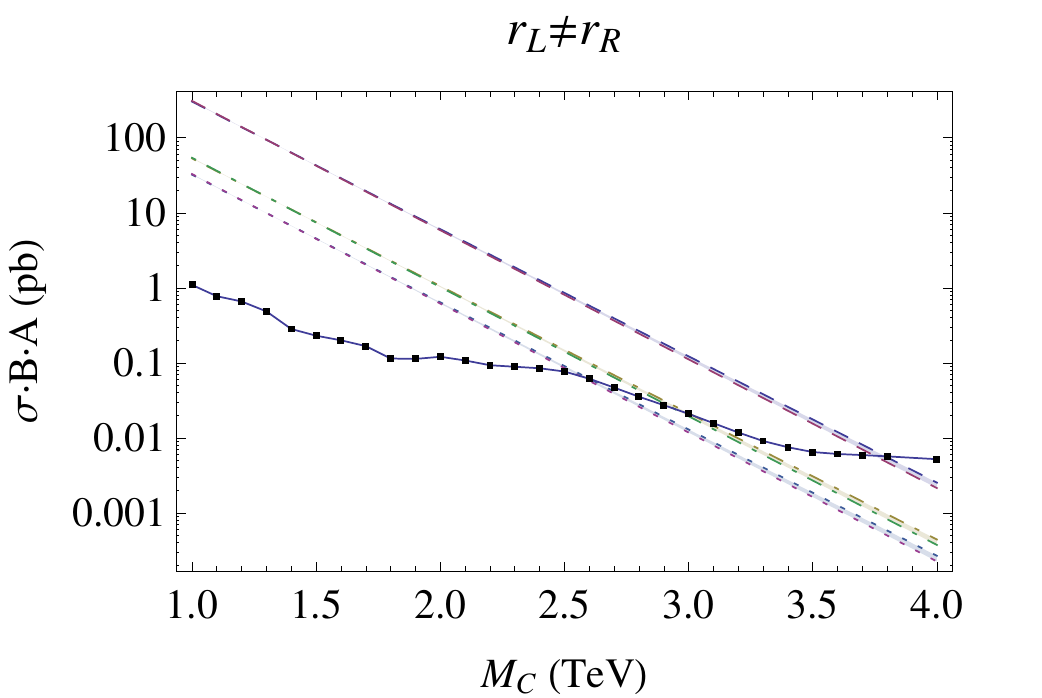}
\includegraphics[width=.8\textwidth,height=2.35in]{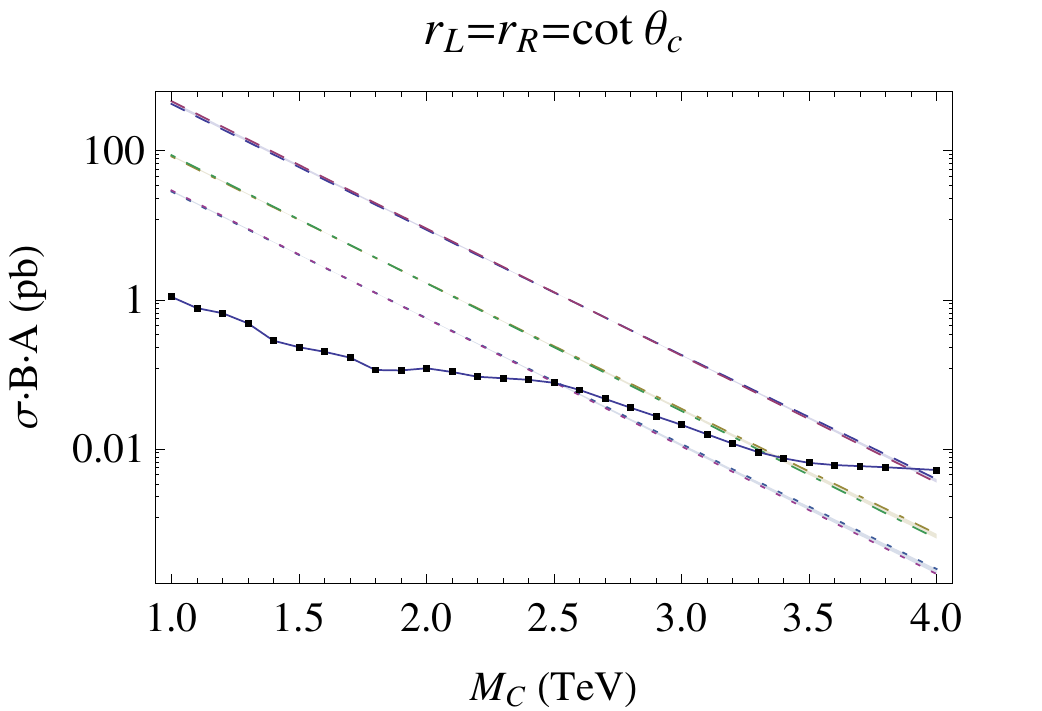}
\caption{NLO cross section times branching ratio to quarks for on-shell coloron production at the LHC ($\sqrt{s}=$ 7 TeV), corrected for acceptance as described in the text. We consider the three possible flavor-universal scenarios for the quark charge assignments, take the renormalization scale $\mu$ to be equal to $M_C$, and plot $\sigma$ for $\sin^2\theta_c\vert_{\mu=M_C}=$ 0.05 (dashed), 0.25 (dot-dashed), and 0.5 (dotted). We plot these cross sections for $\mu_F$ ranging from $M_C/2$ to $2\,M_C$ and,  reflecting the weak dependence of the NLO cross section on the factorization scale, the resulting bands for each $\sin^2\theta_c$ are very narrow. To give a sense of current experimental reach, we plot the CMS \protect\cite{Chatrchyan:2011bd} upper limit (solid line)
on the cross section times dijet branching ratio for a narrow resonance.}
\label{fig:sigma}
\end{center}\end{figure}

\begin{figure}\begin{center}
\includegraphics[width=.8\textwidth,height=2.7in]{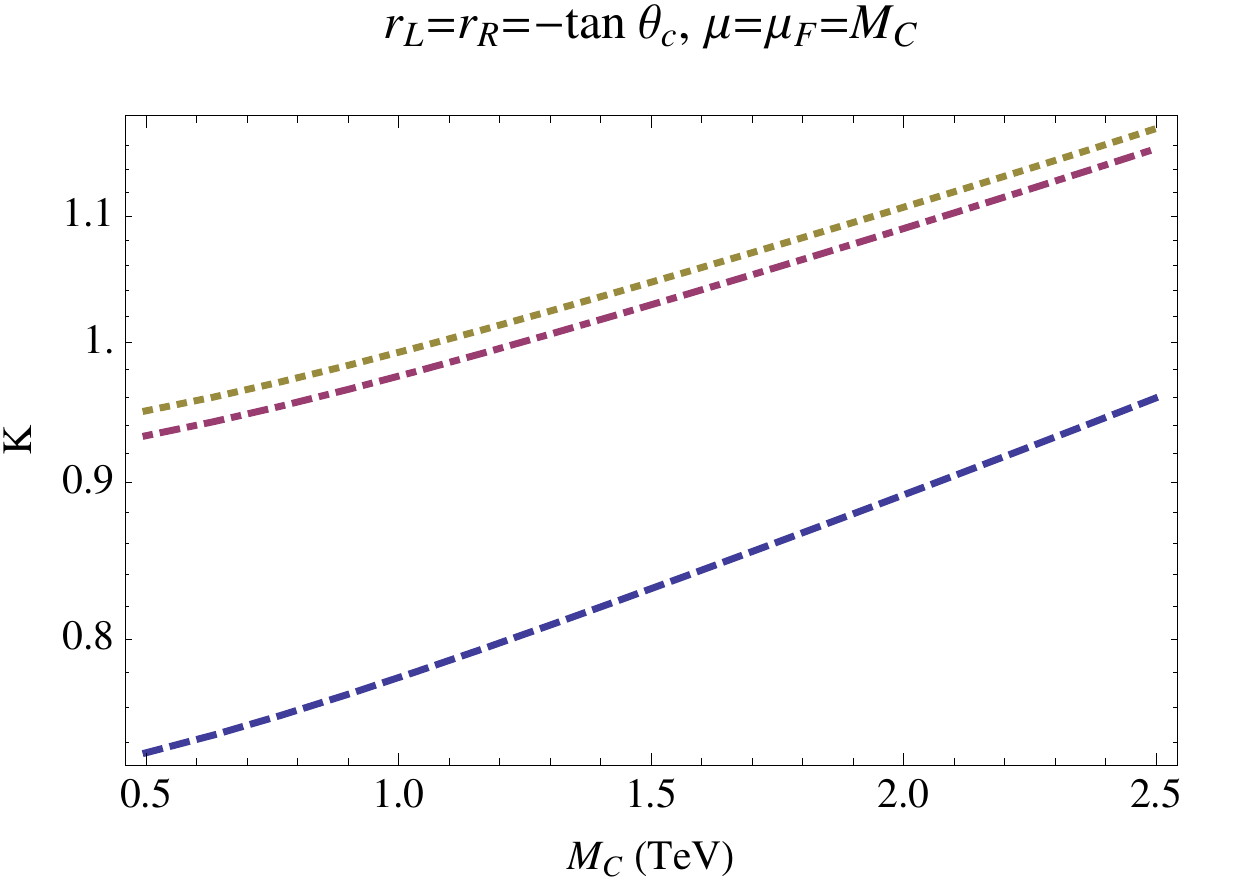}
\includegraphics[width=.8\textwidth,height=2.7in]{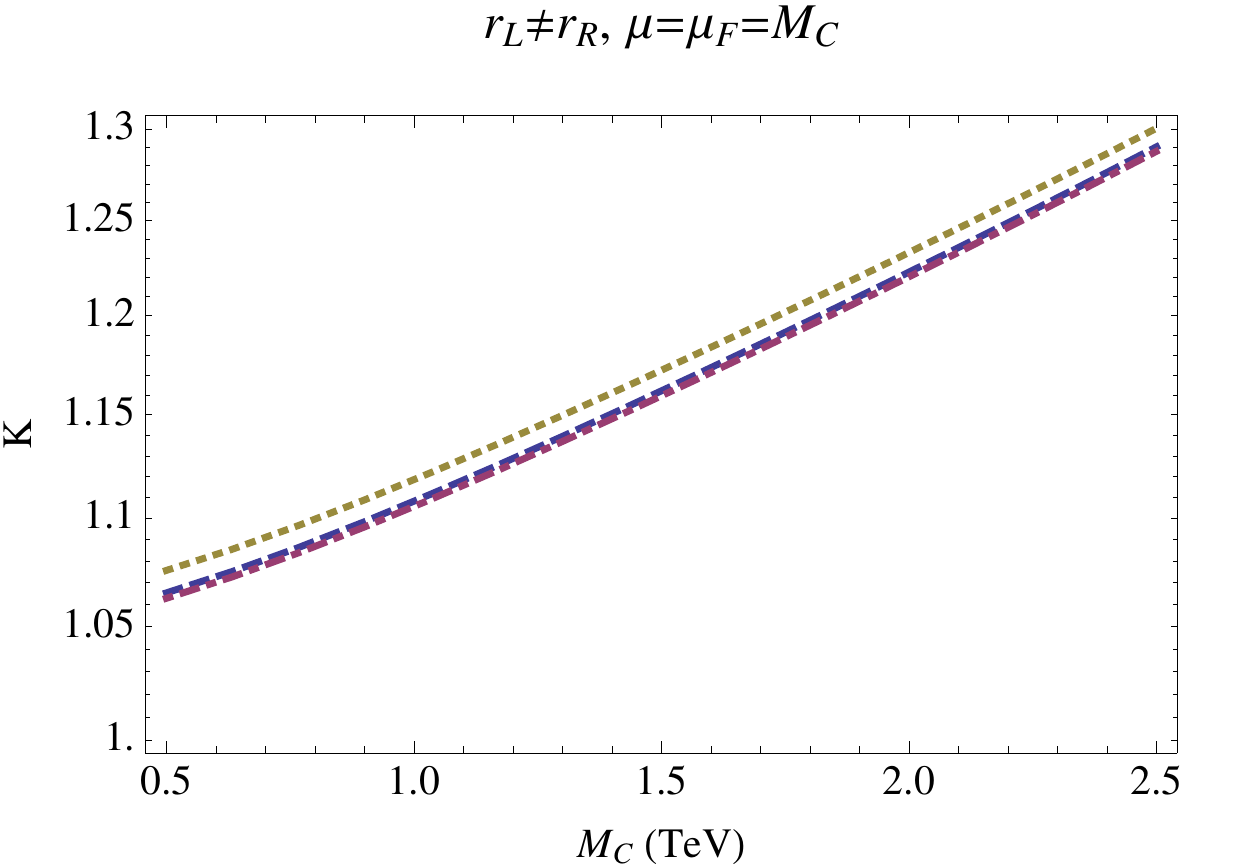}
\includegraphics[width=.8\textwidth,height=2.7in]{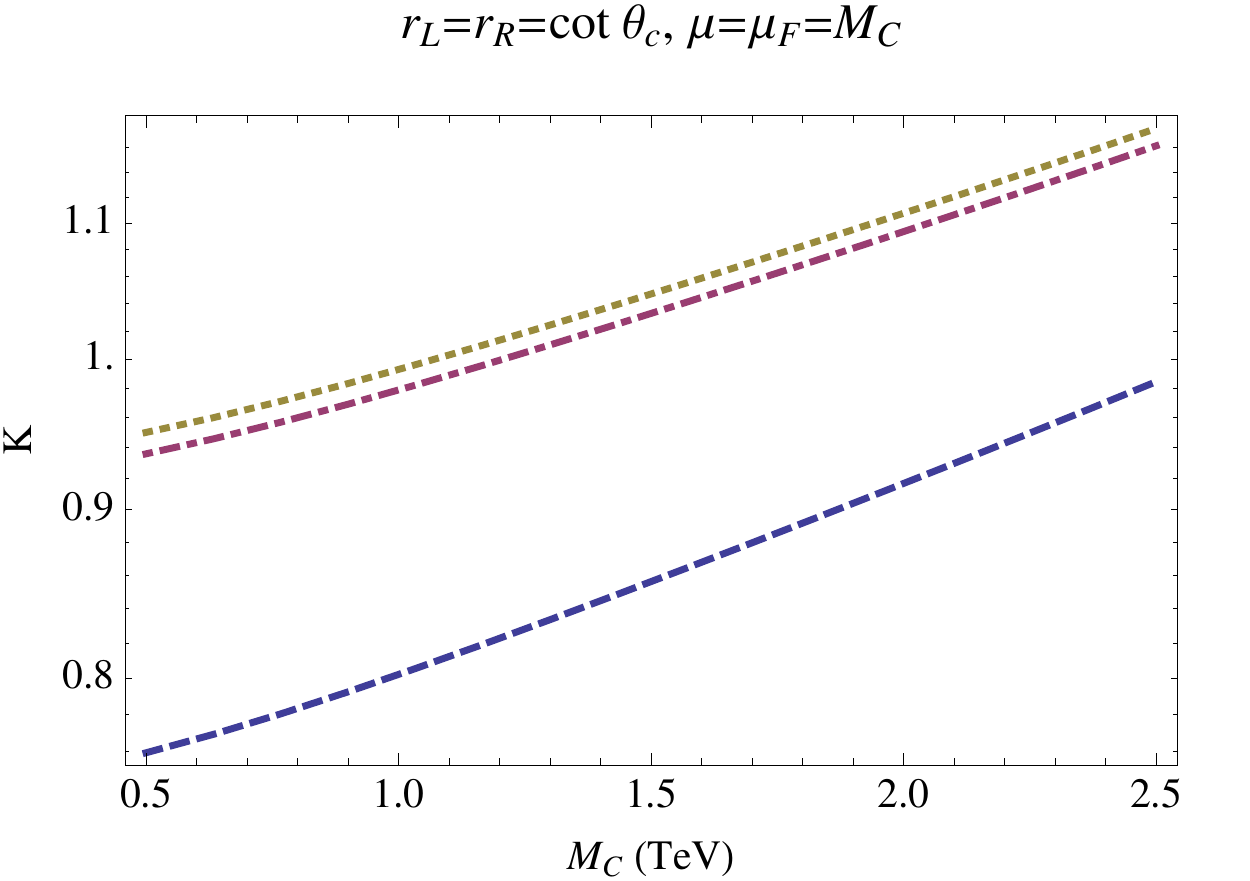}
\caption{``$K$-factor", the ratio of the NLO to LO cross section for coloron production at the LHC ($\sqrt{s}=7$ TeV), plotted as a function of $M_C$ for $\sin^2\theta_c$ = 0.05 (dashed), 0.25 (dot-dashed) and 0.50 (dotted), $\mu_F=M_C$, and the three different quark charge assignments.}
\label{fig:Kfactor}
\end{center}\end{figure}

We now illustrate\footnote{For the purposes of illustration we use  the Mathematica package for CTEQ5 \protect\cite{Lai:2000oa} to evaluate the relevant parton distribution functions.} our results for the NLO coloron production cross section in Figs.~\ref{fig:Q}-\ref{fig:Kfactor}.
In each figure we consider the three possible flavor-universal scenarios for quark charge assignment: $r_L=r_R=-\tan\theta_c$, $r_L\neq r_R$, and $r_L=r_R=\cot\theta_c$. All of the plots refer to coloron production at the LHC with $\sqrt{s}=7$ TeV. 

Notice that the perturbative expansion is only meaningful as long as $\sin\theta_c$ is neither too close to zero (where $g_{2s}\gg g_{1s}$) nor too close to one (where $g_{1s}\gg g_{2s}$). This is clear from Fig.~\ref{fig:Q}, in which we plot the quantity $Q$ defined in Eq.~(\ref{eq:QNLO}), for $\mu=M_C$: the contribution from the virtual corrections to the NLO cross section. The upper curve is for the $r_L\neq r_R$ scenario, whereas the almost identical lower curves are for $r_L=r_R=-\tan\theta_c$, and $r_L=r_R=\cot\theta_c$. For $\sin^2\theta_c\lesssim 0.05$ and $\sin^2\theta_c\gtrsim 0.95$ the virtual corrections become large, and the perturbative expansion in $\alpha_s$ breaks down. Since $\alpha_s\simeq 0.118$ at the $Z$ pole, these boundaries correspond to $g_{2s}\gtrsim 2.7$ and $g_{1s}\gtrsim 2.7$, respectively.

In Fig.~\ref{fig:scale-dependence} we plot the $\mu_F$ dependence of the LO and NLO production cross sections
of a 2.0 TeV coloron (with $\sin^2\theta_c\vert_{\mu=2.0\, \text{TeV}}$=0.25). The scale-dependence of the LO
cross section is of order 30\% while, as expected, the NLO cross section has a much weaker scale-dependence, only of the order of 2\%.

In Fig.~\ref{fig:sigma} we plot the  cross section times branching ratio to quark jets as a function of $M_C$, allowing $\mu_F$ to vary from $M_C/2$ to $2M_C$.
Here, in order to compare to the experimental results of \cite{Chatrchyan:2011bd} (shown as the solid line in the figures), we correct for the acceptance of the
detector by multiplying our partonic-level NLO production cross section by the factor
\begin{equation}
R = \frac{\Big(\sigma(pp \to C)\cdot {\cal B} \cdot A \Big)^\text{CMS}_\text{axigluon}}{\sigma^\text{LO}(pp\to C)_\text{axigluon}} \ .
\end{equation}
In this expression,
$\Big(\sigma(pp \to C)\cdot {\cal B} \cdot A\Big)^\text{CMS}_\text{axigluon}$ is the CMS (LO) prediction for axigluon production cross section, times dijet branching ratio,
times acceptance\footnote{The CMS acceptance for isotropic decays is of order 0.6, independent of resonance mass \protect\cite{Chatrchyan:2011bd}.} reported in \cite{Chatrchyan:2011bd}, and $\sigma^\text{LO}(pp\to C)_\text{axigluon}$ is the  leading
order cross section in Eq.~(\ref{CSfulldetect}) in the case of an axigluon (i.e. $r_L = -r_R = 1$), assuming the branching ratio to quarks ${\cal B}(C \to q\bar{q})=1$.\footnote{It is worth noting that there are examples of models with colorons which do not decay primarily to dijets, e.g. \protect\cite{Kilic:2008xz}.}
The three sets of thin bands correspond to $\sin^2\theta_c\vert_{\mu=M_C}=$ 0.05 (dashed), 0.25 (dot-dashed), and 0.5 (dotted). Here, the weak residual $\mu_F$ dependence is shown by the narrowness of the bands. To give a sense of current experimental reach, we also show the 1 fb$^{-1}$ CMS upper bounds on the cross section times dijet branching ratio for a narrow resonance~\cite{Chatrchyan:2011bd}. Note that the bound on the axigluon \cite{Frampton:1987wo}
corresponds to the $r_L \neq r_R$ plot with $\sin^2\theta_{c}=0.5$ | and, hence, a narrow axigluon resonance is
constrained to have a mass of order 2.6 TeV or higher. The enhancement of the axigluon cross section at NLO
is responsible for the increase in the bound from of order 2.5 TeV as reported in \cite{Chatrchyan:2011bd}.

Next, we compute the ``$K$-factor" for coloron production,
\begin{equation}
K(M_C,\sin\theta_c\vert_{\mu=M_C},\mu_F=M_C) \equiv \frac{\sigma^\text{NLO}(M_C,\sin\theta_c\vert_{\mu=M_C},\mu_F=M_C)}{\sigma^\text{LO}(M_C,\sin\theta_c\vert_{\mu=M_C},\mu_F=M_C)} \ ,
\label{eq:Kfactor}
\end{equation}
shown in Fig.~\ref{fig:Kfactor} for $\sin^2\theta_c$ = 0.05 (dashed), 0.25 (dot-dashed) and 0.50 (dotted).
Again, we see that the NLO corrections are of order 30\%. In Appendix~\ref{sec:Kfactors} we report the numerical
values of the $K$-factors corresponding to Fig.~\ref{fig:Kfactor}, as well as those corresponding to the ATLAS
KK-gluon search reported in \cite{ATLCON2011}.

\begin{figure}
\begin{center}
\includegraphics[width=\textwidth]{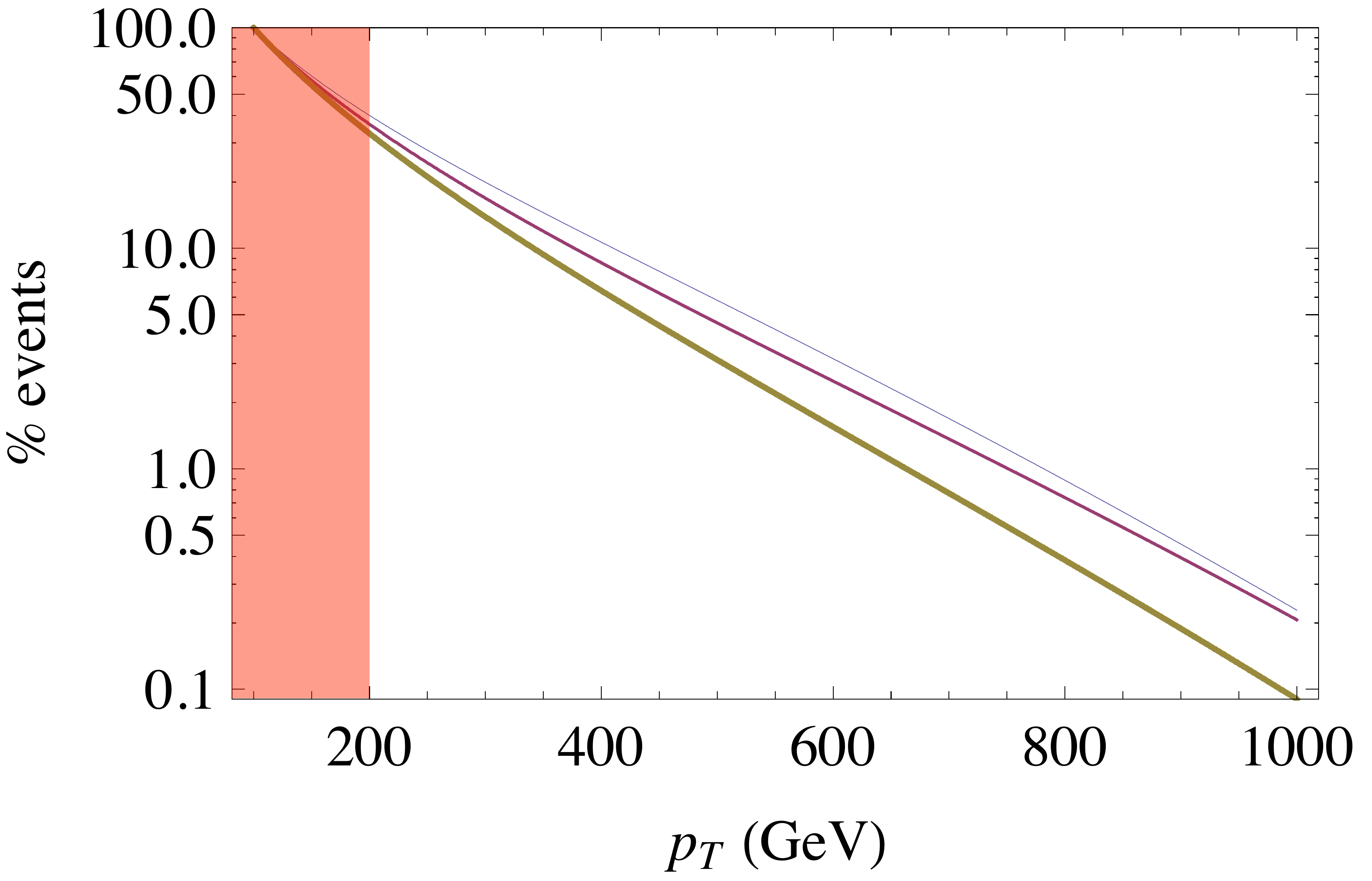}
\end{center}
\caption{Fraction of colorons produced with a $p_T$ greater than $p_{T\, \rm min}$, as
a function of $p_{T\, \rm min}$. The curves are for $M_C=1.2$ (highest, thin blue line), 2.0 (middle, medium purple line), and 3.0 TeV (lowest, thick green line), for the vectorial case $r_L=r_R=-\tan\theta_c$ and $\sin^2\theta_c = 0.05$. Note that of order 30\% of the colorons in
this mass range are produced with $p_T \ge 200$ GeV. As denoted by the red shaded region, below a $p_T$ of 200 GeV the corrections
become larger than 30\%, terms proportional to $\log(M^2_C/p^2_{T\, \rm min})$
become large, and this fixed-order calculation becomes unreliable.}
\label{fig:pTplotINT}
\end{figure}

At leading order, the coloron is produced with zero transverse momentum. 
We may use our results to compute the $p_T$ spectrum in coloron production to leading
nontrivial order from Eq.~(\ref{eq:dsigmadpt}).  Using these formulae, we may compute the \textit{fraction} of colorons produced 
above a momentum
$p_{T\, \rm min}$
\begin{equation} \begin{split}
&{\cal P}(p_T \ge p_{T\, \rm min}, M_C, \sin\theta_c\vert_{\mu=M_C},\mu_F=M_C) \\
&\equiv \frac{1}{\sigma^\text{NLO}(M_C,\sin \theta_c\vert_{\mu=M_C},\mu_F=M_C)} \int_{p_{T\, \rm min}}^{p_{T\, \rm max}} dp_T\, \frac{d\sigma}{dp_T} \ ,
\end{split} \end{equation}
where $p_{T\, \rm max}$ is the kinematic maximum transverse momentum (which depends on the coloron mass). 
For illustration, we plot this fraction for
vectorial colorons ($r_L=r_R=-\tan\theta_c$, with $\sin^2\theta_c=0.05$) with masses of 1.2,
2.0, and 3.0 TeV in Fig.~\ref{fig:pTplotINT}. Note that of order 30\% of the colorons in
this model and mass range are produced with $p_T \ge 200$ GeV. Below a $p_T$ of 200 GeV the corrections
become larger than 30\%, terms proportional to $\log(M^2_C/p^2_{T\, \rm min})$
become large, and this fixed-order calculation becomes unreliable.

In conclusion, we have reported the first complete calculation of QCD corrections to the production of a massive
color-octet vector boson.
Our next-to-leading order calculation includes both virtual corrections as well as corrections arising from the emission of gluons and light quarks, and we have demonstrated the reduction in factorization scale-dependence relative to the leading order
approximation used in previous hadron collider studies. In particular, we have shown that
the QCD NLO corrections to coloron production
are as large as 30\%, and that the residual factorization scale-dependence is reduced to of order 2\%.
We have also calculated the $K$-factor and the $p_T$ spectrum for coloron production, since these are
valuable for comparison with experiment.
Our computation applies
directly to the production of the massive color-octet vector bosons in axigluon, topcolor, and coloron models,
and approximately to the production of KK gluons in extra-dimensional models or colored technivector mesons in
technicolor models.
We look forward to future results from the LHC, and the possible discovery of colorons.

\chapter{CONCLUSION \& DISCUSSION}\label{concl}

\begin{quote}
``\textit{Physics is like sex: sure, it may give some practical results, but that's not why we do it!}"
\begin{flushright}
|Richard Feynman (1918 -- 1988)\\
\end{flushright}
\end{quote}

\vspace{\baselineskip}

\lettrine[lines=1]{I}{n the present Thesis}, we have analyzed two separate extensions of the Standard Model of particle physics~(SM). As explained in Chapter~1, the SM rests on the foundations of an $SU(3)_C \times SU(2)_L \times U(1)_Y$ gauge group, accounting for all three microscopically relevant interactions; namely, the electromagnetism, the strong and the weak forces. On one hand, the field theoretical descriptions of the electromagnetic and the weak interactions are interconnected, forming the electroweak sector of the SM, and are described by the $SU(2)_L \times U(1)_Y$ subgroup. On the other hand, the strong force is contained within the quantum chromodynamics (QCD) $SU(3)_C$ subgroup.

We have seen that the $SU(2)_L \times U(1)_Y$ subgroup is spontaneously broken to $U(1)_\text{EM}$ by means of the Higgs mechanism, which in the SM is facilitated by employing a fundamental scalar doublet. This is the mechanism by which the weak gauge bosons and the fermions acquire mass. Introduction of this scalar doublet predicts the existence of a real degree of freedom, called the Higgs boson, the mass of which appears to be highly sensitive to the high energy behavior of the theory through quadratically divergent quantum corrections. While the actual mass of the Higgs boson, if it exists, needs to lie naturally around the electroweak symmetry breaking energy scale, its sensitivity to ultra high energies necessitates a fine-tuning and causes the Hierarchy problem. Various beyond the Standard Model theories (BSM) have been proposed in order to cancel the quadratic divergences of the Higgs mass in a natural way and to solve the Hierarchy problem.

In Part~I of the Thesis, we explored various formal and phenomenological aspects of one of the proposed solutions to the Hierarchy problem, the Lee-Wick Standard Model~(LW~SM). In this BSM theory, inspired by the Pauli-Villars regulatory scheme, a set of (heavy) auxiliary fields are introduced, which form the Lee-Wick~(LW) partners of the usual SM particles. Unlike the SM fields, however, these LW~partners carry an overall negative sign as part of their description. It is this extra negative sign which induces a cancellation of the quadratically divergent quantum corrections among the contributions originating from the SM particles and their LW~partners. Therefore the LW~SM provides a natural and economical solution to the Hierarchy problem.

We have seen that the SM, as a renormalizable theory, employs a few experimentally determined observables as input in order to make robust predictions regarding the outcome of many other experiments. In the electroweak sector, the SM predictions have been tested to an impressive accuracy through the electroweak precision data. In order to facilitate this comparison between theory and experiment, convenient parametrizations have been introduced. The relatively small deviations of the experimental values of these parameters from their SM predictions can then be utilized as a powerful tool to place tight constraints on the BSM variables possibly influencing those parameters.

Since the LW~particles have not been observed so far in colliders, they must, if they exist, be heavy with masses presumably beyond the previous colliders' production threshold. In Chapter~2 we analyzed the phenomenological consequences of the LW~SM, using the available electroweak precision data in order to set lower bounds on the masses of the LW~particles. We found a lower bound of several TeV at 95\% C.L. for the masses of these hypothetical LW~particles to be consistent with experimental data.

The LW~SM with the auxiliary partners may, equivalently, be expressed in a higher-derivative formulation, in which, instead of introducing new LW~degrees of freedom, one adds higher-derivative terms to various sectors of the original SM Lagrangian. Addition of these higher-derivative terms might raise concerns about the overall renormalizability of the LW~theories and their consideration as reliable BSM alternatives with an arbitrary accuracy in predictions. Previous power counting arguments have exhibited the renormalizability of the LW~theories in the higher-derivative formulation; however, exact translation of this property to the auxiliary-field formulation remained unexplored. In Chapter~3 we investigated the global symmetries and renormalizability of the auxiliary-field formalism, by considering a LW~scalar QED theory as an Abelian toy model. We were able to identify a global $SO(1, 1)$ symmetry, which, together with the $U(1)$ gauge invariance and an $SO(1, 1)$ invariant gauge-fixing condition, allowed us to prove the renormalizability of this class of theories and to clarify the physics involved.

Part~II of the Thesis was dedicated to a separate class of BSM theories, forming an extension to the strong sector of the SM. These strong sector extensions arise naturally, for example as an integral feature of theories in which the Higgs boson is a composite scalar rather than a fundamental degree of freedom, with its constituents held together by a new strong interaction. The SM strong sector gauge group is extended to an $SU(3)_{1C}\times SU(3)_{2C}$ structure, which is spontaneously broken to the ordinary QCD's $SU(3)_C$ group. This introduces, in addition to the usual massless gluon-octet, a massive vector color-octet of states, called colorons.

As is the case in ordinary QCD, a perturbative expansion of the theory in terms of the strong coupling, $\alpha_{s}$, is appropriate at high energies, given its asymptotically free nature. To date, colorons have been studied only to leading order (LO) in perturbation theory, as the colored nature of this final state massive vector boson makes its higher-order non-Abelian analysis rather nontrivial. In Chapter~4 the first complete and consistent calculation of coloron production at next-to-leading order (NLO) in perturbation theory was presented. We provided a finite expression for the production cross section at the LHC, and demonstrated that the NLO effect is as large as 30\%; thereby, dramatically improving upon previous LO results. In addition, we constructed coloron kinematic variables, such as its transverse-momentum distribution, which make a direct comparison with experiment possible.

The research outlined in the present Thesis can be summarized as formal and phenomenological explorations of extensions to two separate sectors of the SM: the electroweak sector and the strong sector. Motivations for going beyond the SM in these sectors, as explained above, are different as they address different issues within the SM. The presented analyses, however, by no means mark an end of theoretical investigations of Lee-Wick and coloron theories, and there is much room left for extending the research in these areas in future studies. Establishing the renormalizability of the non-Abelian LW~theories in the auxiliary-field formalism remains, for example, unexplored and poorly understood, mainly since the discussed $SO(1, 1)$ symmetry is violated by the gauge interactions. A thorough non-Abelian investigation of LW~theories in this formulation promises to be a formidable task, but remains necessary in order to extend the demonstrated Abelian results. The coloron NLO analysis may also be enhanced by taking into account the coloron's finite lifetime and its subsequent decay into quarks and gluons through various channels. This allows for achieving an even higher theoretical accuracy in order to compare with future experimental results. Furthermore, using the NLO results, one may exploit the available experimental data to improve the previously determined theoretical LO lower bounds on coloron masses.

As one might appreciate, we are currently progressing through a particularly exciting era in particle physics research, specifically due to the experiments conducted at the LHC. These experiments might discover the existence of the elusive SM Higgs boson, in which case the SM and its symmetry breaking mechanism would be confirmed, with its last missing piece finally in place. Moreover, in light of a solution to the Hierarchy problem, the LHC may find distinctive signals related to any of the proposed BSM theories, including the LW~SM; thereby, defining the direction of the future theoretical research. A more exciting scenario, however, would involve a physical exclusion of the SM Higgs boson by the LHC data. In that case, theoretical focus will be shifted towards alternative electroweak symmetry breaking mechanisms, such as the strong interaction theories, including the coloron theory. Additionally, one might imagine the discovery of signals not anticipated previously by the proposed BSM theories, which would pave the way for fresh ideas and more advanced theoretical developments. In any case, the LHC findings promise to open the gates to a wealth of knowledge concerning the mysteries of the microscopic world | a world whose precise small-scale exploration might, perhaps not so surprisingly, be connected to the large-scale properties of the macroscopic universe, answering some of the big questions regarding its future, present, and past!

\begin{center}
\noindent\includegraphics[width=\textwidth]{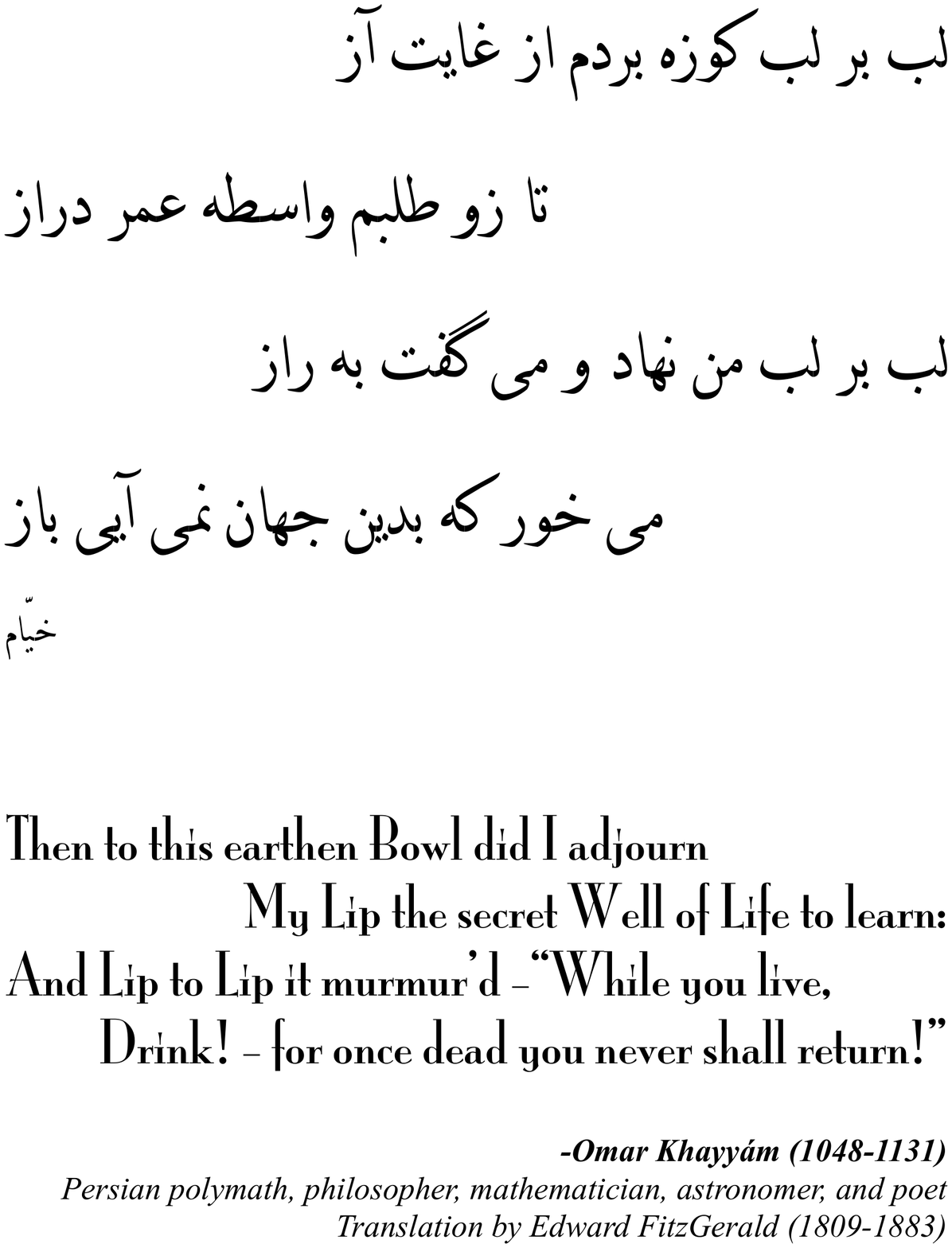}
\end{center}

\appendix

\chapter{EVALUATION OF THE $\phi^0\to b_L\bar{b}_L$ AMPLITUDE}\label{phibb}

The triangle diagram of Fig.~\ref{fig:Zbb} can be easily evaluated once the mass matrix has been diagonalized and the Yukawa couplings have been computed. For an arbitrary theory with heavy replicas of the third generation quarks, and neglecting the bottom Yukawa sector, the interactions with the Nambu-Goldstone bosons eaten by the $W$ and $Z$ boson read
\begin{equation}
\sum_{i,j} -i\,\frac{y_t}{\sqrt{2}}\,\phi^0 \left[\alpha_{ij}\,\bar{t_i} P_R t_j - \alpha_{ji}\,\bar{t_i} P_L t_j\right]
-i\, y_t\, \beta_{ij} \left[\phi^- \bar{b}_i P_R t_j - \phi^+ \bar{t}_j P_L b_i\right] \ ,
\end{equation}
where $t_0$ and $b_0$ are the SM top and bottom, respectively, and where the remaining ones are heavy replicas. From this expression one may extract the Feynman rules. Shifting the momentum of the $\bar{b}_L$ to zero, and omitting the external fermion wavefunctions, the amplitude reads
\begin{equation} \begin{split}
i M = \ & \sum_{i,j}(-)^{N_{ij}} \int\frac{d^4 k}{(2\pi)^4}  (y_t \beta_{0i} P_R) \frac{i(\slashed{k}+\slashed{p}+m_{t_i})}{(k+p)^2-m_{t_i}^2+i\epsilon}\, \frac{y_t}{\sqrt{2}} \\
&\times (\alpha_{ij}P_R-\alpha_{ji}P_L)
\frac{i(\slashed{k}+m_{t_j})}{k^2-m_{t_j}^2+i\epsilon}(-y_t \beta_{0j} P_L)\frac{i}{k^2+i\epsilon} \ , 
\end{split} \end{equation}
where $N_{ij}$ is the number of LW~fermions in the $i,j$ pair. Combining the denominators into a single one, and shifting the loop momentum in the usual way, leads to
\begin{equation} \begin{split}
iM = \ & -\frac{i y_t^3}{\sqrt{2}}\ \slashed{p} P_L \sum_{i,j}(-)^{N_{ij}} \beta_{0i}\beta_{0j}\alpha_{ji} m_{t_j}
\int_0^1 dx \int_0^{1-x} dy\,\int \frac{d^4 l}{(2\pi)^4} \frac{2(1-x)}{(l^2-\Delta)^3} \\
&- \frac{i y_t^3}{\sqrt{2}}\ \slashed{p} P_L \sum_{i,j}(-)^{N_{ij}} \beta_{0i}\beta_{0j}\alpha_{ij} m_{t_i}
\int_0^1 dx \int_0^{1-x} dy\,\int \frac{d^4 l}{(2\pi)^4} \frac{2x}{(l^2-\Delta)^3} \ ,
\end{split} \end{equation}
where
\begin{equation}
\Delta \equiv -x(1-x)p^2+x \, m_{t_i}^2 + y\, m_{t_j}^2 \ .
\end{equation}
Evaluating the integrals in the $p^2\to 0$ limit gives
\begin{equation}\label{iMappendix} \begin{split}
iM = \ & -\frac{1}{16\pi^2} \, \frac{y_t^3}{\sqrt{2}}\ \slashed{p} P_L
\Bigg[\sum_i \frac{\beta_{0i}^2 \alpha_{ii}}{m_{t_i}} \\
&\ + \sum_{i \neq j}(-)^{N_{ij}} \beta_{0i}\beta_{0j}\alpha_{ji} m_{t_j}
\Bigg(-\frac{1}{m_{t_i}^2-m_{t_j}^2}
+\frac{1}{2}\frac{3m_{t_i}^2-m_{t_j}^2}{(m_{t_i}^2-m_{t_j}^2)^2}\log\frac{m_{t_i}^2}{m_{t_j}^2}\Bigg)\Bigg] \ .
\end{split} \end{equation}
Comparing this expression with Eqs.~(\ref{eq:MZbb})~and~(\ref{eq:generalZBB}) gives to leading order in the coupling
\begin{equation} \begin{split}
\delta g_L^{b\bar{b}} = \ & \frac{1}{16\pi^2} \, \frac{y_t^3 v}{2\sqrt{2}}
\Bigg[\sum_i \frac{\beta_{0i}^2 \alpha_{ii}}{m_{t_i}} \\
&+ \sum_{i \neq j}(-)^{N_{ij}} \beta_{0i}\beta_{0j}\alpha_{ji} m_{t_j}
\Bigg(-\frac{1}{m_{t_i}^2-m_{t_j}^2}
+\frac{1}{2}\frac{3m_{t_i}^2-m_{t_j}^2}{(m_{t_i}^2-m_{t_j}^2)^2}\log\frac{m_{t_i}^2}{m_{t_j}^2}\Bigg)\Bigg] \ .
\end{split} \end{equation}

\chapter{FEYNMAN RULES OF THE COLORON THEORY}\label{FR}

\begin{figure}
\begin{center}
\includegraphics[width=\textwidth,height=8in]{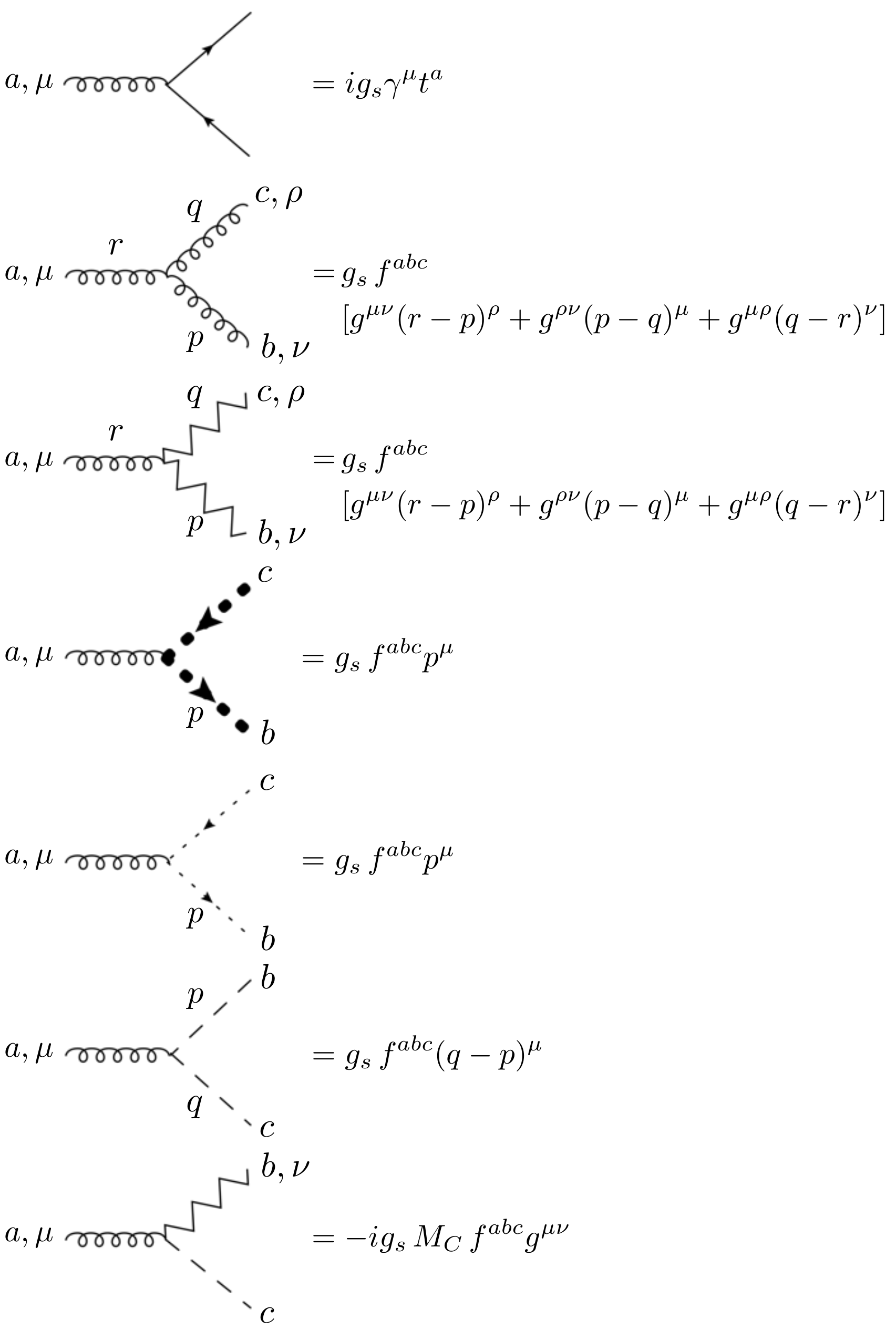}
\caption{Feynman rules for the trilinear vertices. In each diagram the momenta are toward the vertex. A gluon field is, as usual, represented by a coiling line; a coloron field is represented by a zigzag line. The coloron ghost is represented by a sequence of filled circles, and the eaten Goldstone bosons are represented by dashed lines.}
\label{fig:three1}
\end{center}
\end{figure}

\begin{figure}
\begin{center}
\includegraphics[width=\textwidth,height=7.5in]{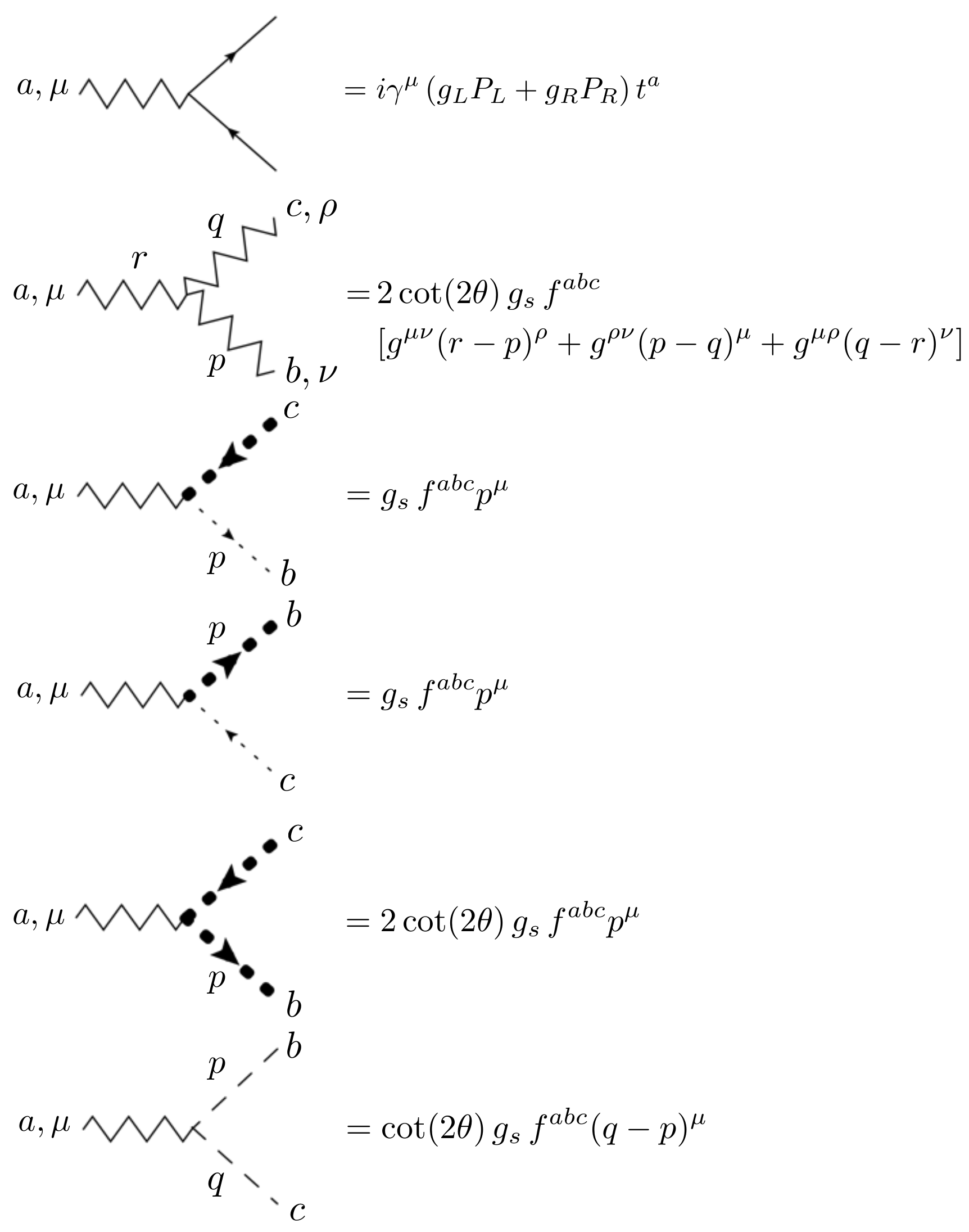}
\caption{Feynman rules for the trilinear vertices (continued).}
\label{fig:three2}
\end{center}
\end{figure}

\begin{figure}
\begin{center}
\includegraphics[width=\textwidth,height=6in]{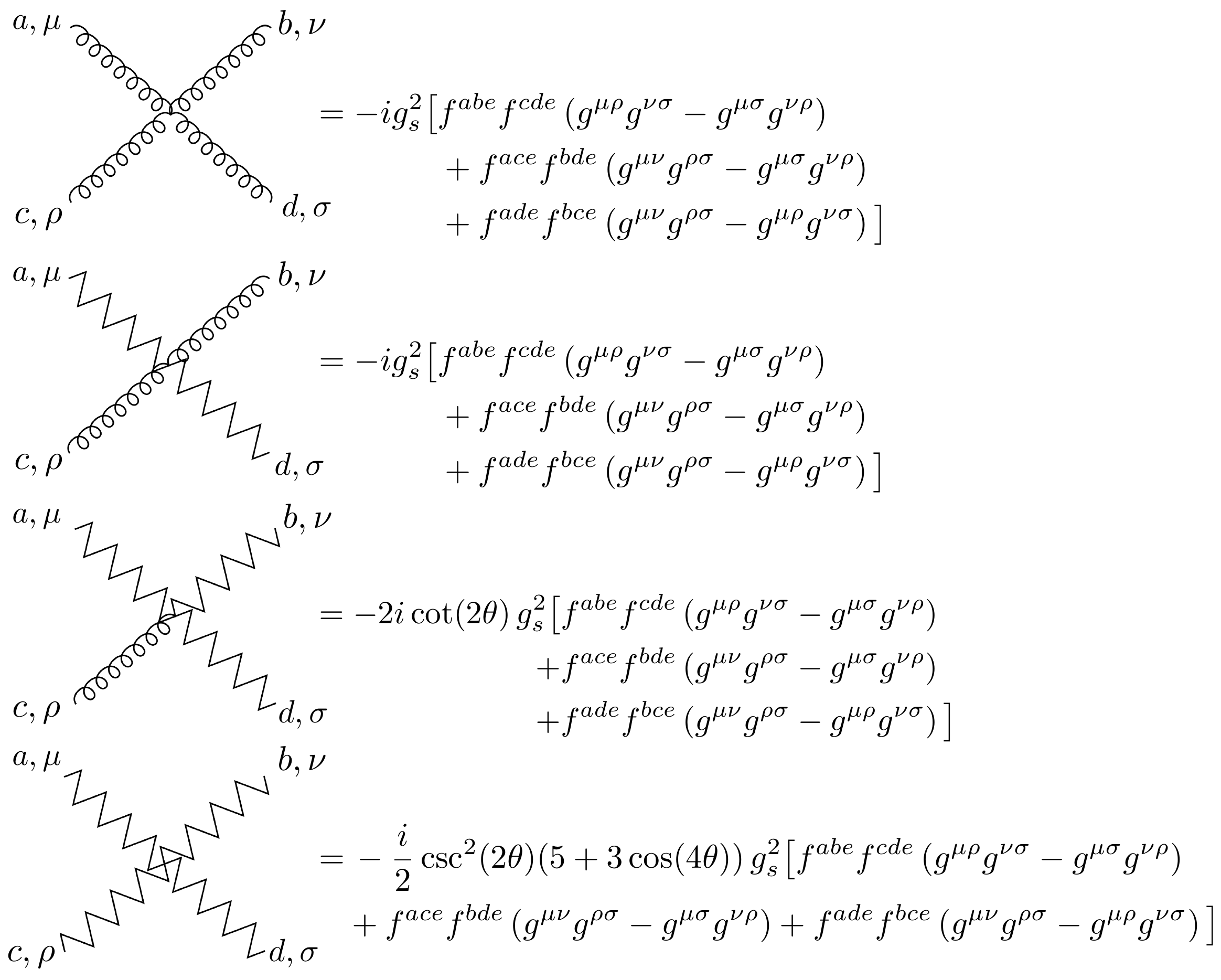}
\caption{Feynman rules for the quartic vertices. A gluon field is, as usual, represented by a coiling line; a coloron field is represented by a zigzag line.}
\label{fig:four1}
\end{center}
\end{figure}

\begin{figure}
\begin{center}
\includegraphics[width=.9\textwidth]{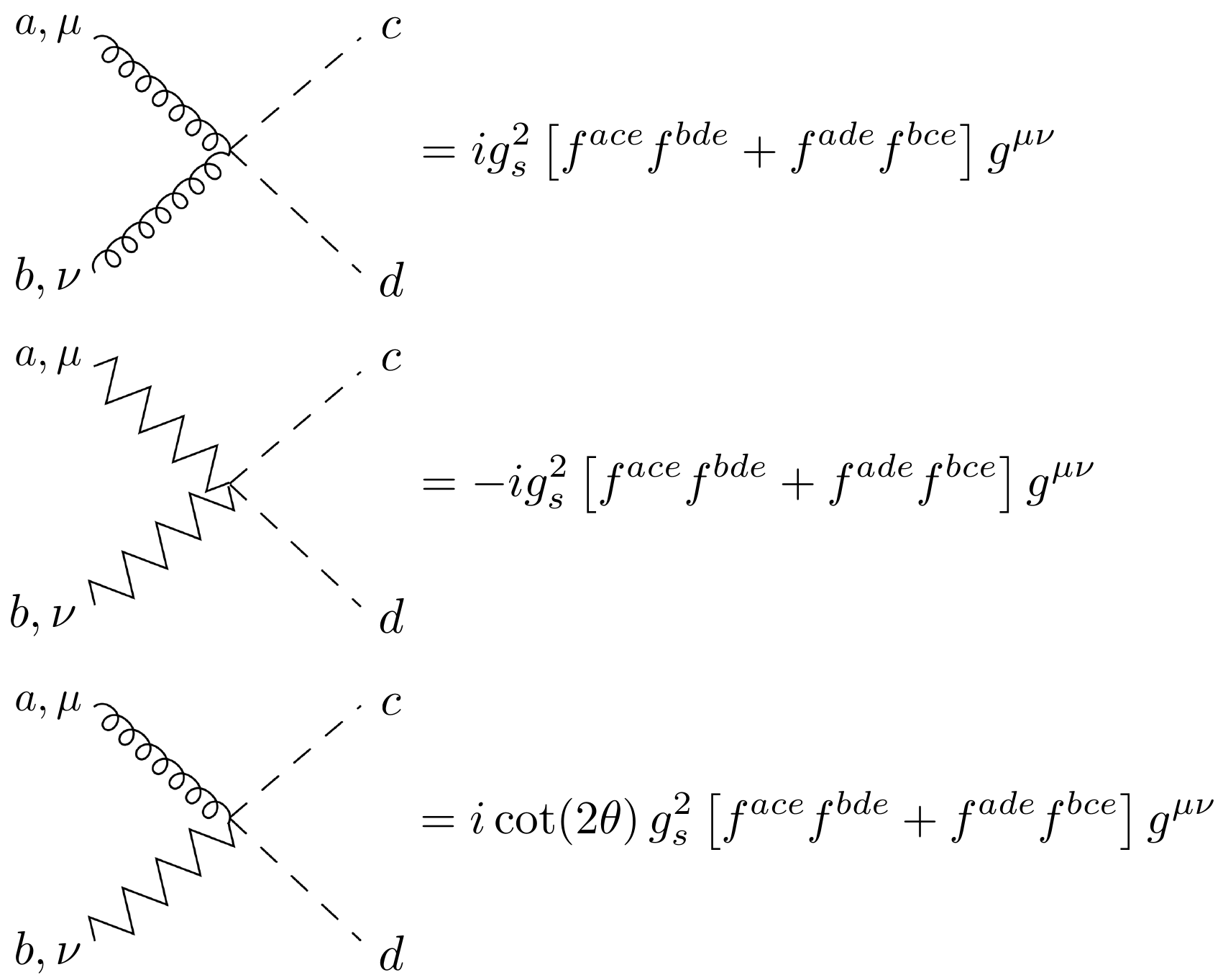}
\caption{Feynman rules for the quartic vertices (continued).}
\label{fig:four2}
\end{center}
\end{figure}

The Feynman rules\footnote{The Feynman rules discussed here are equivalent to those
in \cite{Zhu:2011if}, aside from those for the triple-coloron vertex which is not specified in
that reference.} for the trilinear and quartic vertices are shown in Figs.~\ref{fig:three1}-\ref{fig:four2}. The coloron is represented by a zigzag line, the coloron ghost by a sequence of small circles, and the eaten Goldstone bosons by dashed lines.  All other particles are denoted as in QCD standard notation. Note that a coupling between the eaten Goldstone boson and quarks is absent in the zero quark mass limit.

\chapter{NUMERICAL VALUES OF THE COLORON $K$-FACTOR}\label{sec:Kfactors}

The numerical values of the $K$-factors for various values of the coloron mass and the three
patterns of coloron coupling are shown in Tabs.~\ref{K-factor-1}-\ref{K-factor-3}. Finally, the values of the $K$-factor corresponding to the KK-gluons of \cite{Lillie:2007ca},
corresponding to the experimental search reported in \cite{ATLCON2011}, are shown in Tab.~\ref{KKgluon-K}.

\begin{table}
\begin{center}
\begin{tabular}{|c@{\:\vline\vline\vline\:} c|c|c|}
\hline
$M_{C}$ (GeV)& $\sin^{2}(\theta_{c})=0.05$& $\sin^{2}(\theta_{c})=0.25$& $\sin^{2}(\theta_{c})=0.50$\\ \hline\cline{1-4}
1000&0.780&0.980&0.990\\ \hline
1200&0.800&1.00&1.01\\ \hline
1400& 0.820&1.02&1.04\\ \hline
1600&0.840&1.04&1.06\\ \hline
1800&0.870&1.06&1.08\\ \hline
2000&0.890&1.09&1.11\\ \hline
2200&0.920&1.12&1.13\\ \hline
2400&0.950&1.14&1.16\\ \hline
2600&0.970&1.17&1.19\\ \hline
2800&1.00&1.20&1.22\\ \hline
3000&1.04&1.23&1.25\\ \hline
3200&1.07&1.27&1.28\\ \hline
3400&1.10&1.30&1.32\\ \hline
3600&1.14&1.34&1.36\\ \hline
3800&1.18&1.38&1.39\\ \hline
4000&1.22&1.42&1.43\\ \hline
\end{tabular}
\end{center}
\caption {$K$-factors for colorons of various masses, and $r_L=r_R=-\tan\theta_c$.} \label{K-factor-1}
\end{table}

\begin{table}
\begin{center}
\begin{tabular}{|c@{\:\vline\vline\vline\:} c|c|c|}
\hline
$M_{C}$ (GeV)& $\sin^{2}(\theta_{c})=0.05$& $\sin^{2}(\theta_{c})=0.25$& $\sin^{2}(\theta_{c})=0.50$\\ \hline\cline{1-4}
1000&1.11&1.11&1.12\\ \hline
1200&1.13&1.13&1.14\\ \hline
1400& 1.15&1.15&1.16\\ \hline
1600&1.17&1.17&1.18\\ \hline
1800&1.20&1.20&1.21\\ \hline
2000&1.22&1.22&1.23\\ \hline
2200&1.25&1.25&1.26\\ \hline
2400&1.28&1.27&1.29\\ \hline
2600&1.30&1.30&1.32\\ \hline
2800&1.33&1.33&1.35\\ \hline
3000&1.37&1.36&1.38\\ \hline
3200&1.40&1.40&1.41\\ \hline
3400&1.43&1.43&1.44\\ \hline
3600&1.47&1.47&1.48\\ \hline
3800&1.51&1.51&1.52\\ \hline
4000&1.55&1.55&1.56\\ \hline
\end{tabular}
\end{center}
\caption {$K$-factors for colorons of various masses, and $r_L\neq r_R$. The classic ``axigluon"
\protect\cite{Frampton:1987wo} corresponds to $\sin^2\theta_c=0.50$.} \label{K-factor-2}
\end{table}

\begin{table}
\begin{center}
\begin{tabular}{|c@{\:\vline\vline\vline\vline\:} c|c|c|}
\hline
$M_{C}$ (GeV)& $\sin^{2}(\theta_{c})=0.05$& $\sin^{2}(\theta_{c})=0.25$& $\sin^{2}(\theta_{c})=0.50$\\ \hline\cline{1-4}
1000&0.800&0.980&0.990\\ \hline
1200&0.820&1.00&1.01\\ \hline
1400& 0.840&1.02&1.04\\ \hline
1600&0.870&1.04&1.06\\ \hline
1800&0.890&1.07&1.08\\ \hline
2000&0.920&1.09&1.11\\ \hline
2200&0.940&1.12&1.13\\ \hline
2400&0.970&1.15&1.16\\ \hline
2600&1.00&1.18&1.19\\ \hline
2800&1.03&1.21&1.22\\ \hline
3000&1.06&1.24&1.25\\ \hline
3200&1.09&1.27&1.28\\ \hline
3400&1.13&1.31&1.32\\ \hline
3600&1.17&1.34&1.36\\ \hline
3800&1.20&1.38&1.39\\ \hline
4000&1.24&1.42&1.43\\ \hline
\end{tabular}
\end{center}
\caption {$K$-factors for colorons of various masses, and $r_L= r_R=\cot\theta_c$.} \label{K-factor-3}
\end{table}

\begin{table}
\begin{center}
\begin{tabular}{|c@{\:\vline\vline\vline\:} c|c|c|c|c|}
\hline
$M_{C}$ (GeV)& $-0.20 g_{s}$& $-0.25 g_{s}$& $-0.30 g_{s}$&$-0.35 g_{s}$&$-0.40 g_{s}$\\ \hline\cline{1-6}
500&0.660&0.770&0.830&0.870&0.890\\ \hline
600&0.670&0.780&0.840&0.880&0.900\\ \hline
700&0.670&0.790&0.850&0.890&0.910\\ \hline
800&0.680&0.800&0.860&0.890&0.920\\ \hline
900&0.690&0.810&0.870&0.900&0.930\\ \hline
1000&0.700&0.810&0.880&0.910&0.940\\ \hline
1100&0.710&0.820&0.890&0.920&0.950\\ \hline
1200&0.720&0.840&0.900&0.930&0.960\\ \hline
1300&0.730&0.850&0.910&0.940&0.970\\ \hline
1400&0.740&0.860&0.920&0.960&0.980\\ \hline
1500&0.760&0.870&0.930&0.970&0.990\\ \hline
1600&0.770&0.880&0.940&0.980&1.00\\ \hline
1700&0.780&0.890&0.950&0.990&1.01\\ \hline
1800&0.790&0.900&0.970&1.00&1.03\\ \hline
1900&0.800&0.920&0.980&1.01&1.04\\ \hline
2000&0.820&0.930&0.990&1.03&1.05\\ \hline
2100&0.830&0.940&1.00&1.04&1.06\\ \hline
2200&0.840&0.960&1.02&1.05&1.08\\ \hline
2300&0.860&0.970&1.03&1.07&1.09\\ \hline
2400&0.870&0.980&1.04&1.08&1.10\\ \hline
2500&0.880&1.00&1.06&1.09&1.12\\ \hline
2600&0.900&1.01&1.07&1.11&1.13\\ \hline
2700&0.910&1.03&1.09&1.12&1.15\\ \hline
2800&0.930&1.04&1.10&1.14&1.16\\ \hline
2900&0.940&1.06&1.12&1.15&1.18\\ \hline
3000&0.960&1.07&1.13&1.17&1.19\\ \hline
\end{tabular}
\end{center}
\caption {$K$-factors for KK-gluons of various masses considered in \protect\cite{ATLCON2011}. This calculation is based on the theoretical framework of \protect\cite{Lillie:2007ca}, with the KK-gluon coupling
(specified in the column heading) varying between $-0.20 g_s$ and $-0.40 g_s$.} \label{KKgluon-K}
\end{table}

\end{doublespace}
\clearpage

\bibliography{Thesis}
\bibliographystyle{prsty}


\end{document}